\documentstyle[manuscript,aps,epsf]{revtex}
\tightenlines
\newcommand{\beqa}{\begin{eqnarray}}
\newcommand{\eeqa}{\end{eqnarray}}
\newcommand{\beq}{\begin{equation}}
\newcommand{\eeq}{\end{equation}}

\newcommand{\bc}{\begin{center}}
\newcommand{\ec}{\end{center}}
\begin{document}
%
%
%
\draft
\title{
\vspace*{-2.cm}
\begin{flushright}
{\normalsize UTHEP-484}\\
{\normalsize UTCCP-P-147}\\
\end{flushright}
$I=2$ $\pi\pi$ Scattering Phase Shift 
with two Flavors of $O(a)$ Improved Dynamical Quarks
\\
}
\author{
    T.~Yamazaki\rlap,$^{\rm 1}$
        S.~Aoki\rlap,$^{\rm 1}$
    M.~Fukugita\rlap,$^{\rm 2}$
  K-I.~Ishikawa\rlap,$^{\rm 3}$
    N.~Ishizuka\rlap,$^{\rm 1,4}$
     Y.~Iwasaki\rlap,$^{\rm 1,4}$
      K.~Kanaya\rlap,$^{\rm 1}$
      T.~Kaneko\rlap,$^{\rm 5}$
   Y.~Kuramashi\rlap,$^{\rm 5}$
       M.~Okawa\rlap,$^{\rm 3}$
            A.~Ukawa,$^{\rm 1,4}$  and
          T.~Yoshi\'e$^{\rm 1,4}$\\
(CP-PACS Collaboration)
}
\address{
$^{\rm 1}$
Institute of Physics,
University of Tsukuba, 
Tsukuba, Ibaraki 305-8571, Japan \\
$^{\rm 2}$
Institute for Cosmic Ray Research,
University of Tokyo, 
Kashiwa, Chiba 277-8582, Japan \\
$^{\rm 3}$
Department of Physics,
Hisoshima University, 
Higashi-Hiroshima, Hiroshima 739-8526, Japan \\
$^{\rm 4}$
Center for Computational Physics,
University of Tsukuba, 
Tsukuba, Ibaraki 305-8577, Japan \\
$^{\rm 5}$
High Energy Accelerator Research Organization (KEK), 
Tsukuba, Ibaraki 305-0801, Japan \\
}
\date{\today}
\maketitle
%
%
\begin{abstract}
We present a lattice QCD calculation of phase shift 
including the chiral and continuum extrapolations in two-flavor QCD. 
The calculation is carried out for $I=2$ S-wave $\pi\pi$ scattering.  
The phase shift is evaluated for two momentum systems, 
the center of mass and laboratory systems,
by using the finite volume method proposed by L\"uscher in the center of mass system
and its extension to general systems by Rummukainen and Gottlieb.
The measurements are made
at three different bare couplings $\beta = 1.80$, $1.95$ and $2.10$ 
using a renormalization group improved gauge and a tadpole improved clover fermion action, 
and employing a set of configurations generated for hadron spectroscopy in our previous work.
The illustrative values we obtain for the phase shift in the continuum limit 
are 
$\delta$(deg.)
$= -  3.50(64)$,
$  -  9.5(30)$ and 
$  - 16.9(64)$ 
for 
$\sqrt{s}({\rm GeV})$
$=0.4$, 
$ 0.6$ and 
$ 0.8$,
which are consistent with experiment.
\end{abstract}
\pacs{PACS number(s): 12.38.Gc, 11.15.Ha }
%
%
%
\section{Introduction}
Calculation of scattering phase shift
is an important step for expanding our understanding
of the strong interaction based on lattice QCD beyond the hadron mass spectrum.
For scattering lengths, which are the threshold values of phase shifts,
several studies have already been carried out.
For the simplest case of the two-pion system,
$I=2$ S-wave scattering length has been calculated in detail
\cite{sc-1,sc-2,sc-3,sc-4,sc-5,sc-6,sc-7,CP-PACS:phsh}
including the continuum extrapolation \cite{sc-6,sc-7}.
There is also a pioneering attempt at $I=0$ scattering length~\cite{sc-3},
which is much more difficult 
due to the presence of contributions from disconnected diagrams.

The first pioneering study of $I=2$ S-wave scattering phase shift
was made by Fiebig {\it et al.}~\cite{Fiebig}. 
They estimated the phase shift
from the two-pion effective potential calculated in the lattice simulations.
A direct calculation of phase shift on the lattice without recourse to 
effective potentials is possible if one uses the finite volume method proposed 
by L\"uscher~\cite{FVM:L:1,FVM:L:2}.
In this method phase shift is obtained
from the two-pion energy eigenvalues for finite volume.
However, it is non-trivial to extract the energy eigenvalues 
from the time behavior of the two-pion correlation functions,
because the correlation functions have multi-exponential time behaviors
due to presence of multiple states with the same quantum numbers.
Recently we presented
a direct calculation of the phase shift~\cite{CP-PACS:phsh} in quenched QCD.
In this calculation the diagonalization method proposed by L\"uscher and Wolff~\cite{diag} 
was used to solve the problem of the multi-exponential behavior.
Very recently Kim reported on his preliminary results
on $G$- and $H$-period boundary lattices,
where the problem can be avoided 
by the boundary condition~\cite{Kim}.

In all previous studies of the scattering phase shift
calculations were carried out with quenched approximation
nor the continuum limit was taken.
It is predicted in  chiral perturbation theory 
that unphysical chiral divergences appear
in the scattering length in the quenched theory 
due to the lack of unitarity~\cite{Bernard-Golterman,Colangelo-Pallante}.
The unphysical divergences can also occur in the phase shift.
While the presence of such pathologies has not been numerically confirmed 
in actual lattice calculations with quenched approximation,
we should make our study in full QCD
in order to avoid such uncontrollable quenching problems.

In this paper we present a calculation of the physical scattering phase shift 
for $I=2$ S-wave two-pion system
including the dynamical quark effects and taking the continuum limit.
We use the full QCD configurations previously generated
for a study of light hadron spectrum~\cite{CP-PACS:sprm}
with a renormalization group improved gauge action
and clover fermion action with a tadpole improved clover coefficient.
The phase shift is evaluated by the finite volume method 
as in the previous work~\cite{CP-PACS:phsh}.
In order to obtain the phase shift at several energies from 
one full QCD configuration,
the calculations are carried out for two types of momentum systems.
One of them is the center of mass system, 
where the total momentum of the two-pion system is zero.
In the other system where the total momentum is fixed to a non-zero value,
we use the method proposed by Rummukainen and Gottlieb~\cite{FVM:RG}
which is a simple extension of that by L\"uscher in the center of mass system 
to general momentum systems.
We shall refer to this system as laboratory system in this paper. 

This paper is organized as follows.
In Sec.~\ref{Sec:Method} we describe the finite volume method for both  
momentum systems for evaluating the phase shift from the two-pion energy.
We also explain the diagonalization method to extract the two-pion energy eigenvalues.
The parameters of the calculation in this work are given in Sec.~\ref{Sec:Parameters}.
In Sec.~\ref{Sec:Results} we show data for the pion four-point functions and
the effect of diagonalizations for the two momentum systems.
We then present the results of the scattering length and the phase shift at each $\beta$
and in the continuum limit.
In Sec.~\ref{Sec:Conclusions} we briefly summarize this work.
%
%
\section{Methods}
\label{Sec:Method}
\subsection{Finite volume method}
%
%
%
\subsubsection{Center of mass system}
In the center of mass system 
the total momentum of the two-pion system is zero.
The energy eigenvalue of two pions in a finite periodic box $L^3$ 
without two-pion interactions is given by
\begin{equation}
  E_n = 2 \sqrt{ m_\pi^2 + p_n^2 }, \qquad p_n^2 = ( 2 \pi / L )^2 n,\ \ n \in Z,
\end{equation}
In the interacting case
the total momentum is also zero and
the energy eigenvalue of the $n$-th state is given by
\begin{equation}
  \overline{E}_n   = 2 \sqrt{ m_\pi^2 + \overline{p}_n^2 }, \qquad
  \overline{p}_n^2 = ( 2 \pi / L )^2 \overline{n},\ \ \overline{n} \notin Z.
\end{equation}
The energy eigenvalue is written as that of the free two-pion case
with momentum 
$ \overline{\bf p}_n$ and 
$-\overline{\bf p}_n$, but the quantity 
$\overline{n } =  ( L / 2 \pi )^2 \cdot \overline{p }_n^2$
is no longer an integer.
L\"uscher~\cite{FVM:L:1,FVM:L:2} 
found that the momentum $ \overline{p }_n $ satisfies the relation
\begin{equation}
  \tan \delta ( \overline{p}_n )
        = \frac{ \pi^{3/2} \sqrt{ \overline{n} } }
               { Z_{00} ( 1 ; \overline{n} ) },
  \label{eq:FVM:L}
\end{equation}
where $\delta ( \overline{p}_n )$ is the S-wave scattering phase shift 
in the infinite volume and
\begin{equation}
   Z_{00} ( s ; \overline{n} ) 
   = \frac{ 1 }{ \sqrt{ 4 \pi } }
     \sum_{ { \bf n } \in Z^3 } ( n^2 - \overline{n} )^{ -s } .
\label{eq:Z00}
\end{equation}
The calculation method of $Z_{ 00 } ( s ; \overline{n } )$ 
is discussed in Appendix~\ref{appe:zeta}.
Using Eq.(\ref{eq:FVM:L}), 
we can obtain the phase shift from the energy eigenvalue calculated 
in the lattice simulations.
%
%
\subsubsection{Laboratory system}
Let us consider a two-pion system 
with a non-zero total momentum ${\bf P} \ne 0$ in a periodic box $L^3$.
We shall refer to this system as the laboratory system in the following.
In the laboratory system
the energy eigenvalue for the $n$-th energy state $E_n^{\bf P}$ 
without interaction is given by
\begin{equation}
  E_n^{\bf P} =      \sqrt{ m_{\pi}^2 + (p_{ 1, n }^{\bf P})^2 } 
             + \sqrt{ m_{\pi}^2 + (p_{ 2, n }^{\bf P})^2 } , \qquad
  L/(2\pi) {\bf p}_{1,n}^{\bf P} \in Z^3 , \ \ 
  L/(2\pi) {\bf p}_{2,n}^{\bf P} \in Z^3 ,  
\label{eq:energy_L}
\end{equation}
where ${\bf p}_{i,n}^{\bf P}$ is the $i$-th pion momentum of the $n$-th energy state 
in the laboratory system,
which takes discrete values due to the periodic boundary condition 
in finite volume.
The two-pion interaction shifts $E_n^{\bf P}$ to $\overline{E}_n^{\bf P}$
as in the center of mass system.
By the Lorentz transformation with a boost factor
\begin{equation}
\gamma = \overline{E}_n^{\bf P} / \sqrt{ ( \overline{E}_n^{\bf P} )^2 - {\bf P}^2 },
\label{eq:gamma}
\end{equation}
the energy in the center of mass system can be obtained as
\begin{equation}
    \overline{E}_n = \overline{E}_n^{\bf P} / \gamma .
\label{eq:E_CM}
\end{equation}
It is noted that the total momentum ${\bf P}$ is not shifted
by the two-pion interaction.

The center of mass momentum $\overline{p}_n$
is determined from $\overline{E}_n$ by
\begin{equation}
    \overline{E}_n   = 2 \sqrt{ m_\pi^2 + \overline{p}_n^2 }, \qquad
    \overline{p}_n^2 = ( 2 \pi / L )^2 \overline{m },
                                     \ \ \overline{m } \notin Z,
\label{eq:p_CM}
\end{equation}
where $\overline{m }$ is also not an integer.
Rummukainen and Gottlieb~\cite{FVM:RG} found that 
the momentum $\overline{p}_n$
is related to the phase shift in the infinite volume through the relation
\begin{equation}
    \tan \delta ( \overline{p}_n ) 
  = \frac{ \gamma \pi^{3/2} \sqrt{ \overline{m} } }
                                 { Z^{\bf d}_{00} ( 1 ; \overline{m } ) },
\label{eq:FVM:RG}
\end{equation}
where 
\begin{equation}
     Z^{\bf d}_{00} ( s ; \overline{m} ) 
   = \frac{ 1 }{ \sqrt{4\pi} }
     \sum_{ {\bf r} \in P^{\bf d} } ( r^2 - \overline{m} )^{ -s } .
\label{eq:Z00d}
\end{equation}
The summation for ${\bf r}$ is carried out over the set 
\begin{equation}
  P^{\bf d} = \{ {\bf r} | \ {\bf r} = \hat{\gamma}^{ -1 } ( {\bf n} + {\bf d}/2 ), {\bf n} \in Z^3 \}  
\end{equation}
where ${ \bf d } = L / ( 2 \pi ) { \bf P }$. 
The operation $\hat{ \gamma }^{ -1 }$ is the inverse Lorentz transformation : 
$\hat{\gamma}^{ -1 } {\bf n} = 1/\gamma \cdot {\bf n}_{||} + {\bf n}_{\bot}$ ,
where $ {\bf n}_{||} = ( {\bf n} \cdot {\bf d} ) {\bf d} / d^2$
is the parallel component and 
${\bf n}_{\bot} = {\bf n} - {\bf n}_{||}$ 
the perpendicular component of ${\bf n}$ in the direction ${\bf d}$.
The calculation method of $Z^{ \bf d }_{ 00 } ( s ; \overline{m } )$ 
is discussed in Appendix~\ref{appe:zeta}.
Substituting ${\bf d} = {\bf 0}$ and $\gamma = 1$ into Eq.(\ref{eq:FVM:RG}),
we obtain the formula for the center of mass system Eq.(\ref{eq:FVM:L}).
%
%
\subsection{Extraction of energy eigenvalues of the two-pion system}
\label{SubSec:Diagonalization}
In order to extract the two-pion energy eigenvalues 
in the system with a total momentum ${\bf P}$,
we construct the pion four-point function by 
\begin{equation}
  G_{nm}^{(N_R)}( t ) 
  = \langle 0 | \Omega_n ( t ) \Omega_m^{(N_R)} ( t_S ) | 0 \rangle,
\label{eq:prop}
\end{equation}
where we omit the index for the total momentum.
The operator $\Omega_n ( t )$ is a two-pion operator at time $t$
for the $n$-th energy eigenstate which is defined by 
\begin{equation}
    \Omega_n ( t ) 
    = \sum_{ {\bf p}_1, {\bf p}_2 \in S_n } 
      \pi( { \bf p}_1, t ) \pi( {\bf p}_2, t )  
      \ / \ \bigl[  \sum_{ {\bf p}_1, {\bf p}_2 \in S_n } 1 \ \bigr] \ ,
\label{eq:Omega}
\end{equation}
where $\pi ({\bf p},t)$ is the pion operator at time $t$ with momentum ${\bf p} $.
The ${\bf p}_1$ and ${\bf p}_2$
are the momenta on the lattice
and take discrete values,
{\it i.e.}, $L/(2\pi) {\bf p}_1 \in Z^3$, $L/(2\pi) {\bf p}_2 \in Z^3$.
The summation for these momenta is taken over the set
\begin{equation}
  S_n = 
    \{ 
           {\bf p}_1, {\bf p}_2 | \ \ 
               \sqrt{ m_\pi^2 + {\bf p}_1^2 } 
             + \sqrt{ m_\pi^2 + {\bf p}_2^2 } = E_n^{\bf P} , \ \ 
           {\bf p}_1 + {\bf p}_2 = {\bf P}
    \} ,
\label{eq:Sdn}
\end{equation}
where $E_n^{\bf P}$ is fixed at the energy of the $n$-th energy state
with the total momentum ${\bf P}$ in the free two-pion case.
This summation of momenta
projects out the $A^+$ representation of the rotation group on the lattice, 
which equals the S-wave states in the continuum,
ignoring the states with higher angular momentum.
For example,
states with angular momentum
$l \ge 4$ are ignored for $L/(2\pi){\bf P} = (0,0,0)$, and
$l \ge 2$ for $L/(2\pi){\bf P} = (1,0,0)$ and $(1,1,0)$.

For the source we use a different operator $\Omega_n^{ ( N_R ) } (t)$ defined by 
\begin{equation}
  \Omega_n^{ (N_R) }( t ) = 
  \frac{ 1 }{ N_R } \sum_{j=1}^{N_R} \pi( {\bf p}_{1,n }, t, \xi_j ) 
                                     \pi( {\bf p}_{2,n }, t, \eta_j ),
  \label{eq:Omega_source}
\end{equation}
where 
\begin{equation}
  \pi( {\bf p}, t, \xi_j ) = \frac{ 1 }{ L^3 } 
    \biggl[
      \sum_{\bf x} 
          \overline{q}( {\bf x}, t ) 
                         e^{ i {\bf p} \cdot {\bf x} } 
                         \xi_j^{\dagger}( {\bf x} )
    \biggr] \gamma_5
    \biggl[
        \sum_{ \bf y } q ( {\bf y}, t ) 
        \xi_j ( {\bf y} )
    \biggr]  .
\end{equation}
Here ${\bf p}_{ 1, n }, {\bf p}_{ 2, n }$ are fixed to one of the values 
in the set $S_n$ for the sink operator $\Omega_n(t)$ 
defined in Eq.(\ref{eq:Omega}).
The functions $\xi_j({\bf x})$ and $\eta_j({\bf x})$
are complex random numbers in three-dimensional space, 
whose property is
\begin{equation}
  \lim_{ N_R \rightarrow \infty } \frac{ 1 }{ N_R } \sum_{ j = 1 }^{ N_R }
  \xi_j^{ \dagger } ( { \bf x } ) \xi_j ( { \bf y } ) 
  =
  L^3 \delta_{ \bf x, y }.
\end{equation}
The pion propagator is also calculated with the random number as
\begin{equation}
    G_n^{ \pi ( N_R ) }( t ) 
    = \frac{ 1 }{ N_R } 
      \sum_{ j = 1 }^{ N_R } 
          \langle 0 | \pi(  { \bf p }_n, t ) 
                      \pi( -{ \bf p }_n, t_S, \xi_j ) | 0 \rangle.
\end{equation}

When the number of random noise sources $N_R$ is taken large 
or the number of gauge configurations are large,
we expect 
\begin{eqnarray}
&&  G_{ nm }^{ ( N_R ) }( t ) \sim G_{nm} (t) 
               = \langle 0 | \Omega_n( t ) \Omega_m( t_S ) | 0 \rangle \cr
&&  G_n^{ \pi ( N_R )  }( t ) \sim G_n^\pi(t) 
               = \langle 0 | \pi( {\bf p}_n, t ) \pi( -{\bf p}_n, t_S ) | 0 \rangle 
\end{eqnarray}
and the four-point function will be symmetric 
under the exchange of the sink and the source indices $n$ and $m$.
We fix $N_R = 2$ in all calculations.  
The number of the gauge configurations is from 380 to 725
depending on the lattice spacing as shown in Sec.~\ref{Sec:Parameters}.

The four-point function can be written 
in terms of the energy eigenstates $| \overline{\Omega}_n \rangle$ 
with the total momentum ${\bf P}$ as follows:
\begin{equation}
   G_{nm}( t ) = \sum_j V_{ nj }^{ T } V_{ jm } e^{ - \overline{E}_j^{\bf P} ( t - t_S ) },\ \ 
   V_{jm}      = \langle \overline{\Omega }_j | \Omega_m ( 0 ) | 0 \rangle,
\label{eq:multiexp}
\end{equation}
where $\overline{E}_n^{\bf P}$ is the energy eigenvalues with the two-pion interaction.
Since the matrix $V_{jm}$ is not diagonal generally,
the four-point function contains many exponential terms and 
is not diagonal with respect to $n$ and $m$.

A simple method for extracting the energy eigenvalues
is diagonalization of $G(t)$ at each $t$.
L\"uscher and Wolff~\cite{diag} found that the eigenvalue is given by 
$\lambda_n ( t ) = \exp ( - \overline{E }_n^{\bf P} t ) \times \{ 1 + O( \exp [ - \Delta_n t ] ) \}$
with 
$\Delta_n = { \mathrm{ min } }_{ n \ne m } | \overline{E }_n^{\bf P} - \overline{E }_m^{\bf P} |$.
In order to extract $\overline{E}_n^{\bf P}$ from the eigenvalue 
by a single exponential fit,
we have to analyze $G(t)$ in the large $t$ region
where $O( \exp [ - \Delta_n t ] )$ terms can be neglected.
However, it may be very difficult to employ the fitting with this method
due to loss of statistics of $G(t)$ in the large $t$ region.

L\"uscher and Wolff also proposed another method~\cite{diag},
which is a diagonalization of the matrix $M( t, t_0 )$ 
constructed from $G(t)$ by
\begin{equation}
  M( t, t_0 ) = G ( t_0 )^{ - 1 / 2 } G ( t ) G ( t_0 )^{ - 1 / 2 },
  \label{eq:matrix}
\end{equation}
where $t_0$ is a reference time.
The eigenvalues of $ M( t, t_0 ) $ equals
\begin{equation}
 \lambda_n ( t, t_0 ) = e^{ -\overline{E}_n^{\bf P} ( t - t_0 ) }.
\end{equation}
without $O(\exp [ -\Delta_n t ] )$ terms.
Therefore after the diagonalization of $M(t,t_0)$
we can extract the two-pion energy eigenvalue 
by a single exponential fitting of $\lambda_n (t,t_0)$.
In this work we adopt the second method.

In the diagonalization method 
we assume that the two-pion states dominate 
and effects from other states can be neglected
in $G(t)$ for the considered range of $t$.
For example, our analysis loses its validity 
when the center of mass two-pion energy is over the inelastic scattering limit,
e.g., $\overline{E} \ge 4 m_\pi$.
In this case $G(t)$ is dominated by the four-pion ground state
rather than the two-pion state in the large $t$ region.
Further we should pay attention 
to the states contained with the excited pion state, such as $\pi \pi^*(1300)$.
The existence of such undesirable states 
is discussed in Sec.~\ref{Sec:Parameters}.

Since we cannot calculate all components of $G(t)$
in the actual lattice calculations,
a cut-off of the state $N$ has to be introduced.
We expect that the components of $G_{nm}(t)$ with $n,m \le k$ 
dominate for the $k$-th eigenvalue $\lambda_k (t)$
in the large $t$ and $t_0$ region,
while the components $n,m > k$ are less important.
We set $t$ and $t_0$ large and investigate 
the dependence for the cut-off $N \ge k$.
%
%
\section{Parameters}
\label{Sec:Parameters}
We calculate the scattering phase shift on the gauge configurations
previously generated including two flavors of dynamical quark effects 
for the study of the light hadron spectrum~\cite{CP-PACS:sprm}.
This work employed a renormalization group improved gauge action
and clover fermion action with
a tadpole improved clover coefficient $c_{SW}$, which we also use in the 
present study.
The gauge action is constructed in terms of the $1 \times 1$ and $1 \times 2$ 
Wilson loops $W^{ 1 \times 1 }$ and $W^{ 1 \times 2 }$,
\begin{equation}
  S_G = \frac{ \beta }{ 6 } 
        \left( 
                c_0 \sum_{ x, \mu < \nu } W^{ 1 \times 1 }_{ \mu \nu }( x ) 
              + c_1 \sum_{ x, \mu, \nu  } W^{ 1 \times 2 }_{ \mu \nu }( x ) 
        \right).
\end{equation}
The coefficient $c_1$ is fixed to $c_1=-0.331$
by an approximate renormalization group analysis
and $c_0 = 1 - 8 c_1 = 3.648$ by the normalization condition~\cite{Iwasaki_action} .
The bare gauge coupling $\beta$ is defined by $\beta = 6 / g^2$.
For clover fermion action~\cite{clover} a mean-field improved clover coefficient 
$c_{SW} = ( W^{ 1 \times 1 } )^{-3/4}$ is adopted,
where the plaquette $W^{ 1 \times 1 } = 1 - 0.8412 / \beta$
is determined by 
one-loop perturbation theory~\cite{Iwasaki_action}.

The parameters for the configuration generation 
are summarized in Table~\ref{tab:param}.
The configurations were generated
at $\beta =1.80$ and $1.95 $ and $2.10$, and for 
four hopping parameters $\kappa$ 
corresponding to $m_\pi / m_\rho \approx 0.6 - 0.8$ in each $\beta$.
The lattice spacing is estimated from the $\rho$ meson mass and
equals $a = 0.2150(22)$, $0.1555(17)$ and $0.1076(13)\ {\rm fm}$, respectively.
The lattice size $L^3 \times T$ at each $\beta$ is 
$12^3 \times 24$, $16^3 \times 32$ and $24^3\times 48$,
which correspond to a $2.5^3\ {\rm fm}^3$ lattice.
The periodic boundary conditions are imposed 
both in the spatial and time directions.

The quark propagators are calculated 
with the periodic boundary condition in the spatial directions,   
and the Dirichlet boundary condition in the temporal direction. 
The source operator $\Omega_n^{ ( N_R ) } ( t_S )$ is set at 
$t_S = 4$ for $\beta = 1.80$ and $\beta=1.95$, and $t_S = 6$ for $\beta = 2.10$
to reduce effects from the temporal boundary.
In order to avoid effects from excited states
the reference time for the diagonalization introduced in Eq.(\ref{eq:matrix})
is fixed to large value;  
$t_0=10$, $12$ and $16$ at $\beta = 1.80$, $1.95$ and $2.10$, respectively. 
At $\beta = 1.80$ for the lightest quark mass
we carry out extra measurements to reduce statistical errors 
in which the source operator is located at $t_S+T/2$ 
and the Dirichlet boundary condition in the temporal direction is imposed at $T/2$.
We average over the two measurements for the analysis of 
the pion four-point functions and the pion propagator.

In order to extract the phase shift at various momenta 
from a single full QCD configuration,
the calculations are carried out in three momentum systems, 
the center of mass (CM) and two laboratory systems (L1 and L2),
whose total momenta are
\begin{equation}
  \begin{array}{llllllll}
                 &                                & \qquad & L/(2\pi) {\bf P} \\
     \mbox{ CM } & \mbox{ Center of mass system } &        & ( 0, 0, 0 ) \\
     \mbox{ L1 } & \mbox{ Laboratory system     } &        & ( 1, 0, 0 ) \\
     \mbox{ L2 } & \mbox{ Laboratory system     } &        & ( 1, 1, 0 ) \ . \\
  \end{array}
\end{equation} 
In Table~\ref{tab:momsetup}
we show the momenta chosen
for the source operator $\Omega_n^{ ( N_R ) }$ defined in Eq.(\ref{eq:Omega_source}).
All elements in $S_n$,
which appear in the summation over momenta in the sink operator $\Omega_n$
defined in Eq.(\ref{eq:Omega}), 
can be obtained from the source momenta by 
cubic, tetragonal and orthorhombic rotations on the lattice for the CM, L1 and L2 systems, respectively.

\begin{figure}
\begin{center}
\leavevmode \epsfxsize=7cm\epsfbox{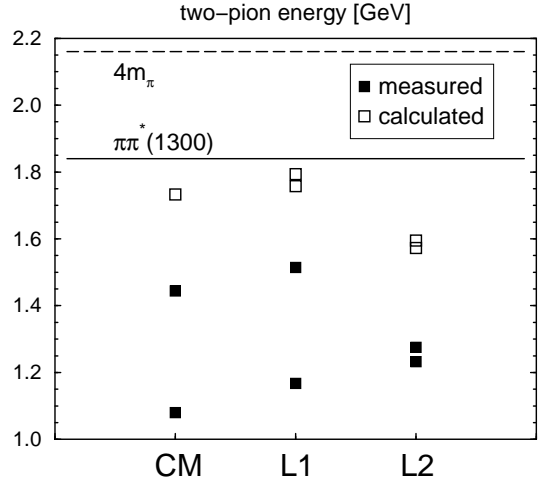}
\end{center}
\caption{
Center of mass energy of two-pion system
for $m_\pi/m_\rho \approx 0.6$ at $\beta = 2.10$ without the two-pion interaction.
We measure the scattering phase shifts at the energies referred by filled symbols.
States denoted by open symbols are used only to examine
the effect of the cut-off of the state for the diagonalization.
Solid and dashed lines denote the $\pi\pi^*(1300)$ state energy 
and the inelastic scattering limit, respectively.
\label{fig:non-interact}
}
\end{figure}

The center of mass energies of the free two-pion system $E_n$
in this work are plotted in Fig.~\ref{fig:non-interact},
where the smallest pion mass at $\beta=2.10$ is assumed.
In all systems the phase shift 
is evaluated at the ground ($n=0$) and first excited states ($n=1$), 
which are denoted by closed symbols.
Other higher energy states ($n \ge 2$) plotted by open symbols
are used to investigate the effects of the cut-off $N$ introduced 
in the diagonalization.
In the two laboratory systems, L1 and L2, we also calculate the $n=3$ state, 
because the energies of these states are very close to the $n=2$ state 
as shown in the figure and effects from these states can be comparable.
We also plot the location of the inelastic scattering threshold, 
$E \ge 4 m_\pi$, and the energy of state $\pi \pi^*(1300)$, 
which can appear in the pion four-point function 
and causes loss of validity of the diagonalization method
as discussed in Sec.~\ref{SubSec:Diagonalization}.
We neglect the quark mass dependence of $\pi^*(1300)$ in the estimation.
As shown in Fig.~\ref{fig:non-interact} 
the energies of these undesirable states
are very far from the measured point of the scattering phase shift.

We estimate the errors of the four-point functions 
and the pion propagator by the jackknife method.
In our study of the light hadron spectrum~\cite{CP-PACS:sprm},
we have shown that the separation of $50$ trajectories 
covers all the autocorrelations of the configuration,
so that in the present analysis we also use bins of $50$ trajectories 
in the jackknife method.
In the actual analysis a bin size of $5$ or $10$ 
($\beta = 2.10$ at the lightest quark mass) 
measurements are employed,
because we skip 10 or 5 trajectories between successive measurements.
%
%
\section{Results}
\label{Sec:Results}
\subsection{Effect of diagonalization}
\label{Sec:Results:diag}
\begin{figure}
\begin{center}
\begin{tabular}{cc}
\leavevmode  \epsfxsize=7cm \epsfbox{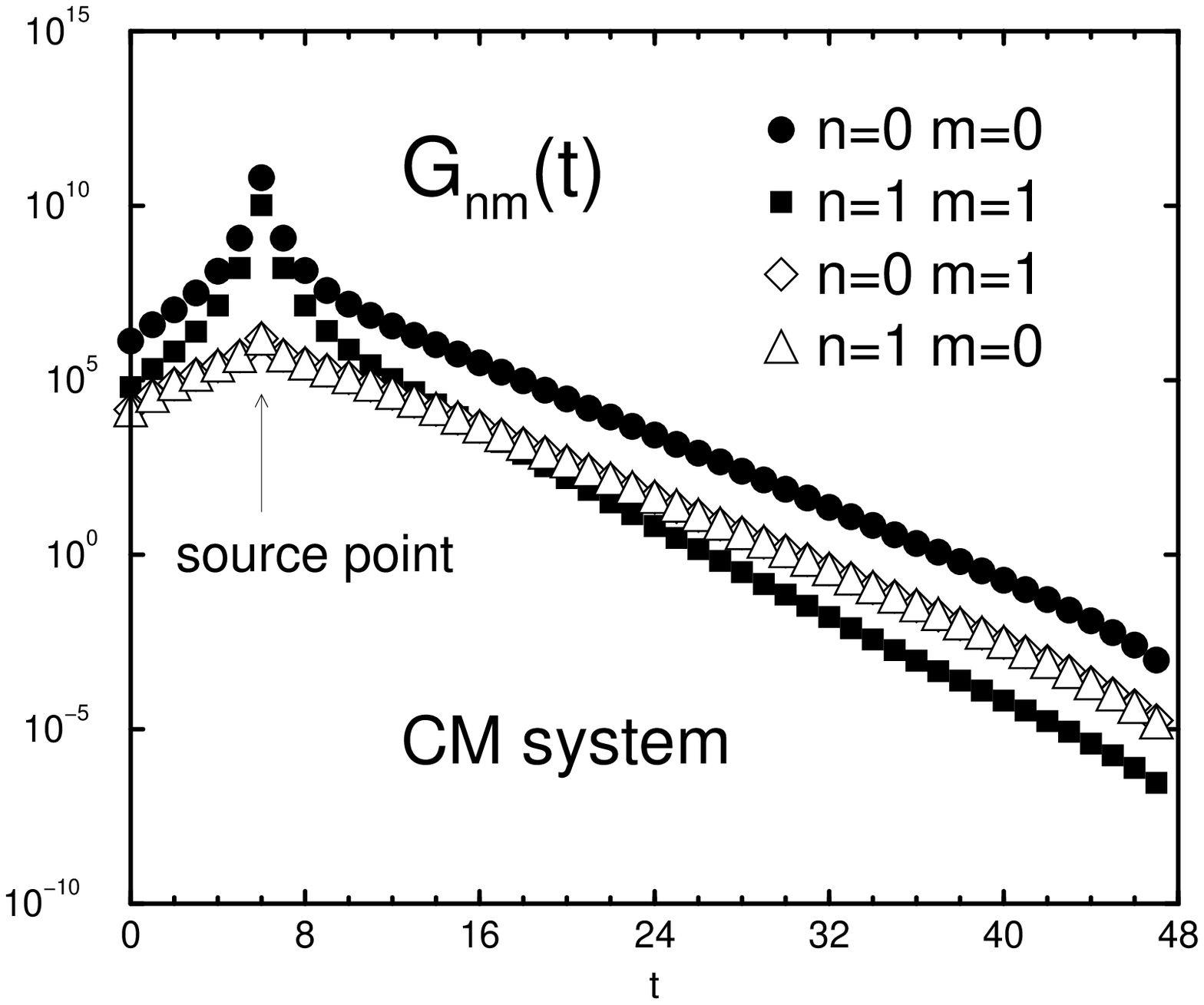} \\
\leavevmode  \epsfxsize=7cm \epsfbox{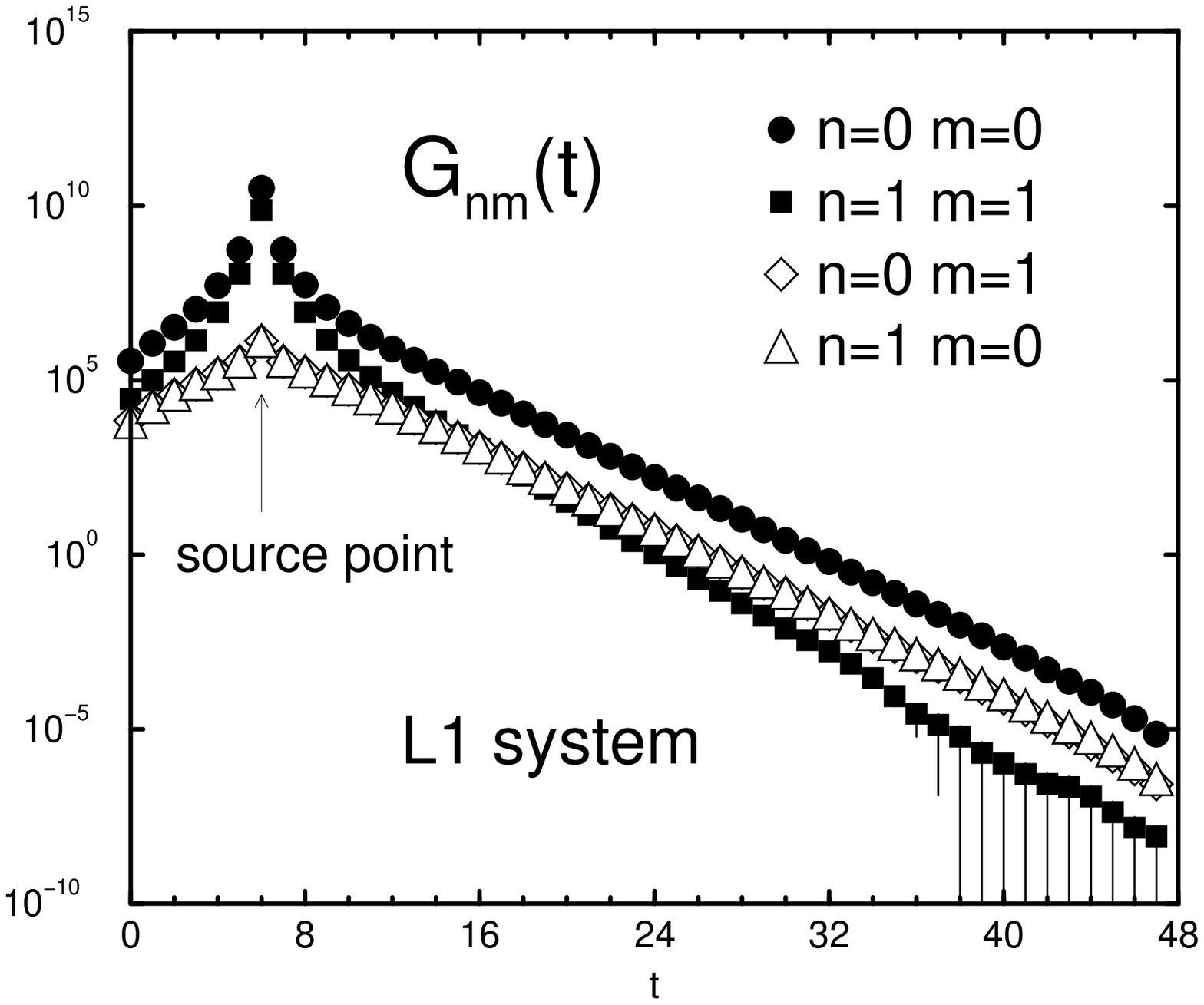} \\
\leavevmode  \epsfxsize=7cm \epsfbox{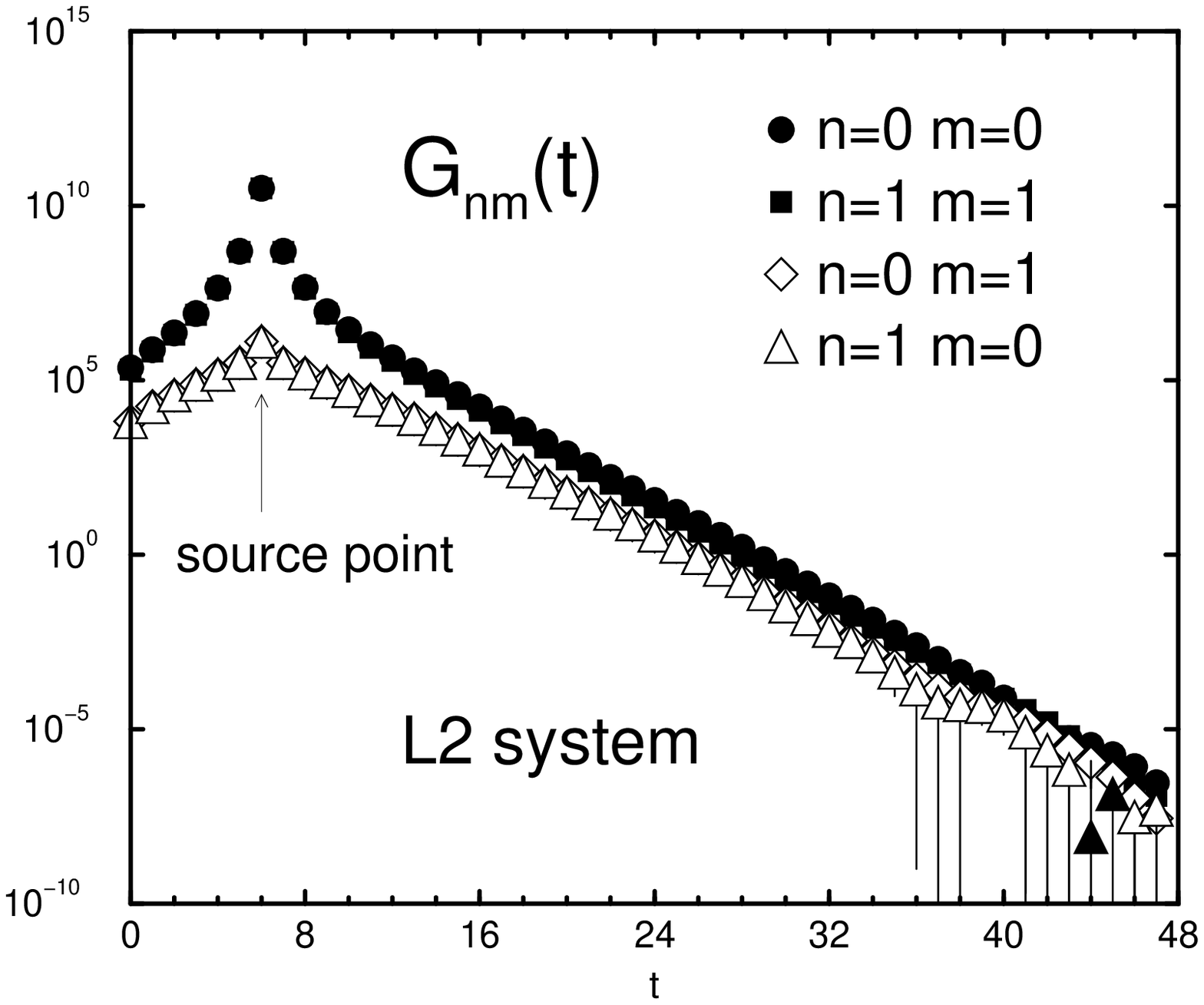}
\end{tabular}
\end{center}
\caption{
Examples of the pion four-point function $G_{nm}(t)$ 
in the center of mass system CM
and two laboratory systems L1 and L2 
for $m_\pi / m_\rho \approx 0.6$ at $\beta=2.10$.
Filled and open symbols indicate positive and negative value.
\label{fig:prop}
}
\end{figure}

In Fig.~\ref{fig:prop}
we plot the absolute value of the pion four-point function $G_{nm}(t)$
defined by Eq.(\ref{eq:prop}) at several values of $n$ and $m$ 
for the three momentum systems.
This figure corresponds to $m_\pi / m_\rho \approx 0.6$ at $\beta=2.10$.
Filled and open symbols indicate positive and negative values.
In all systems
the off-diagonal components of $G_{nm}(t)$ ($n \ne m$)
are not negligible compared with the diagonal components ($n=m$).
We also observe that
the shape of $G_{nm}(t)$ in the CM and L1 systems are very similar.
In the L2 system $G_{00}(t)$ is almost the same as $G_{11}(t)$ for all $t$.
This is attributed to the fact that these energies are very close
in the free two-pion case.
The figure also shows that $G_{nm}(t)$ is symmetric
under the exchange of the indices as expected.
In the following analysis  
we assume the symmetry of $G_{nm}(t)$
and we average values of the symmetric components.

In order to investigate the effect of diagonalization 
we define two ratios $R_n(t)$ and $D_n(t)$ as follows,
\begin{eqnarray}
  R_n ( t ) &=& \frac{ G_{nn} ( t ) }{ G^{ \pi }_{ 1, n }( t ) G^{ \pi }_{ 2, n }( t ) }, 
\label{eq:before_diag} \\
  D_n ( t ) &=& \lambda_n ( t, t_0 )
                \frac{ G^{ \pi }_{ 1, n }( t_0 ) G^{ \pi }_{ 2, n }( t_0 ) }
                { G^{ \pi }_{ 1, n }( t )   G^{ \pi }_{ 2, n }( t )   }, 
\label{eq:after_diag}
\end{eqnarray}
where $G^{\pi}_{i, n}(t)$ is the $i$-th pion propagator for the $n$-th energy state, 
and $\lambda_n (t,t_0)$ is the $n$-th eigenvalue of
the matrix $M(t,t_0)$ defined by Eq.(\ref{eq:matrix})
calculated with a finite cut-off of the energy state $N$.
If the pion four-point function contains only a single exponential term,
i.e., $G_{nm}(t) \propto \delta_{nm} \cdot \exp [ - \overline{E}_n^{\bf P} ( t - t_S )]$, 
then the ratio $R_n(t)$ behaves as
\begin{equation}
  R_n ( t ) 
  = A \cdot \exp ( -\Delta \overline{E }_n^{\bf P} ( t - t_S ) ),
\end{equation}
where $\Delta \overline{E}_n^{\bf P} = \overline{E}_n^{\bf P} - E_n^{\bf P}$ and $A$ is a constant.
If the cut-off of the state $N$ for the diagonalization is sufficiently large,
then the ratio $D_n(t)$ behaves as
\begin{equation}
  D_n ( t ) = \exp ( -\Delta \overline{E}_n^{\bf P} ( t - t_0 ) ).
\end{equation}
In these cases
we can extract the $n$-th energy shift $\Delta \overline{E }_n^{\bf P}$ 
from the ratios $R_n(t)$ or $D_n(t)$
by a single exponential fitting.

\begin{figure}
\begin{center}
\hspace{-26mm}
\begin{tabular}{rccc}
& $\beta = 1.80$ & $\beta = 1.95$ & $\beta = 2.10$ 
\\
\raisebox{25mm}{ $\displaystyle{ \frac{m_{\pi}}{m_{\rho}} \approx 0.6 }$ }
& \leavevmode \epsfxsize=5.1cm \epsfbox{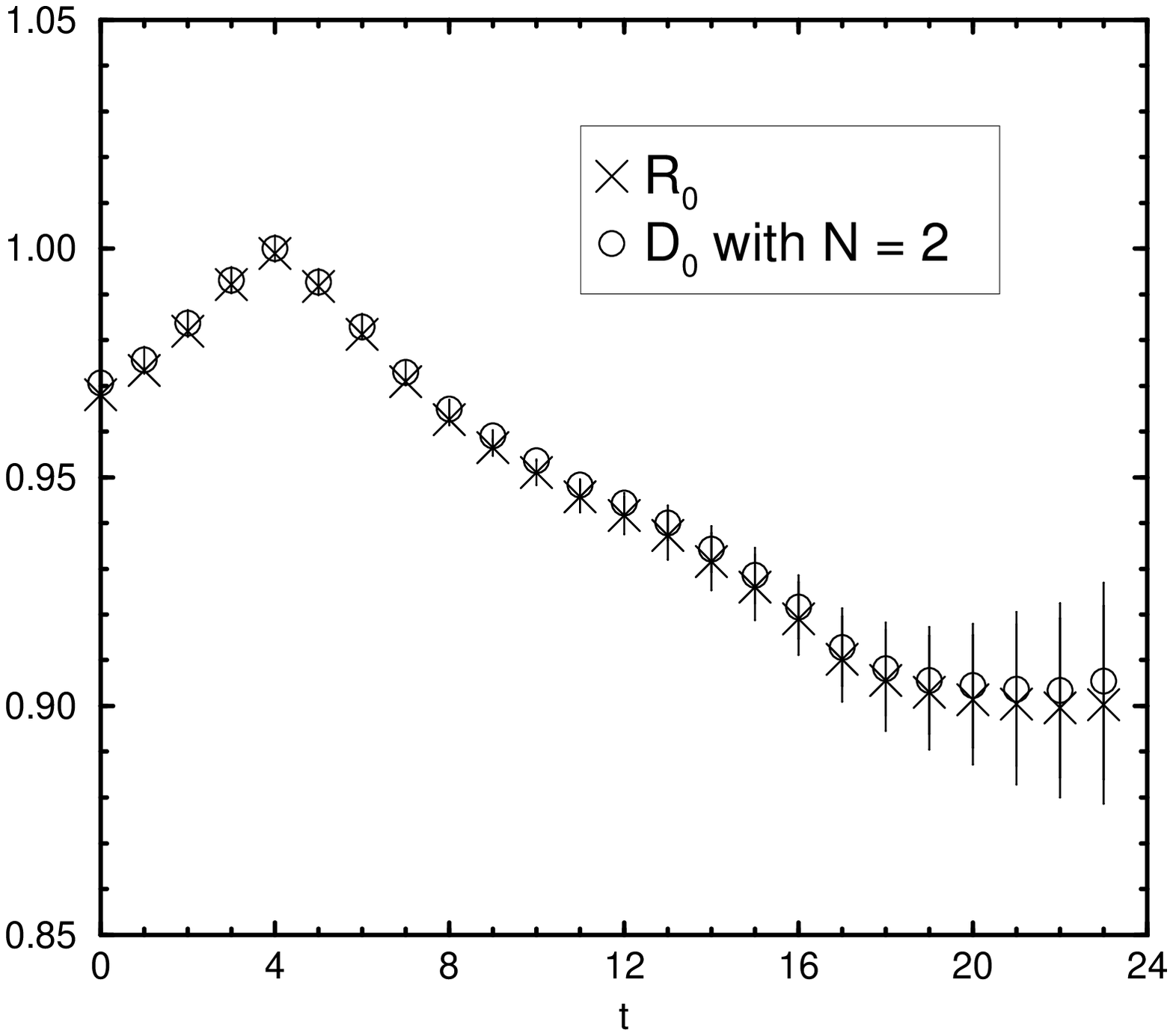}
& \leavevmode \epsfxsize=5cm   \epsfbox{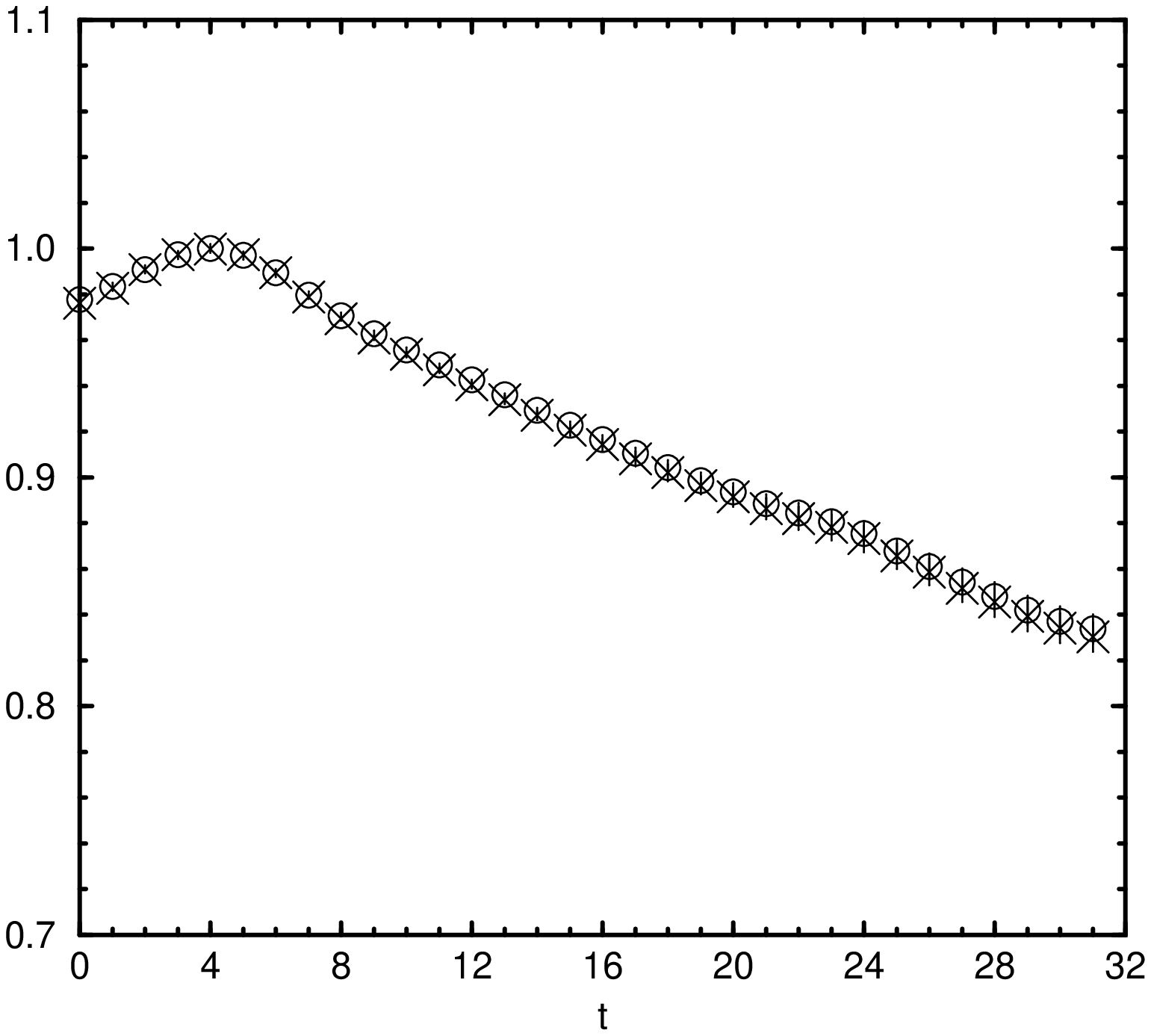}
& \leavevmode \epsfxsize=5cm   \epsfbox{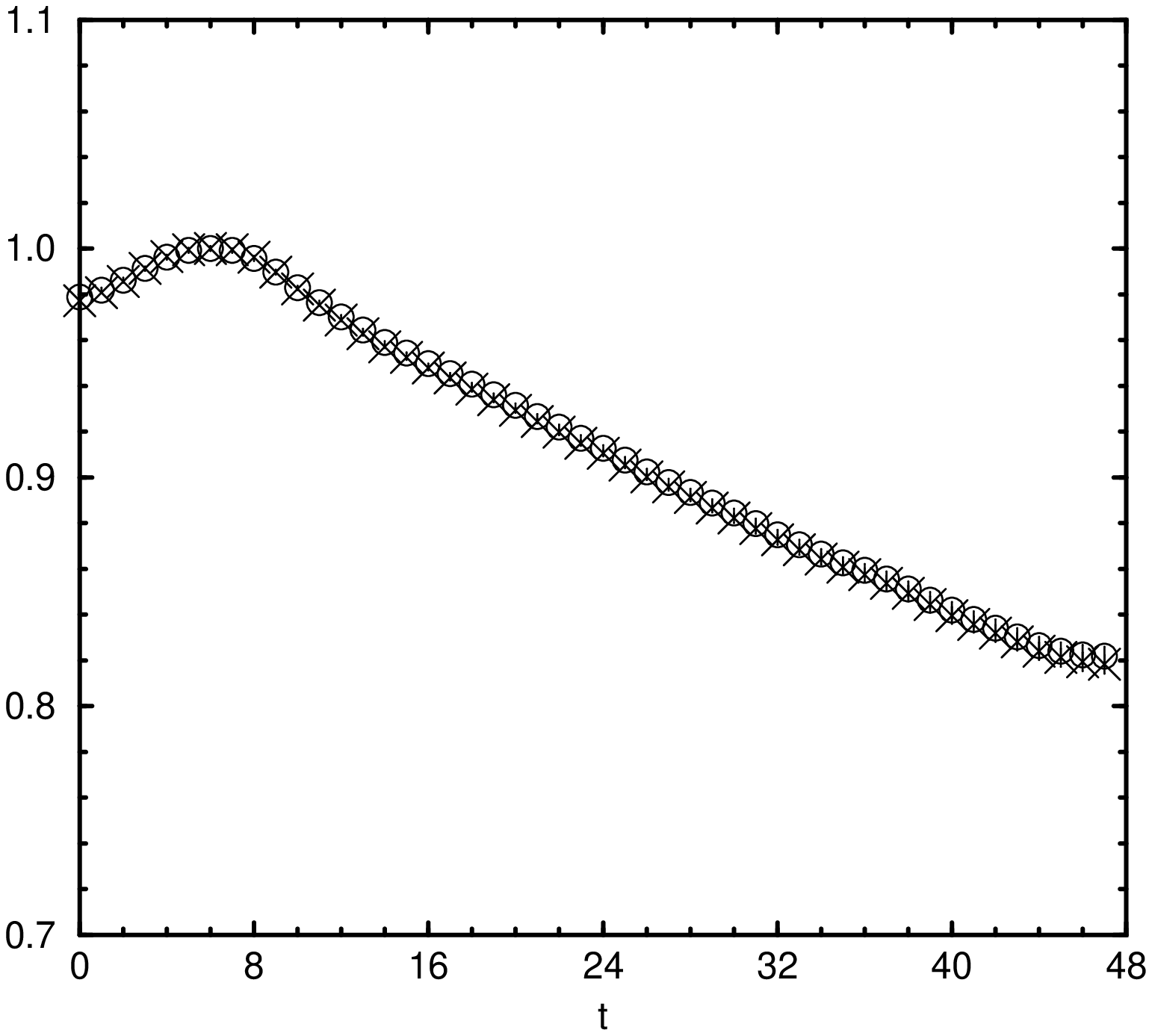} 
\\ 
\raisebox{25mm}{ $\displaystyle{ \frac{m_{\pi}}{m_{\rho}} \approx 0.7 }$ }
& \leavevmode \epsfxsize=5.1cm \epsfbox{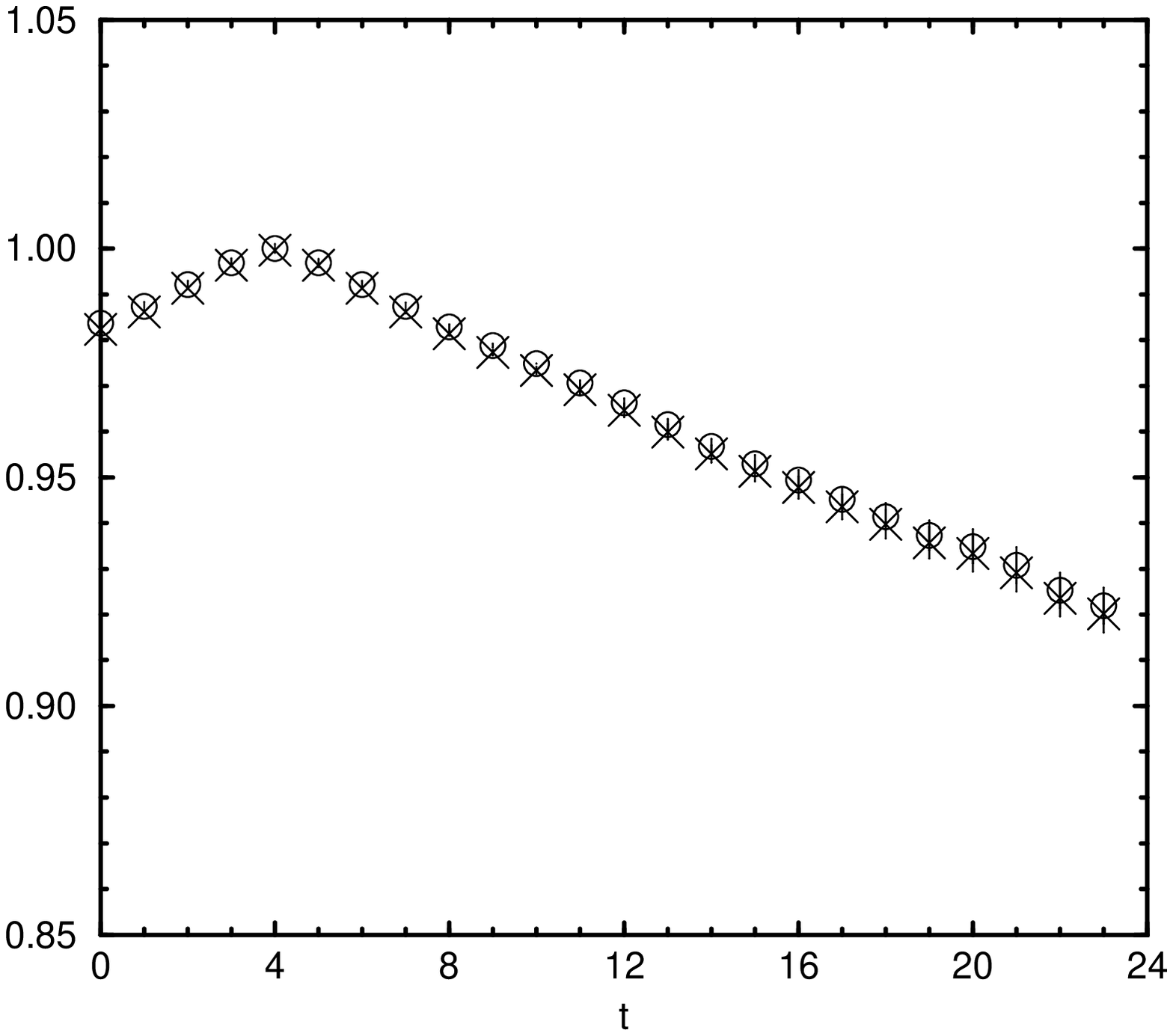}
& \leavevmode \epsfxsize=5cm   \epsfbox{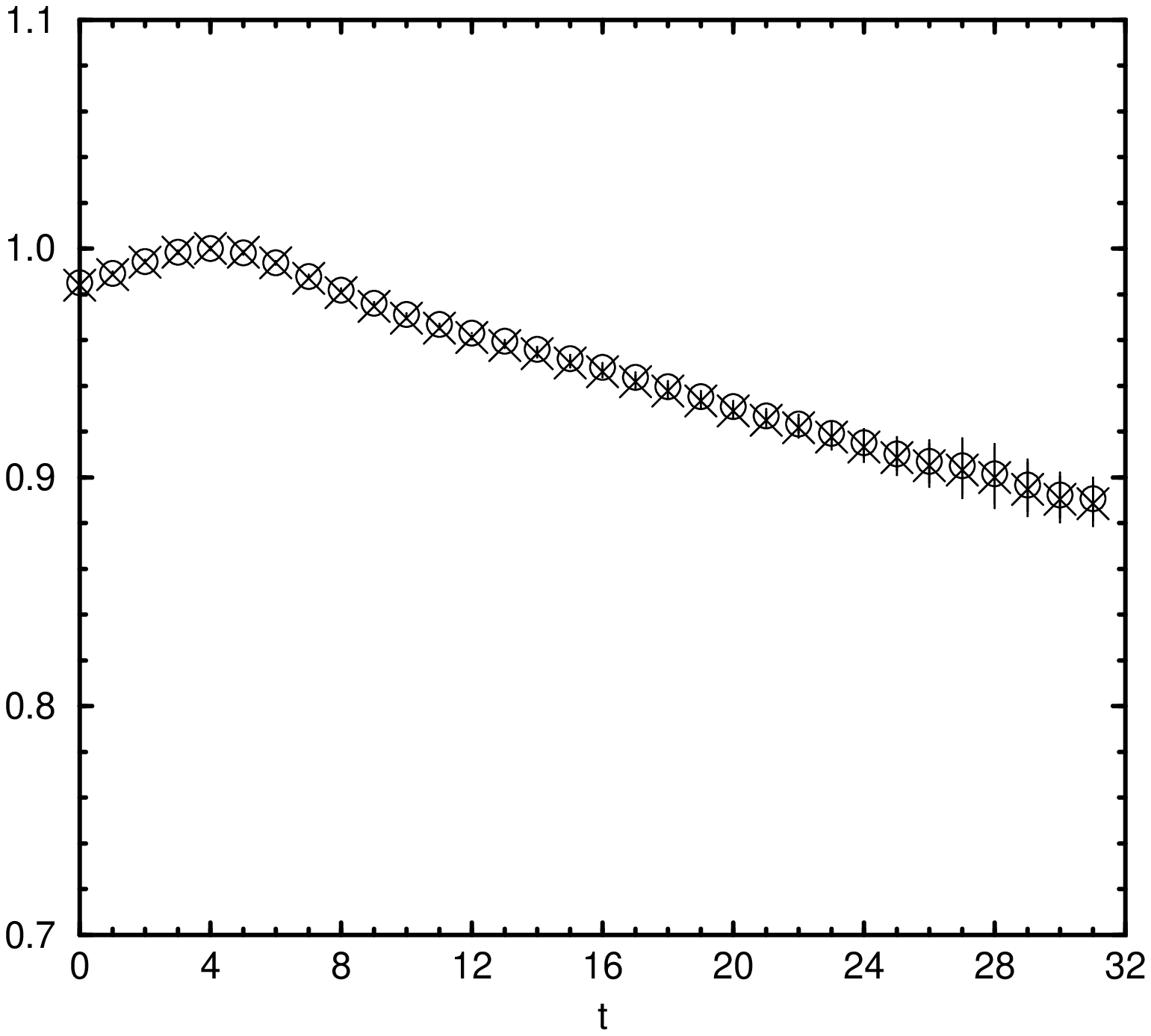}
& \leavevmode \epsfxsize=5cm   \epsfbox{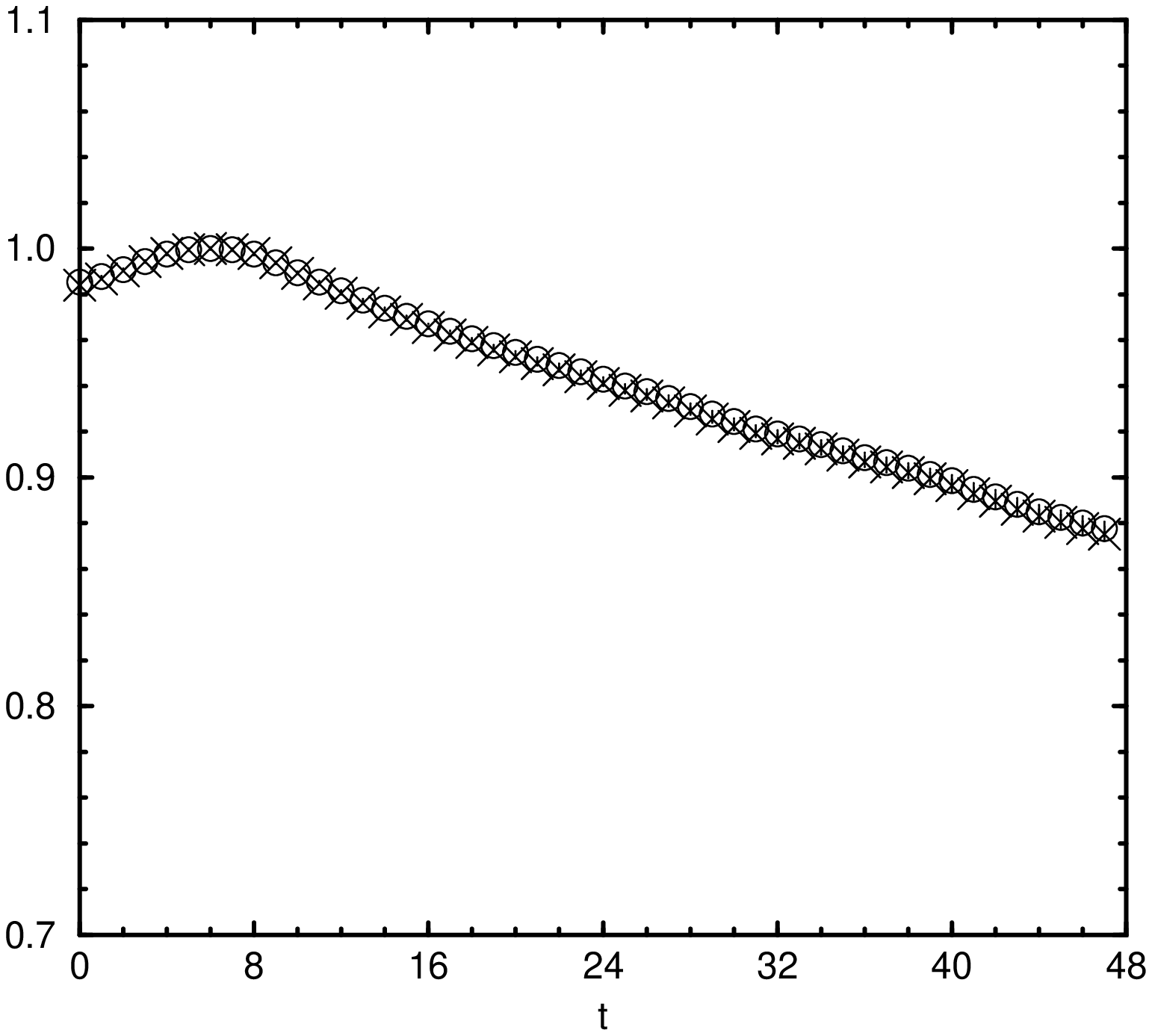} 
\\
\raisebox{25mm}{ $\displaystyle{ \frac{m_{\pi}}{m_{\rho}} \approx 0.75 }$ }
& \leavevmode \epsfxsize=5.1cm \epsfbox{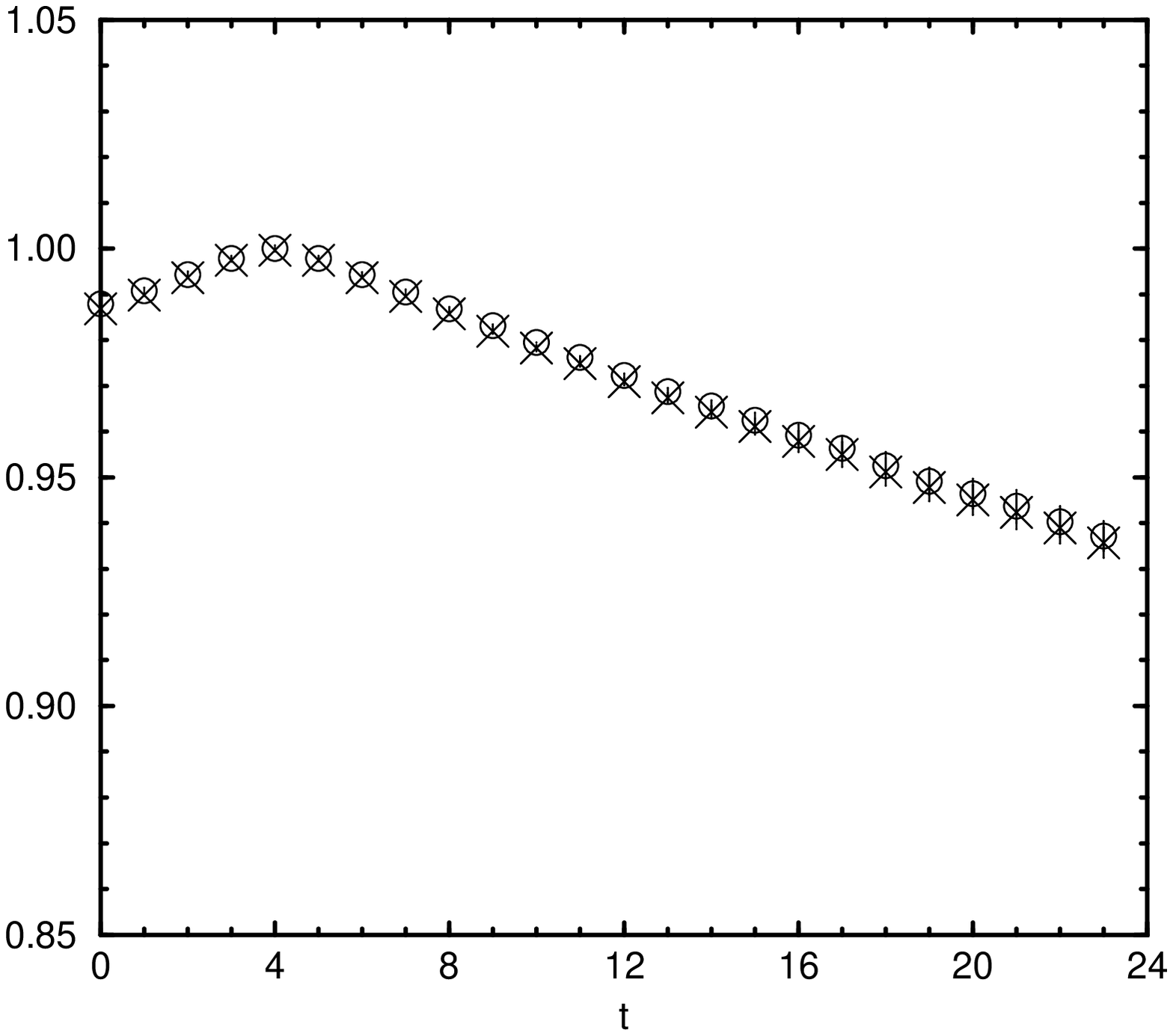}
& \leavevmode \epsfxsize=5cm   \epsfbox{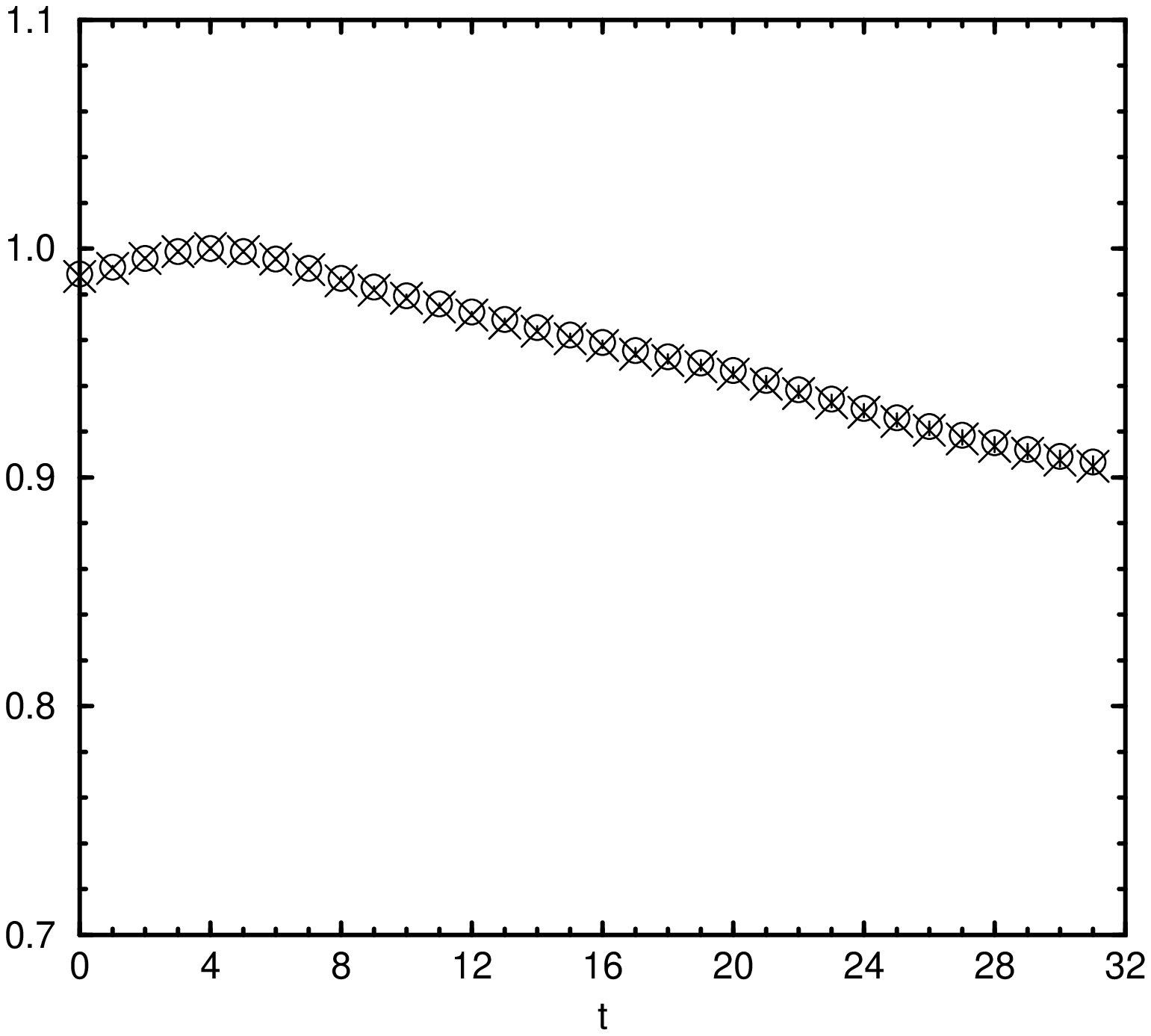}
& \leavevmode \epsfxsize=5cm   \epsfbox{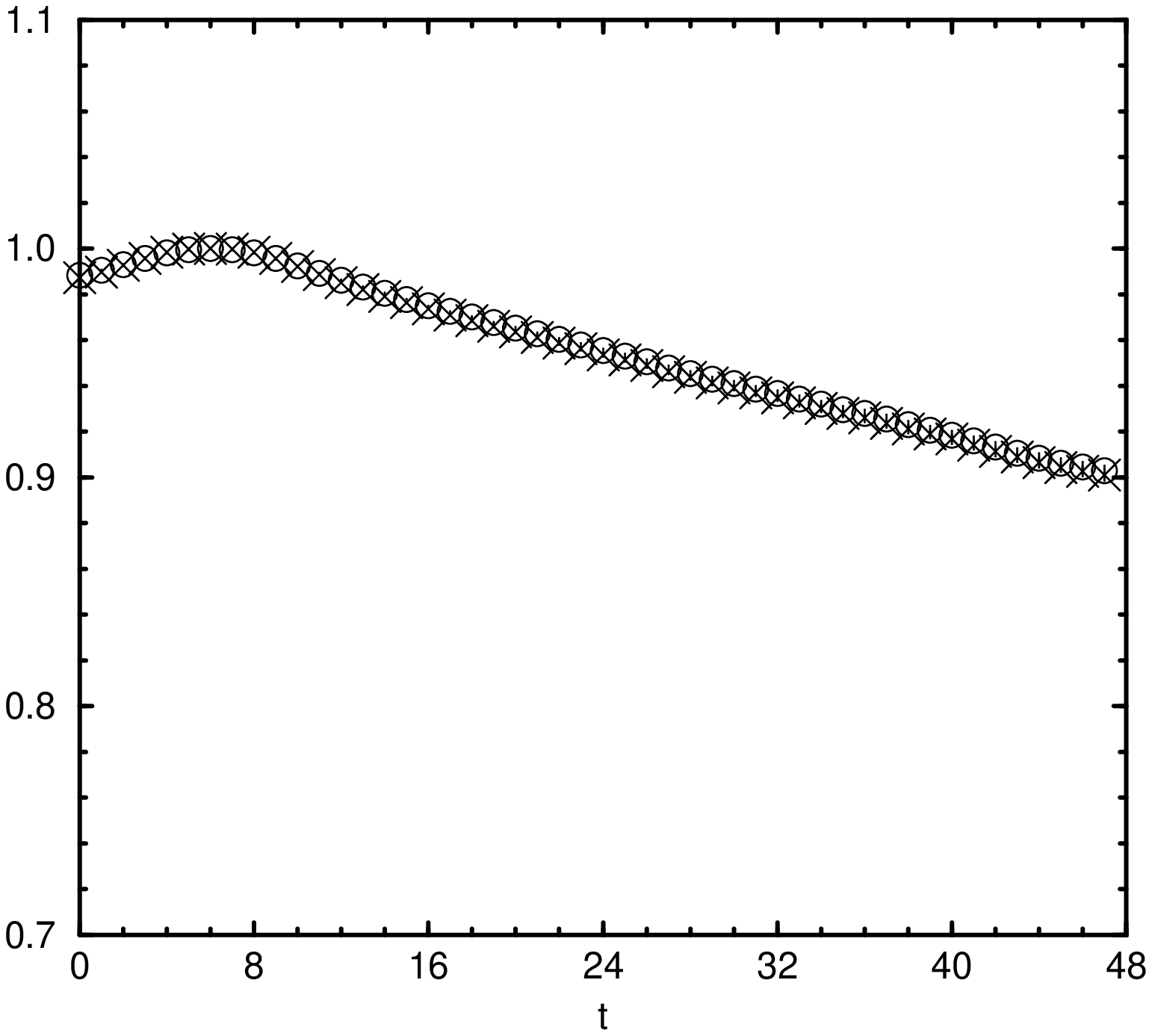} 
\\
\raisebox{25mm}{ $\displaystyle{ \frac{m_{\pi}}{m_{\rho}} \approx 0.8 }$ }
& \leavevmode \epsfxsize=5.1cm \epsfbox{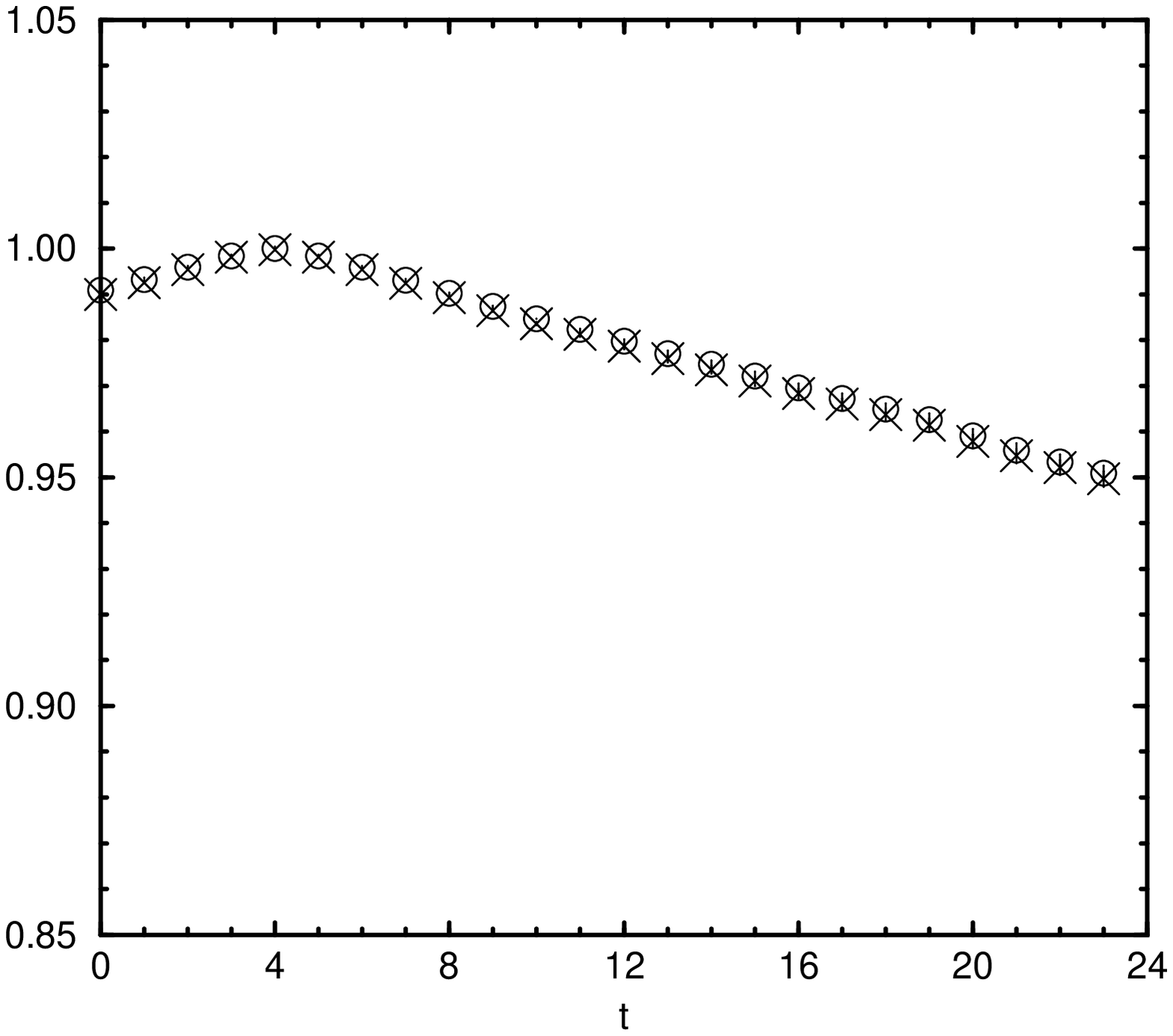}
& \leavevmode \epsfxsize=5cm   \epsfbox{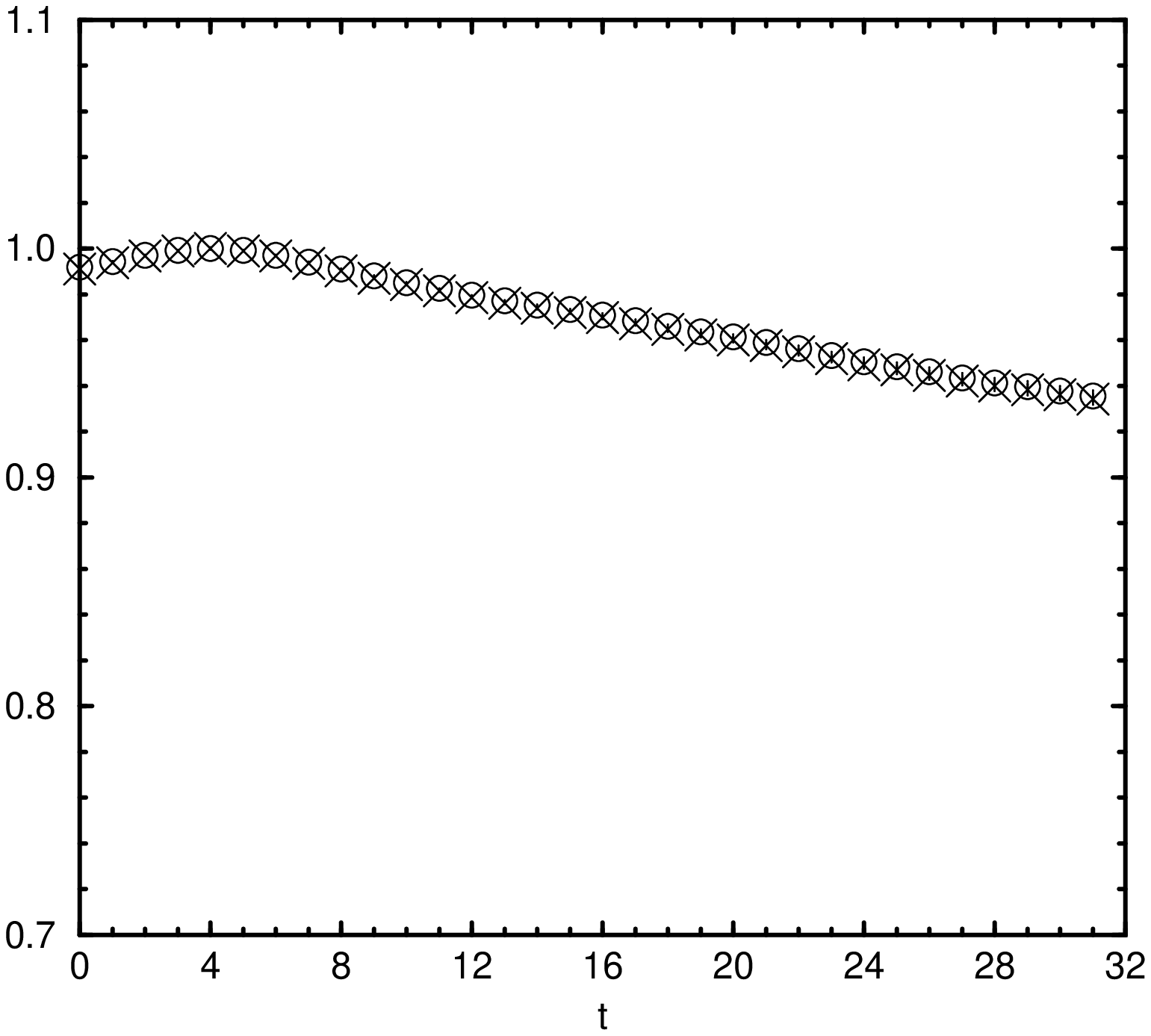}
& \leavevmode \epsfxsize=5cm   \epsfbox{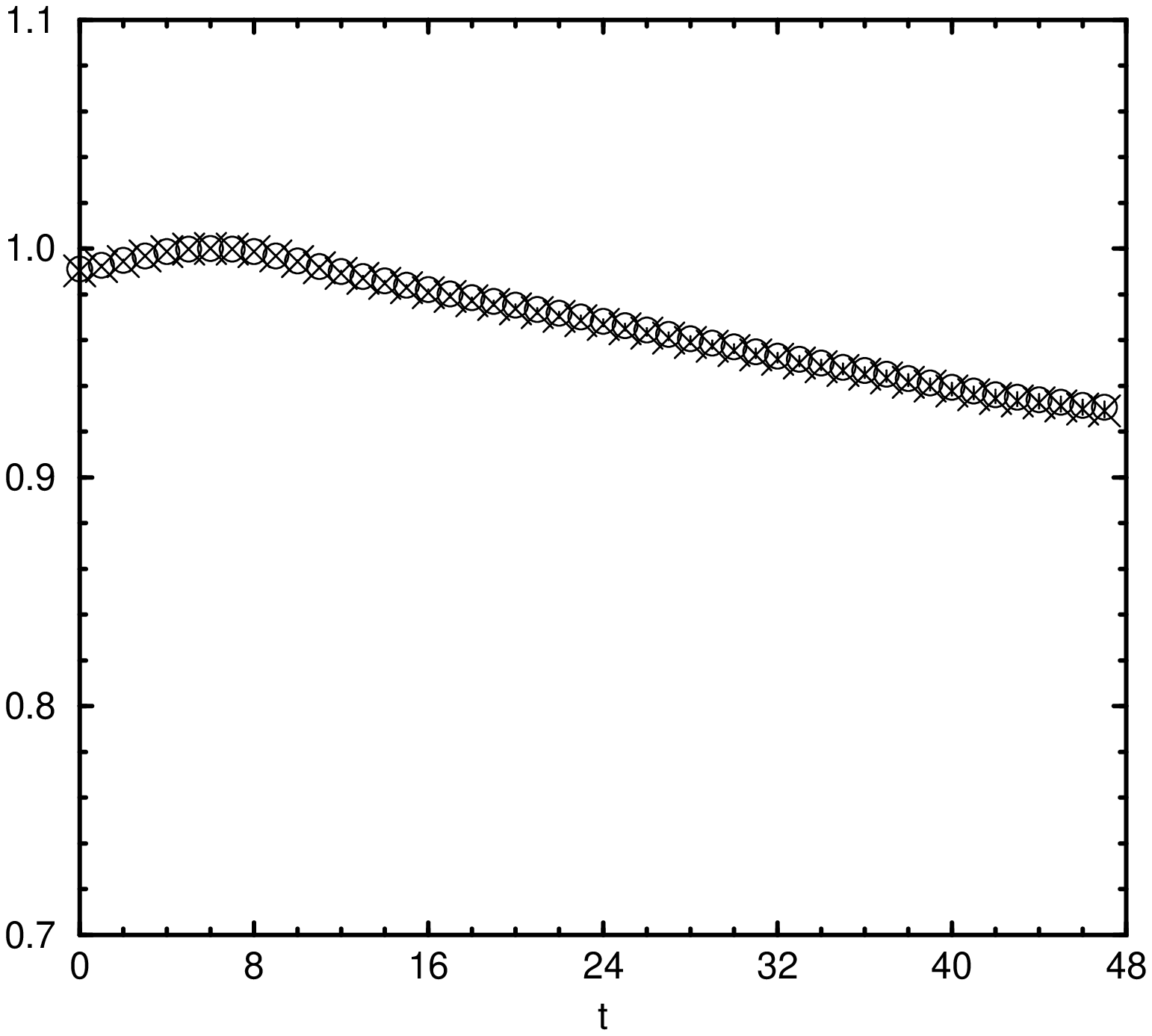}
\end{tabular}
\end{center}
\caption{
Ratios $R_n(t)$ and $D_n(t)$ for $n = 0$ in the center of mass system CM
with energy state cut-off $N=2$.
$m_\pi / m_\rho$ increases from top to bottom, 
while $\beta$ increases from left to right.
\label{fig:diag:CM_0}
}
\end{figure}

First we focus on the results in the CM system.
For the ground state ($n = 0$)
the ratios $R_0(t)$ and $D_0(t)$
for all $m_\pi / m_\rho$ and $\beta$
are presented in Fig.~\ref{fig:diag:CM_0}.
We divide the ratio $D_0(t)$ by $D_0(t_S)$ for comparison with $R_0(t)$.
The cut-off of the state $N$ for the diagonalization is set at $N=2$.
We also check the cut-off dependence by taking $N=1$ and
confirm that it is negligible.
From the figures we find that
the diagonalization does not affect the result of the ground state
and the energy shift
can be extracted from the ratio $R_0(t)$
without the diagonalization.

\begin{figure}
\begin{center}
\hspace{-26mm}
\begin{tabular}{rccc}
& $\beta = 1.80$ & $\beta = 1.95$ & $\beta = 2.10$ 
\\
\raisebox{25mm}{ $\displaystyle{ \frac{m_{\pi}}{m_{\rho}} \approx 0.6 }$ }
& \leavevmode \epsfxsize=5.1cm \epsfbox{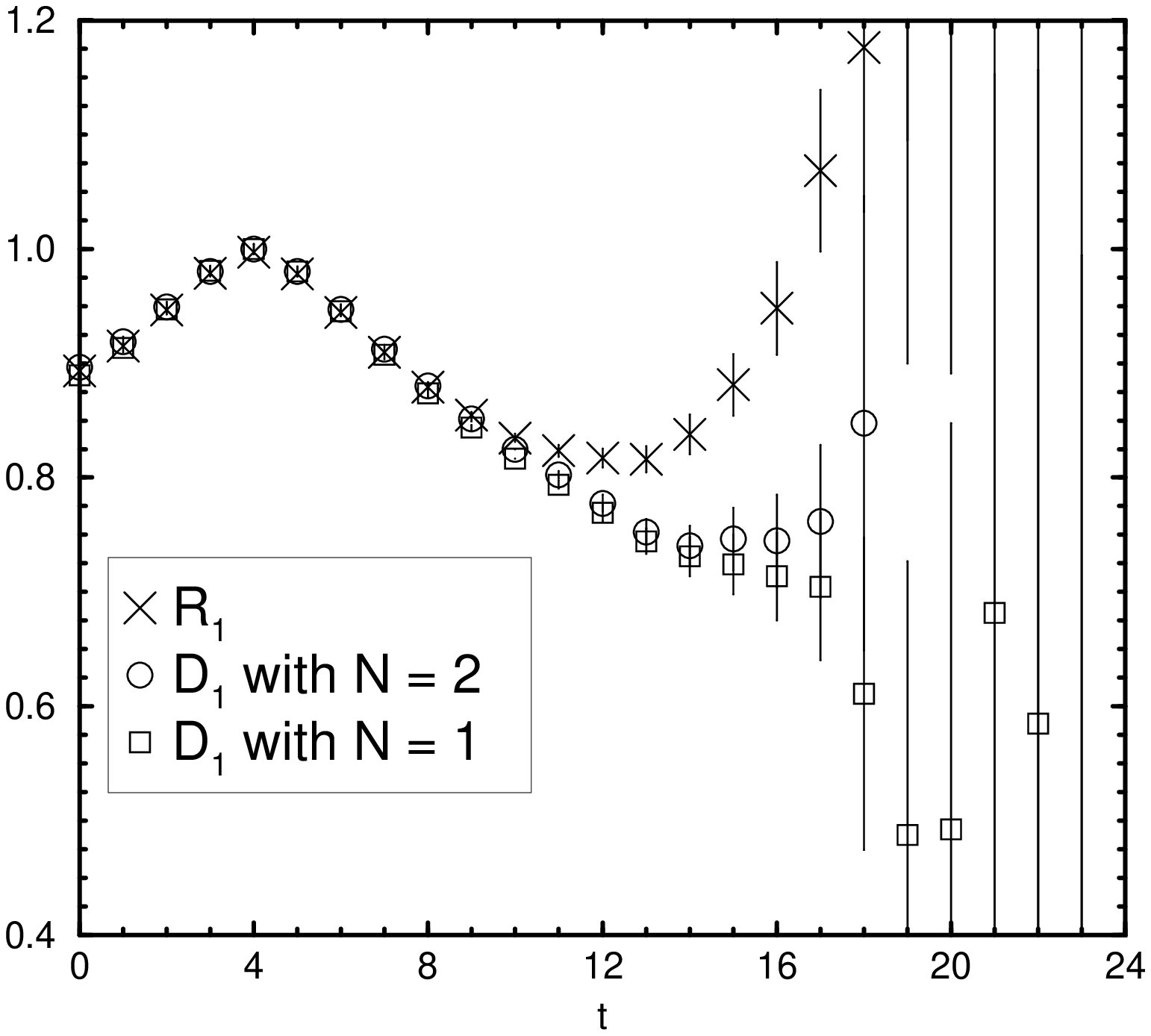}
& \leavevmode \epsfxsize=5cm   \epsfbox{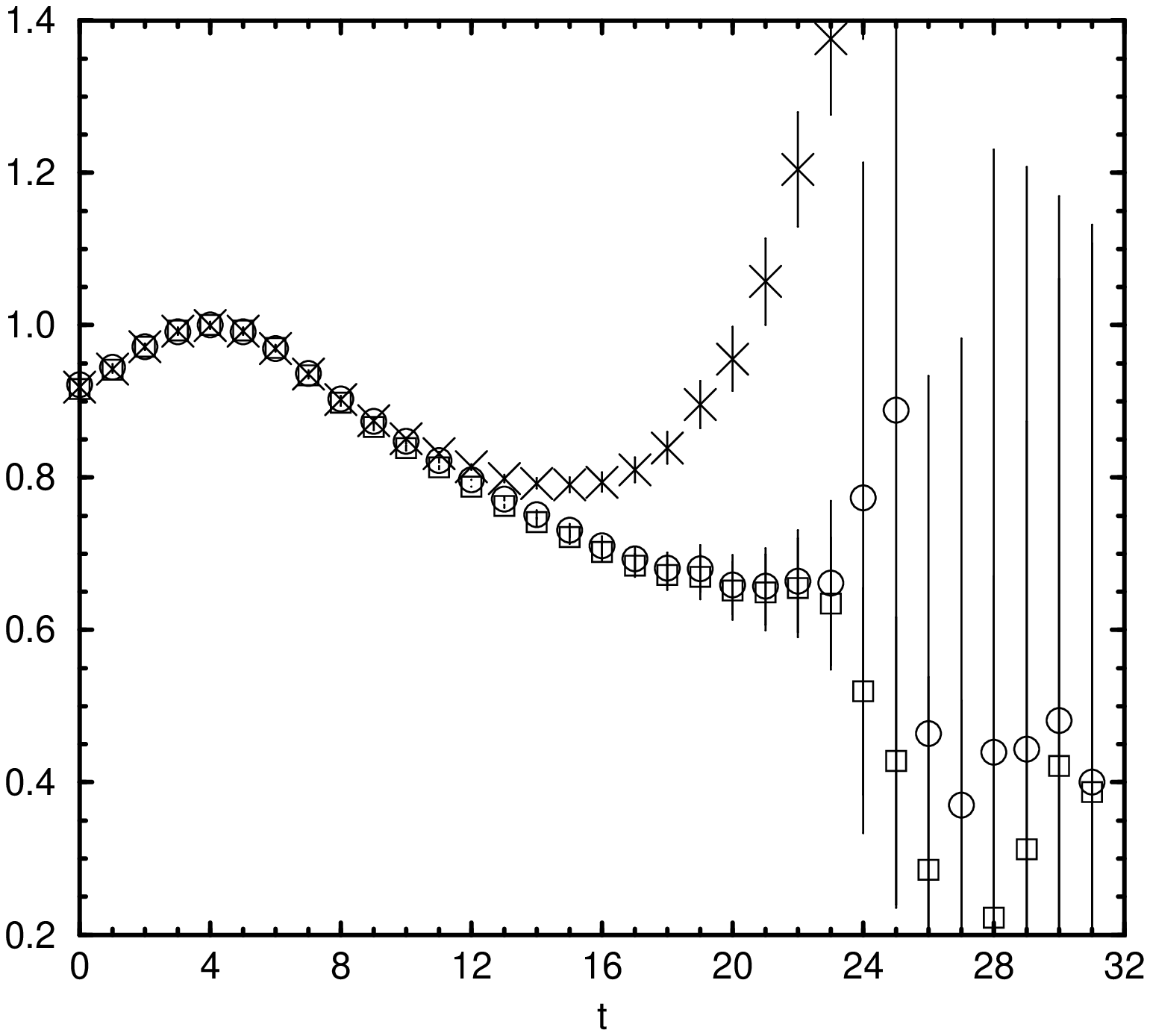}
& \leavevmode \epsfxsize=5cm   \epsfbox{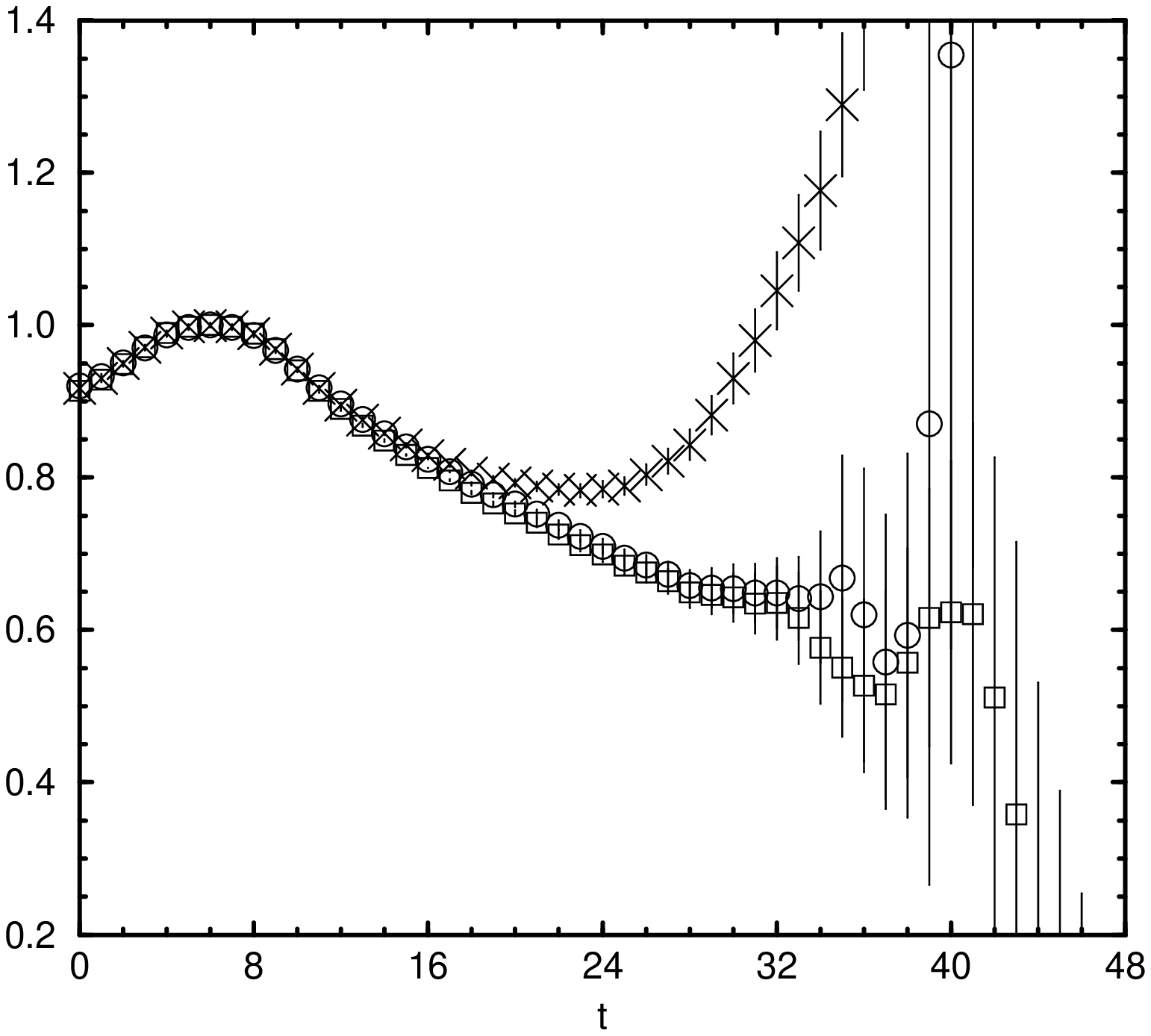} 
\\
\raisebox{25mm}{ $\displaystyle{ \frac{m_{\pi}}{m_{\rho}} \approx 0.7 }$ }
& \leavevmode \epsfxsize=5.1cm \epsfbox{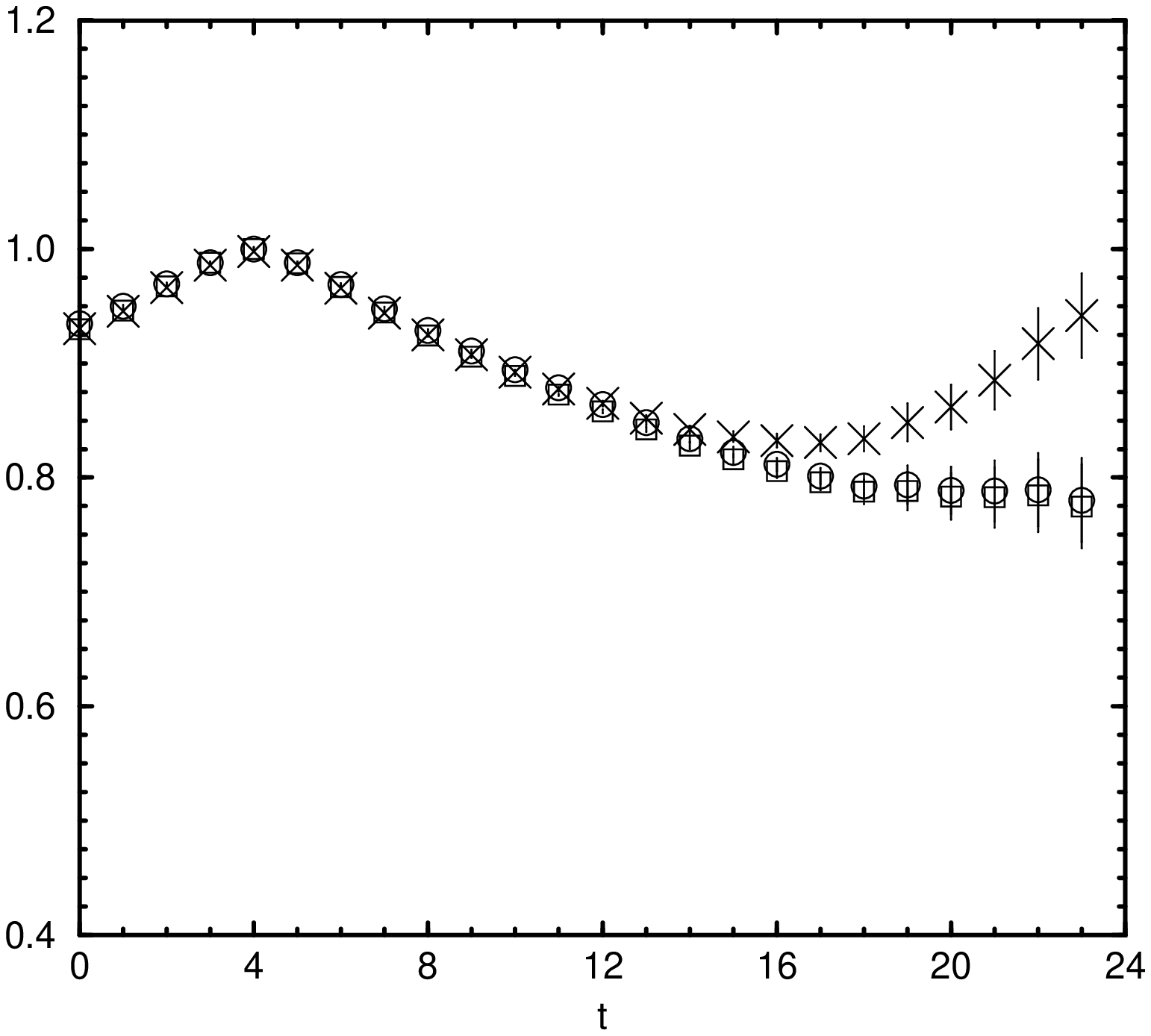}
& \leavevmode \epsfxsize=5cm   \epsfbox{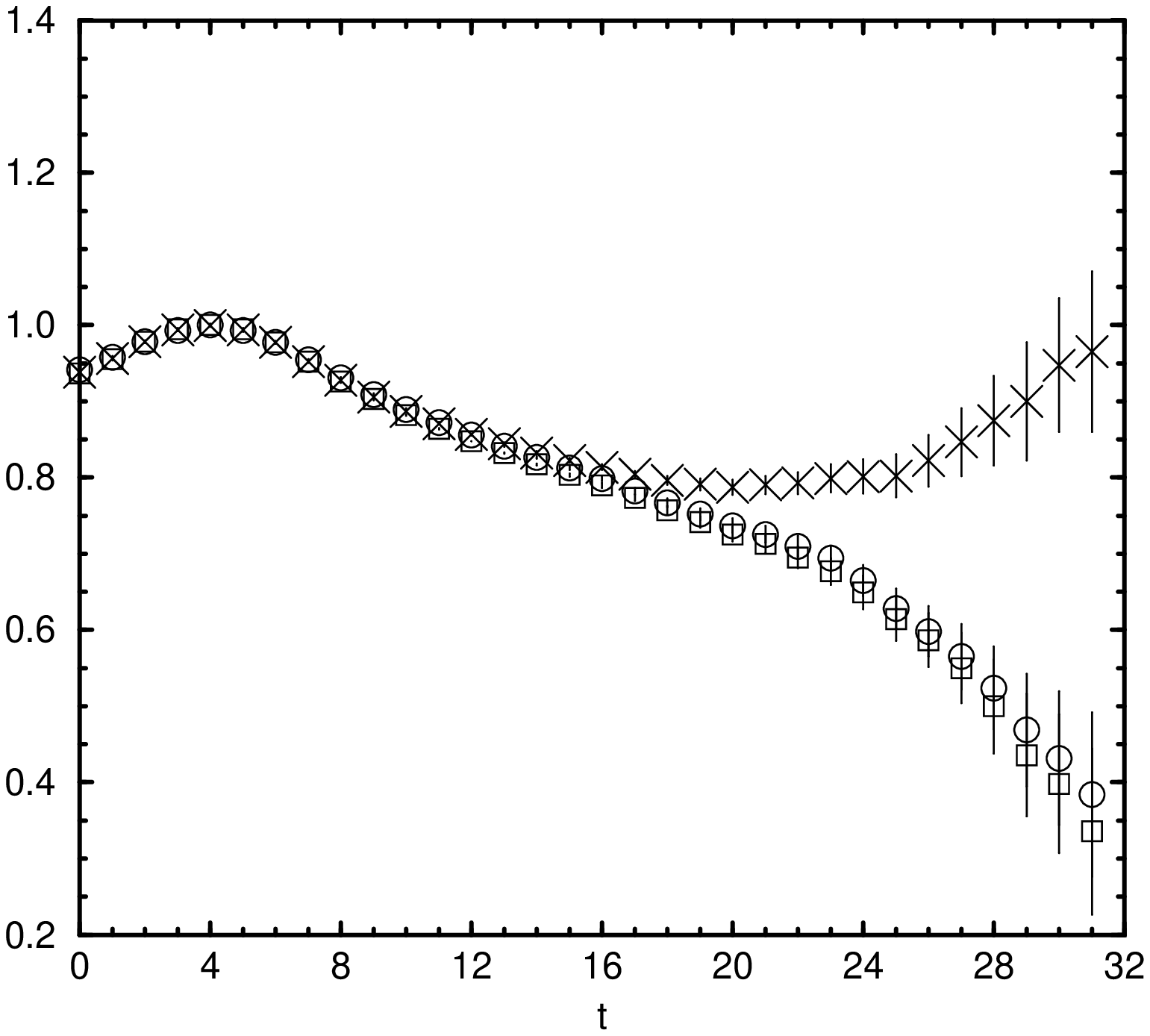}
& \leavevmode \epsfxsize=5cm   \epsfbox{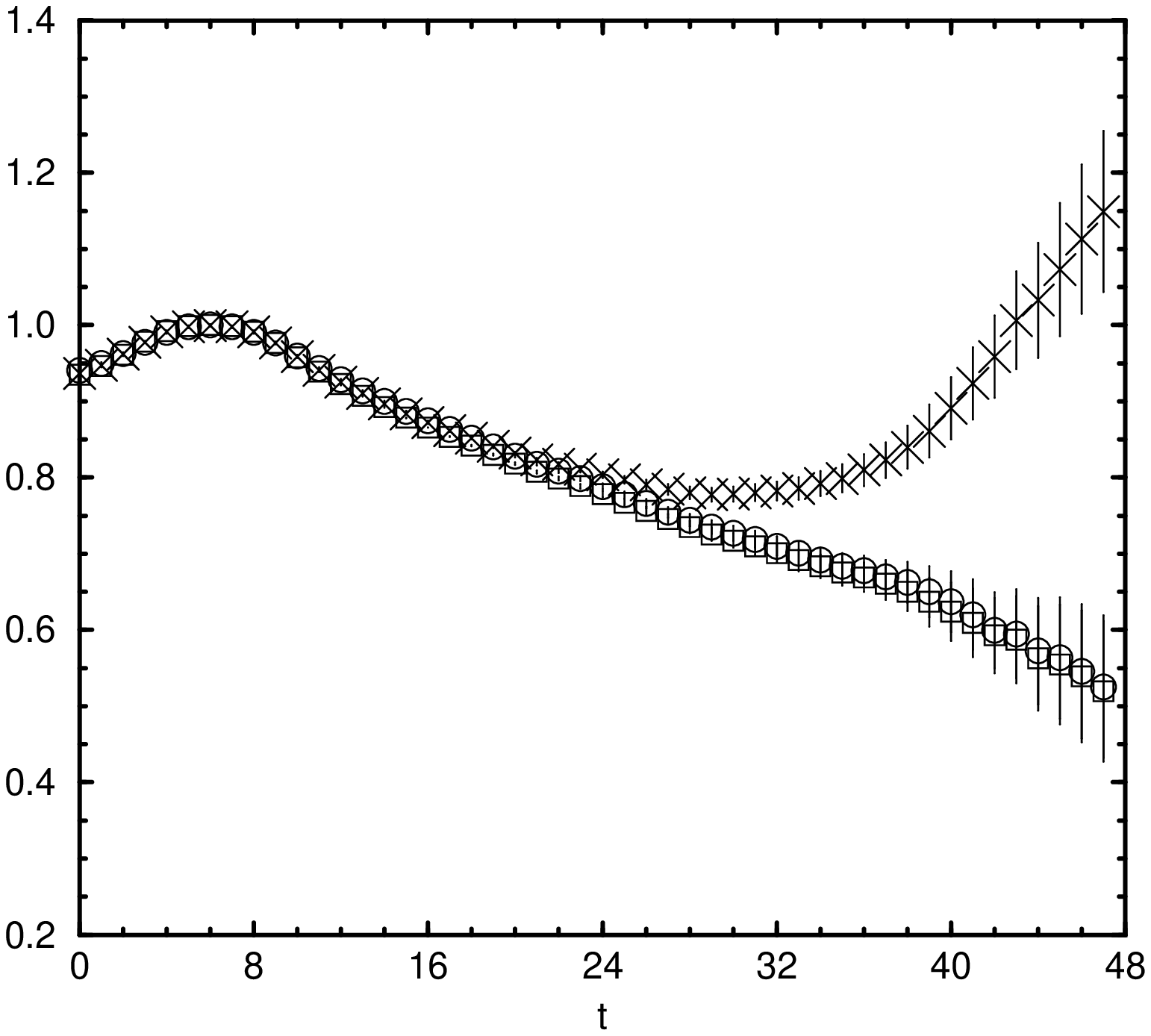}
\\
\raisebox{25mm}{ $\displaystyle{ \frac{m_{\pi}}{m_{\rho}} \approx 0.75 }$ }
& \leavevmode \epsfxsize=5.1cm \epsfbox{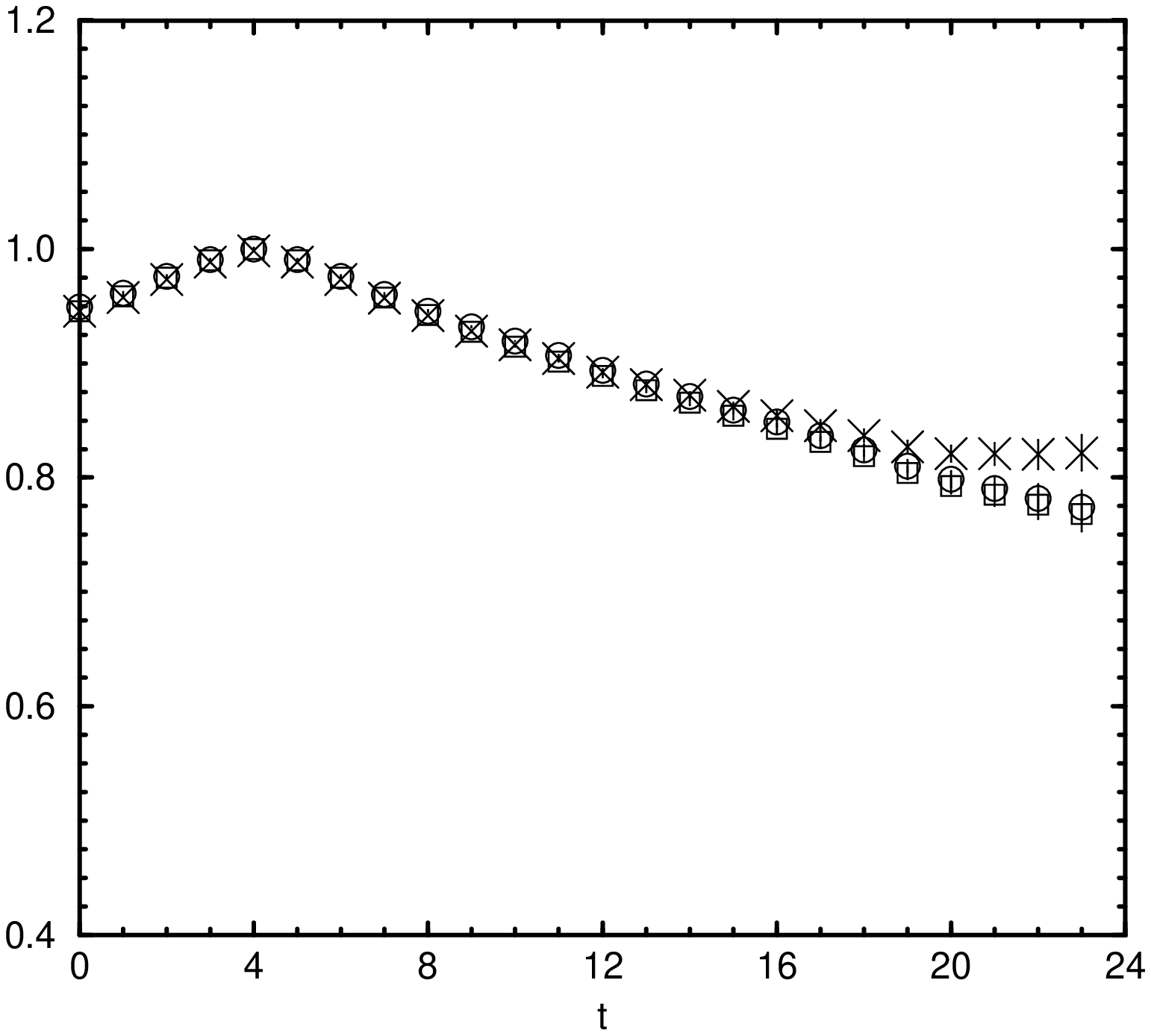}
& \leavevmode \epsfxsize=5cm   \epsfbox{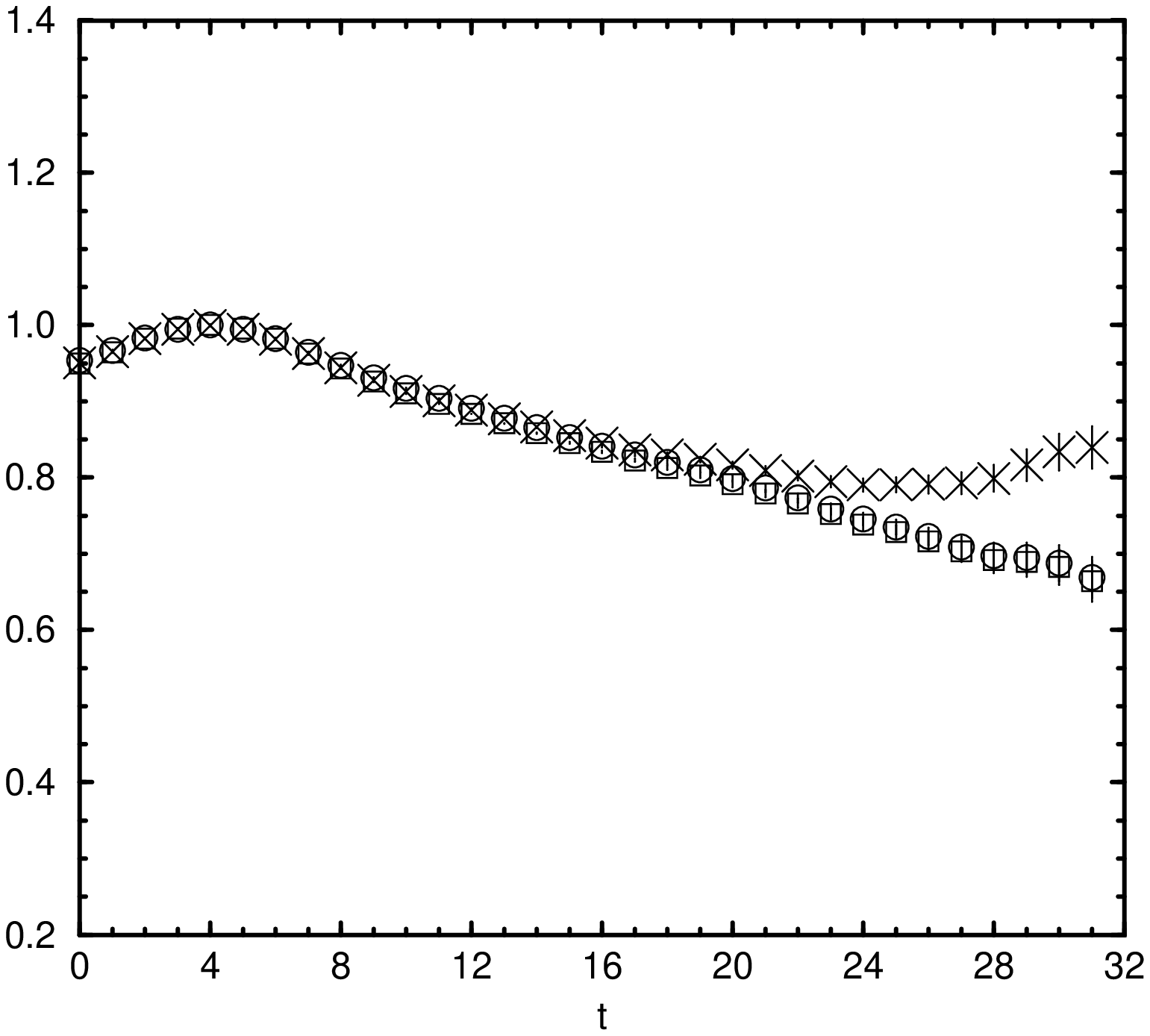}
& \leavevmode \epsfxsize=5cm   \epsfbox{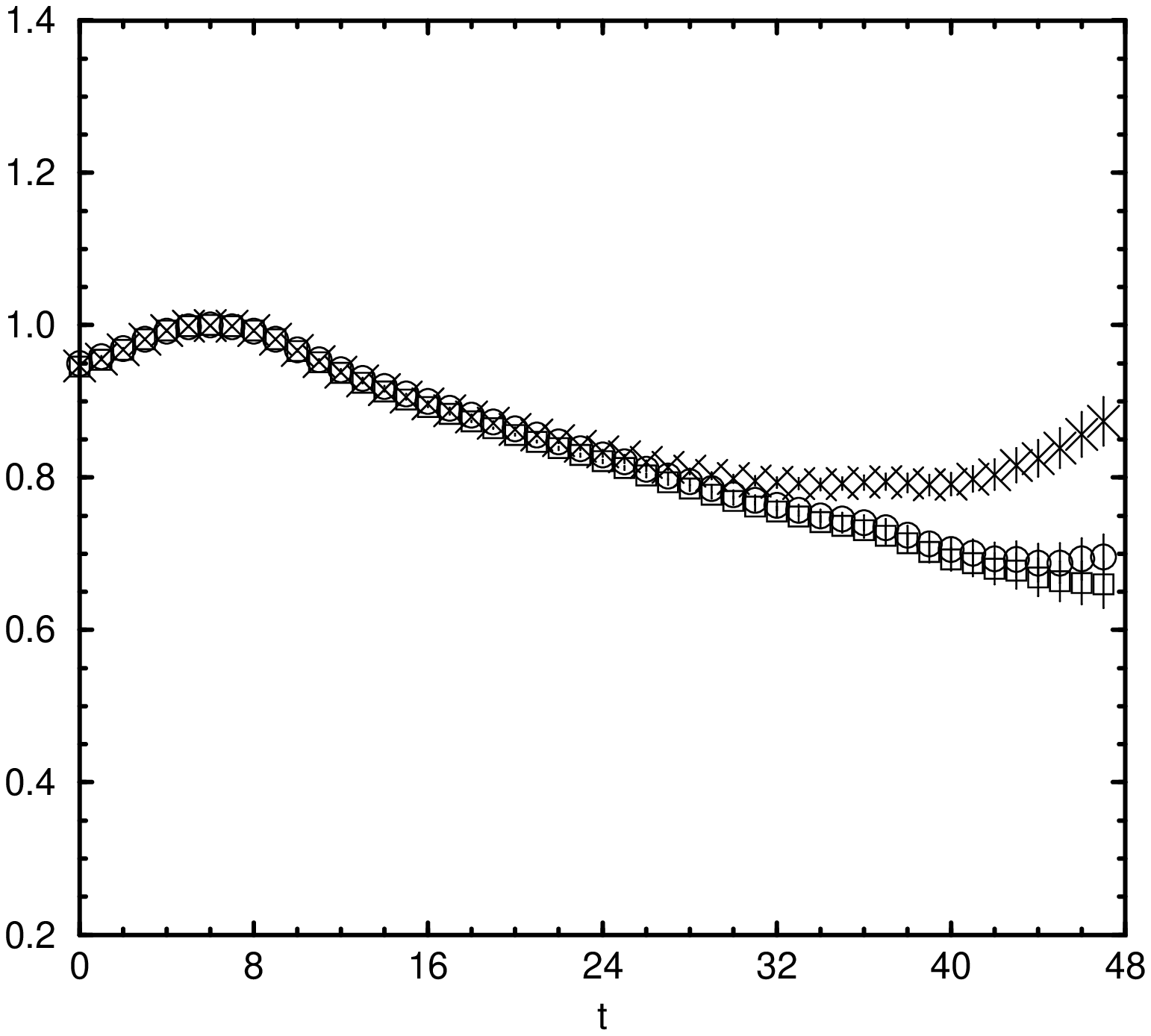}
\\
\raisebox{25mm}{ $\displaystyle{ \frac{m_{\pi}}{m_{\rho}} \approx 0.8 } $ }
& \leavevmode \epsfxsize=5.1cm \epsfbox{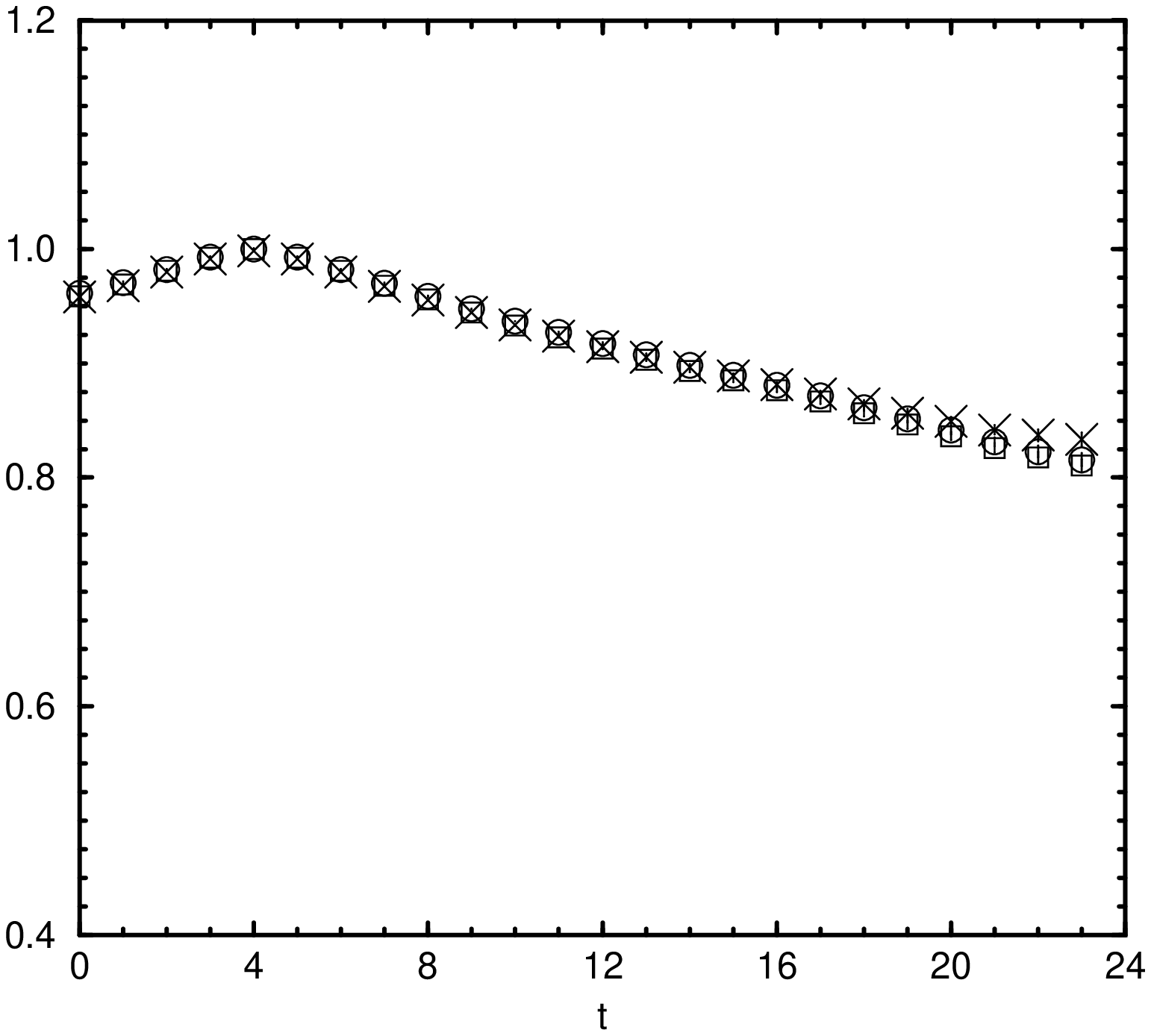}
& \leavevmode \epsfxsize=5cm   \epsfbox{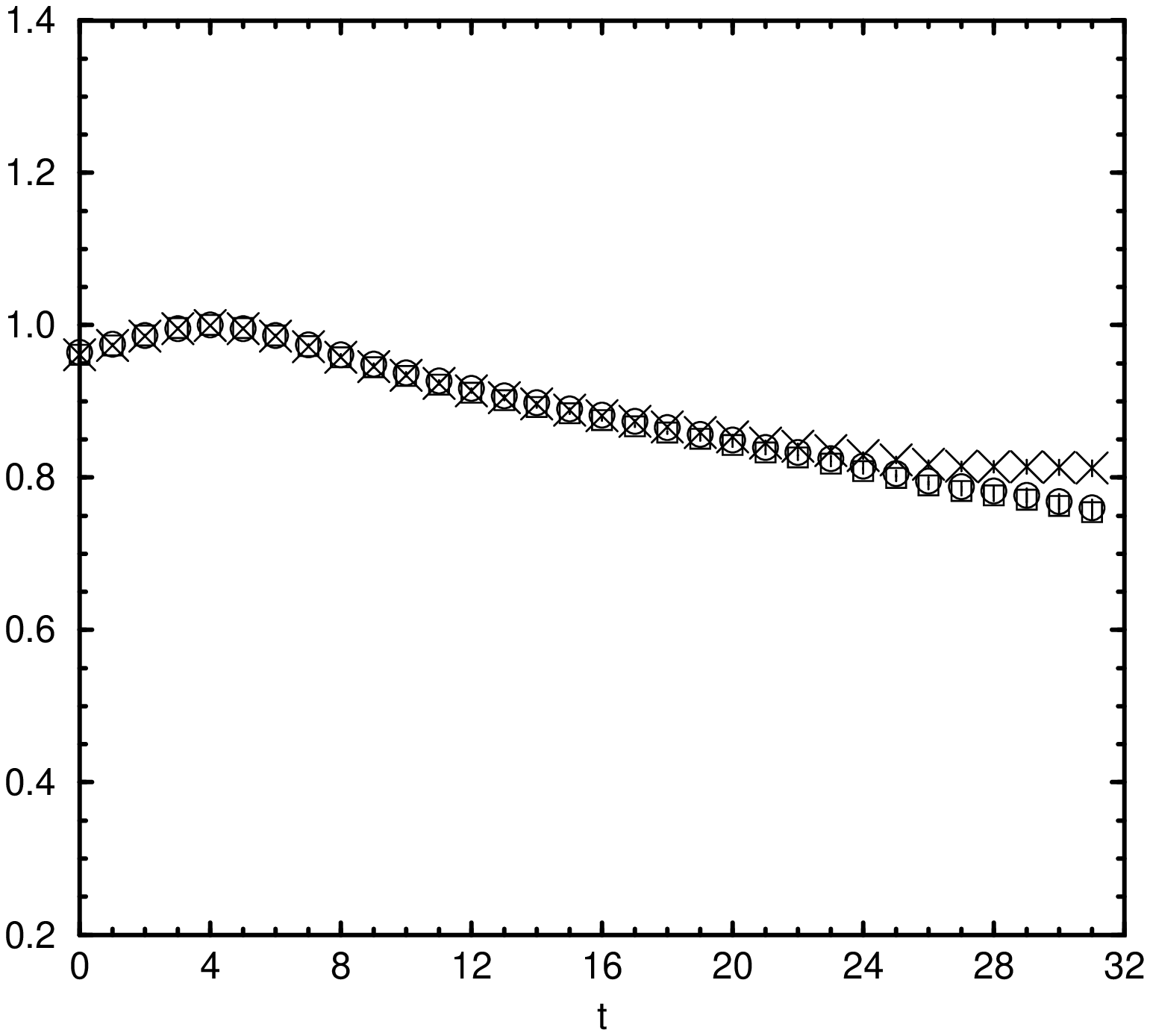}
& \leavevmode \epsfxsize=5cm   \epsfbox{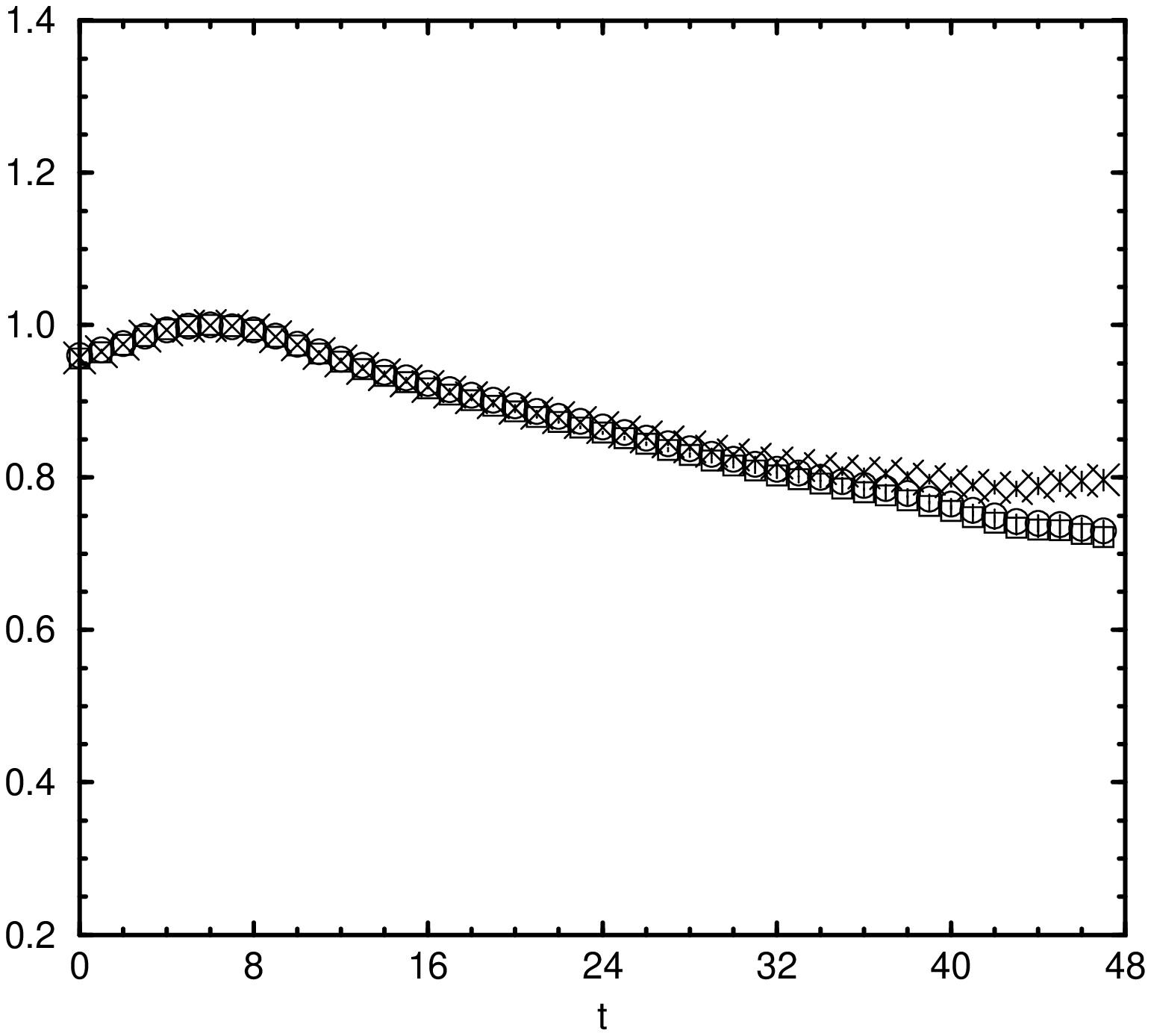}
\end{tabular}
\end{center}
\caption{
Ratio $R_n(t)$ and $D_n(t)$ for $n=1$ in the center of mass system CM
with energy state cut-off $N=1,2$.
$m_\pi / m_\rho$ increases from top to bottom, 
while $\beta$ increases from left to right.
\label{fig:diag:CM_1}
}
\end{figure}

We compare the ratios for the first excited state ($n=1$) 
in the CM system in Fig.~\ref{fig:diag:CM_1},
where the cut-off is set at $N=1$ and $2$.
Here we divide $D_1(t)$ by $D_1(t_S)$ as for $n=0$. 
In contrast to the case of the ground state 
the diagonalization is very effective for the smaller quark masses.
The ratio $R_1(t)$ for smaller masses rapidly increases,
while such behavior cannot be seen in $D_1(t)$.
The cut-off dependence of $D_1(t)$
is negligible for whole parameter regions as shown in the figure.
Hence we can extract the energy shift
from the ratio $D_1(t)$ by a single exponential fitting.

\begin{figure}
\begin{center}
\hspace{-26mm}
\begin{tabular}{rccc}
& $ \beta = 1.80 $ & $ \beta = 1.95 $ & $ \beta = 2.10 $ \\
\raisebox{25mm}{ $\displaystyle{ \frac{m_{\pi}}{m_{\rho}} \approx 0.6 }$ }
& \leavevmode \epsfxsize=5.1cm \epsfbox{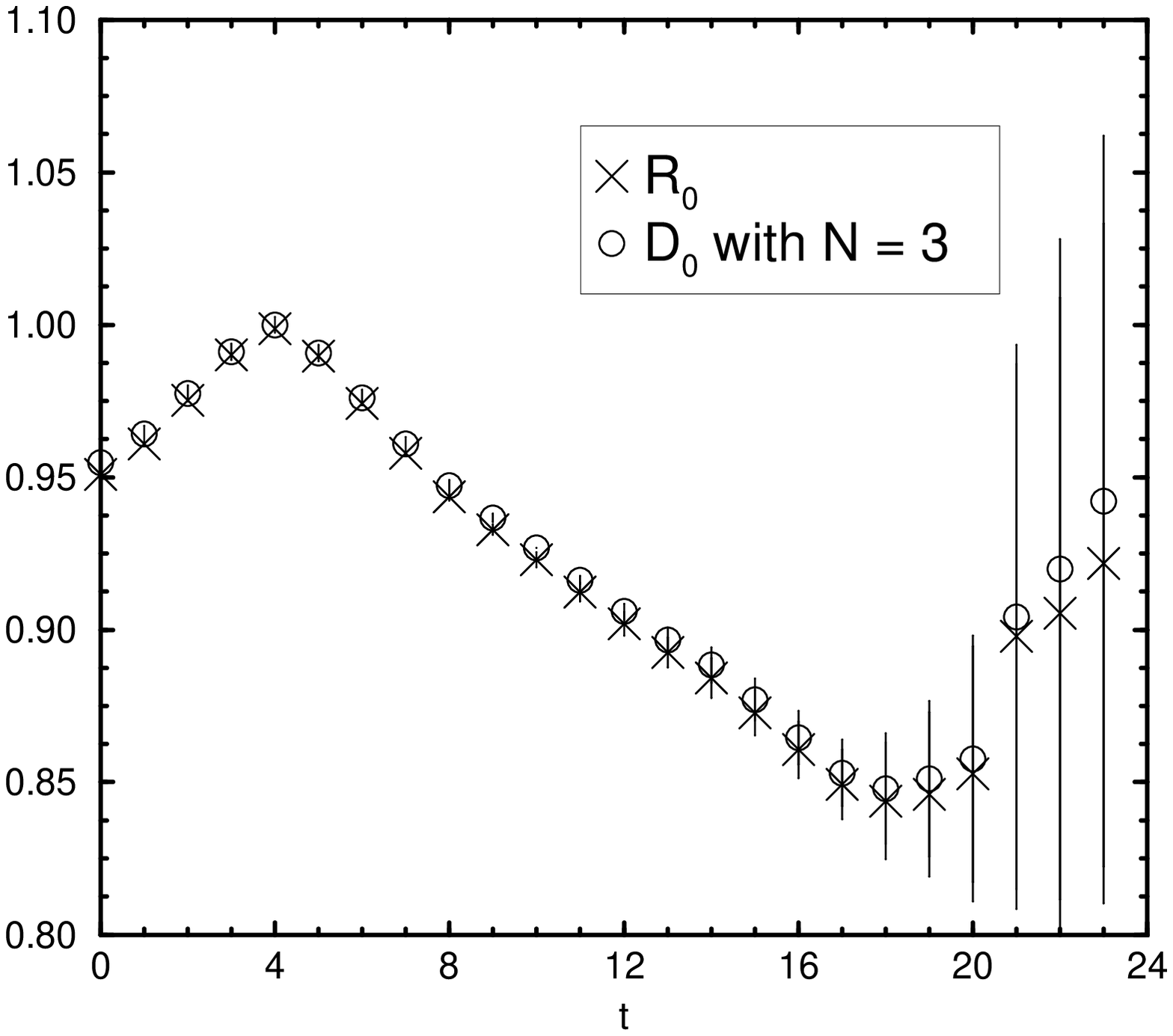}
& \leavevmode \epsfxsize=5cm   \epsfbox{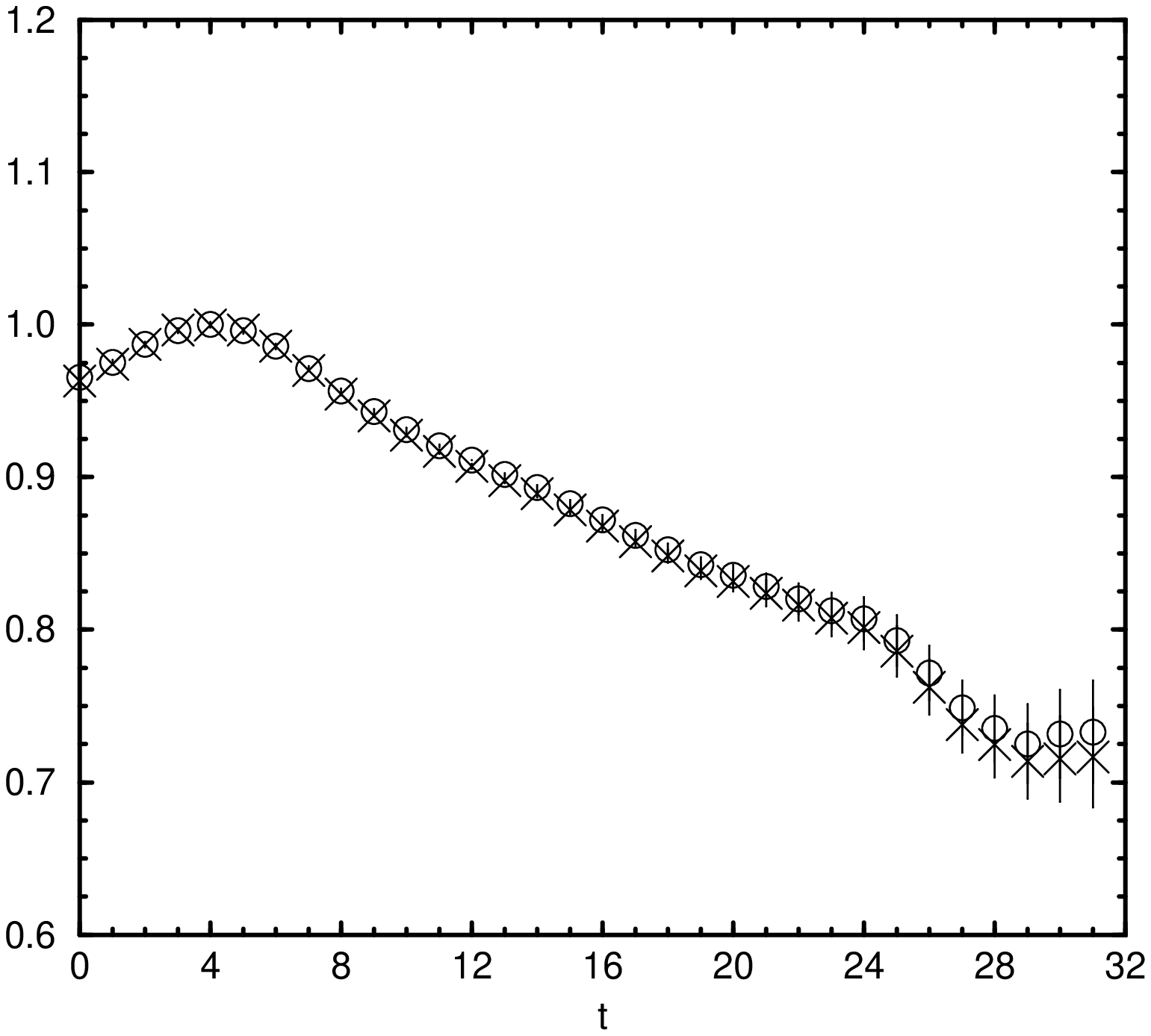}
& \leavevmode \epsfxsize=5cm   \epsfbox{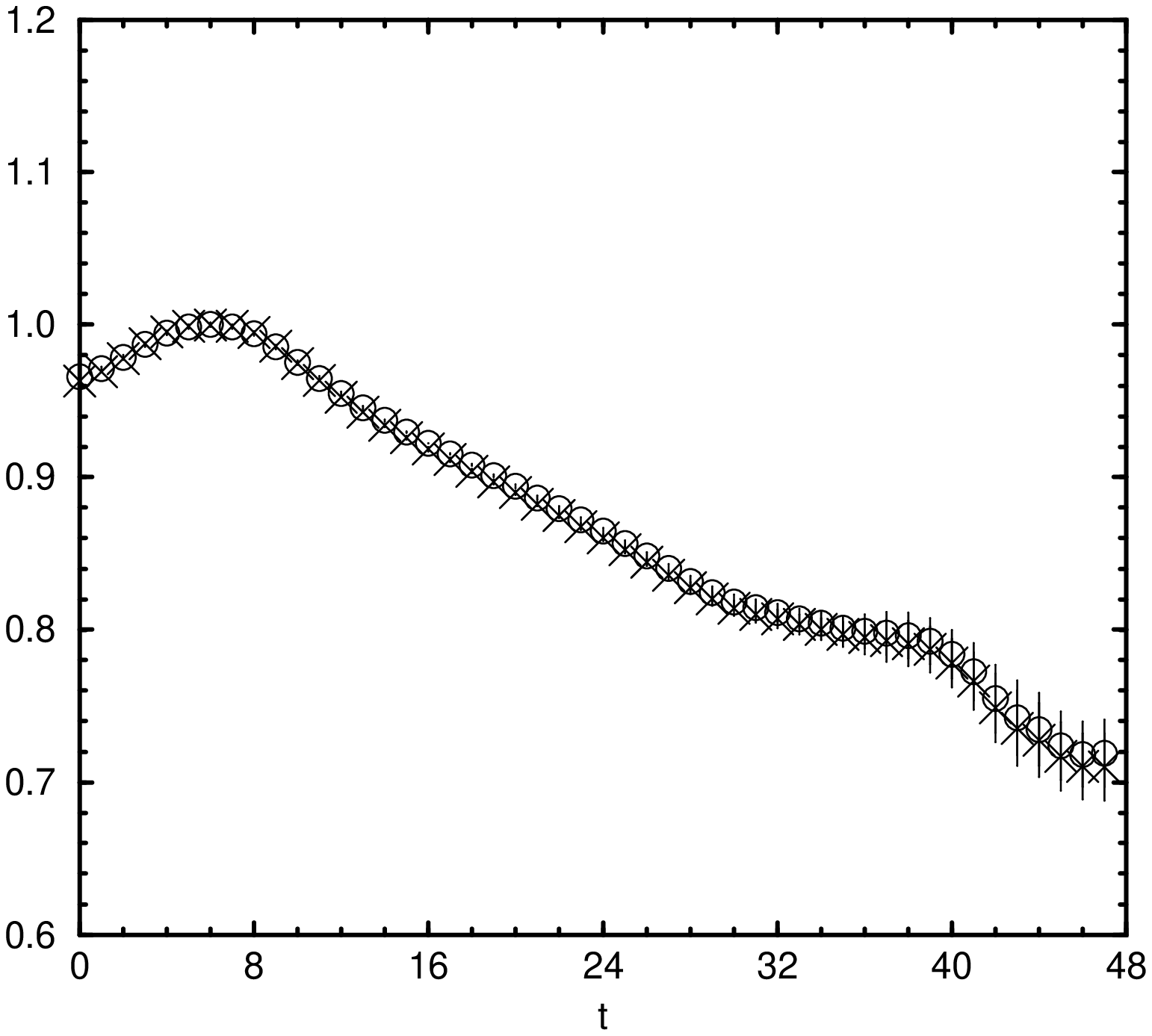}
\\
\raisebox{25mm}{ $\displaystyle{ \frac{m_{\pi}}{m_{\rho}} \approx 0.7 }$ }
& \leavevmode \epsfxsize=5.1cm \epsfbox{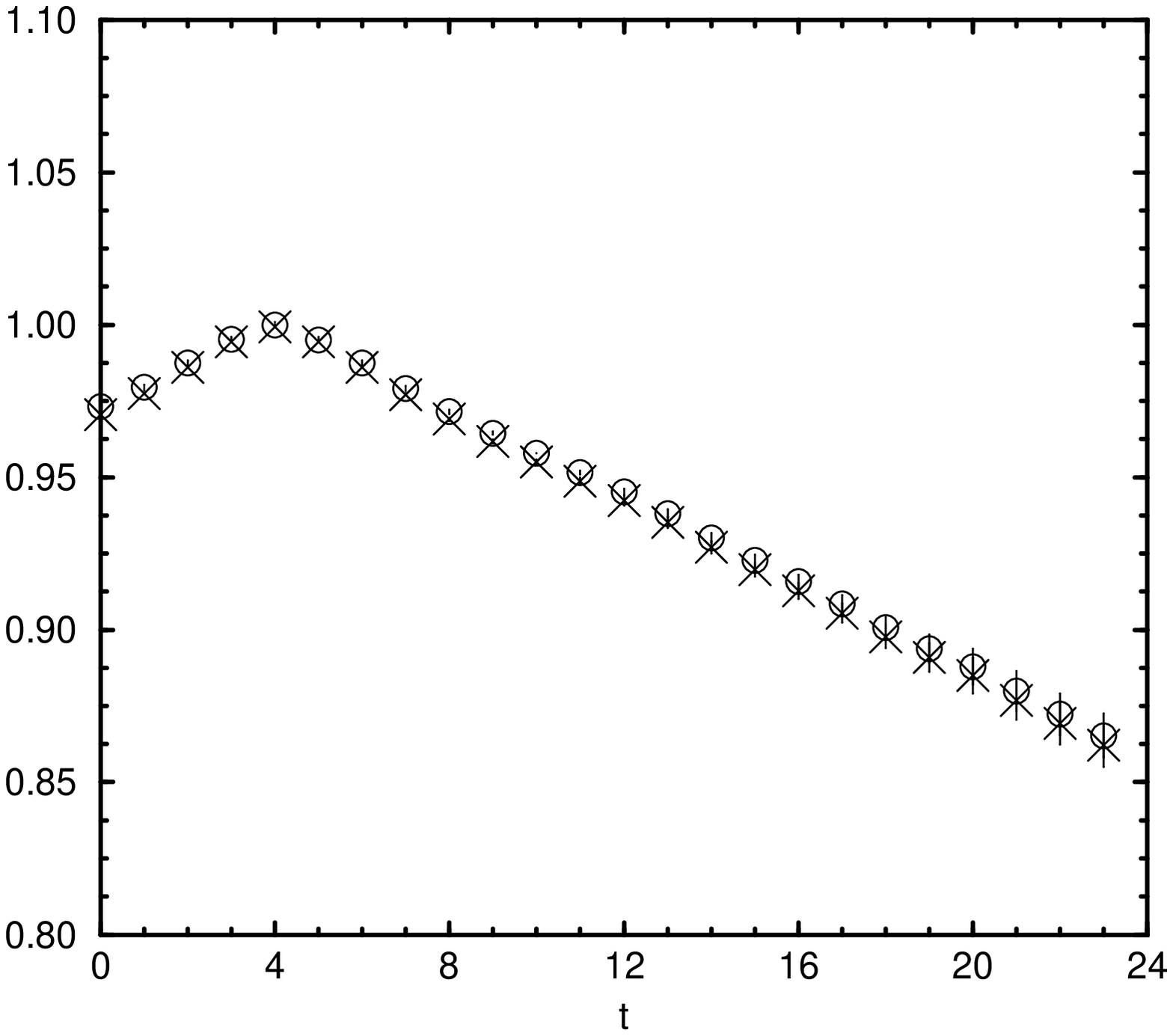}
& \leavevmode \epsfxsize=5cm   \epsfbox{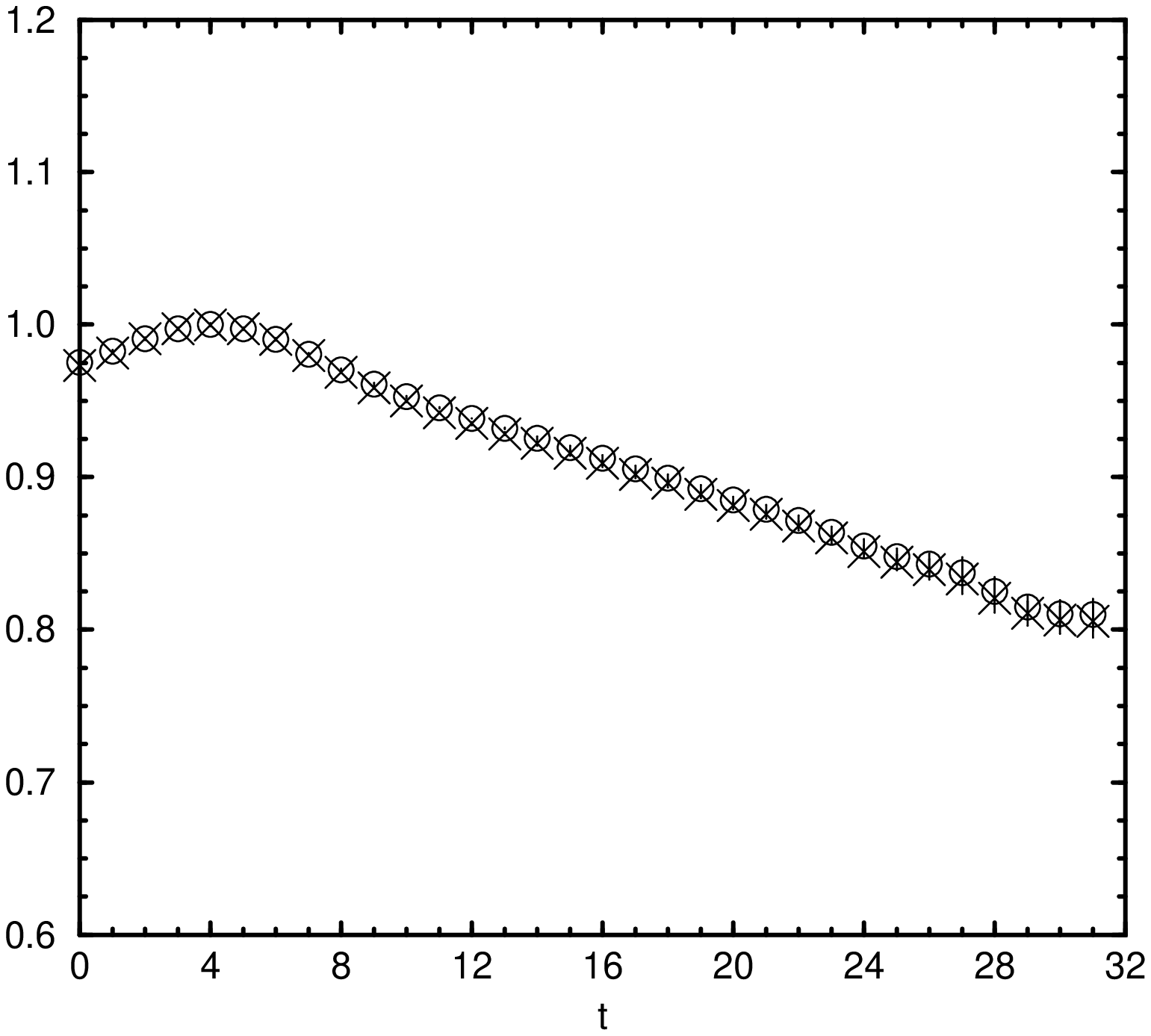}
& \leavevmode \epsfxsize=5cm   \epsfbox{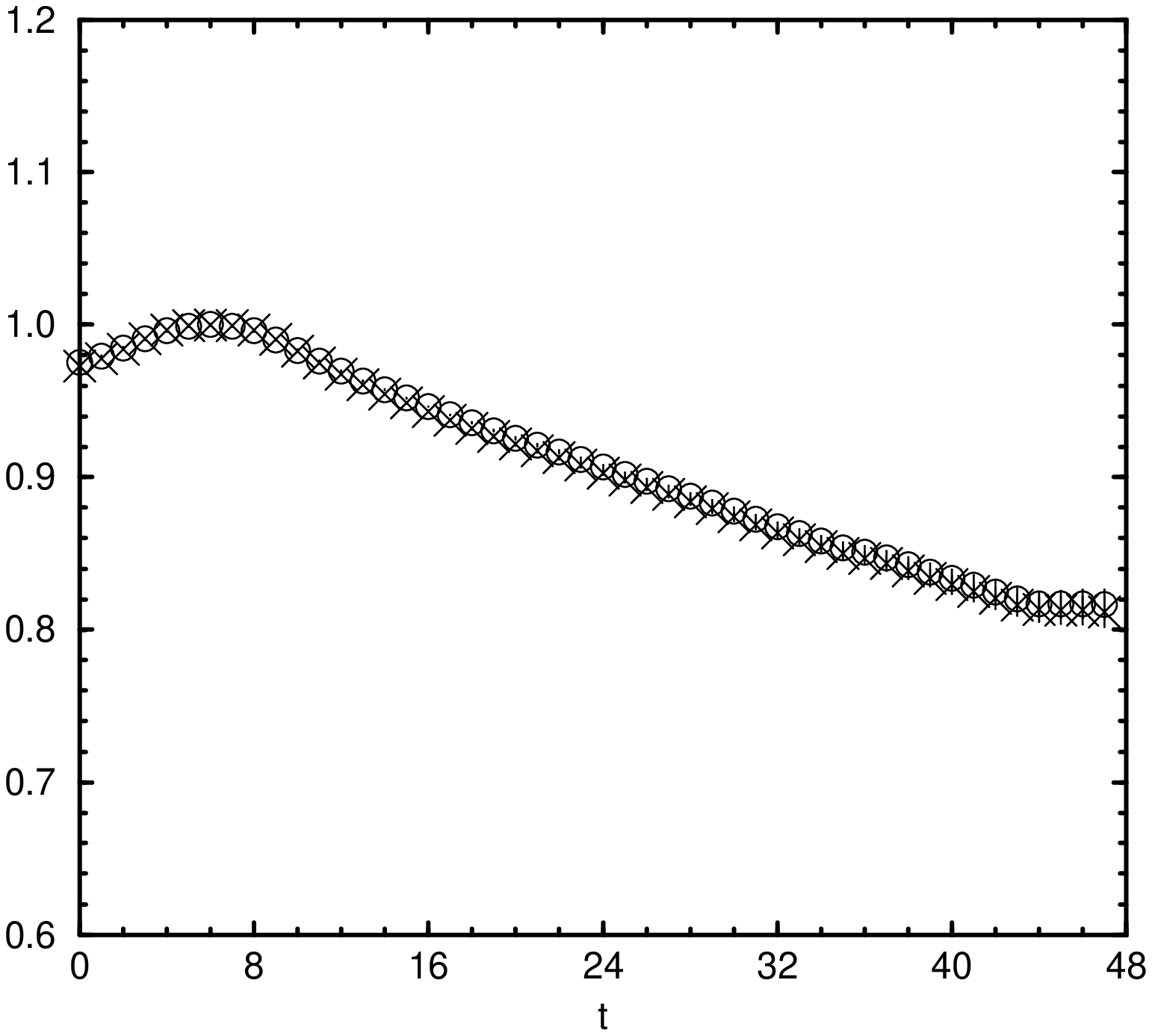}
\\
\raisebox{25mm}{ $\displaystyle{ \frac{m_{\pi}}{m_{\rho}} \approx 0.75 }$ }
& \leavevmode \epsfxsize=5.1cm \epsfbox{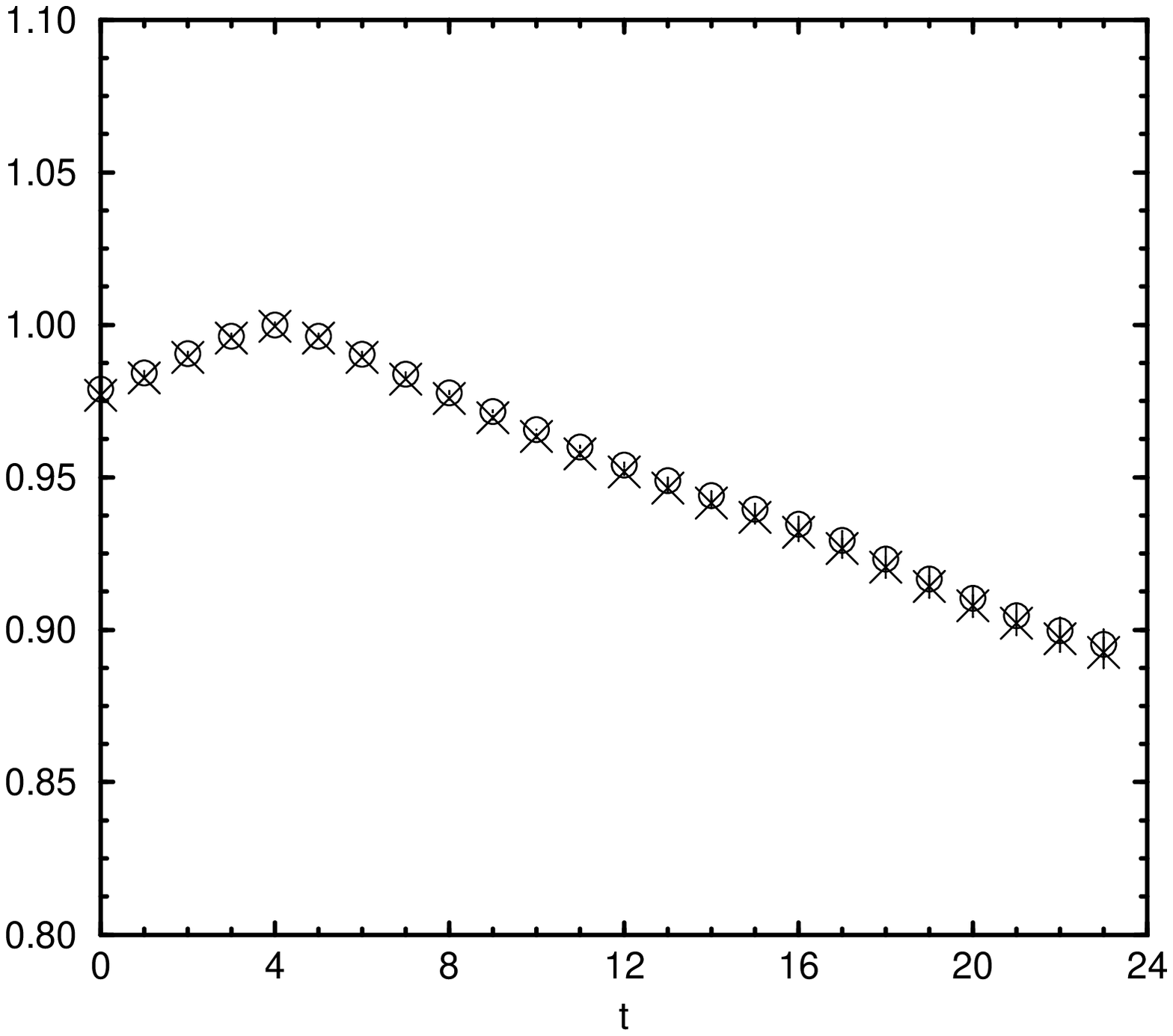}
& \leavevmode \epsfxsize=5cm   \epsfbox{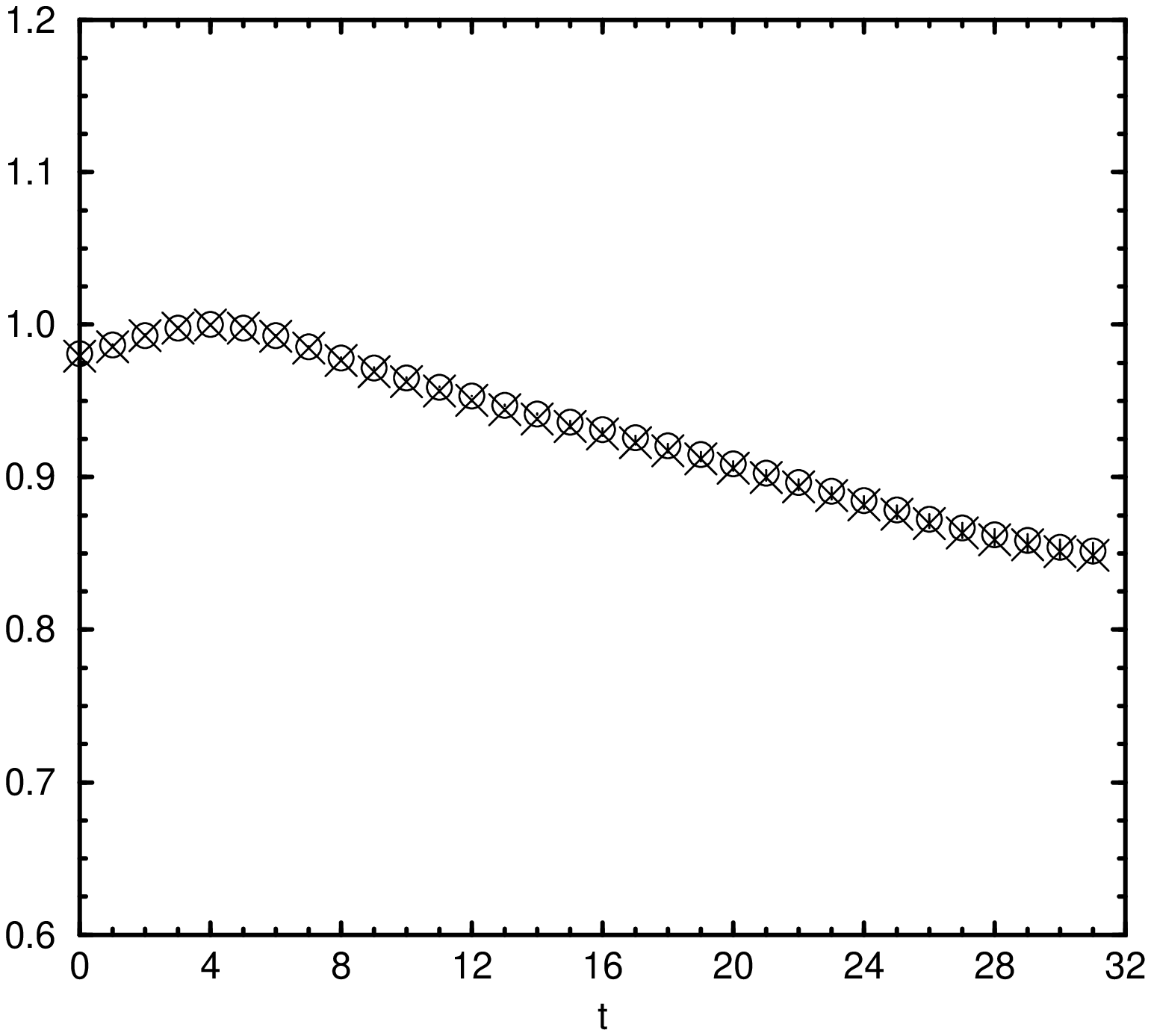}
& \leavevmode \epsfxsize=5cm   \epsfbox{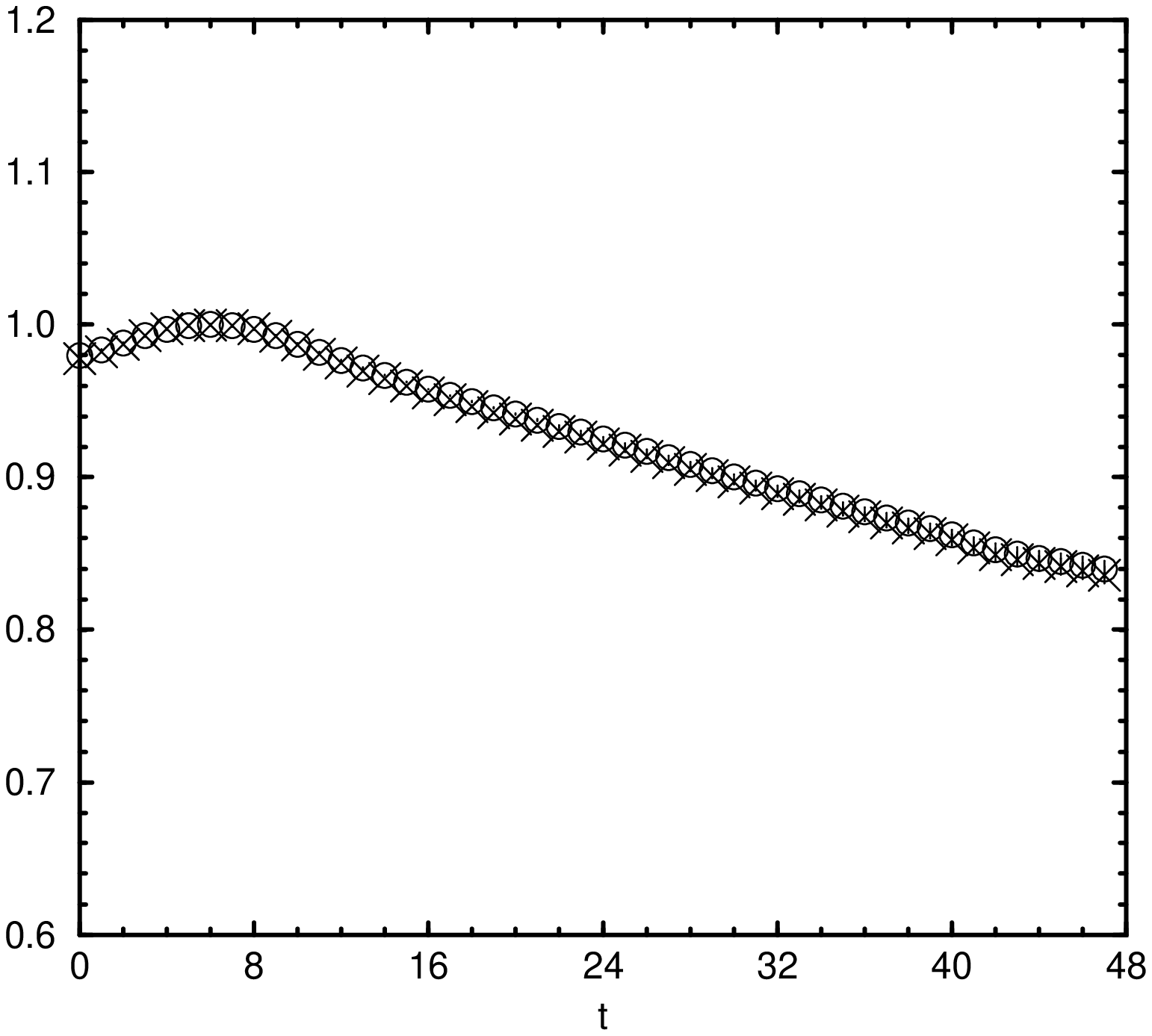}
\\
\raisebox{25mm}{ $\displaystyle{ \frac{m_{\pi}}{m_{\rho}} \approx 0.8 }$ }
& \leavevmode \epsfxsize=5.1cm \epsfbox{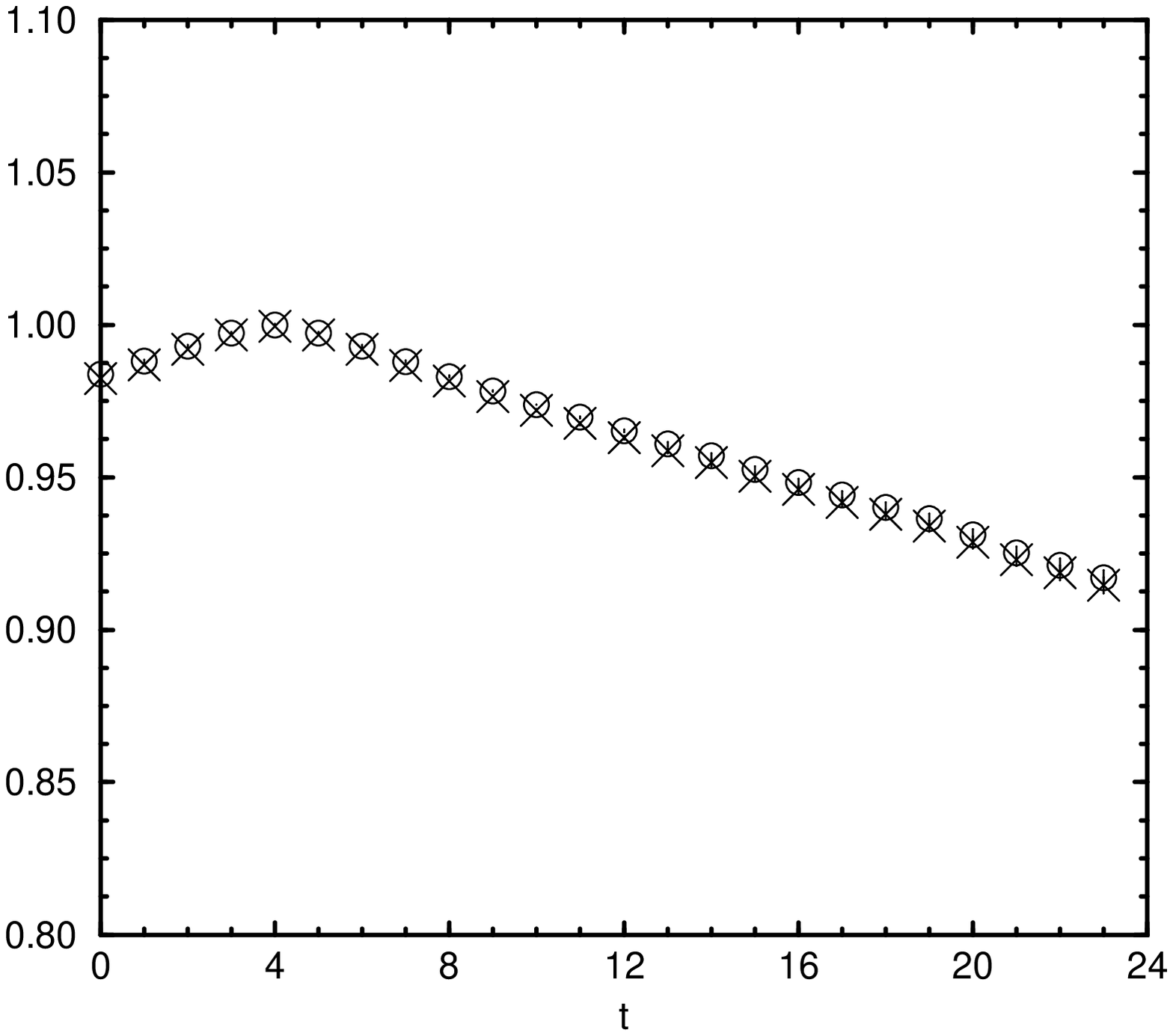}
& \leavevmode \epsfxsize=5cm   \epsfbox{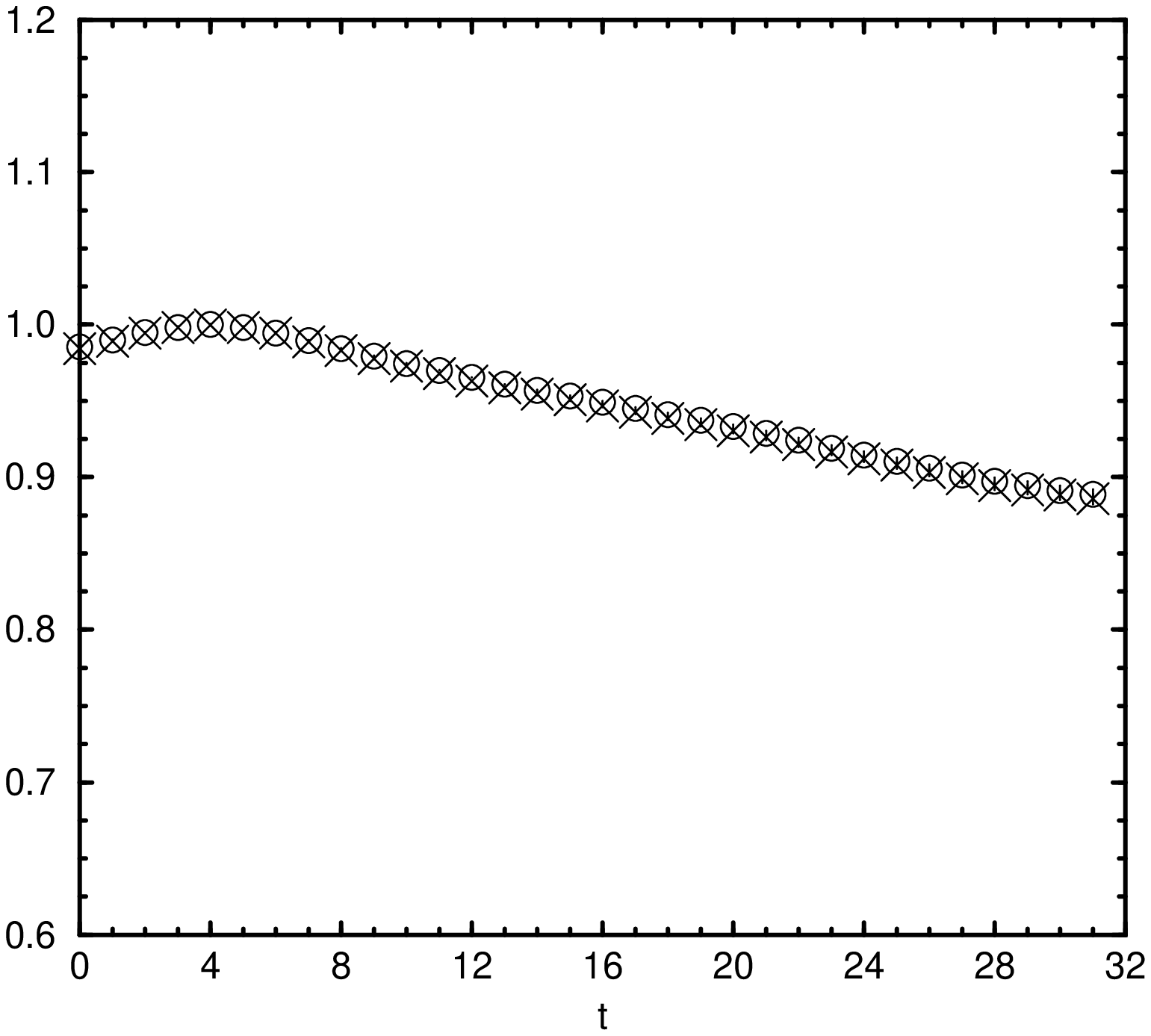}
& \leavevmode \epsfxsize=5cm   \epsfbox{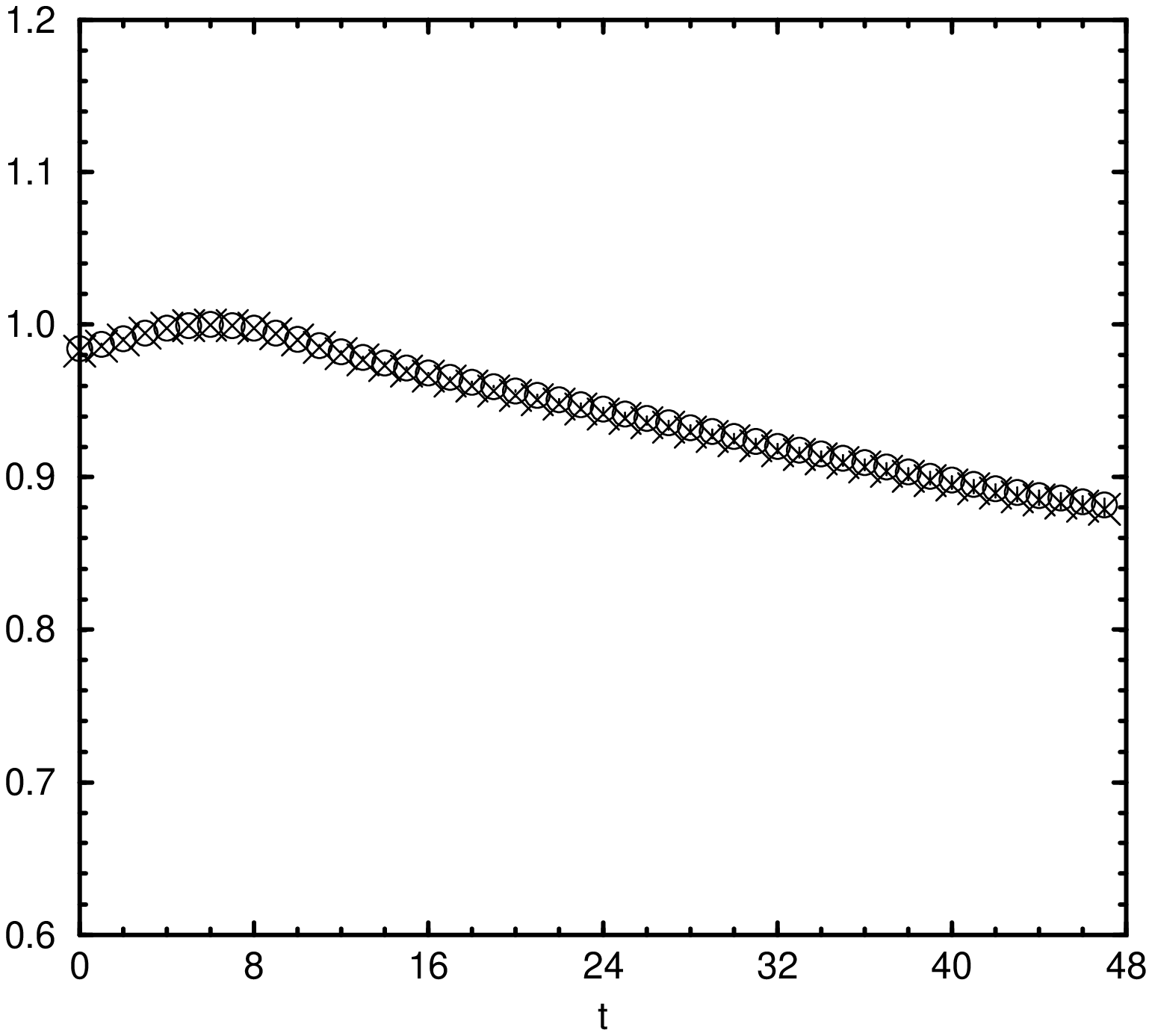}
\end{tabular}
\end{center}
\caption{
Ratio $R_n ( t )$ and $D_n ( t )$ for $n = 0$ state in the laboratory system L1
with energy state cut-off $N = 3$.
$m_\pi / m_\rho$ increases from top to bottom, 
while $\beta$ increases from left to right.
\label{fig:diag:L1_0}
}
\end{figure}

\begin{figure}
\begin{center}
\hspace{-26mm}
\begin{tabular}{rccc}
& $\beta = 1.80 $ & $\beta = 1.95$ & $\beta = 2.10$ 
\\
\raisebox{25mm}{ $\displaystyle{ \frac{m_{\pi}}{m_{\rho}} \approx 0.6 }$ }
& \leavevmode \epsfxsize=5.1cm \epsfbox{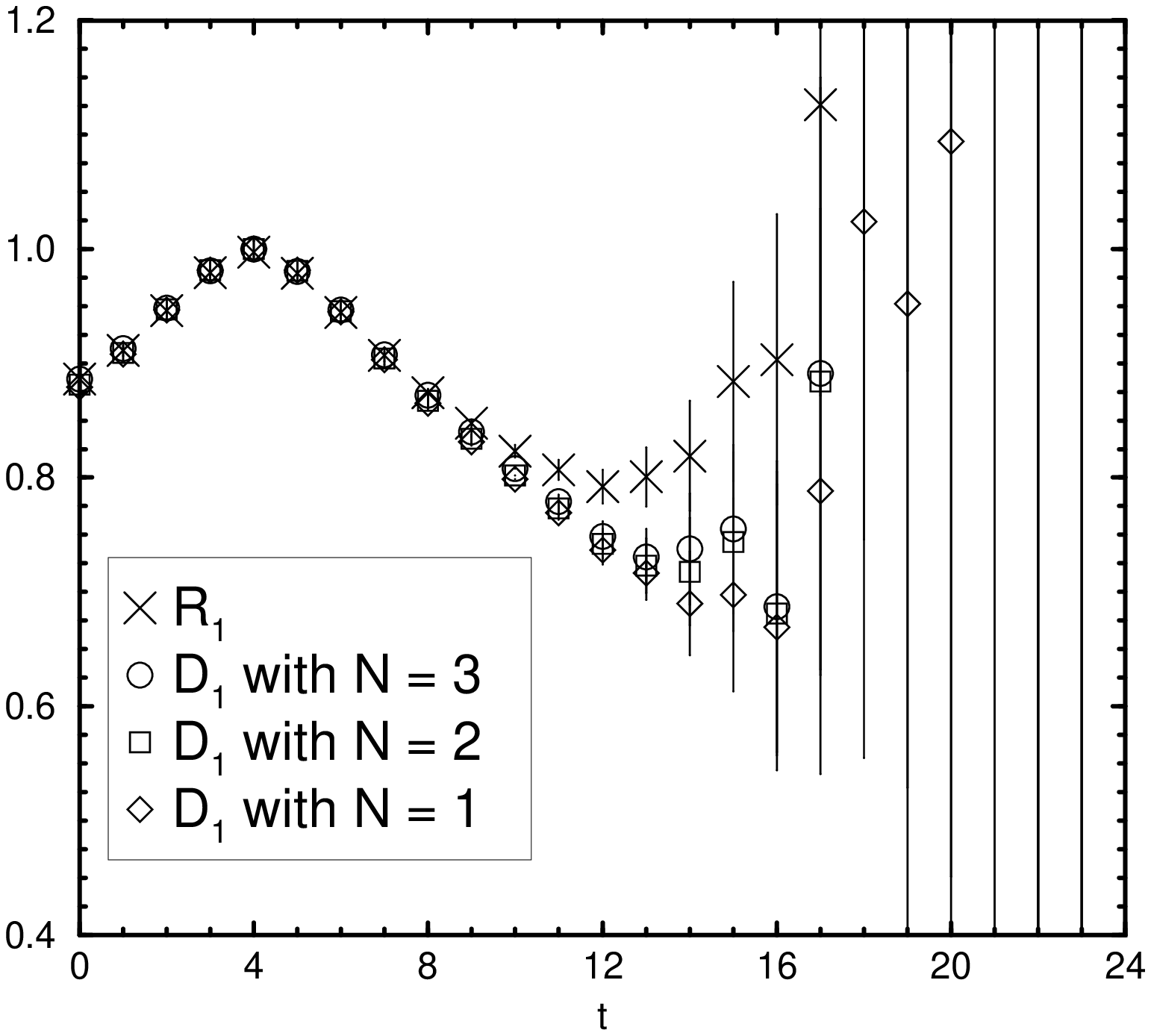}
& \leavevmode \epsfxsize=5cm   \epsfbox{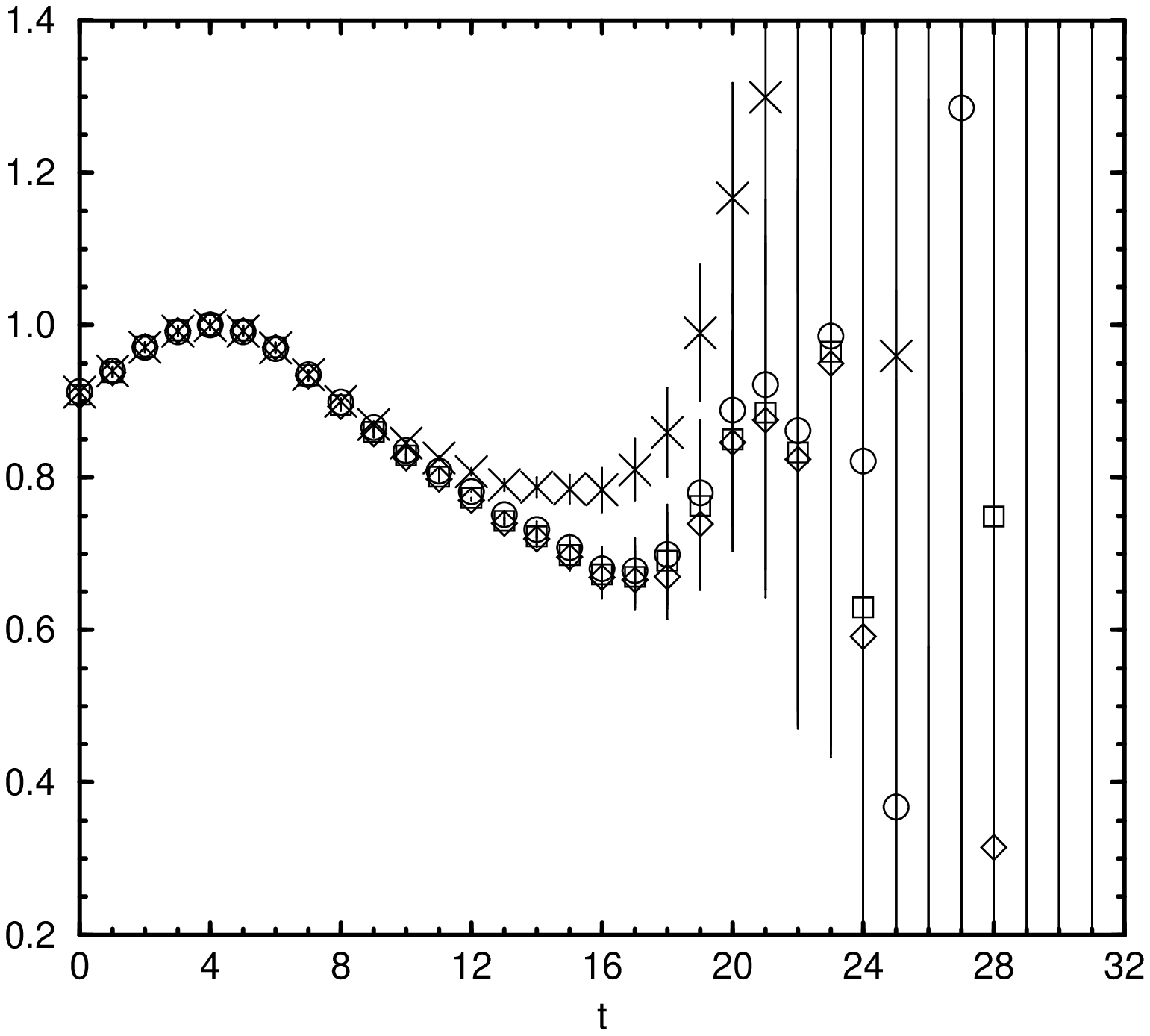}
& \leavevmode \epsfxsize=5cm   \epsfbox{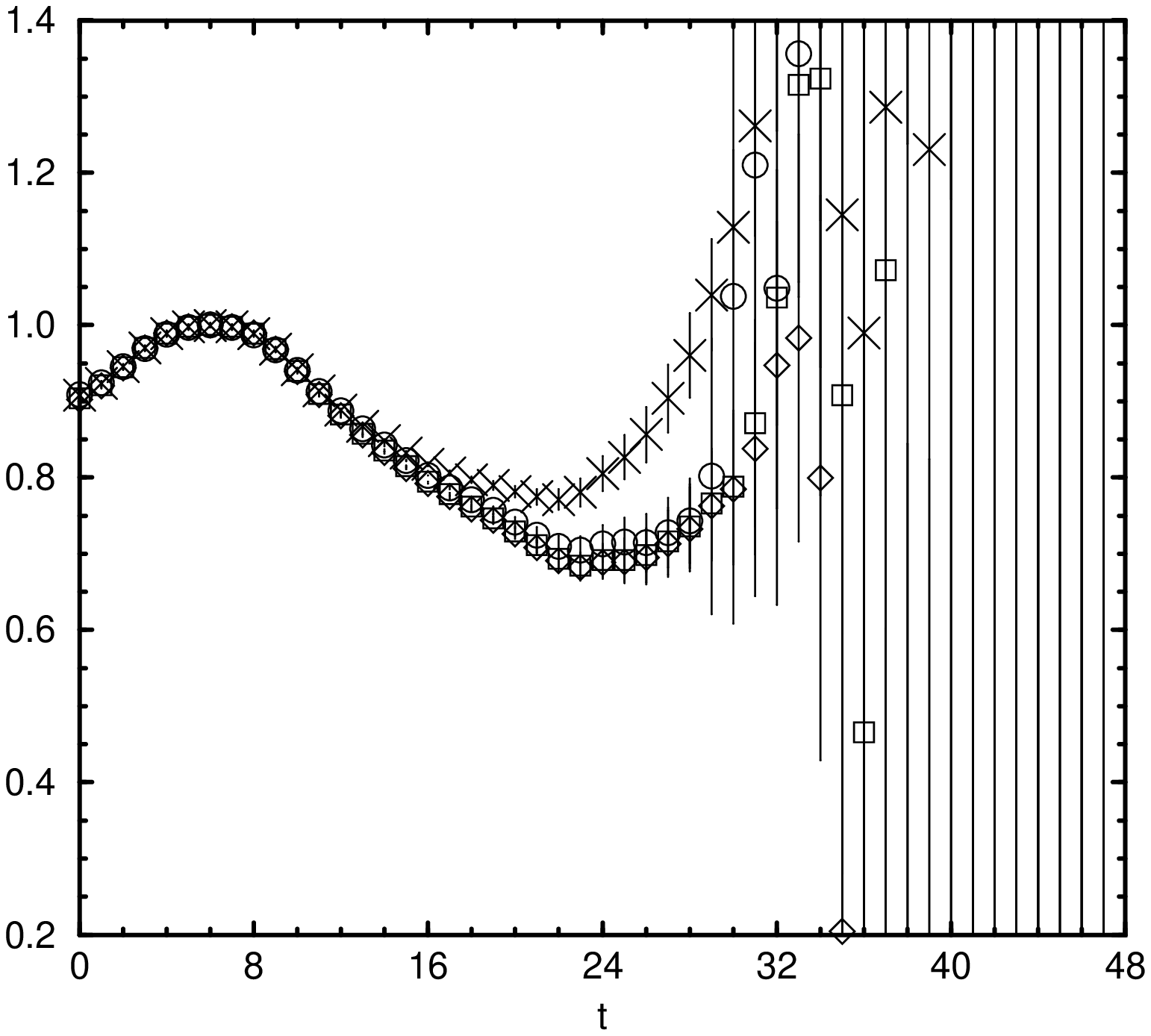}
\\
\raisebox{25mm}{ $\displaystyle{ \frac{m_{\pi}}{m_{\rho}} \approx 0.7 }$ }
& \leavevmode \epsfxsize=5.1cm \epsfbox{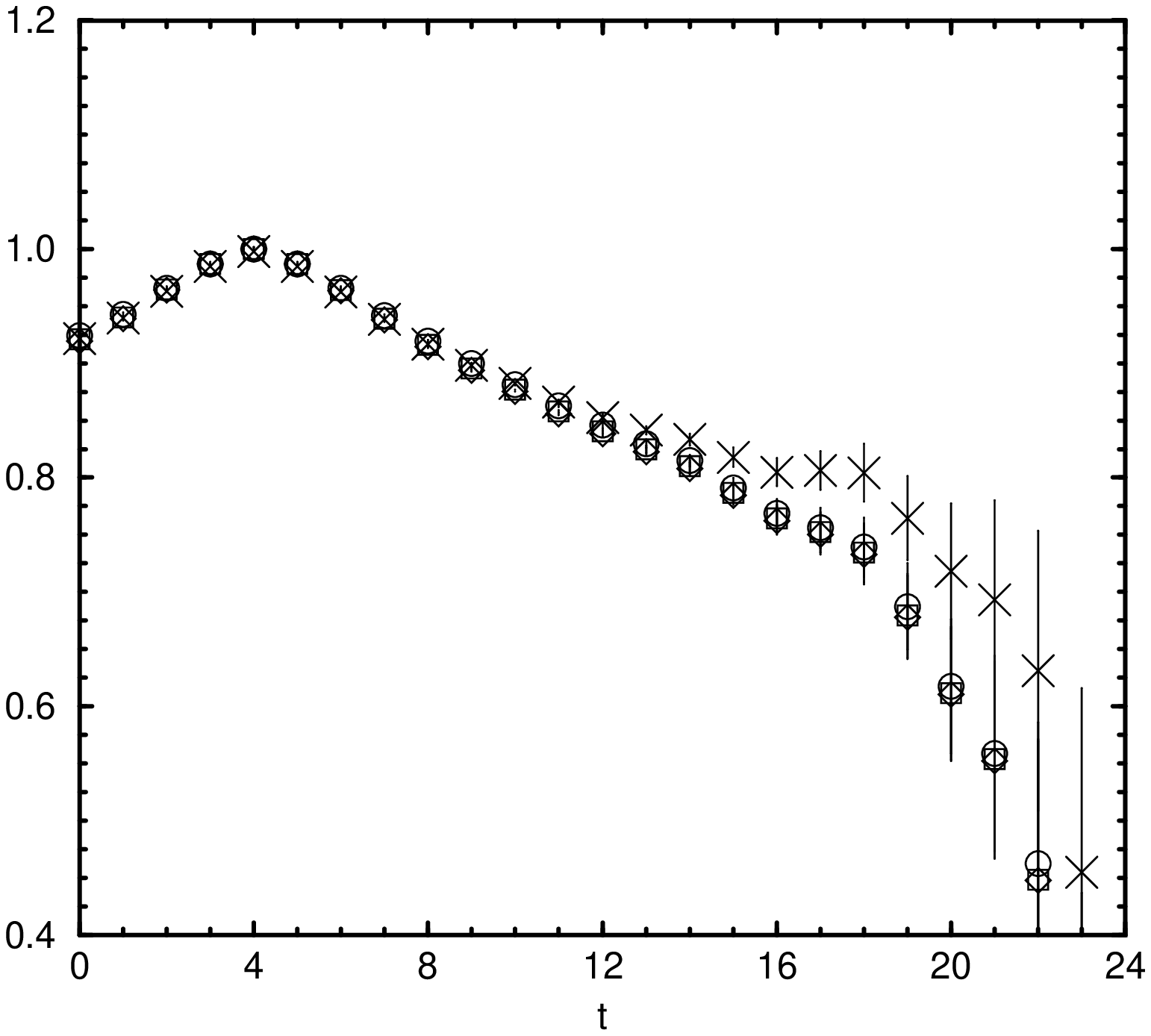}
& \leavevmode \epsfxsize=5cm   \epsfbox{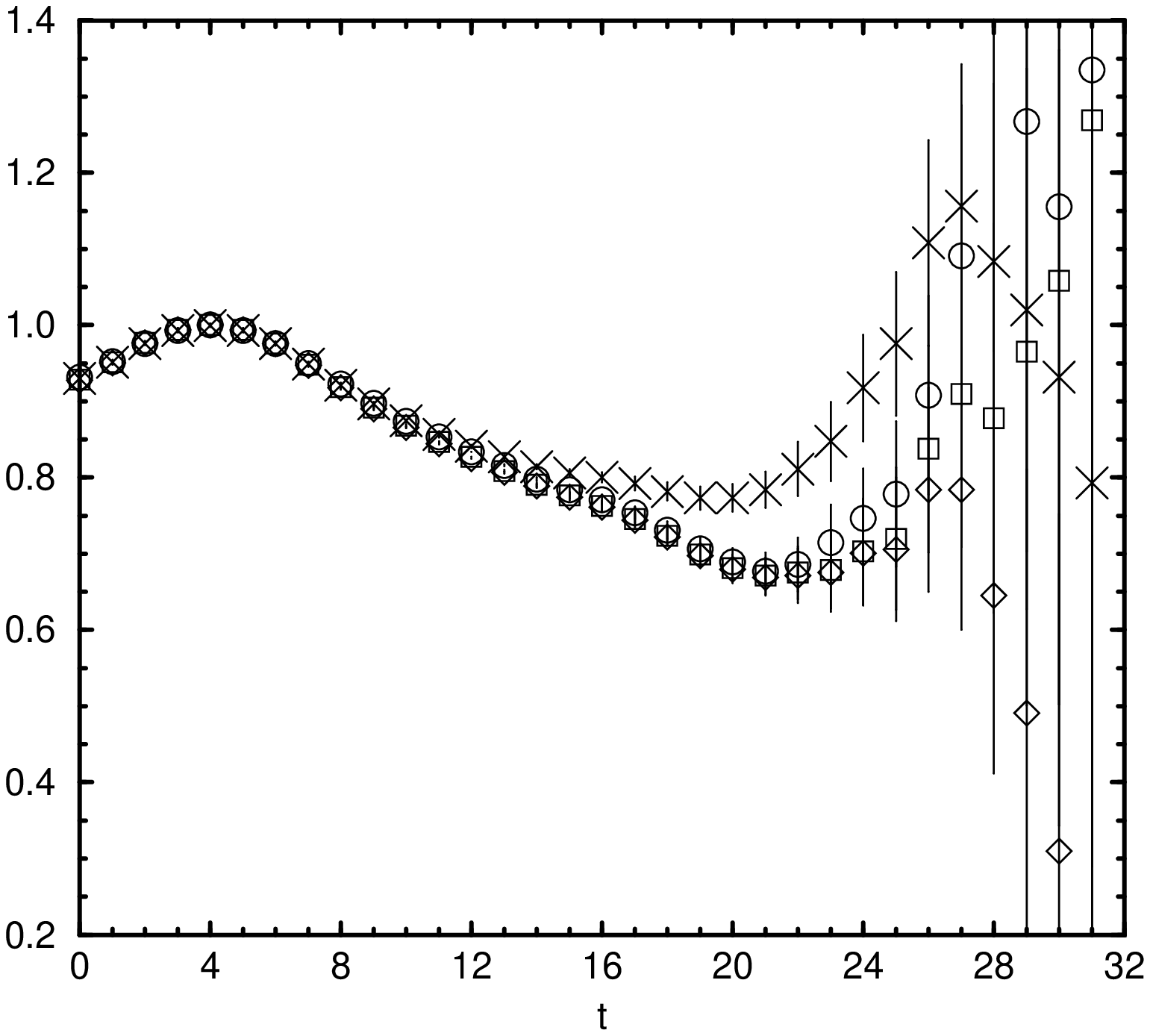}
& \leavevmode \epsfxsize=5cm   \epsfbox{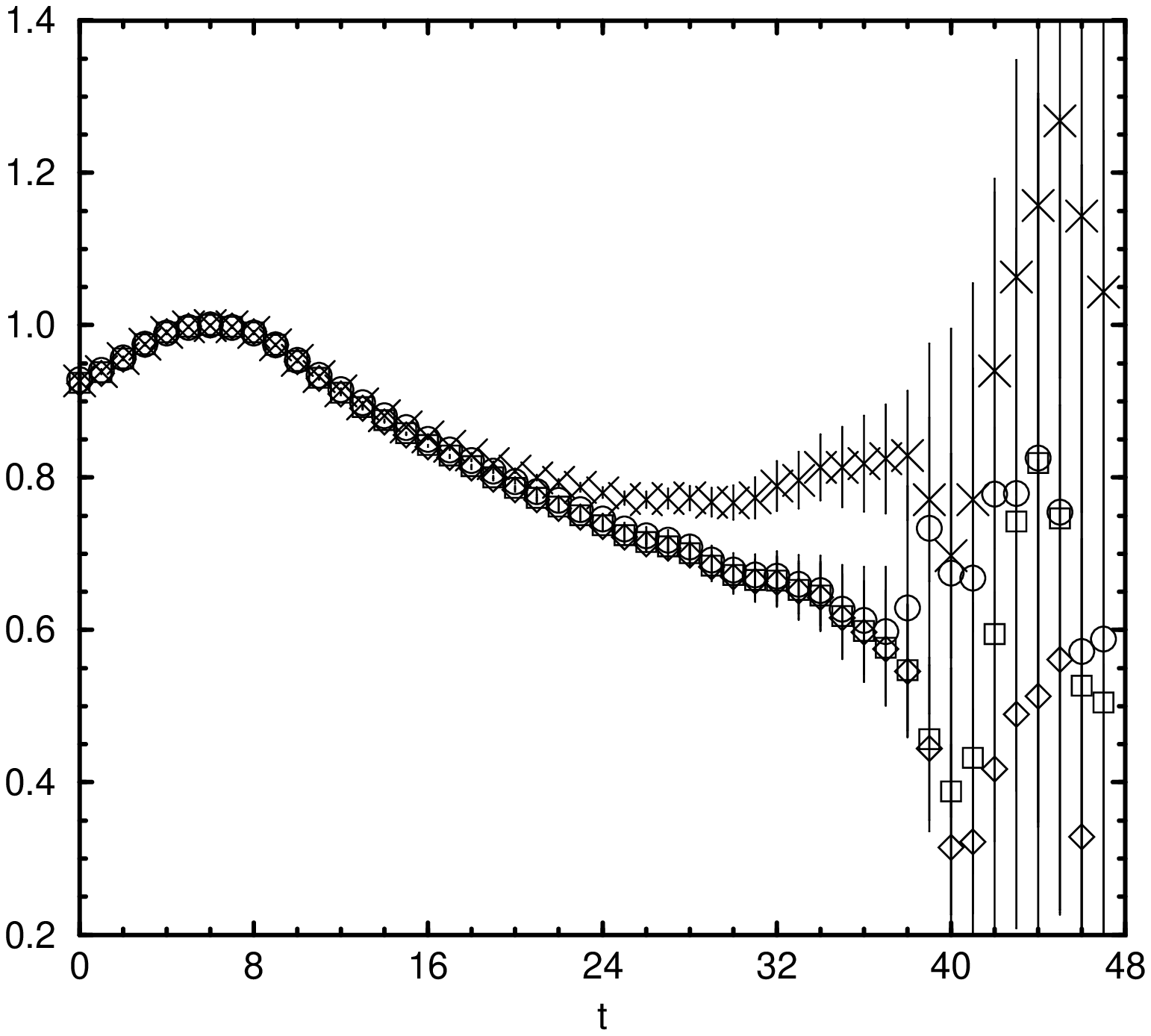}
\\
\raisebox{25mm}{ $\displaystyle{ \frac{m_{\pi}}{m_{\rho}} \approx 0.75}$ }
& \leavevmode \epsfxsize=5.1cm \epsfbox{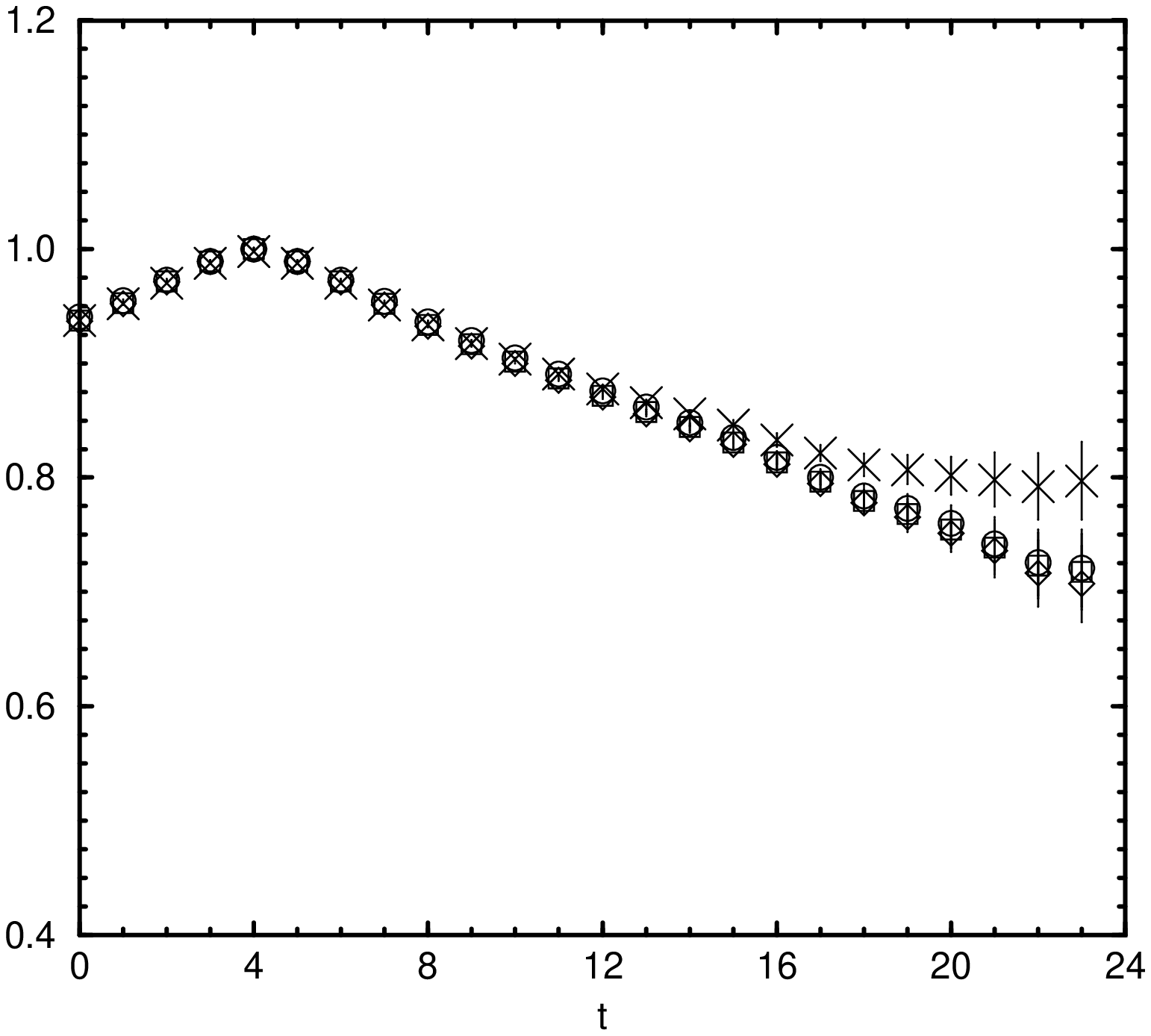}
& \leavevmode \epsfxsize=5cm   \epsfbox{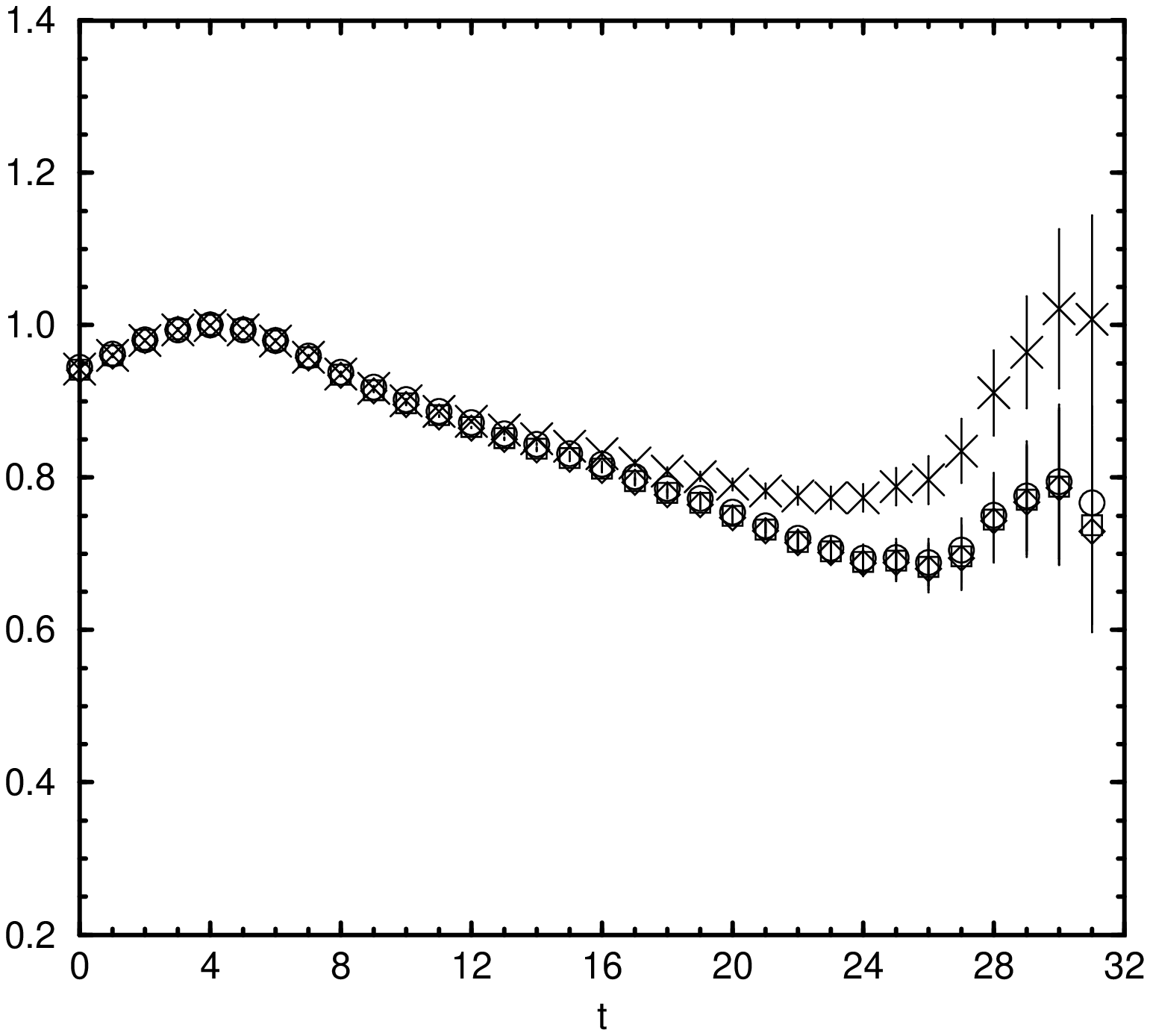}
& \leavevmode \epsfxsize=5cm   \epsfbox{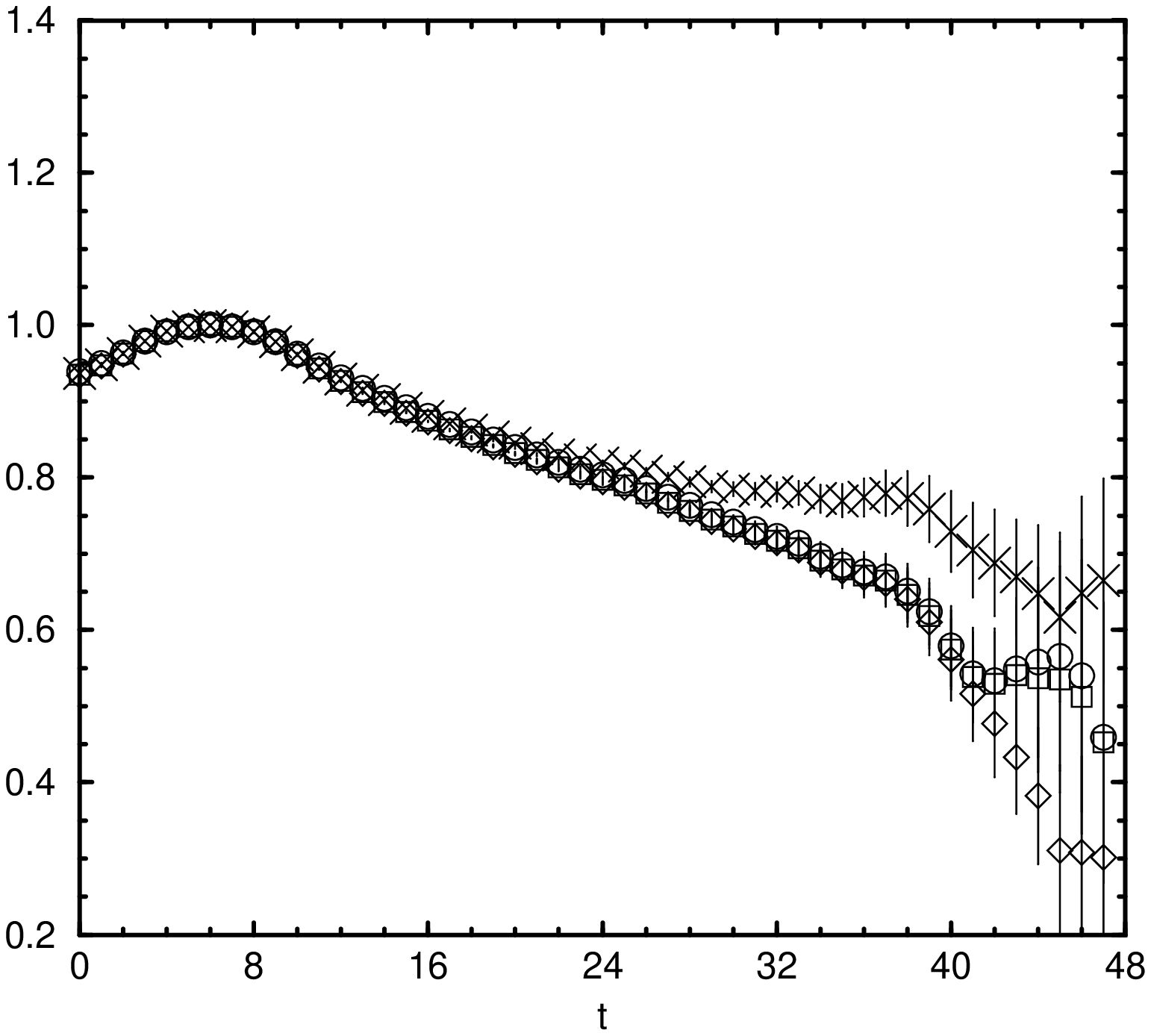}
\\
\raisebox{25mm}{ $\displaystyle{ \frac{m_{\pi}}{m_{\rho}} \approx 0.8 }$ }
& \leavevmode \epsfxsize=5.1cm \epsfbox{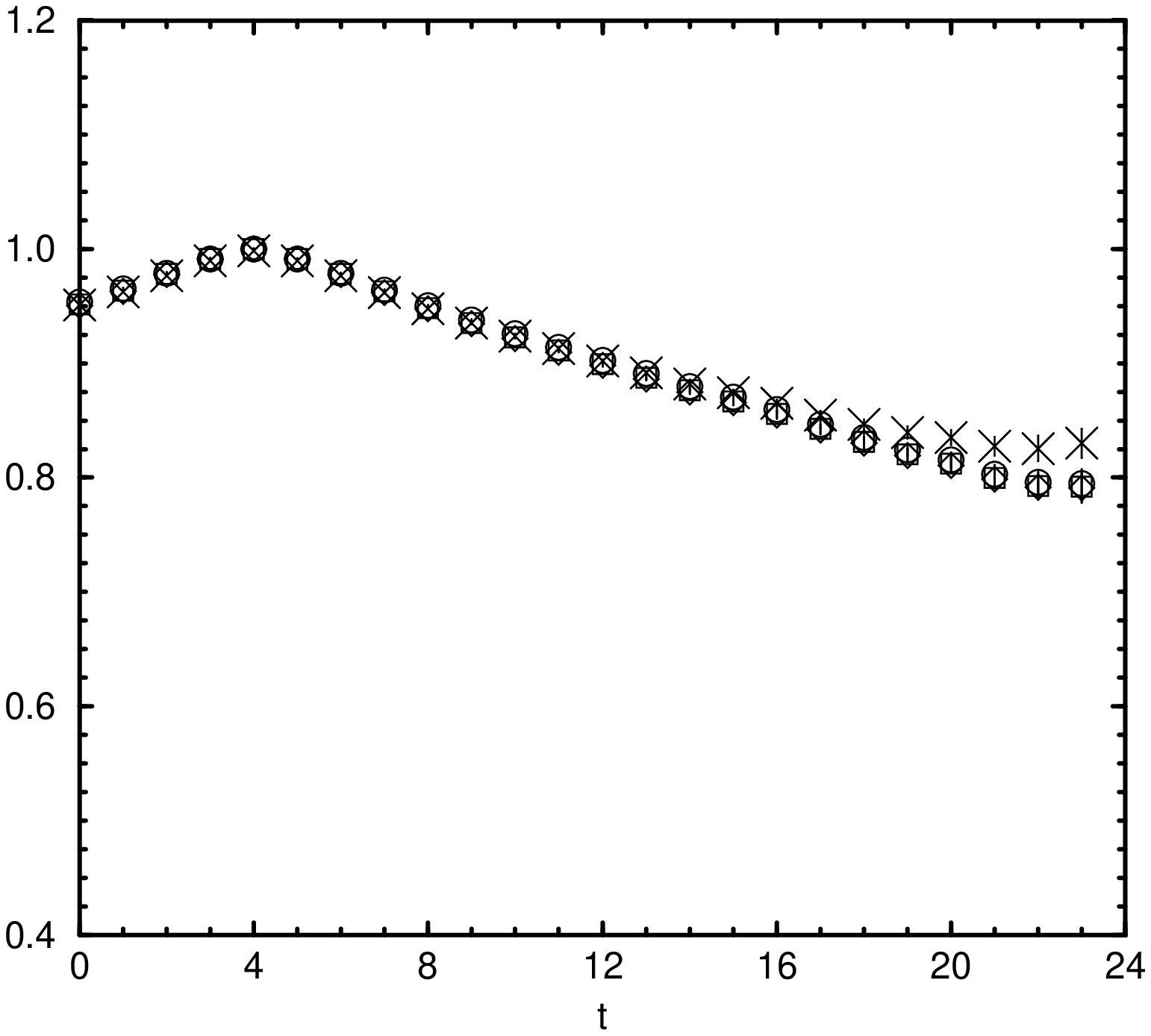}
& \leavevmode \epsfxsize=5cm   \epsfbox{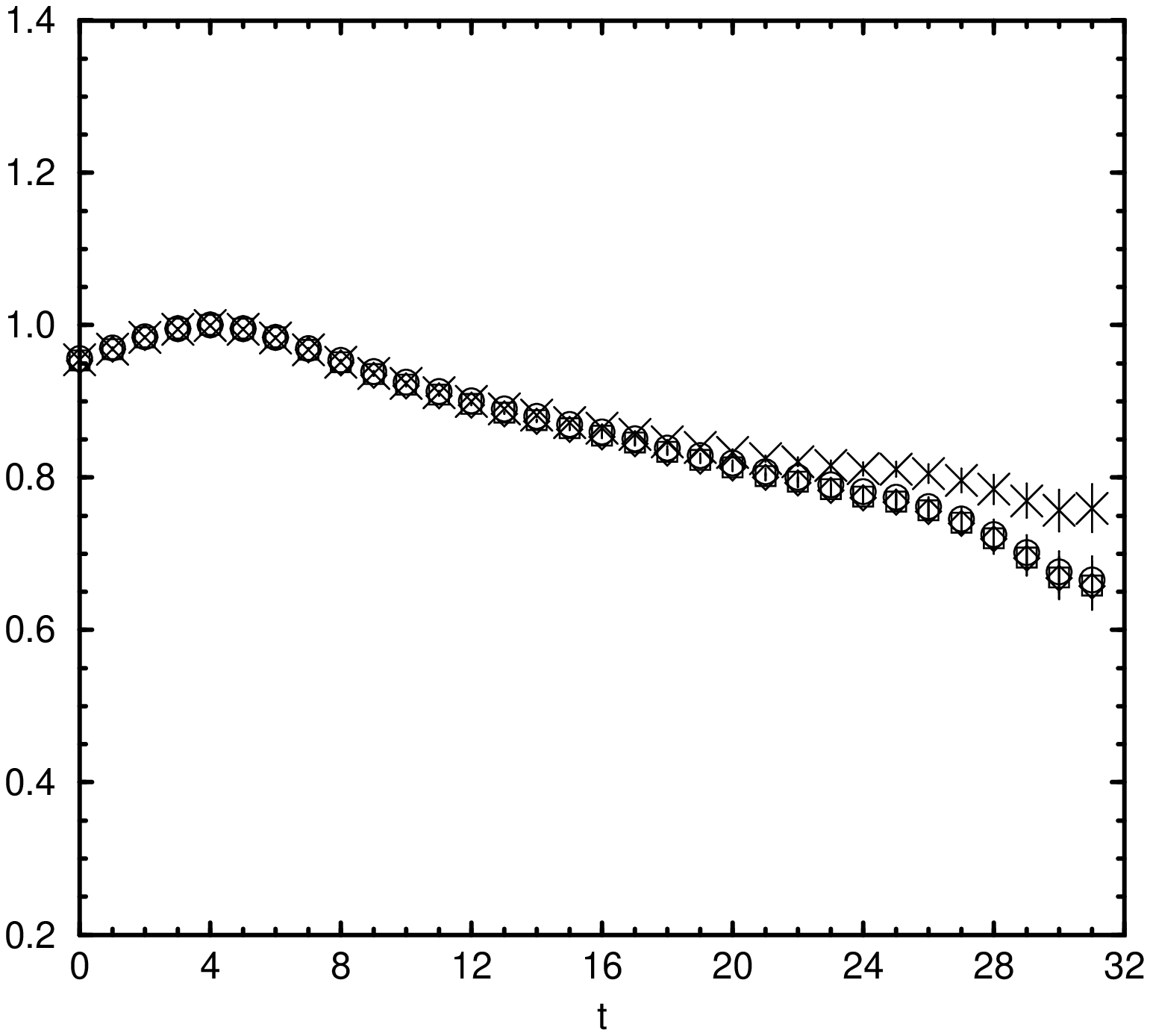}
& \leavevmode \epsfxsize=5cm   \epsfbox{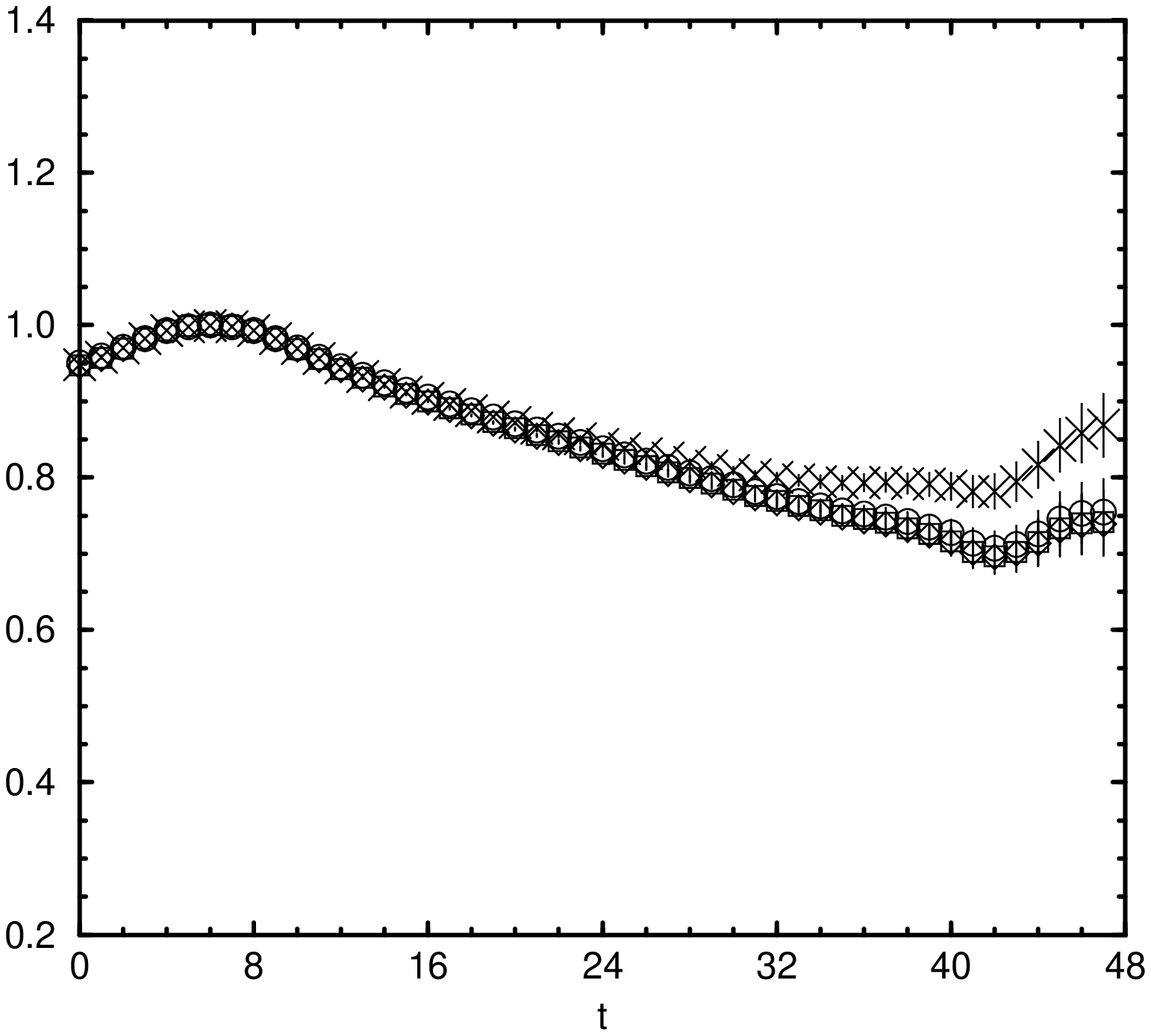}
\end{tabular}
\end{center}
\caption{
Ratio $R_n ( t )$ and $D_n ( t )$ for $n=1$ state in the laboratory system L1
with energy state cut-off $N=1$, $2$, and $3$.
$m_\pi / m_\rho$ increases from top to bottom, 
while $\beta$ increases from left to right.
\label{fig:diag:L1_1}
}
\end{figure}

The ratios in the L1 system are plotted 
in Fig.~\ref{fig:diag:L1_0} for the ground state ($n=0$);
we also divide $D_n(t)$ by $D_n(t_S)$ as for the CM system. 
We find that $R_0(t)$ agrees with $D_0(t)$ with $N=3$.
We also check the cut-off dependence by taking $N=1,2$ 
and confirm that it is negligible.
This indicates that the energy shift 
can be extracted without the diagonalization 
as for the ground state in the CM system.
For the first excited state ($n=1$), however, the diagonalization is effective 
as shown in Fig.~\ref{fig:diag:L1_1}.
We also find that the effect of the cut-off for $D_1(t)$ is negligible
by comparing the results with $N=1$, $2$ and $3$.
The ratio $D_1(t)$ at $\beta = 1.95$ and $2.10$ 
for $m_\pi / m_\rho \approx 0.6$
do not show good exponential behaviors.
We consider that this is due to insufficient statistics 
and do not include these data in the following analysis.

\begin{figure}
\begin{center}
\hspace{-26mm}
\begin{tabular}{rccc}
& $\beta = 1.80$ & $\beta = 1.95$ & $\beta = 2.10$ 
\\
\raisebox{25mm}{ $\displaystyle{ \frac{m_{\pi}}{m_{\rho}} \approx 0.6 }$ }
& \leavevmode \epsfxsize=5.1cm \epsfbox{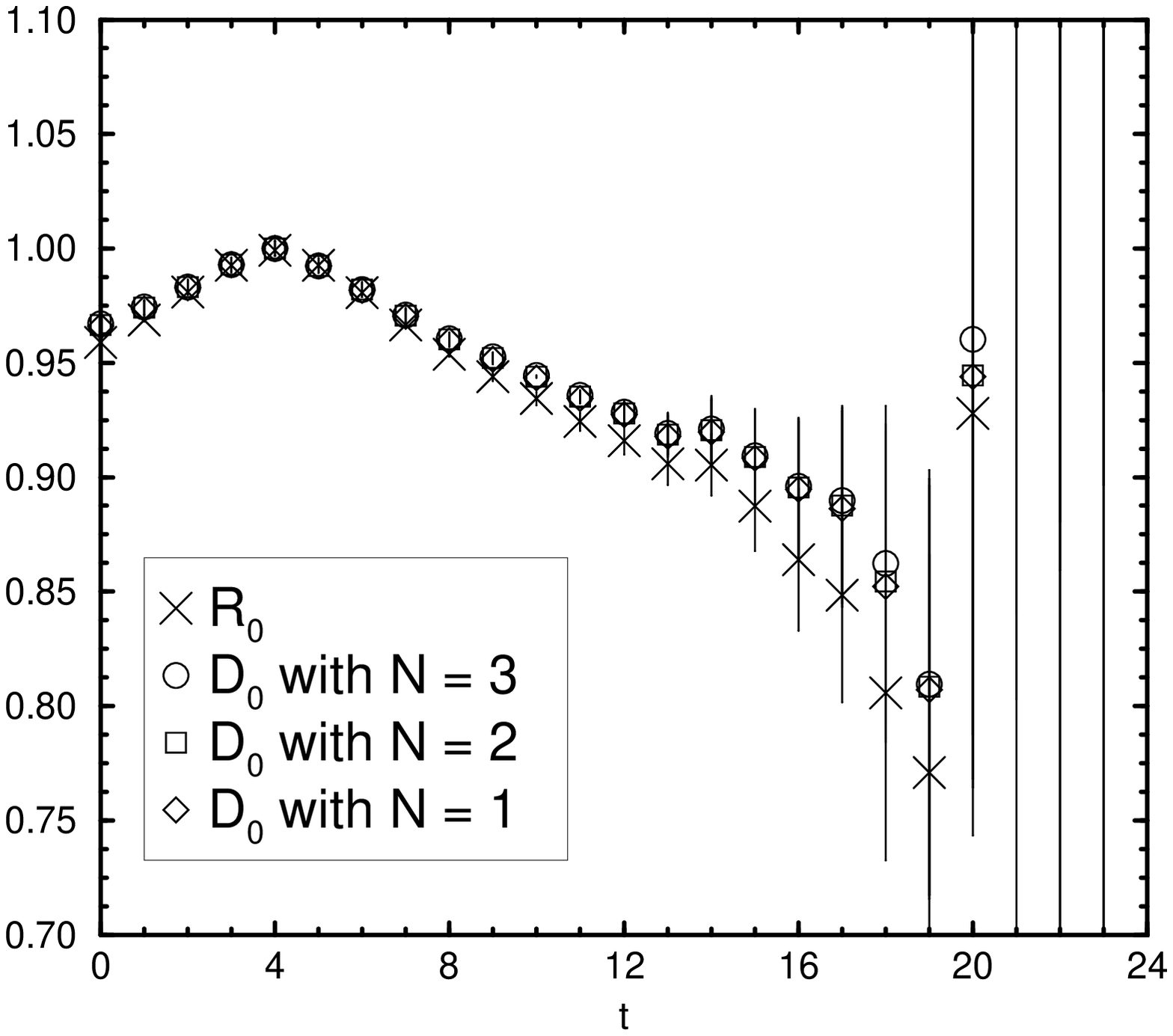}
& \leavevmode \epsfxsize=5cm   \epsfbox{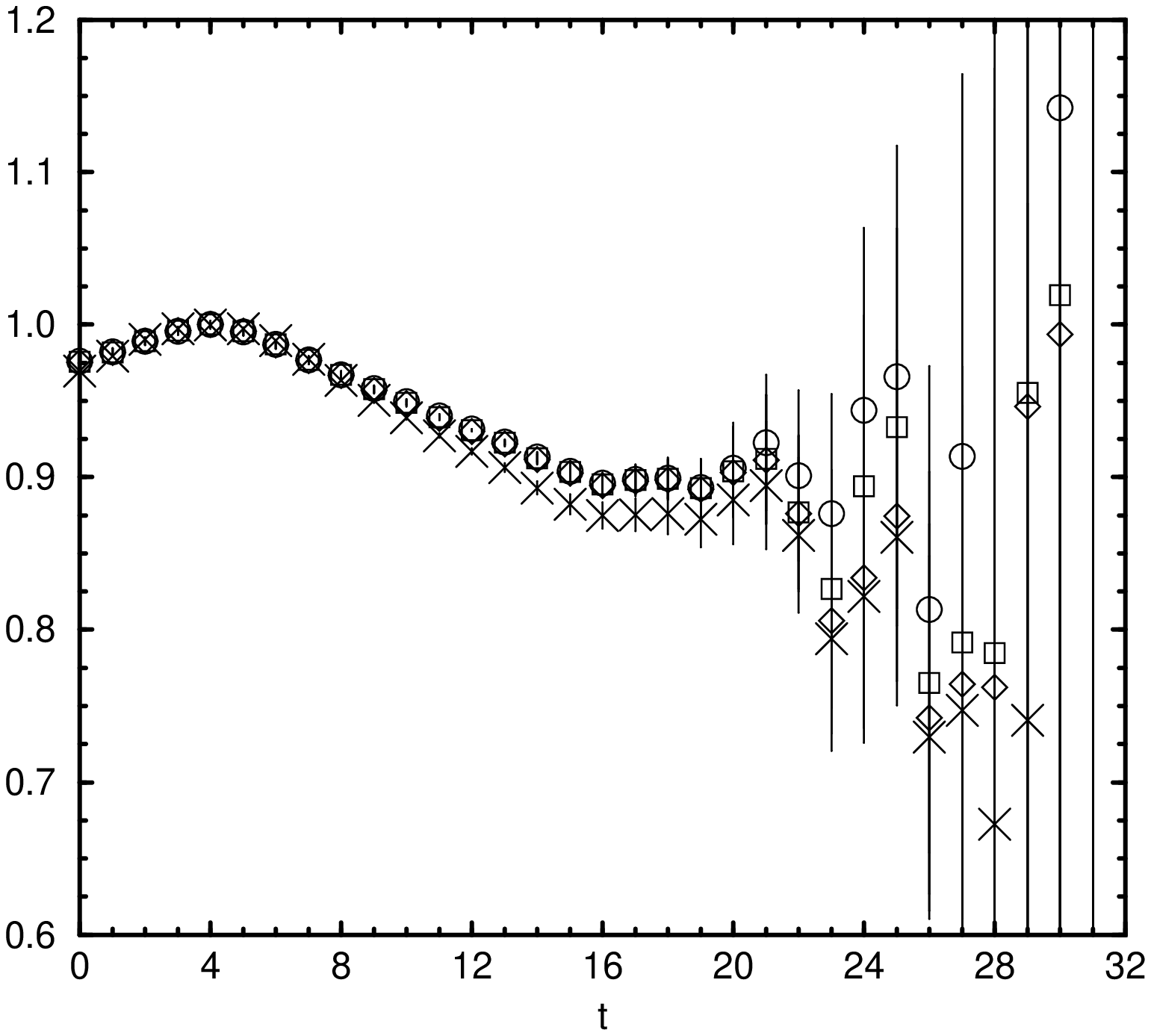}
& \leavevmode \epsfxsize=5cm   \epsfbox{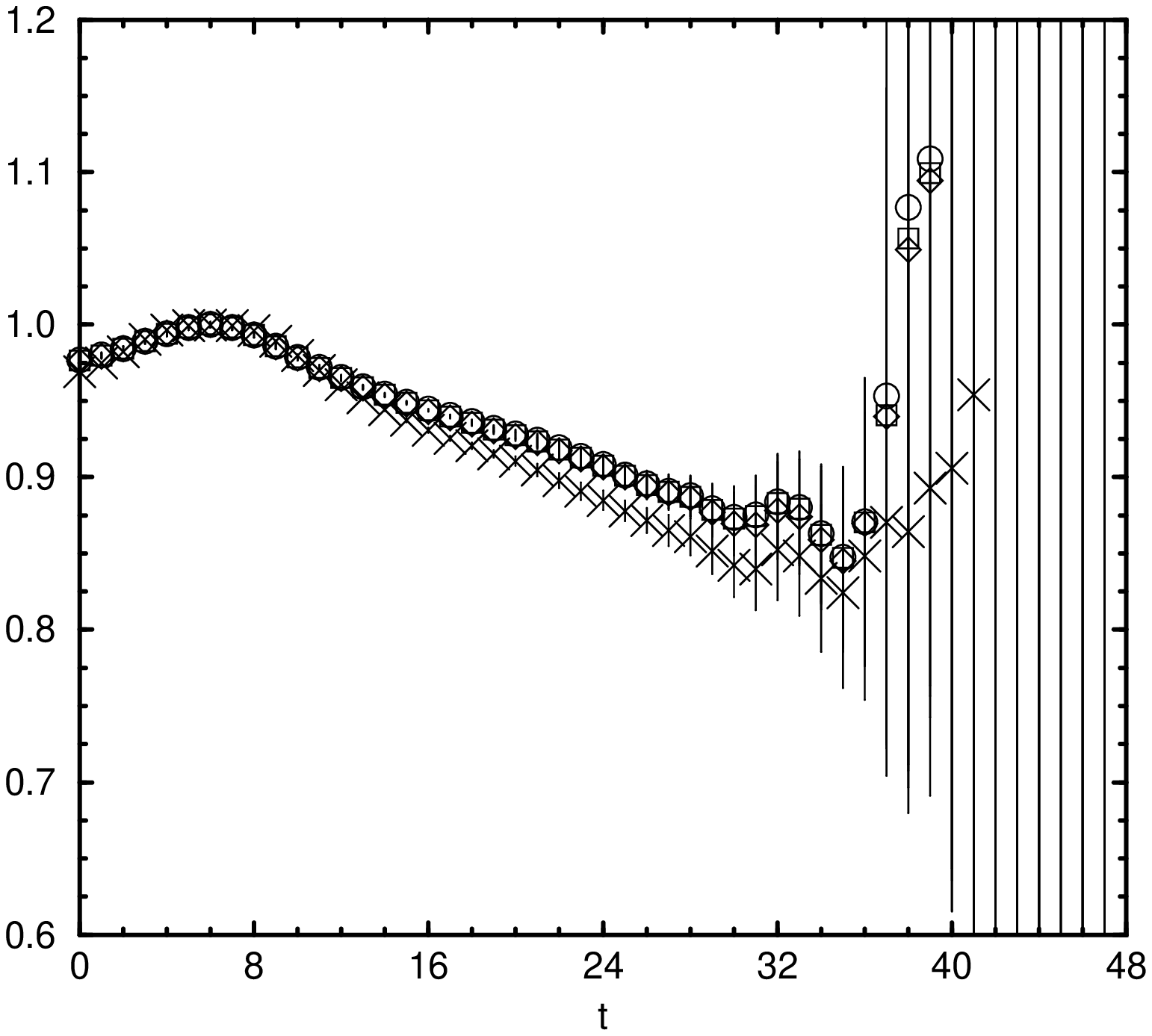}
\\
\raisebox{25mm}{ $\displaystyle{ \frac{m_{\pi}}{m_{\rho}} \approx 0.7 }$ }
& \leavevmode \epsfxsize=5.1cm \epsfbox{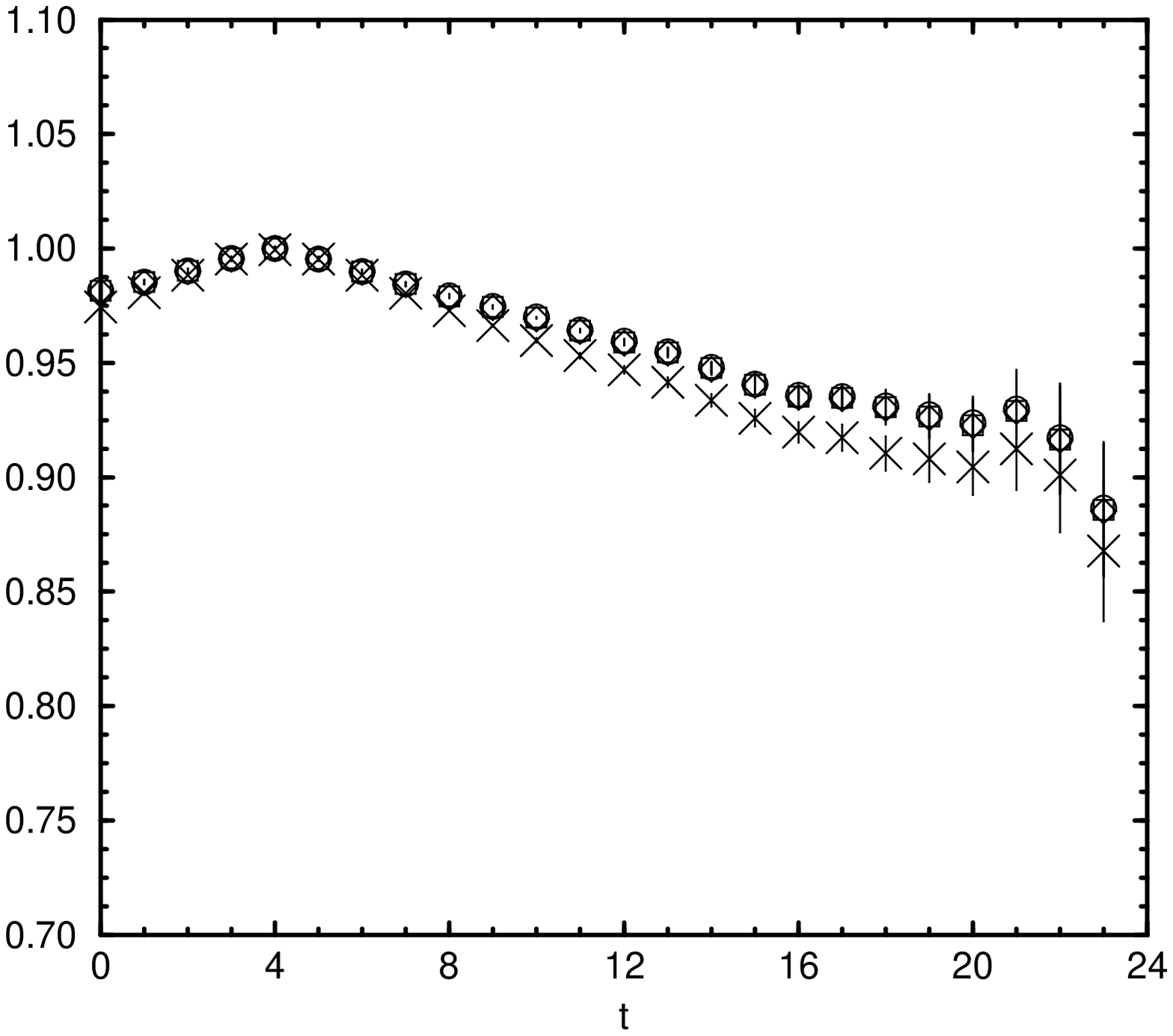}
& \leavevmode \epsfxsize=5cm   \epsfbox{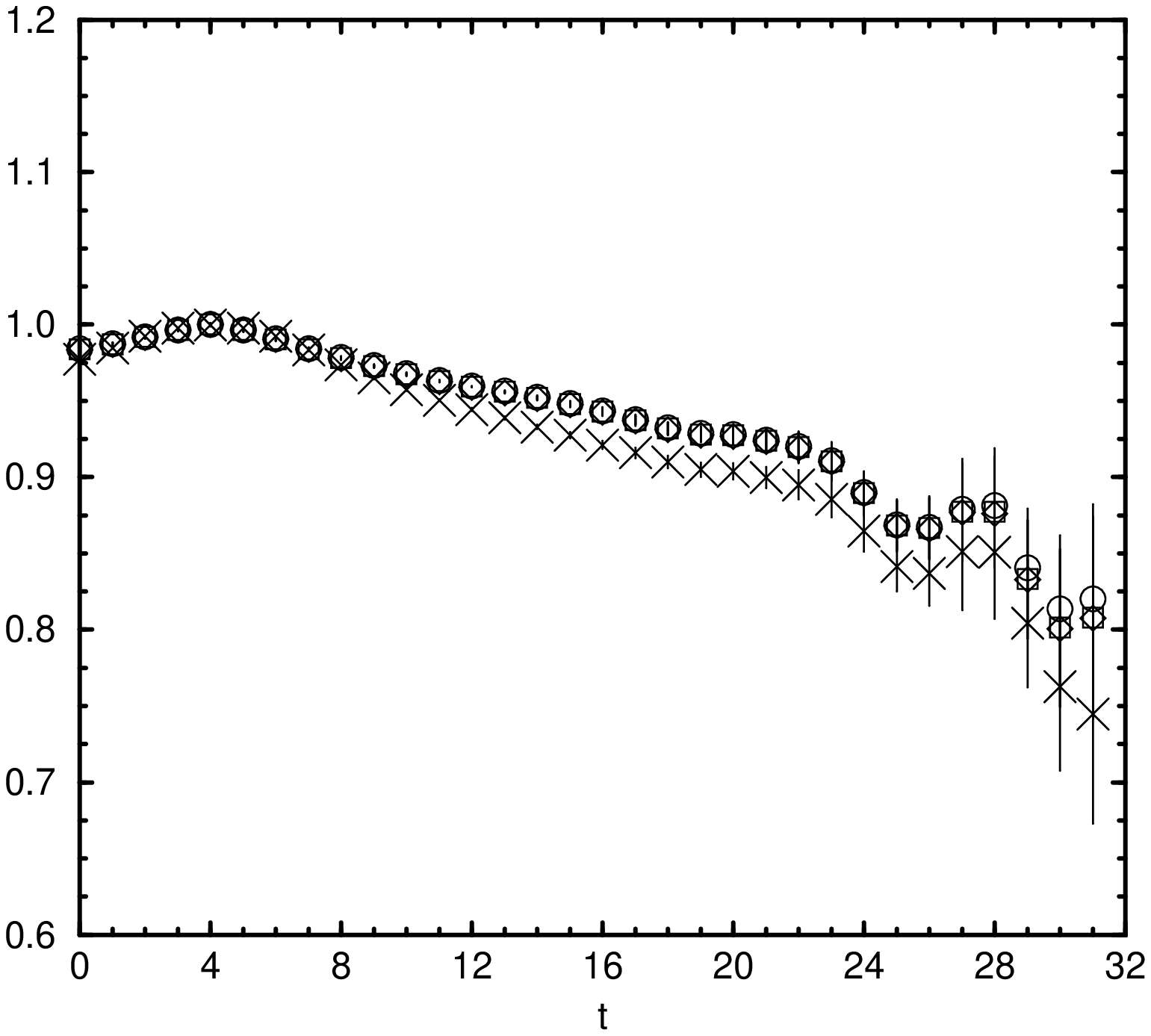}
& \leavevmode \epsfxsize=5cm   \epsfbox{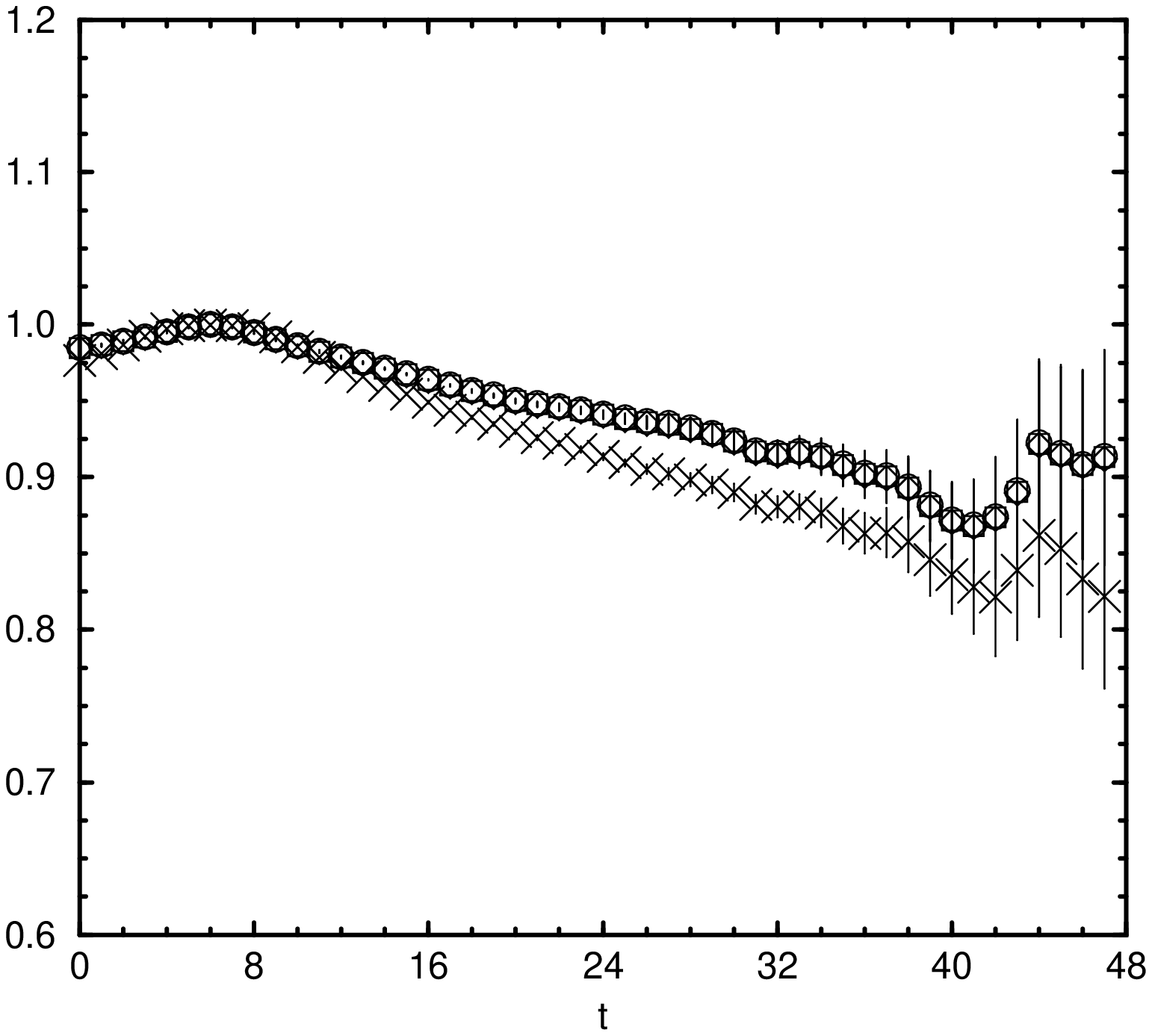}
\\
\raisebox{25mm}{ $\displaystyle{ \frac{m_{\pi}}{m_{\rho}} \approx 0.75 }$ }
& \leavevmode \epsfxsize=5.1cm \epsfbox{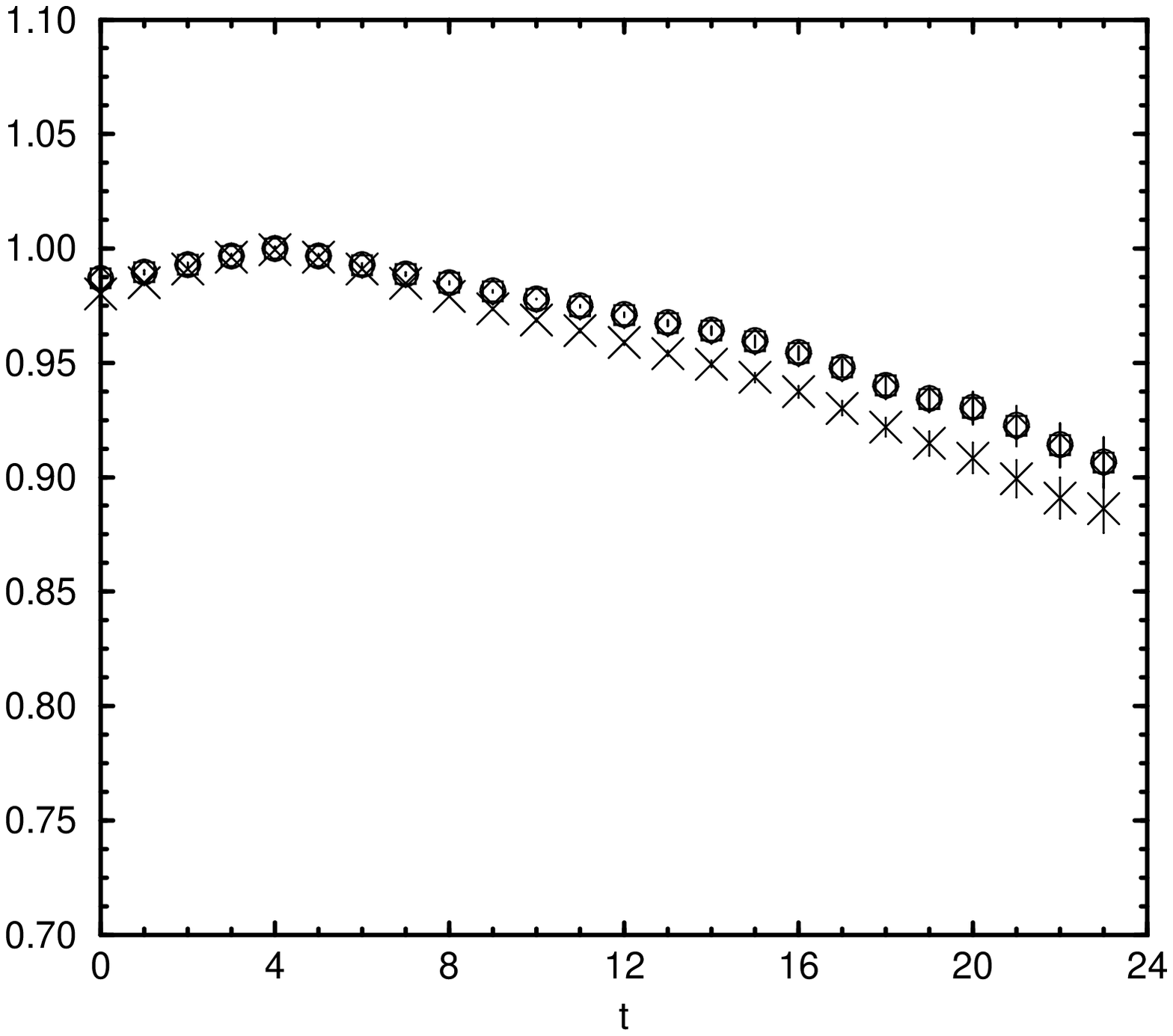}
& \leavevmode \epsfxsize=5cm   \epsfbox{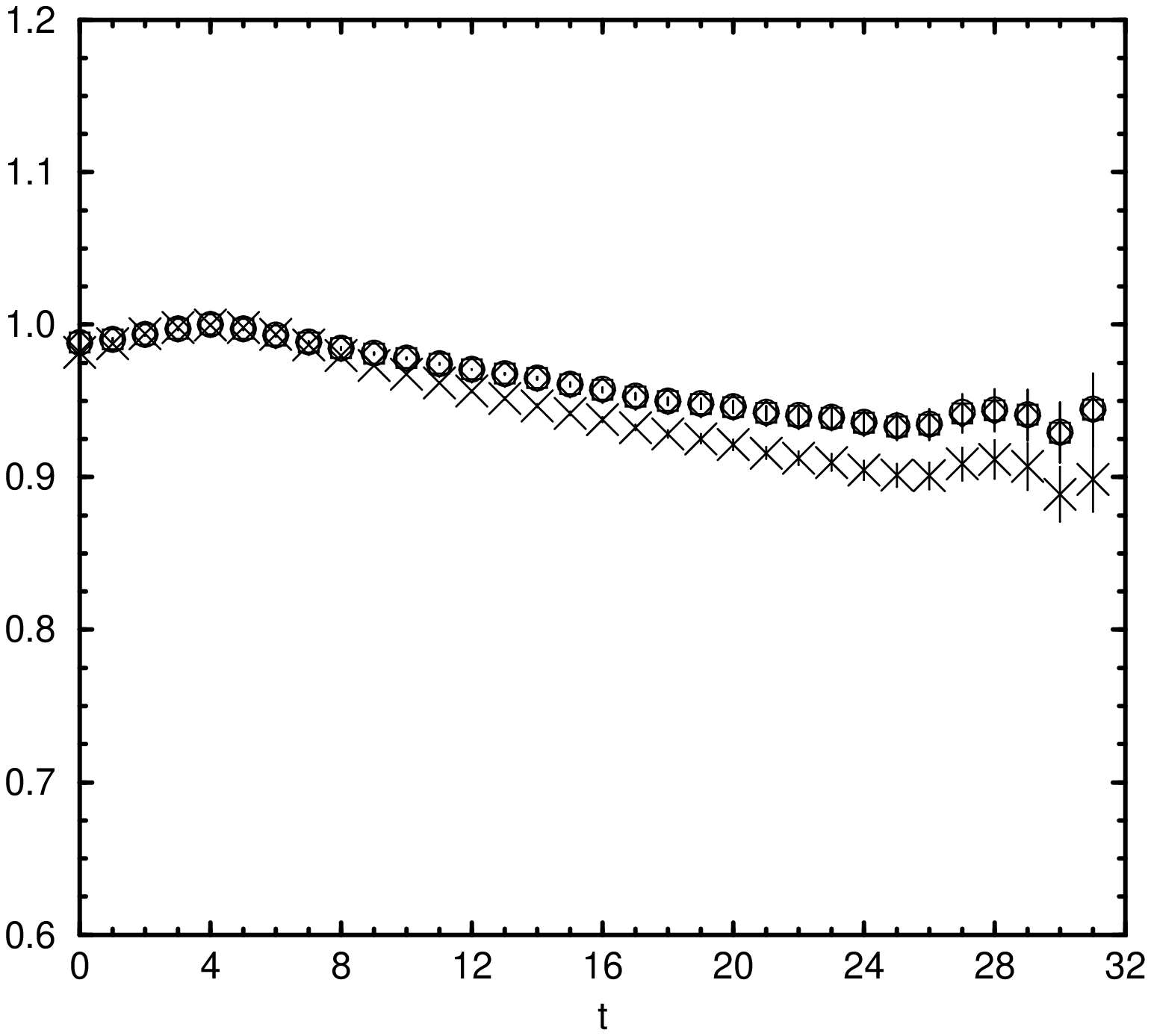}
& \leavevmode \epsfxsize=5cm   \epsfbox{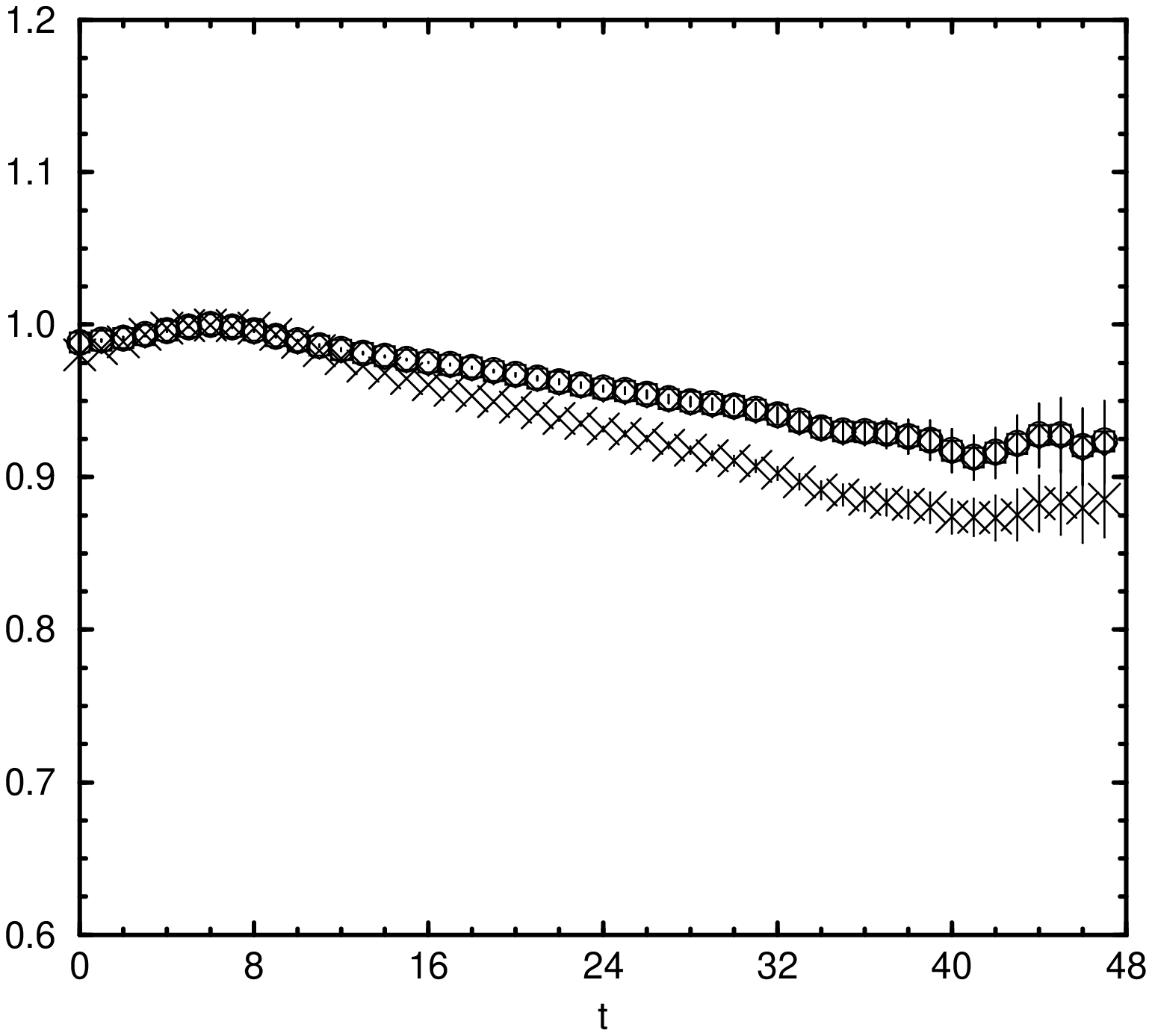}
\\
\raisebox{25mm}{ $\displaystyle{ \frac{m_{\pi}}{m_{\rho}} \approx 0.8 }$ }
& \leavevmode \epsfxsize=5.1cm \epsfbox{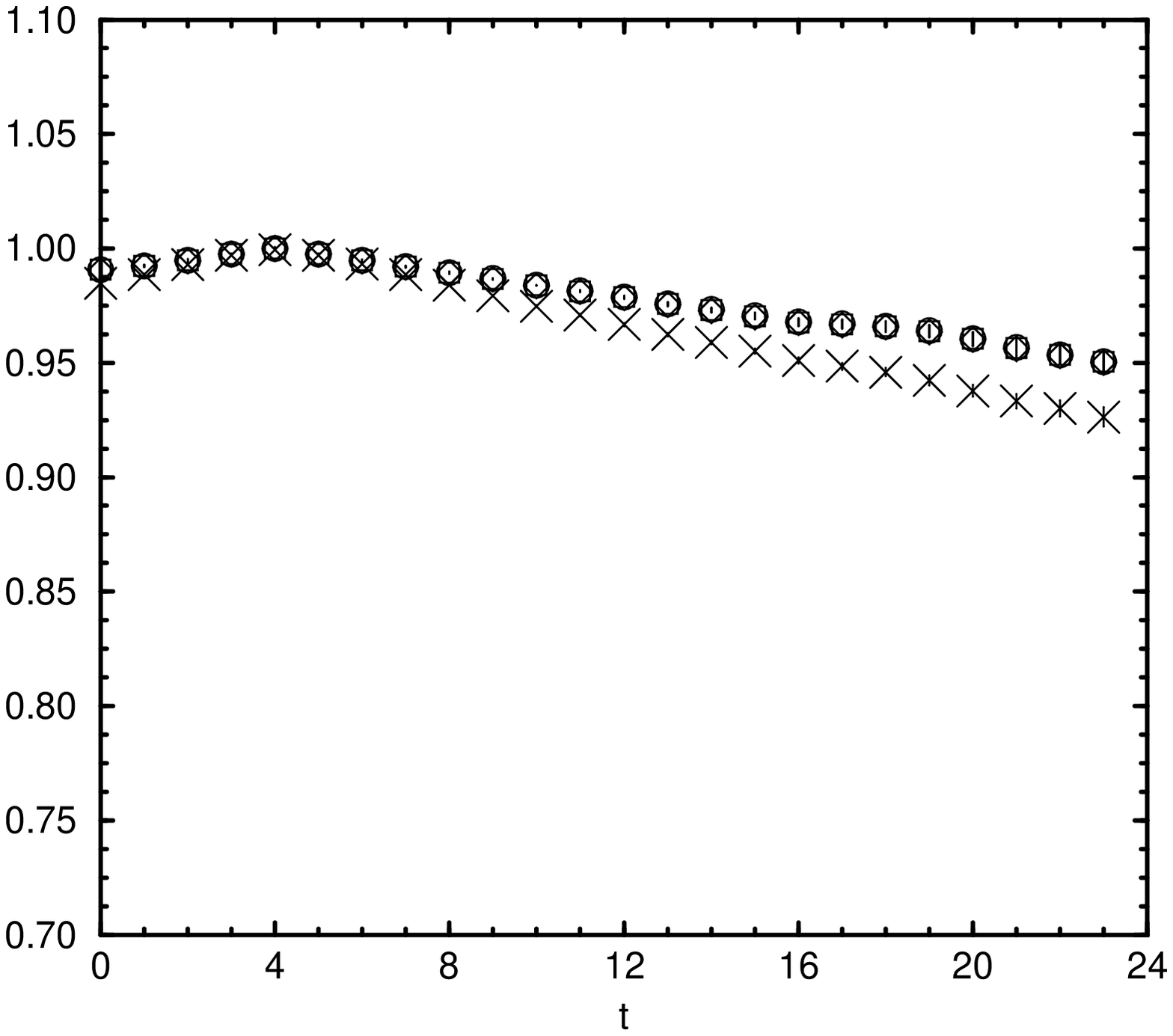}
& \leavevmode \epsfxsize=5cm   \epsfbox{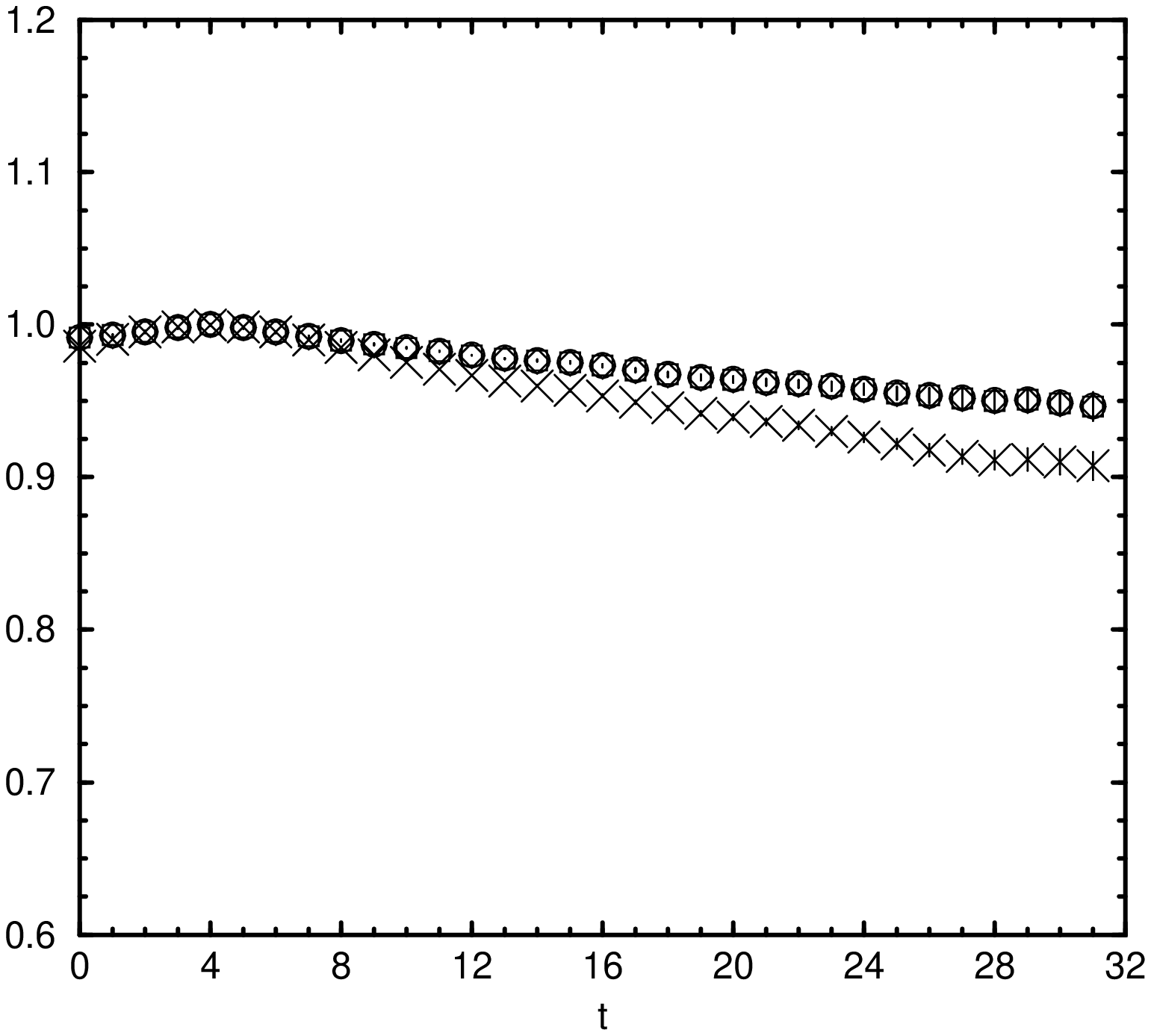}
& \leavevmode \epsfxsize=5cm   \epsfbox{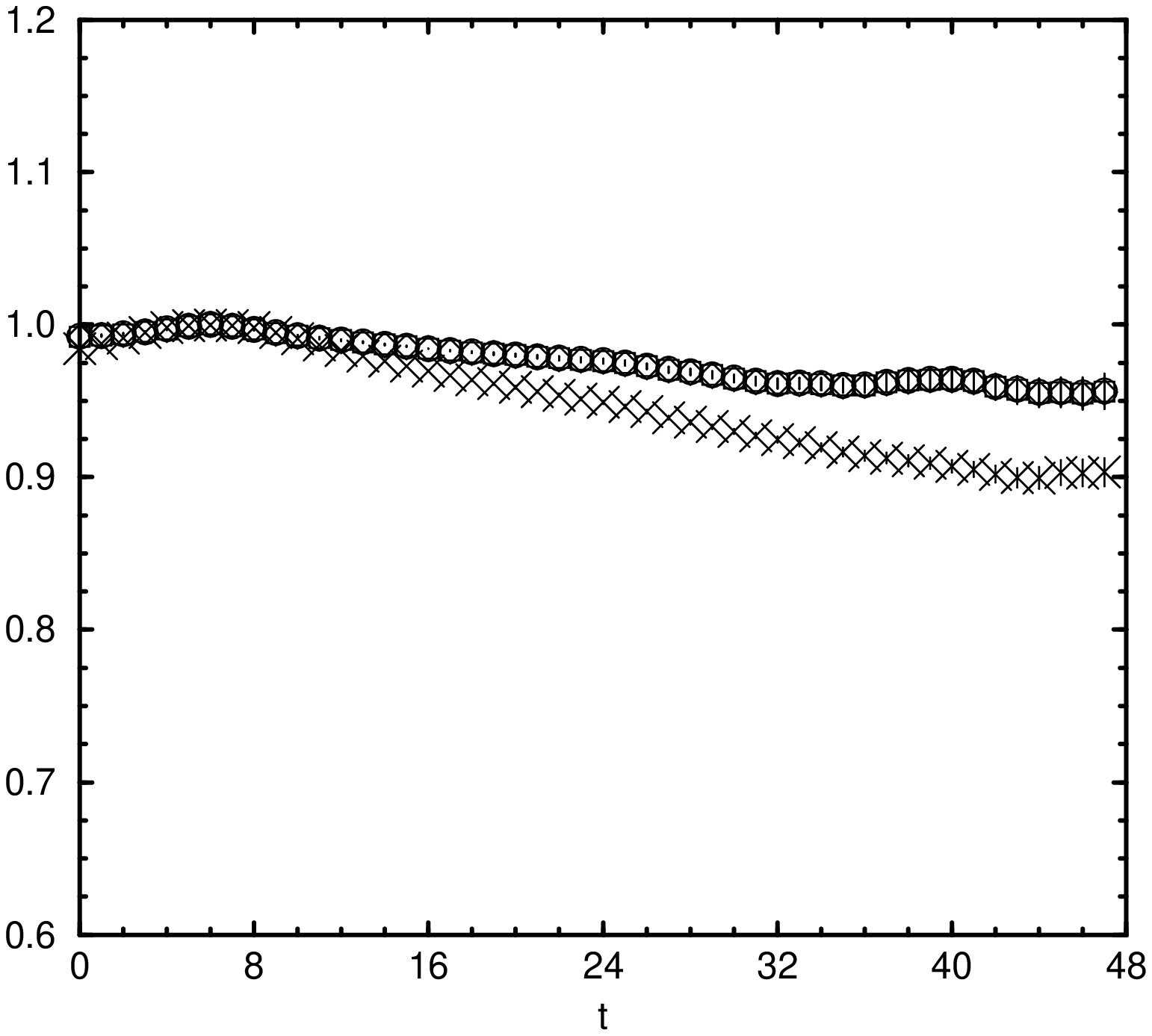}
\end{tabular}
\end{center}
\caption{
Ratio $R_n(t)$ and $D_n(t)$ for $n=0$ state in the laboratory system L2
with energy state cut-off $N = 3$.
$m_\pi / m_\rho$ increases from top to bottom, 
while $\beta$ increases from left to right.
\label{fig:diag:L2_0}
}
\end{figure}

\begin{figure}
\begin{center}
\hspace{-26mm}
\begin{tabular}{rccc}
& $\beta = 1.80$ & $\beta = 1.95$ & $\beta = 2.10$ 
\\
\raisebox{25mm}{ $\displaystyle{ \frac{m_{\pi}}{m_{\rho}} \approx 0.6 }$ }
& \leavevmode \epsfxsize=5.1cm \epsfbox{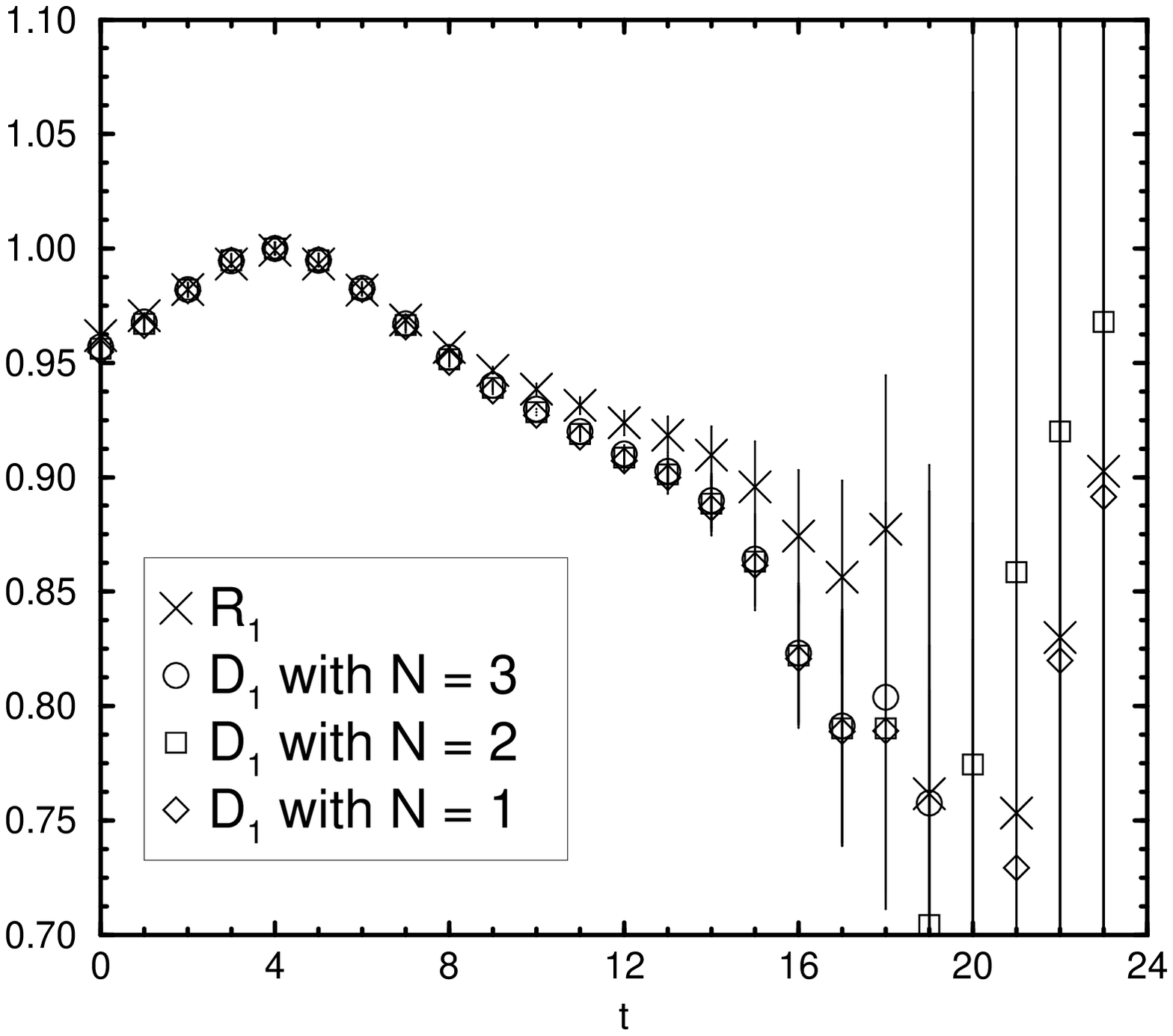}
& \leavevmode \epsfxsize=5cm   \epsfbox{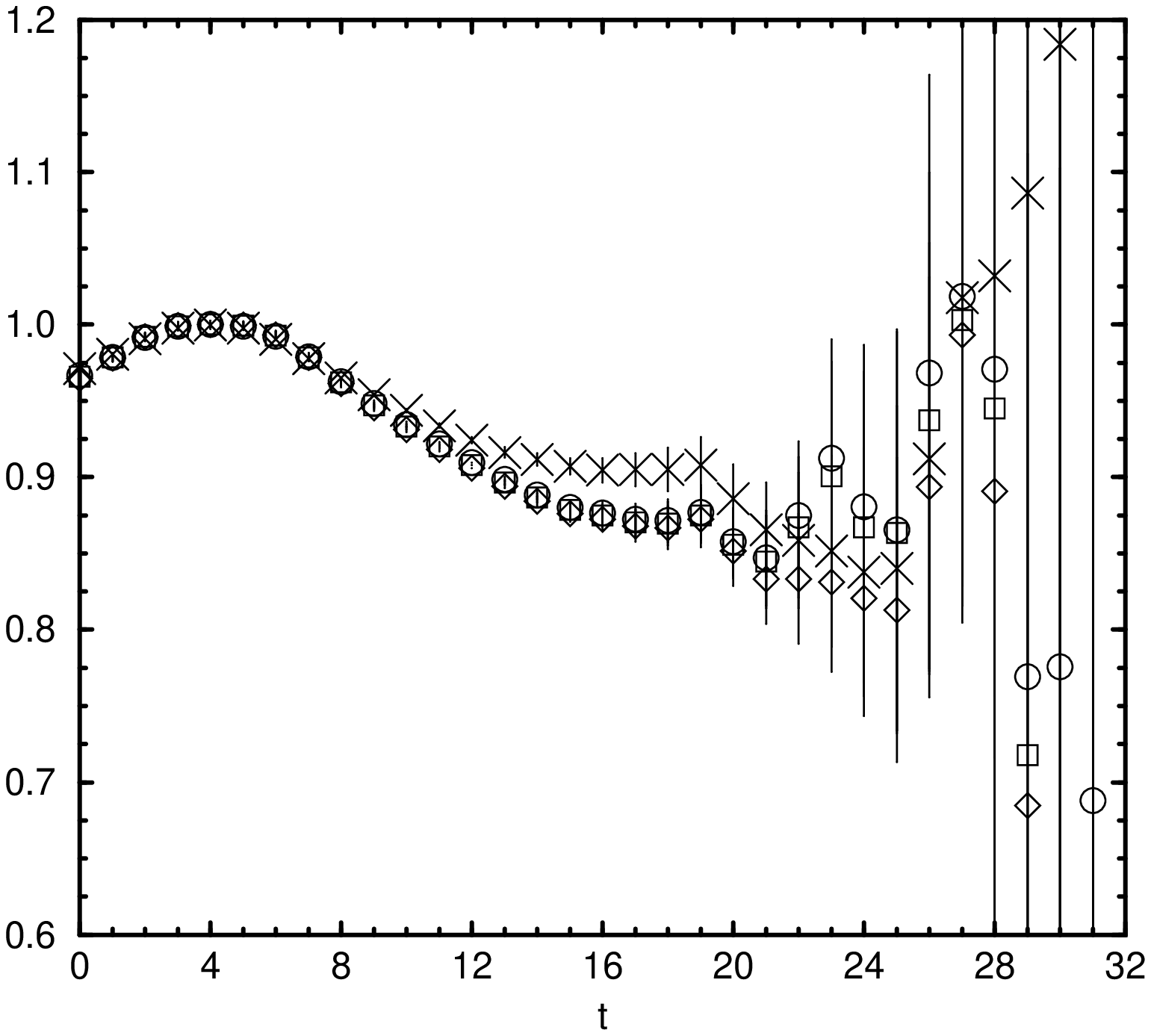}
& \leavevmode \epsfxsize=5cm   \epsfbox{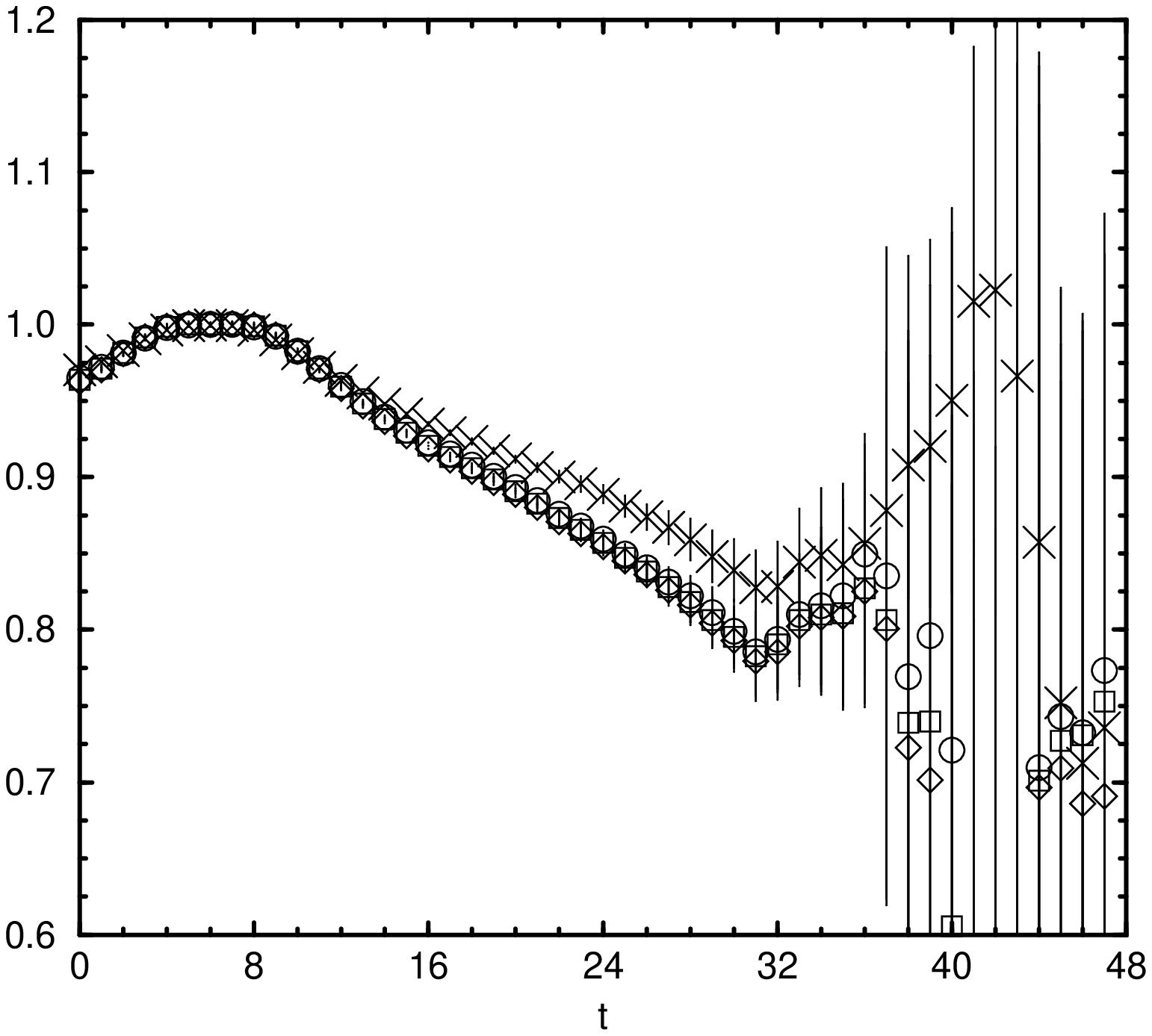}
\\
\raisebox{25mm}{ $\displaystyle{ \frac{m_{\pi}}{m_{\rho}} \approx 0.7 }$ }
& \leavevmode \epsfxsize=5.1cm \epsfbox{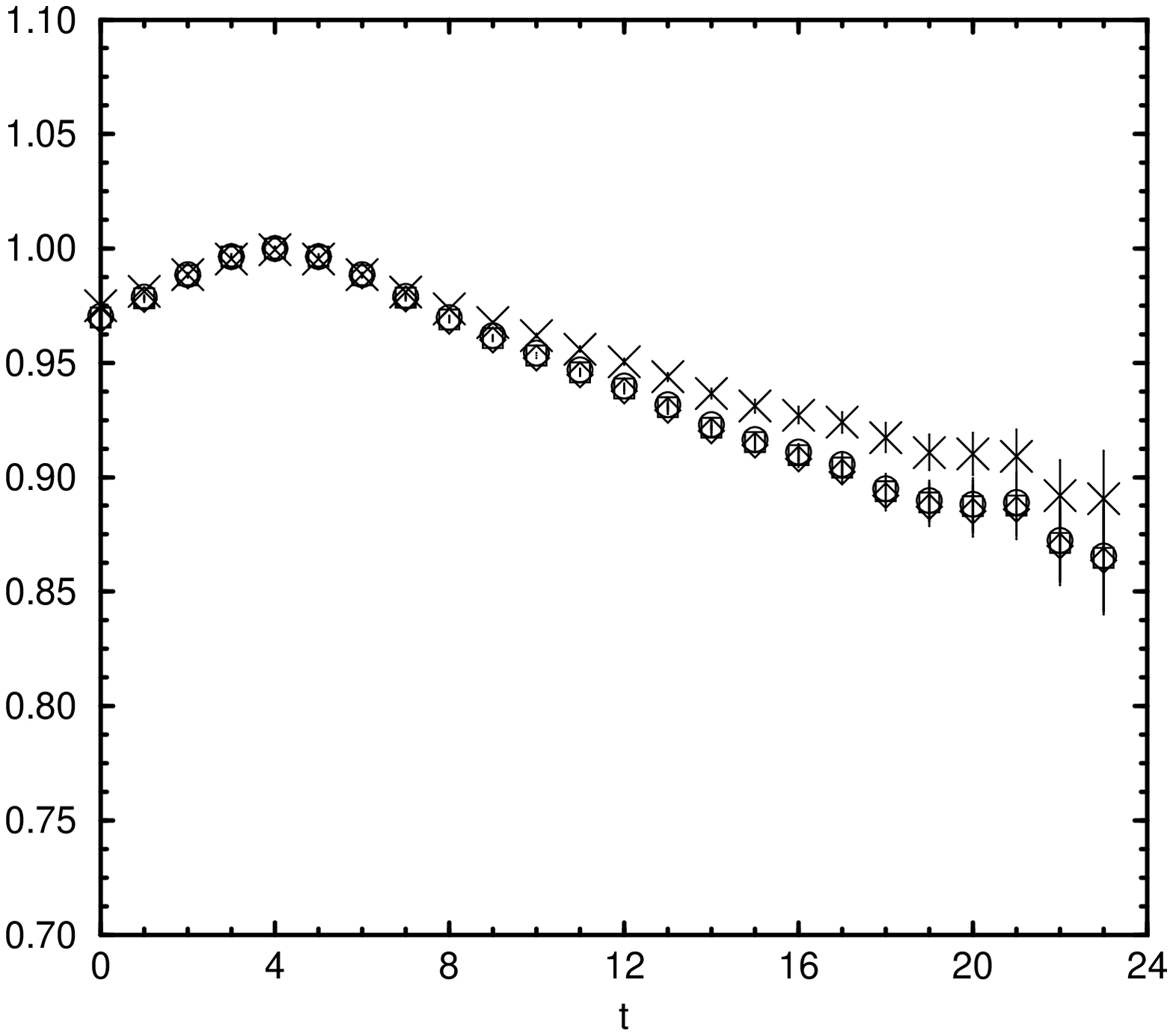}
& \leavevmode \epsfxsize=5cm   \epsfbox{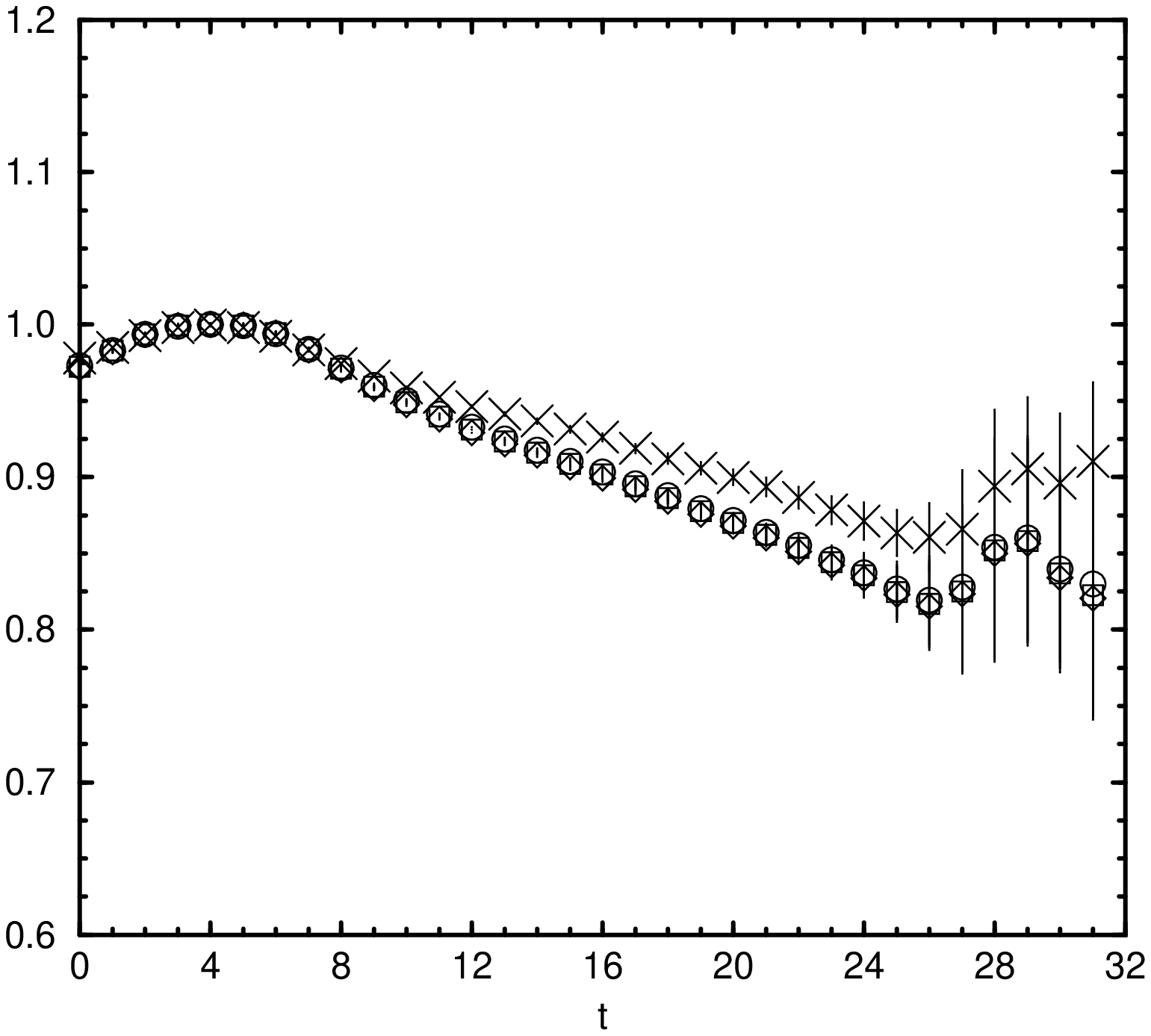}
& \leavevmode \epsfxsize=5cm   \epsfbox{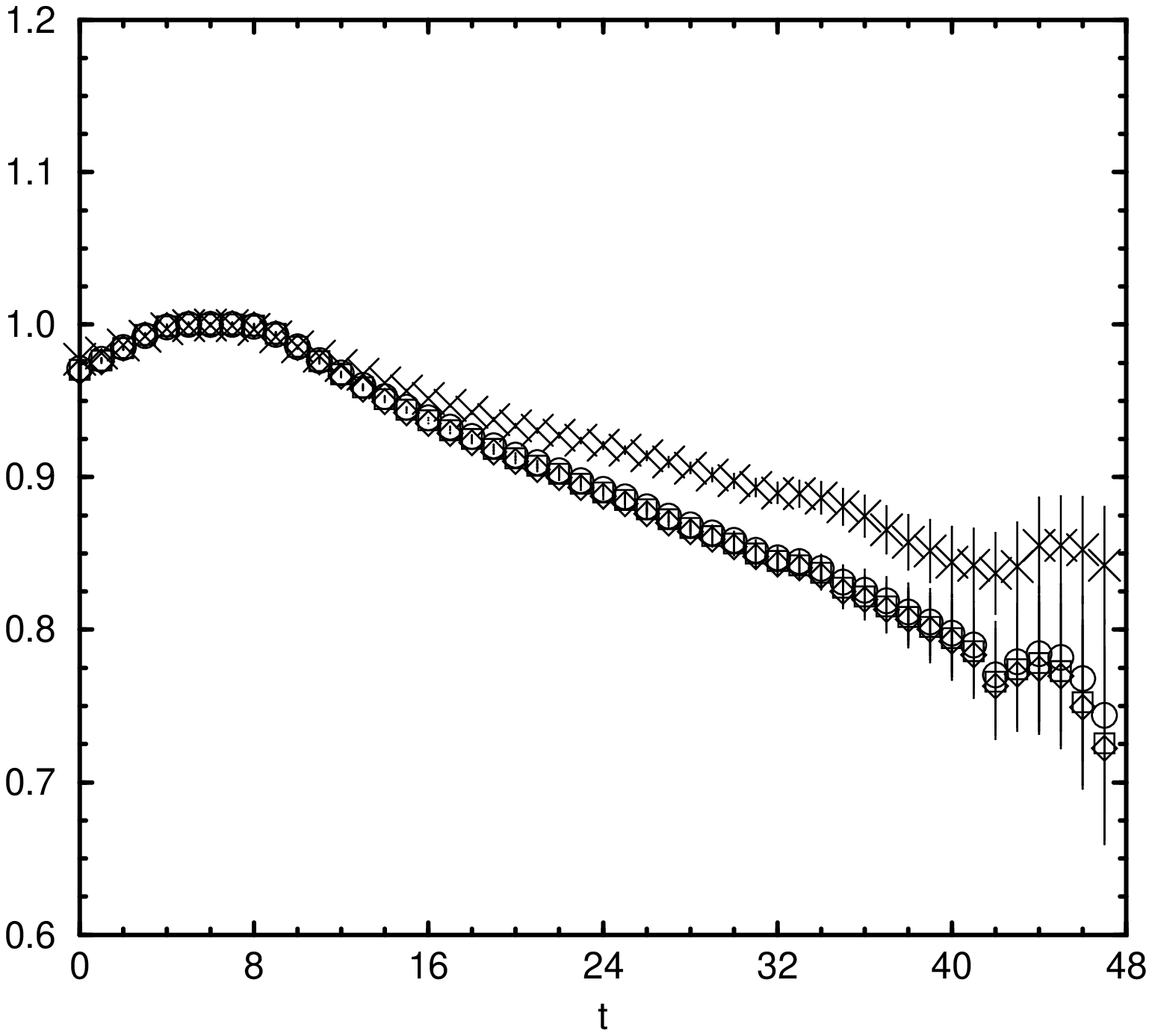}
\\
\raisebox{25mm}{ $\displaystyle{ \frac{m_{\pi}}{m_{\rho}} \approx 0.75 }$ }
& \leavevmode \epsfxsize=5.1cm \epsfbox{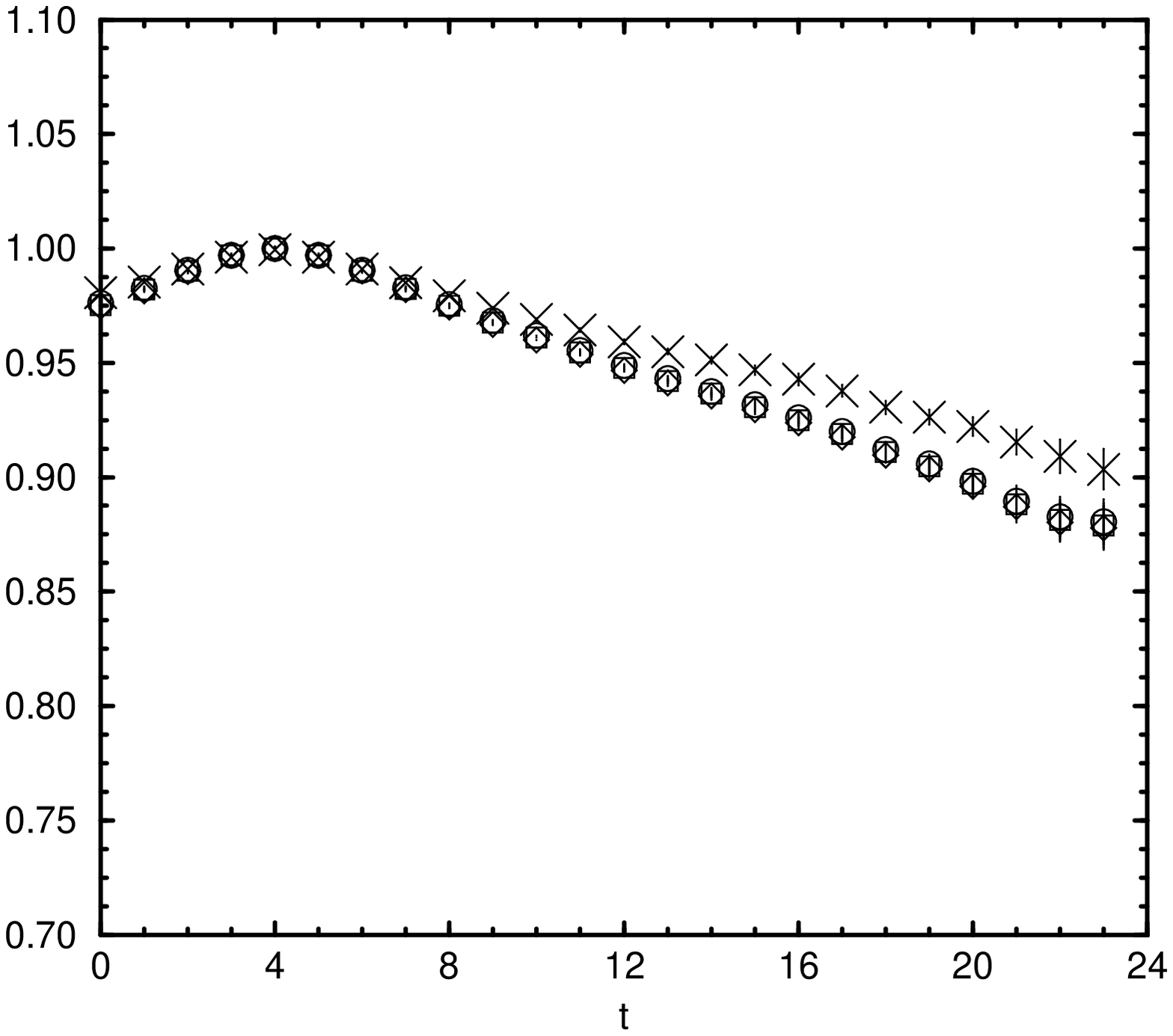}
& \leavevmode \epsfxsize=5cm   \epsfbox{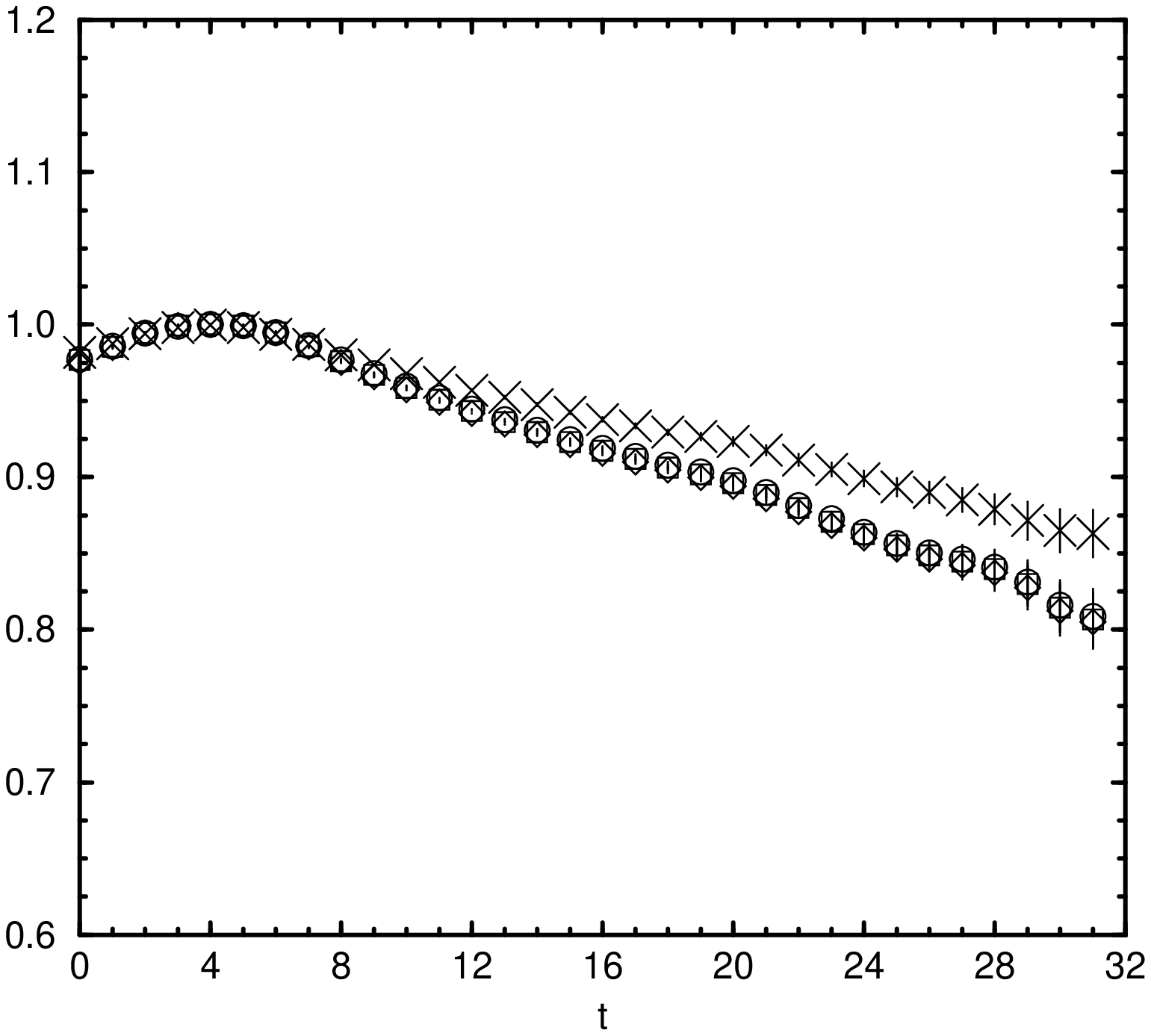}
& \leavevmode \epsfxsize=5cm   \epsfbox{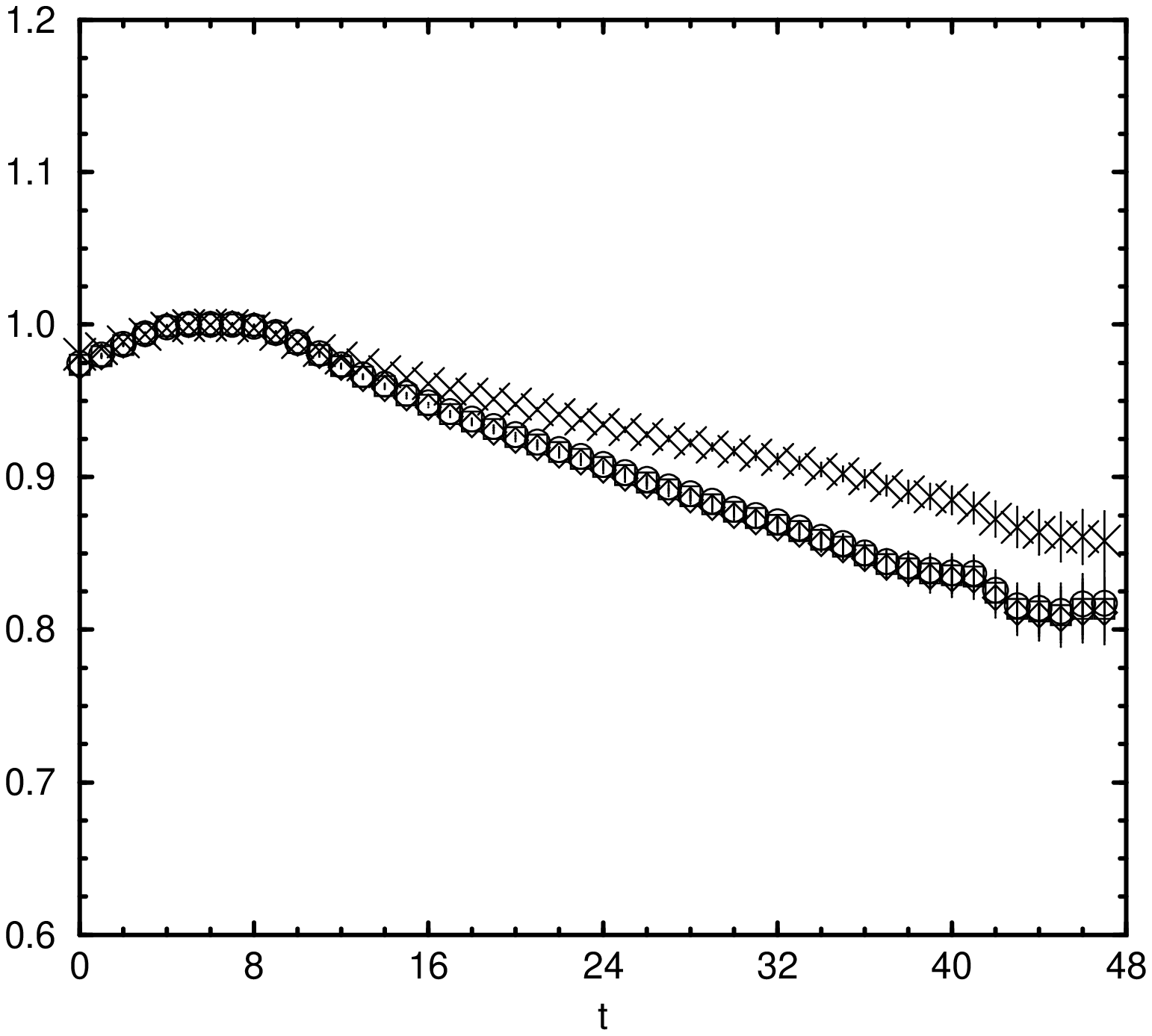}
\\
\raisebox{25mm}{ $\displaystyle{ \frac{m_{\pi}}{m_{\rho}} \approx 0.8 }$ }
& \leavevmode \epsfxsize=5.1cm \epsfbox{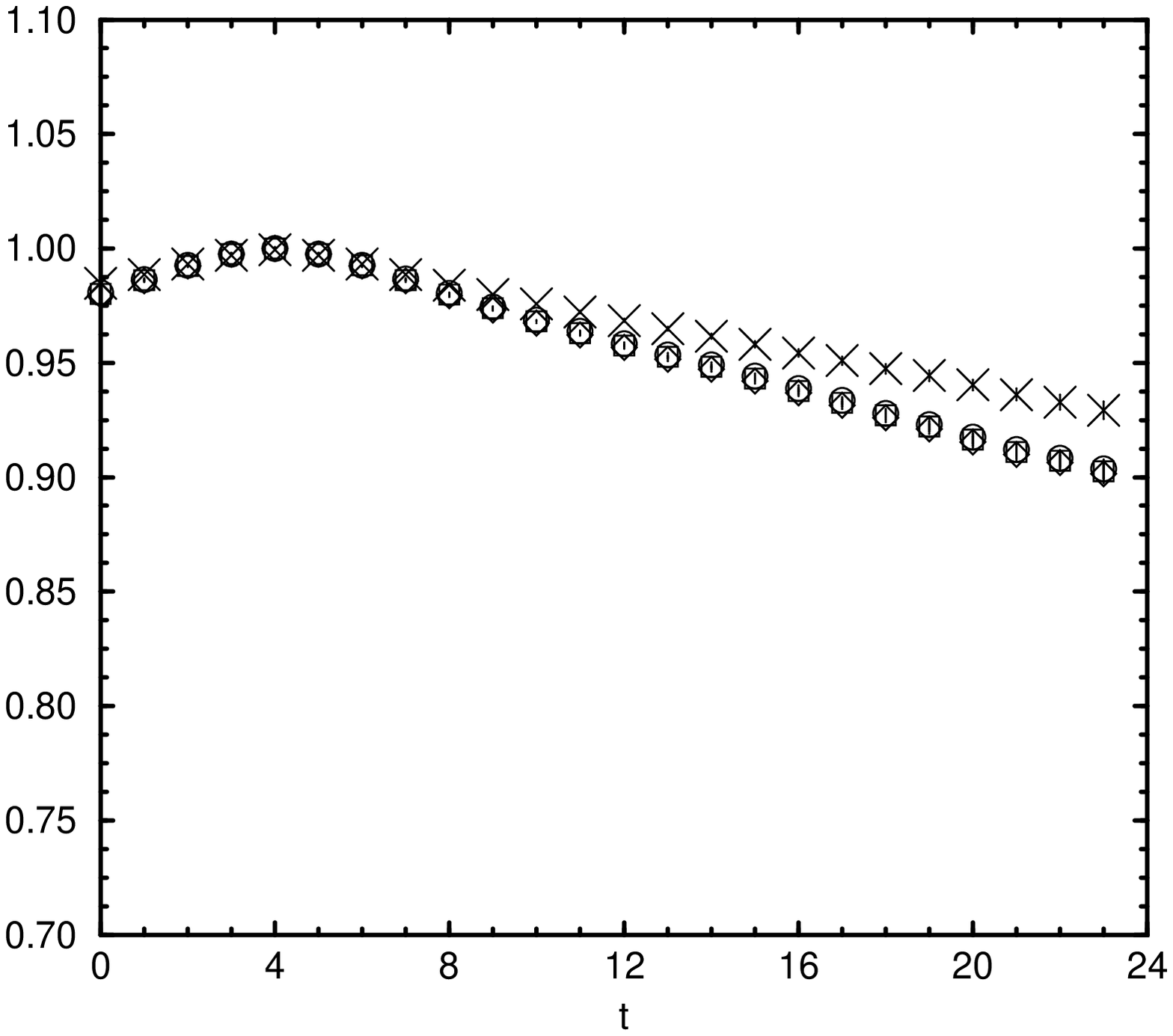}
& \leavevmode \epsfxsize=5cm   \epsfbox{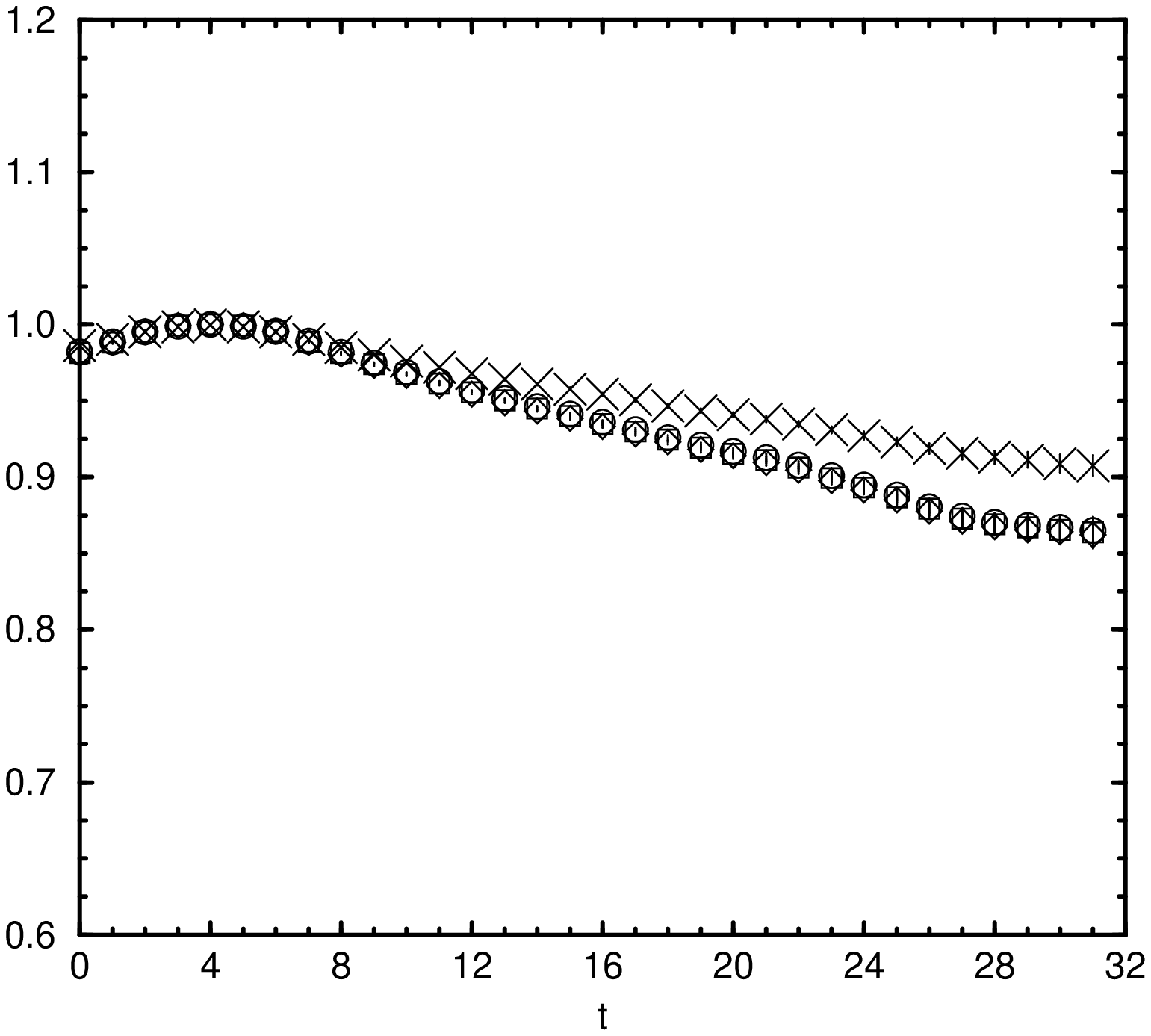}
& \leavevmode \epsfxsize=5cm   \epsfbox{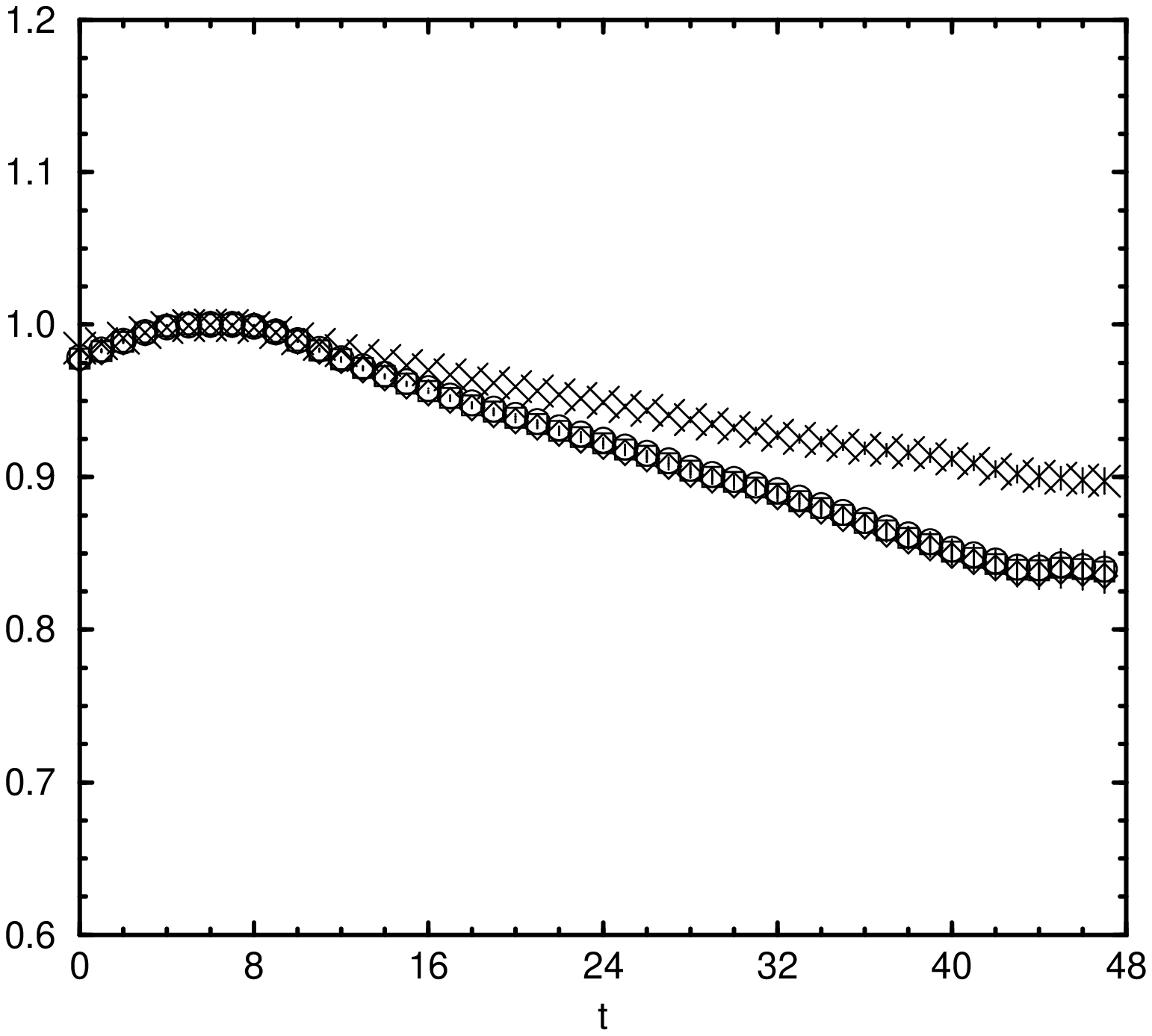}
\end{tabular}
\end{center}
\caption{
Ratio $R_n(t)$ and $D_n(t)$ for $n=1$ state in the laboratory system L2
with energy state cut-off $N = 1$, $2$, and $3$.
$m_\pi / m_\rho$ increases from top to bottom, 
while $\beta$ increases from left to right.
\label{fig:diag:L2_1}
}
\end{figure}

Finally we focus on the L2 system.
The behavior of the ratios in this system is essentially different
from those in the other systems.
The ratios $R_n(t)$ and $D_n(t)$ are shown 
in Fig.~\ref{fig:diag:L2_0} for the ground ($n=0$) and
in Fig.~\ref{fig:diag:L2_1} for the first excited states ($n=1$).
In these figures $D_n(t)$ is divided by $D_n(t_S)$. 
Since the energies of $n=0$ and $1$ states are very close 
in the free two-pion case,
not only the ratio for $n=1$
but also that for $n=0$ are affected by the diagonalization.
We also observe that the cut-off dependence is negligible 
by comparing the results with $N=1$, $2$ and $3$.
The ratios $D_n(t)$ for $n =0,1$
for $m_\pi / m_\rho \approx 0.6$ at $\beta = 1.95$
are not clearly  exponential in behavior.
These results are not included in the following analysis 
considering that these defects are probably caused by insufficient statistics.

\begin{figure}
\begin{center}
\leavevmode \epsfxsize=7cm\epsfbox{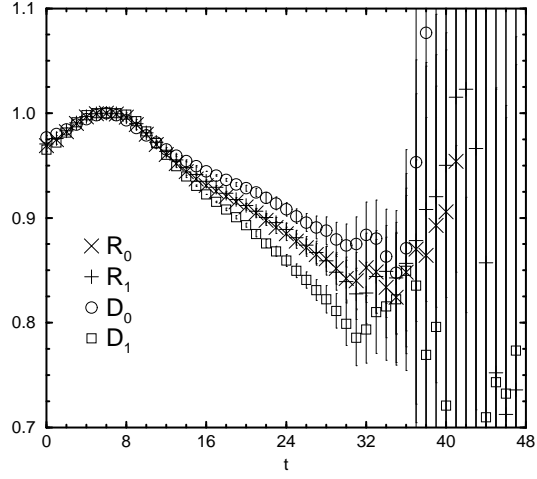}
\end{center}
\caption{
Ratios $R_n(t)$ and $D_n(t)$
for $n=0$ and $1$ states in the laboratory system L2
with energy state cut-off $N=3$ at $\beta = 2.10$ 
for $m_\pi / m_\rho \approx 0.6$.
\label{fig:diag:split}
}
\end{figure}

In order to show the effect of diagonalization in the L2 system clearly,
we gather both ratios $R_n(t)$ and $D_n(t)$ for $n=0$ and $1$ 
in Fig.~\ref{fig:diag:split}.
The data at $\beta=2.10$ for $m_\pi / m_\rho \approx 0.6$ is shown in the figure.
Before the diagonalization the two ratios $R_n(t)$ have almost the same shape.
After the diagonalization, however, the slope of the ratios $D_n(t)$ for $n=0$ decreases
and that for $n=1$ increases.
This behavior can be understood
by considering a simple eigenvalue problem for two degenerate states.
We assume that 
the energies for $n=0$ and $1$ states in the L2 system 
have the same energy $E$ in the free two-pion case.
Further we neglect the effects from higher energy states.
This assumption is supported by the independence on the cut-off $N$.
In the interacting case
the Hamiltonian $H$ for this system can be written as 
\begin{equation}
  H = 
    \left(
        \begin{array}{cc}
            E   + \Delta  &  \beta            \\
            \beta         &  E + \Delta       \\
        \end{array}
    \right),
\end{equation}
where the components of the Hamiltonian are defined by 
$H_{nm} \equiv \langle \Omega_n | H | \Omega_m \rangle$ 
with the non-interacting two-pion state $| \Omega_n \rangle$.
The $\Delta$ and $\beta$ are unknown constants
induced by the two-pion interaction.
The constant $\Delta$ corresponds to the slopes of $R_n(t)$ up to $O( (\beta t)^2 )$.
The eigenvalues $\overline{E}$ of $H$ are given by
\begin{equation}
  \overline{E} = E + \Delta \pm \beta.
\end{equation}
The values of $\Delta \pm \beta$ correspond to the slopes of $D_n (t)$ 
for $n=0$ and $1$ in the figure.
Here we can see 
that the existence of the off-diagonal component of the Hamiltonian $\beta$
causes a separation of $D_n (t)$.

Up to now we have shown that 
the ratios $D_n(t)$ for any $n$ in all the momentum systems
behave as single exponential functions in $t$ and 
it is possible to extract the energy shift $\Delta \overline{E}_n^{\bf P}$
by a single exponential fitting of the ratios.
Our choice of the fitting range and the results of $\Delta \overline{E}_n^{\bf P}$ 
are summarized in Appendix~\ref{tab:result}.
The procedure of calculating
the scattering phase shift $\delta( \overline{p}_n )$
from $\Delta \overline{E}_n^{\bf P}$ is the following.
First we construct the two-pion energy eigenvalue $\overline{E}_n^{\bf P}$
by $\overline{E}_n^{\bf P} = E_n^{\bf P} + \Delta \overline{E}_n^{\bf P}$,
where $E_n^{\bf P}$ is the two-pion energy in the free two-pion case.
Then we evaluate the Lorentz boost factor $\gamma$,
the center of mass energy $\overline{E}_n$ and momentum $\overline{p}_n^2$
from $\overline{E}_n^{\bf P}$ by Eqs.(\ref{eq:gamma}), (\ref{eq:E_CM}) and (\ref{eq:p_CM}).
Finally 
we obtain the phase shift $\delta( \overline{p}_n )$ 
by substituting $\overline{p}_n^2$
into the finite volume formulae given by Eqs.(\ref{eq:FVM:L}) and (\ref{eq:FVM:RG}).
The results are tabulated in Appendix~\ref{tab:result}.
In the appendix
we also quote the 'scattering amplitude' defined by 
\begin{equation}
  A( m_\pi, \overline{p} ) = 
   \frac{ \tan \delta( \overline{p} ) }{ \overline{p} }
   \cdot
   \frac{ \overline{E} }{ 2 }  , 
\label{eq:amplitude}
\end{equation}
where the amplitude is normalized as
\begin{equation}
  \lim_{ \overline{p} \to 0 } A( m_\pi, \overline{p} ) = a_0 m_\pi,
\end{equation}
with the scattering length $a_0$.
%
%
\subsection{Result for scattering length}
\label{Sec:Results:sl}
\begin{figure}
\begin{center}
\leavevmode \epsfxsize=7cm\epsfbox{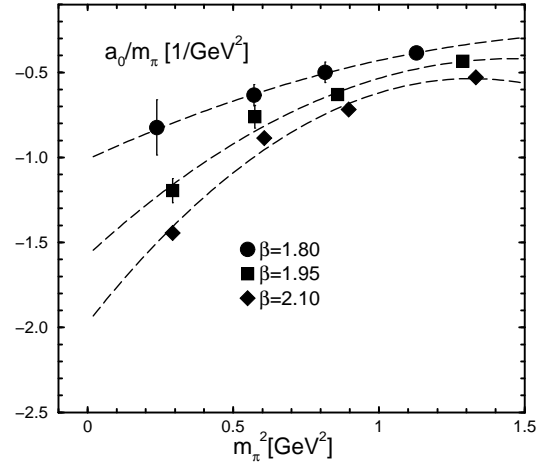}
\end{center}
\caption{
Results for $a_0/m_\pi$ at each $\beta$ and 
polynomial fits (dashed lines).
\label{fig:a0:fit-pow}
}
\end{figure}

In the CM system the momentum $\overline{p}^2$ for the ground state ($n=0$) 
is very small as shown in Appendix~\ref{tab:result}.
Thus the scattering length $a_0$ can be evaluated
from the scattering amplitude defined in Eq.(\ref{eq:amplitude})
by $A(m_\pi,\overline{p}) \approx a_0 m_\pi$.
In Fig.~\ref{fig:a0:fit-pow}
we plot $a_0/m_\pi$ as a function of $m_\pi^2$ at each $\beta$, 
which are also tabulated in Appendix~\ref{tab:result} denoted by $A(m_\pi,\overline{p})/m_\pi^2$. 
A significant curvature in the $m_\pi^2$ dependence at large $\beta$ is seen.
This has not been clearly observed in the previous studies of 
the scattering length.
The existence of a large curvature
renders the chiral extrapolation of $a_0/m_\pi$ very difficult.
In this work we attempt to fit the data 
with various fitting assumptions 
to obtain the scattering length at the physical pion mass $m_\pi = 0.14$ GeV.

The prediction of chiral perturbation theory (ChPT)
for the $m_\pi^2$ dependence 
has been worked out by Gasser and Leutwyler~\cite{ChPT}.
In one loop order it is given by
\begin{equation}
  \frac{ a_0 }{ m_\pi } = -\frac{ 1 }{ 16 \pi F^2 }
  \left\{ 1 - \frac{ m_\pi^2 }{ 16 \pi^2 F^2 } 
  \left[ L( \mu ) - C_L \cdot \log\frac{ m_\pi^2 }{ \mu^2 } \right] \right\},
\label{eq:a0:fit-ChPT}
\end{equation}
where $F$ is the pseudoscalar decay constant in the chiral limit,
$L(\mu)$ is a low energy constant at a scale $\mu$, and $C_L = 7/2$.
It is clear that this one-loop formula cannot be naively applied 
to our results of the scattering length.
As shown in Fig.~\ref{fig:a0:fit-pow}
the dependence of $a_0/m_\pi$ on the lattice spacing is comparable 
to that of $m_\pi^2$.
This implies that we should extrapolate our data
with the formula including the $O(a)$ effect.
Such a formula for the pseudo scalar mass and the decay constant 
have been obtained in one-loop order of chiral perturbation theory 
for the Wilson type fermions~\cite{WChPT},
but that for the scattering length is not yet available.
Further, even if the $O(a)$ effect is negligible, 
it is not clear whether the formula of ChPT is applicable
for our data calculated in a heavy pion mass region,
$m_\pi = 0.5 - 1.1\ {\rm GeV}$.
Here, as a trial,
we fit our data with the fitting assumption given by Eq.(\ref{eq:a0:fit-ChPT})
with three unknown parameters $F$, $L(\mu)$ and $C_L$, and
consider the dependence of fitted parameter values on the lattice spacing.
The results are summarized in Table~\ref{tab:a0:fit-ChPT},
where
the scale $\mu$ is fixed at $\mu=1\ {\rm GeV}$ in the analysis.
We find a significant difference between $C_L$ obtained by the fitting
and that of the prediction from ChPT ($C_L = 7/2$) at all $\beta$.

It is expected that
the chiral breaking effect of the Wilson type fermion
causes a divergence in the chiral limit 
as $a_0/m_\pi \propto 1/m_\pi^2$~\cite{kawamoto}.
The large curvature of the scattering length may originate from this effect.
In Table~\ref{tab:a0:fit-diverge}
we tabulate the results of the fitting with
\begin{equation}
  \frac{ a_0 }{ m_\pi } = \frac{ A_{00} }{ m_\pi^2 } + A_{10} + A_{20} m_\pi^2 .
\label{eq:a0:div}
\end{equation}
The coefficient of the divergent term $A_{00}$, 
which comes from the chiral breaking effect,
is expected to vanish in the continuum limit.
However, our results for $A_{00}$ increases 
toward the continuum limit as opposed to the expectation.
We consider that the effect of chiral breaking 
is not separated from the regular mass dependence.
It is a very important future work
to detect the divergence term through simulations with much higher statistics
and closer to the chiral limit.
We assume that the effect of chiral breaking is small 
in our simulation points in the following analysis.

Next we attempt to fit our data 
assuming the following polynomial function in $m_\pi^2$,
\begin{equation}
   \frac{ a_0 }{ m_\pi } = A_{10} + A_{20} m_\pi^2 + A_{30} m_\pi^4 .
\label{eq:a0:pow}
\end{equation}
The results of the fitting are plotted in Fig.~\ref{fig:a0:fit-pow}
and summarized in Table~\ref{tab:a0:fit-pow}.
At $\beta = 2.10$ the value of $\chi^2 /{\rm d.o.f.}$ is large.
This indicates
that the fitting including higher order terms, such as $m_\pi^6$ term or higher, 
is necessary to obtain a more precise value at the physical pion mass.
Such fitting cannot be carried out in this work, 
because the number of our simulation points is only four.

\begin{figure}
\begin{center}
\leavevmode \epsfxsize=7cm\epsfbox{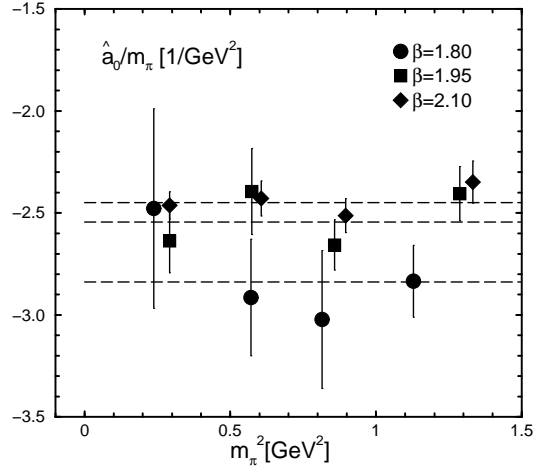}
\end{center}
\caption{
Results for $\hat{a}_0 / m_\pi = ( f_\pi^{lat} / f_\pi )^2 \cdot a_0 / m_{\pi}$
with $f_\pi = 93$ MeV at each $\beta$
and constant fits (dashed lines).
\label{fig:a0:fit-f_pi-a0}
}
\end{figure}

In Refs.~\cite{sc-2,sc-3,sc-4}
the authors found that the mass dependence of the ratio 
$( f_\pi^{lat} )^2 \cdot a_0 / m_\pi$ was very small,
where $f_\pi^{lat}$ is the decay constant measured on the lattice at each $m_\pi$.
In Fig.~\ref{fig:a0:fit-f_pi-a0}
we plot the normalized scattering length defined by
\begin{equation}
   \frac{ \hat{a}_0 }{ m_\pi } 
                 = \biggl( \frac{ f_\pi^{lat} }{ f_\pi } \biggr)^2 \cdot
                   \frac{ a_0  }{ m_\pi },
\label{eq:nom_SCL}
\end{equation}
where $f_\pi^{lat}$ is 
the decay constants measured in Ref.~\cite{CP-PACS:sprm},
which are tabulated in Table~\ref{tab:f_pi},
and 
$f_\pi=93\ {\rm MeV}$.
The values of $\hat{a}_0 / m_\pi$
are also tabulated in Appendix~\ref{tab:result};
they are written under the column for $\hat{A}( m_\pi, \overline{p} ) / m_\pi^2$ 
defined by
\begin{equation}
   \hat{A}( m_\pi, \overline{p} )
          = \biggl( \frac{ f_\pi^{lat} }{ f_\pi } \biggr)^2 \cdot
               A( m_\pi, \overline{p} ).
\label{eq:NOR_amplitude}
\end{equation}
In the calculation of $\hat{a}_0 / m_\pi$
the statistical errors of $f_\pi^{lat}$ are not included,
since they are small compared with those of the scattering length.
We observe in Fig.~\ref{fig:a0:fit-f_pi-a0} that $\hat{a}_0 / m_\pi$
is almost independent of $m_\pi^2$ within the statistical errors.
This fact implies a strong correlation 
between the scattering length and the decay constant.
The results of a constant fitting for $\hat{a}_0/m_\pi$
are also plotted in the figure 
and tabulated in Table~\ref{tab:a0:fit-f_pi-a0}.
The values of $\chi^2 /$d.o.f. are reasonably small. 
Especially at $\beta=2.10$
it is much smaller than that of the polynomial fitting for $a_0/m_\pi$.

\begin{figure}
\begin{center}
\leavevmode \epsfxsize=7cm\epsfbox{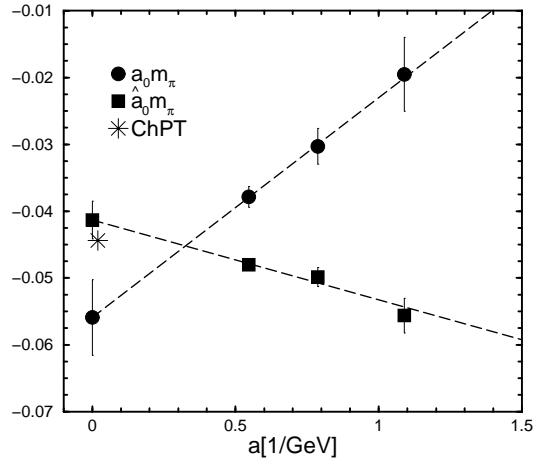}
\end{center}
\caption{
Results for $a_0 m_\pi$ and 
$\hat{a}_0 m_\pi = ( f_\pi^{lat} / f_\pi )^2 \cdot a_0 / m_\pi$
at the physical pion mass as functions of lattice spacing.
\label{fig:a0:fit.cont}
}
\end{figure}

In Fig.~\ref{fig:a0:fit.cont} 
we present $a_0 m_\pi$ at the physical pion mass
obtained by the polynomial fitting defined by Eq.(\ref{eq:a0:pow})
and $\hat{a}_0 m_\pi$ by the constant fitting
as a function of the lattice spacing.
We also plot the results of the continuum extrapolations,
which are summarized in Table~\ref{tab:a0:fit.cont}.
The prediction of ChPT~\cite{cola}:
$a_0 m_\pi = -0.0444(10)$ is denoted by the star symbol in the figure.
We see large $O(a)$ effects in both $a_0 m_\pi$ and $\hat{a}_0 m_\pi$.
Furthermore the $O(a)$ effect for $\hat{a}_0 m_\pi$ is opposite 
to that of $a_0 m_\pi$ 
due to large $O(a)$ effects in $f_\pi^{lat}$.
The difference between them decreases
going toward the continuum limit.

As shown in Fig.~\ref{fig:a0:fit.cont} 
$\hat{a}_0 m_\pi$ in the continuum limit is 
closer to the prediction of ChPT than that of $a_0 m_\pi$.
We should note, however, that the decay constant is introduced 
in  $\hat{a}_0/m_\pi$ to compensate the mass dependence of the scattering 
length. Extrapolations only with the scattering length 
are difficult as discussed before.

The decay constant 
measured in the previous work of Ref.~\cite{CP-PACS:sprm}, 
is tabulated in Table~\ref{tab:f_pi}.
As shown in the table, the value in the continuum limit 
seems not consistent with the experiment $f_\pi=93\ {\rm MeV}$.
One of possible reasons for this discrepancy is an uncertainty 
of the renormalization factor for the axial vector current; 
in Ref.~\cite{CP-PACS:sprm} the perturbative renormalization factor was used.
This causes a large uncertainty 
in the prediction of the scattering length in the continuum limit.
In order to obtain more reliable predictions for 
the physical scattering length,
a non-perturbative determination of the renormalization factor and 
higher statics calculations of the scattering length
closer to the physical pion mass and the continuum limit
are needed.
Here we present the two values estimated by $a_0 m_\pi$ 
and $\hat{a}_0 m_\pi = ( f_\pi ^{lat} / f_\pi )^2 \cdot a_0 m_\pi$
as our results for the scattering length in this work
with the provisions noted above:

\begin{eqnarray}
   &&  a_0       m_\pi = - 0.0558(56),  \label{eq:a_0} \\ 
   &&  \hat{a}_0 m_\pi = - 0.0413(29).  \label{eq:a_0^f} 
\end{eqnarray}
%
%
\subsection{Result for scattering phase shift}
\begin{figure}
\begin{center}
\hspace{-26mm}
\begin{tabular}{rccc}
& $\beta = 1.80$ & $\beta = 1.95$ & $\beta = 2.10$
\\
\raisebox{25mm}{ $\displaystyle{ \frac{m_{\pi}}{m_{\rho}} \approx 0.6 }$ }
& \leavevmode \epsfxsize=5.2cm \epsfbox{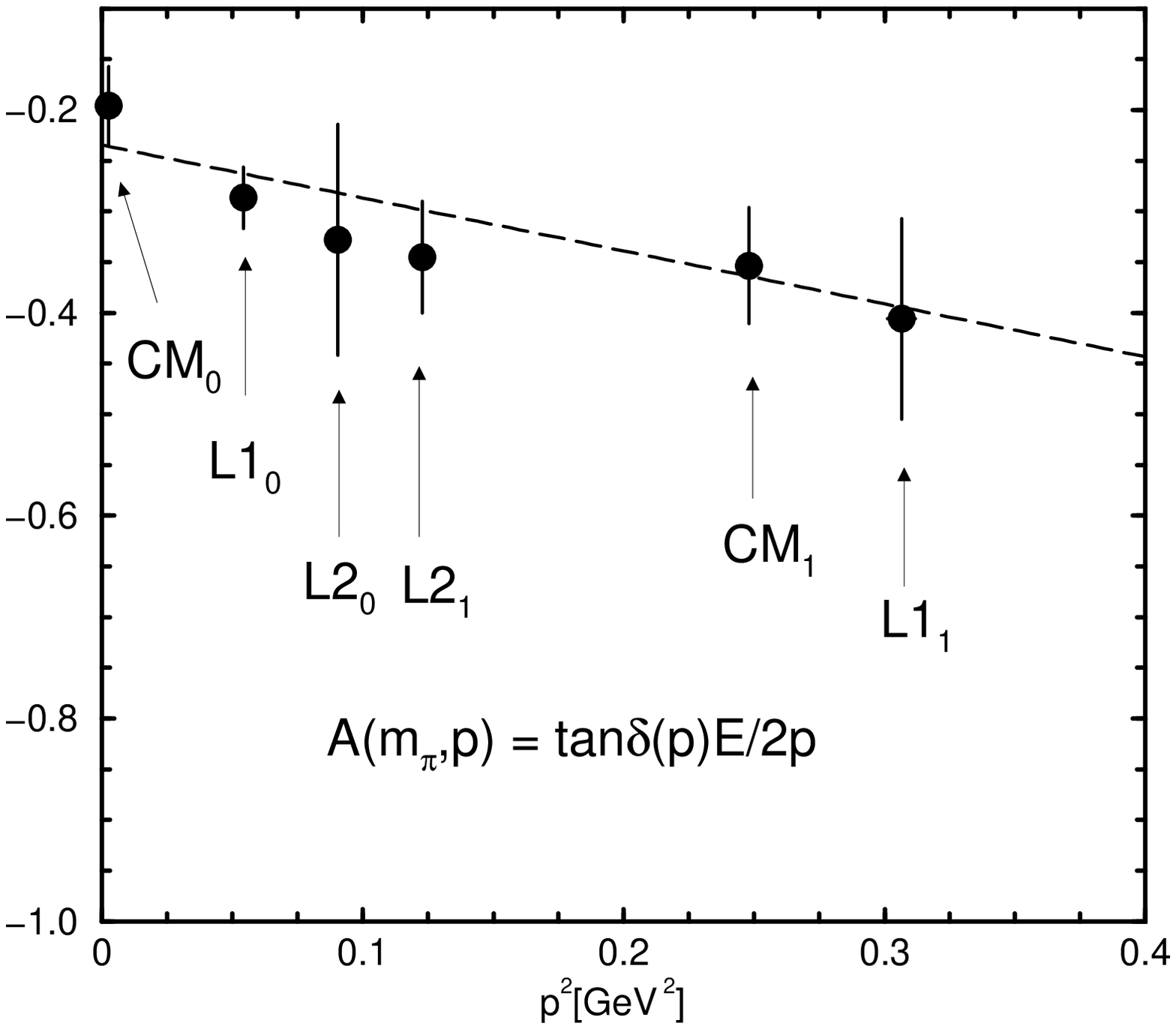}
& \leavevmode \epsfxsize=5.2cm \epsfbox{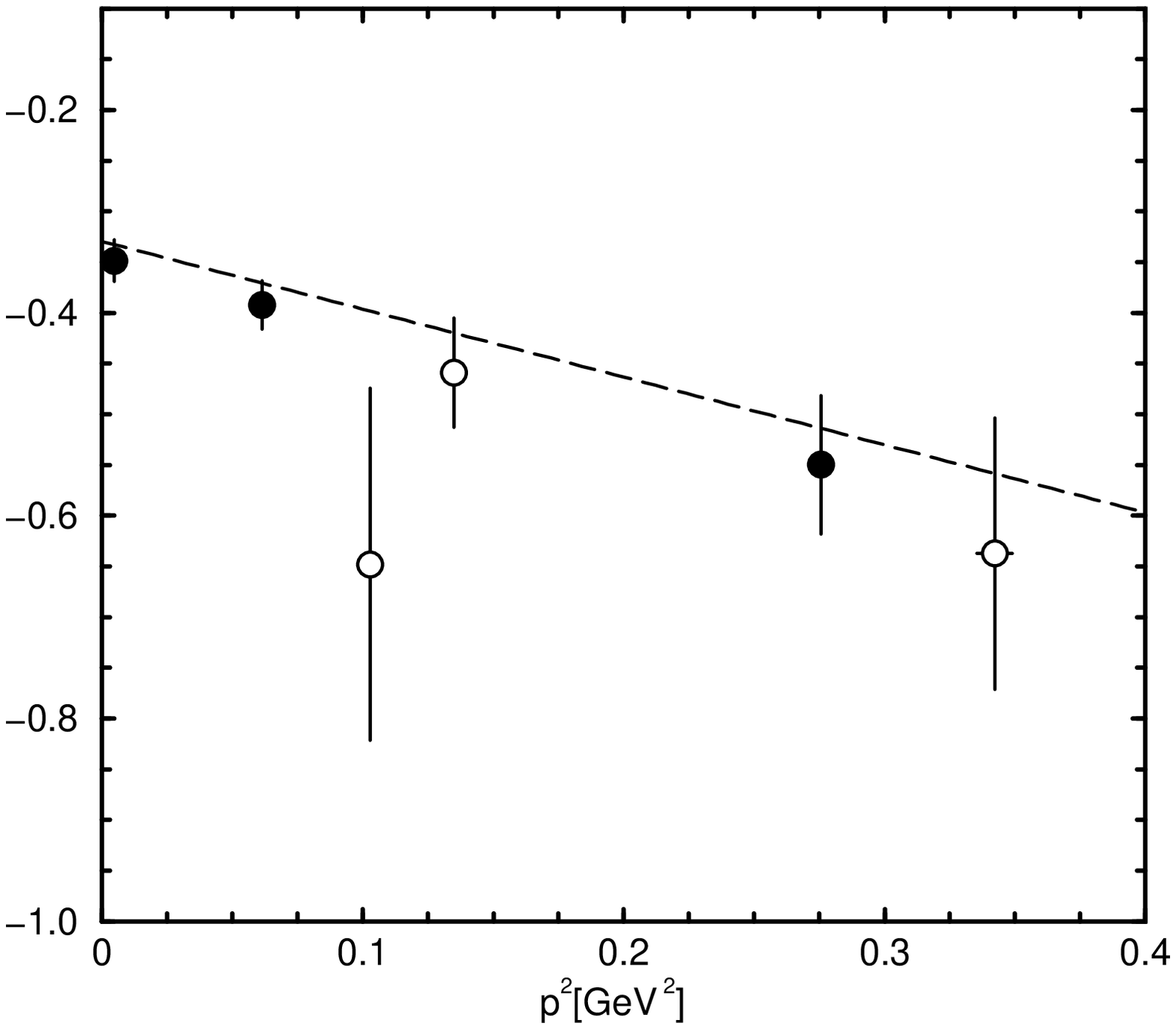}
& \leavevmode \epsfxsize=5.2cm \epsfbox{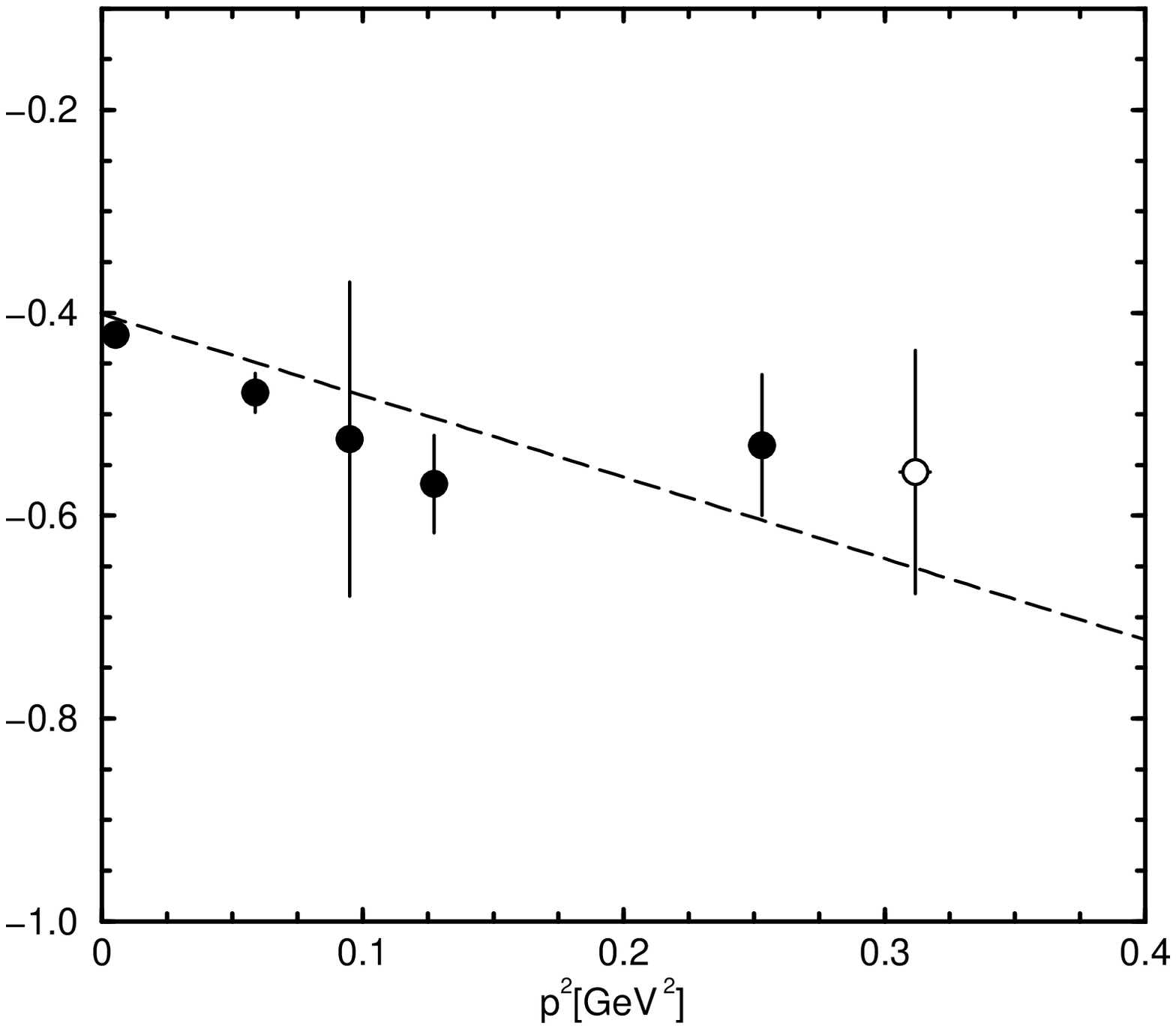}
\\
\raisebox{25mm}{ $\displaystyle{ \frac{m_{\pi}}{m_{\rho}} \approx 0.7 }$ }
& \leavevmode \epsfxsize=5.2cm \epsfbox{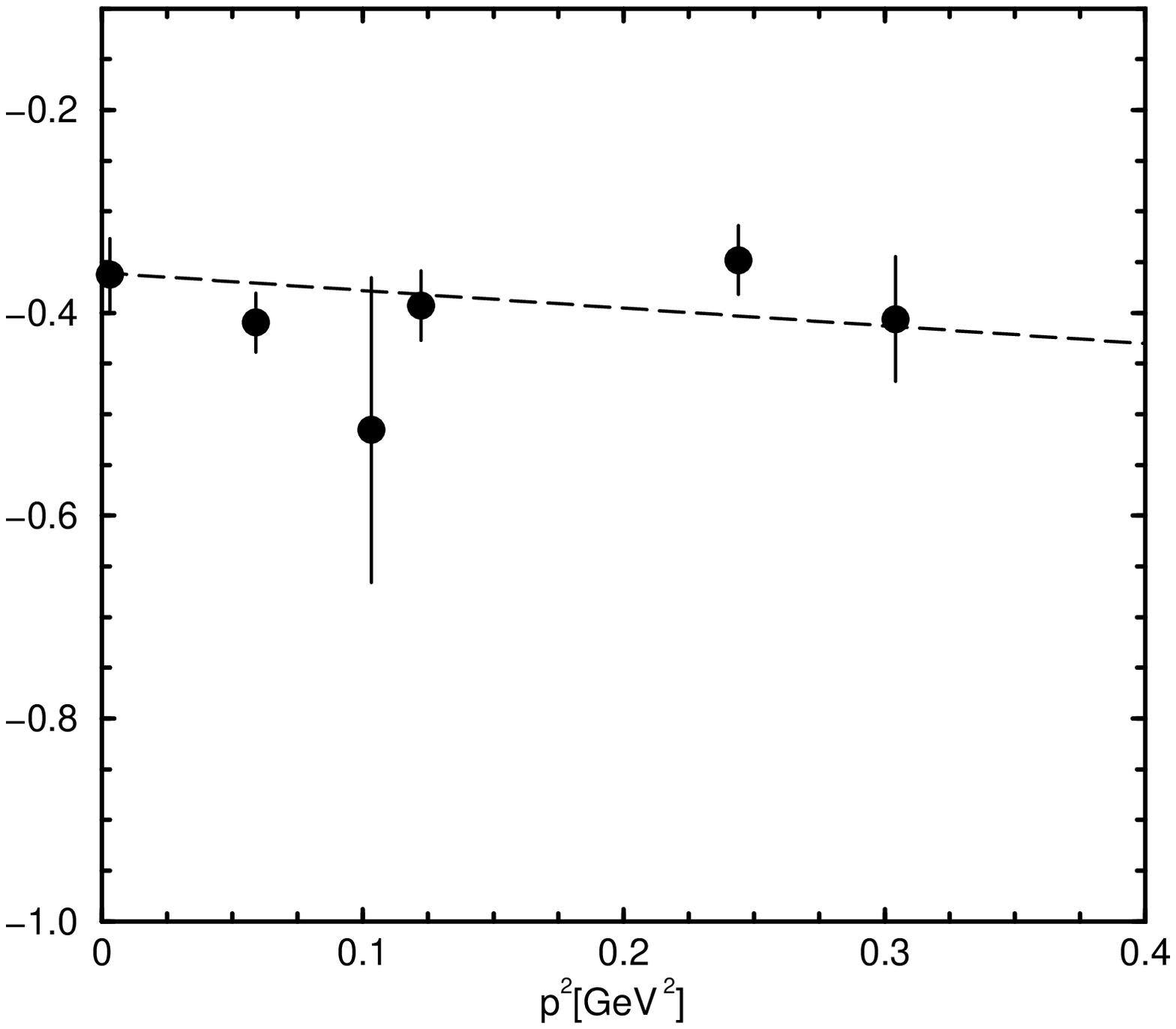}
& \leavevmode \epsfxsize=5.2cm \epsfbox{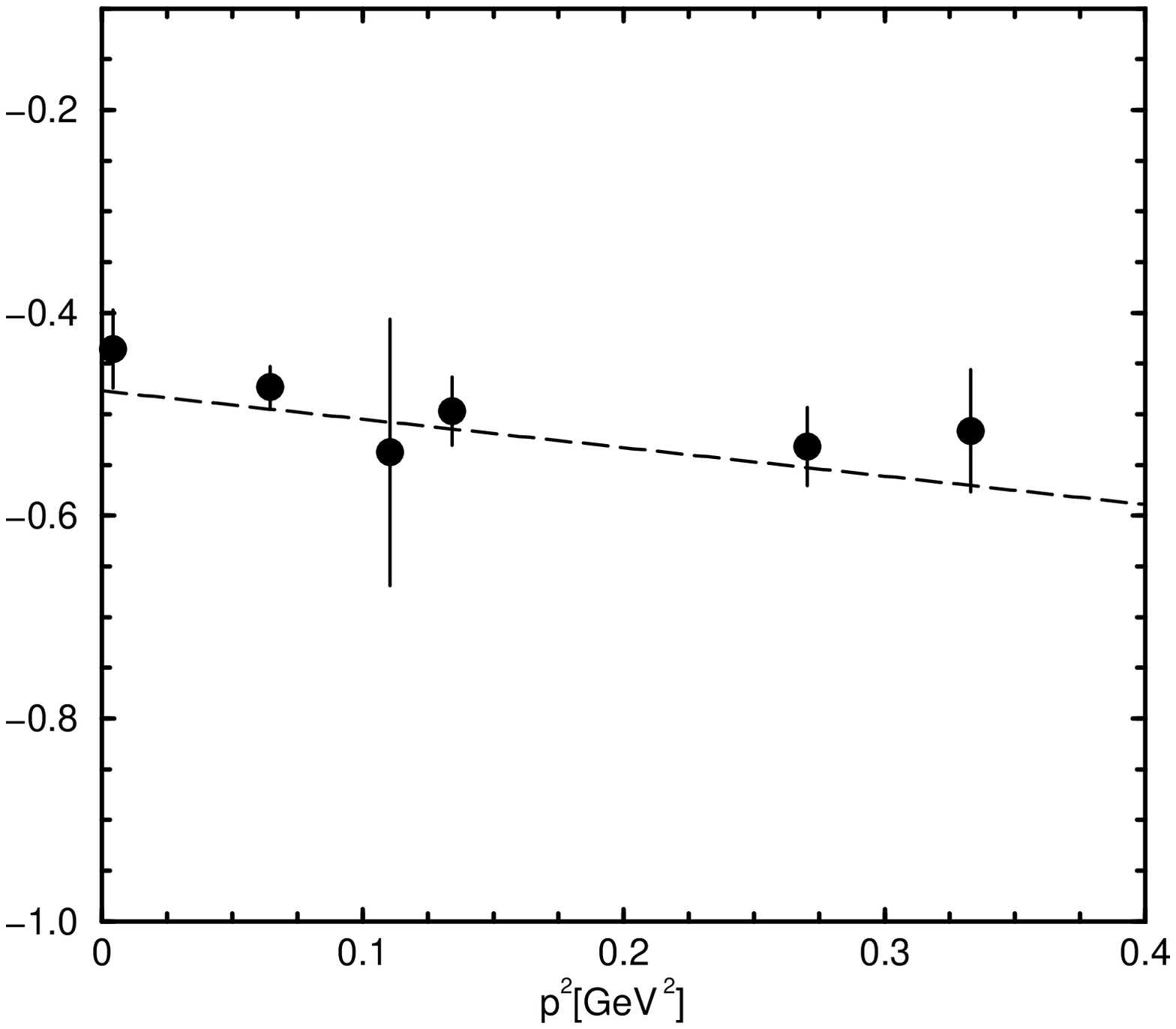}
& \leavevmode \epsfxsize=5.2cm \epsfbox{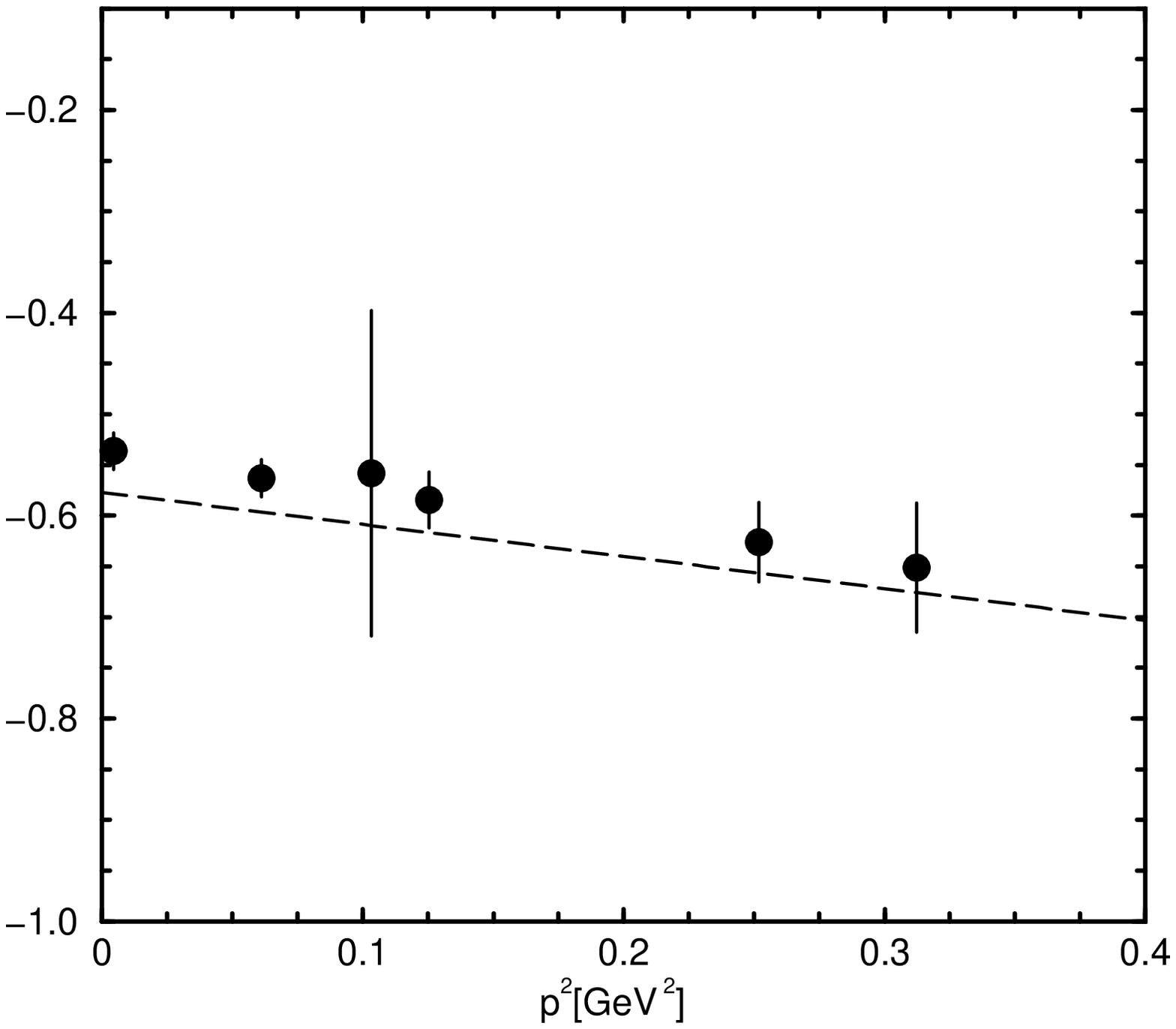}
\\
\raisebox{25mm}{ $\displaystyle{ \frac{m_{\pi}}{m_{\rho}} \approx 0.75 }$ }
& \leavevmode \epsfxsize=5.2cm \epsfbox{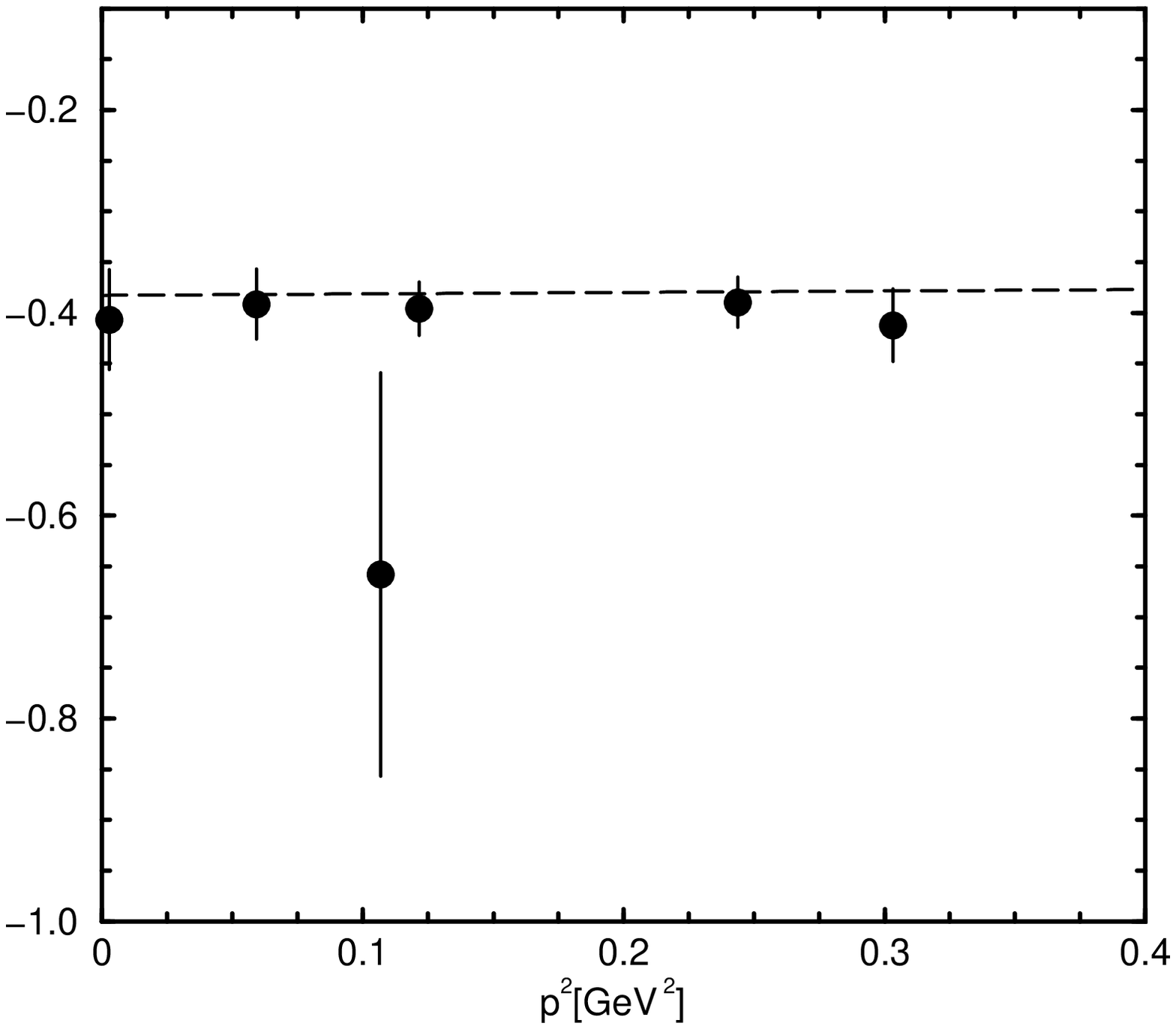}
& \leavevmode \epsfxsize=5.2cm \epsfbox{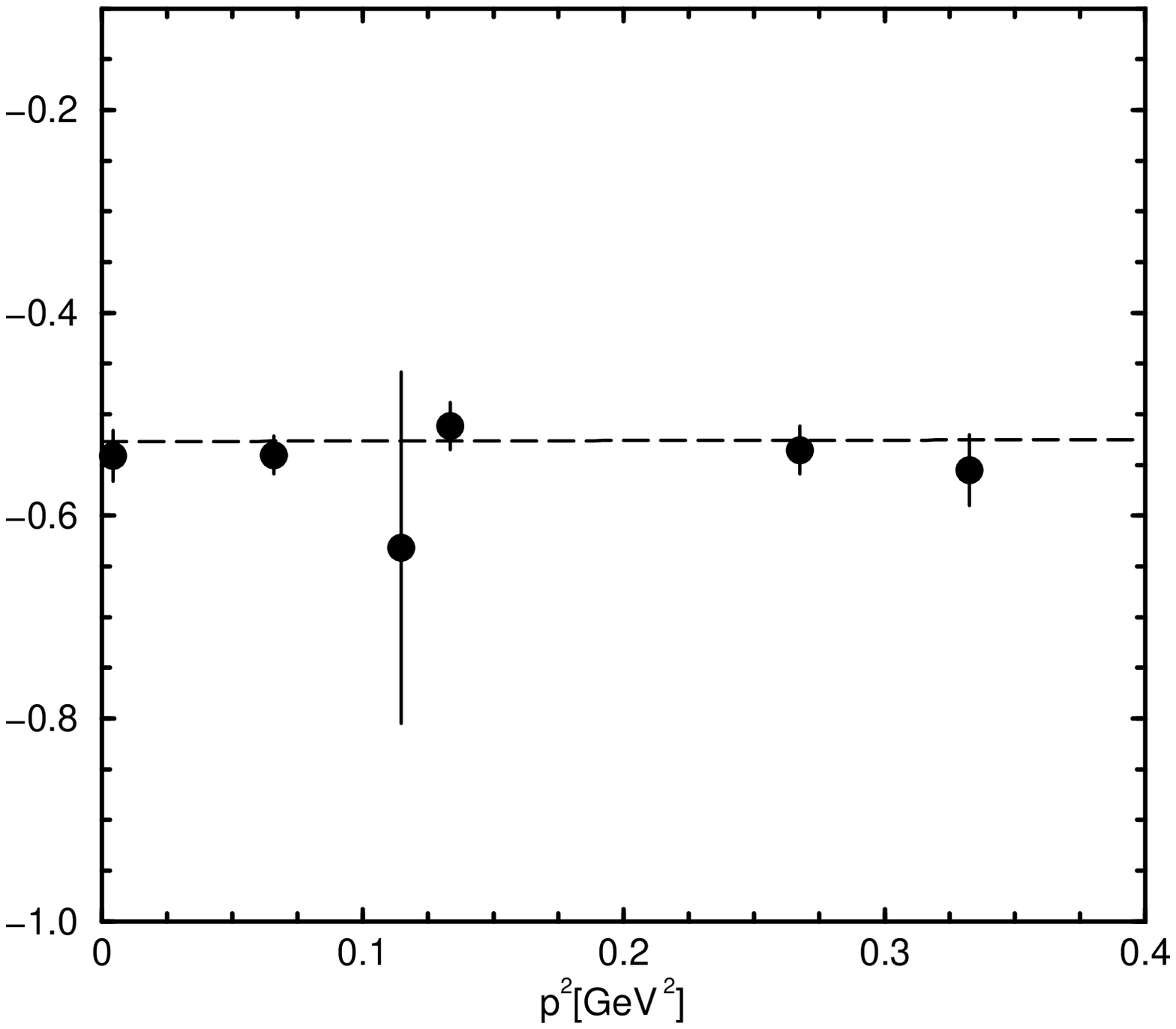}
& \leavevmode \epsfxsize=5.2cm \epsfbox{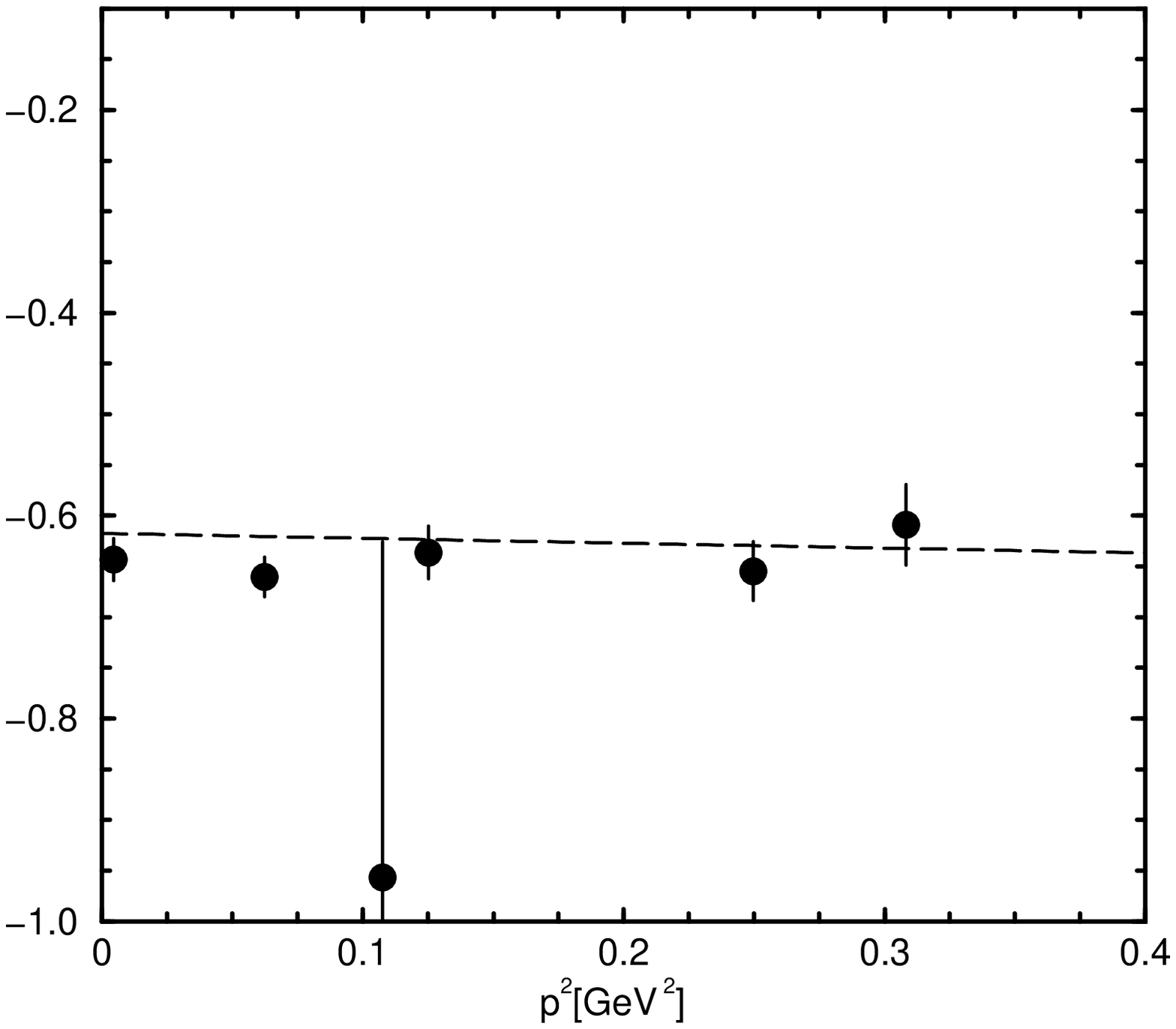}
\\
\raisebox{25mm}{ $\displaystyle{ \frac{m_{\pi}}{m_{\rho}} \approx 0.8 }$ }
& \leavevmode \epsfxsize=5.2cm \epsfbox{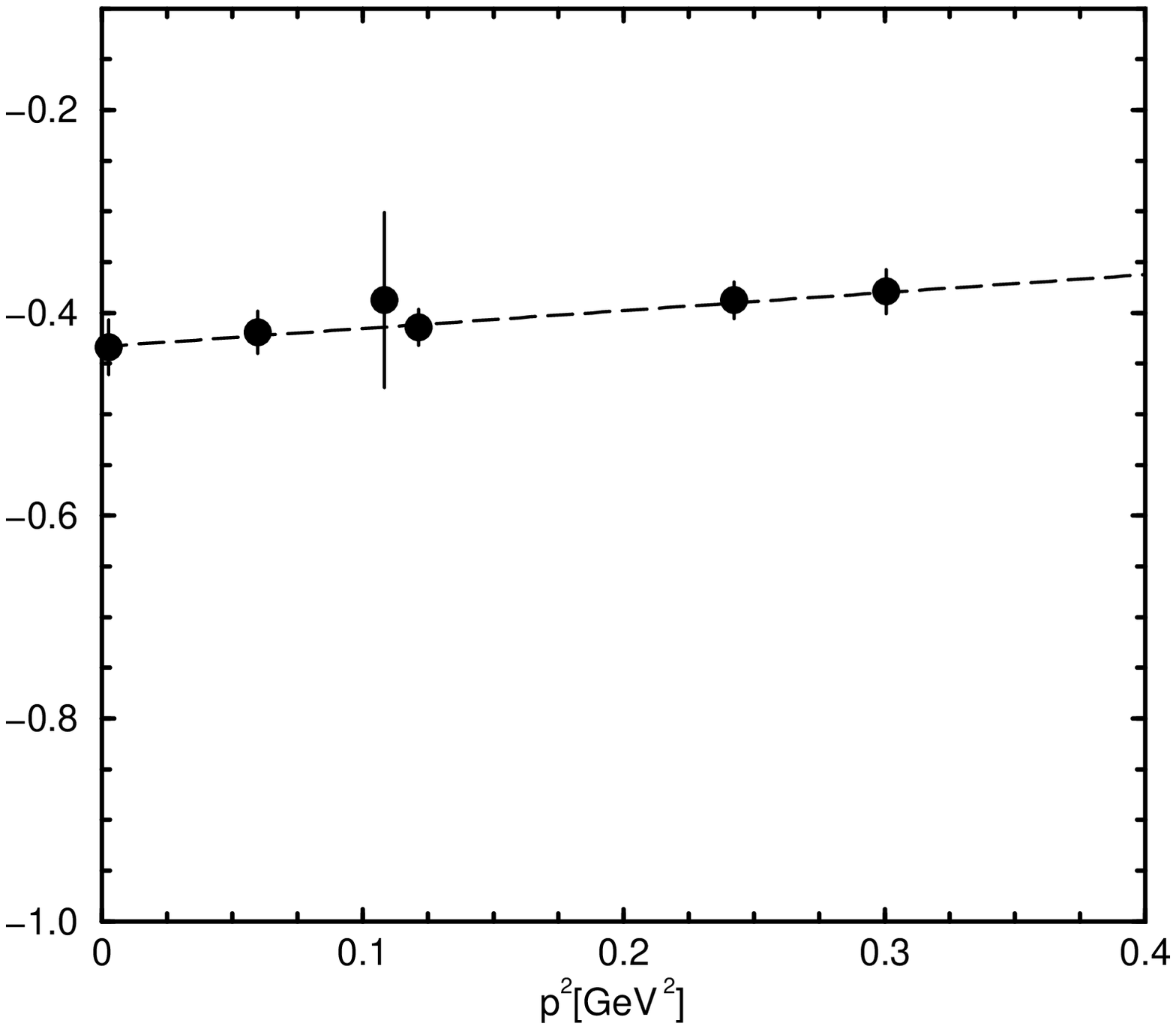}
& \leavevmode \epsfxsize=5.2cm \epsfbox{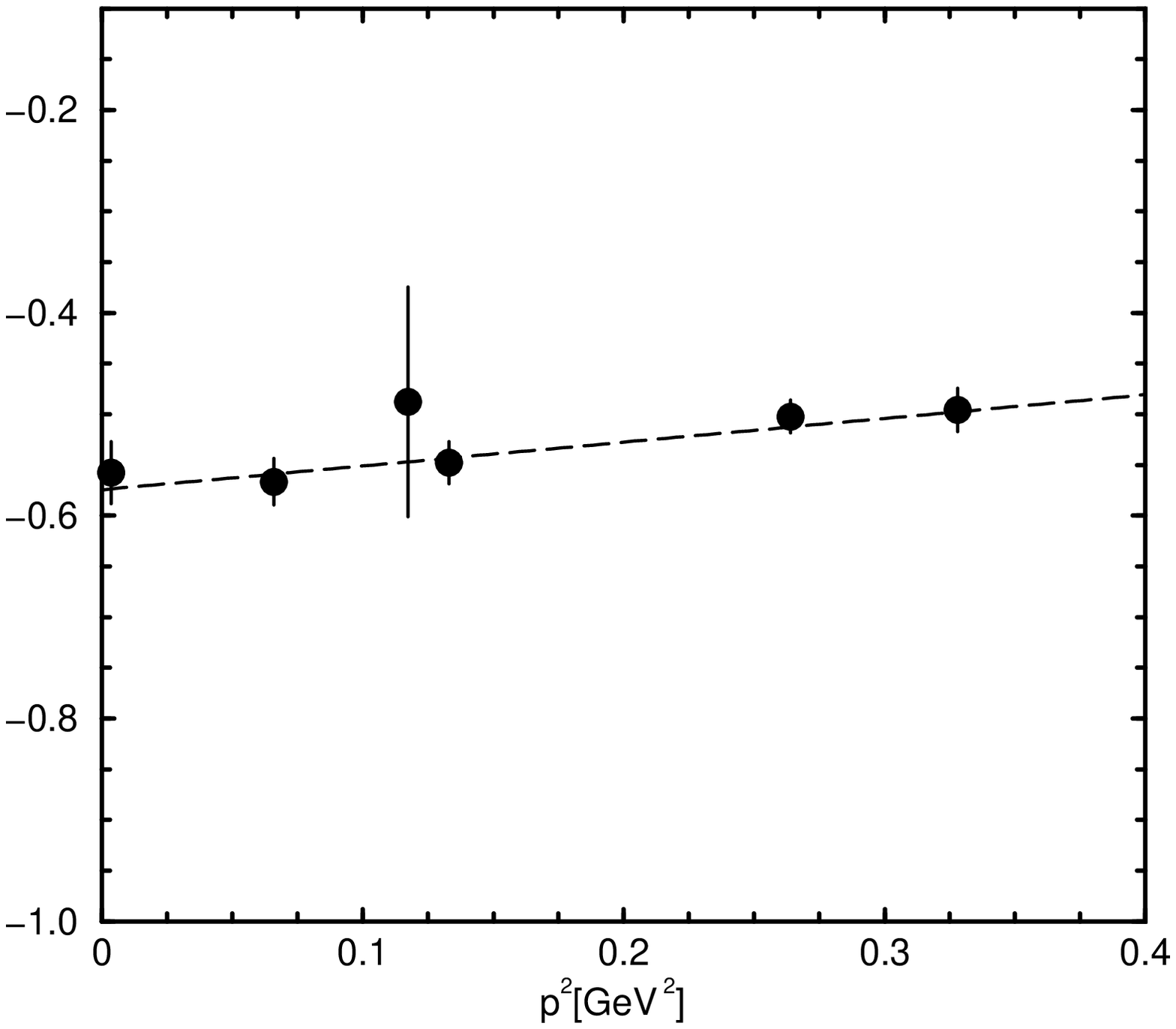}
& \leavevmode \epsfxsize=5.2cm \epsfbox{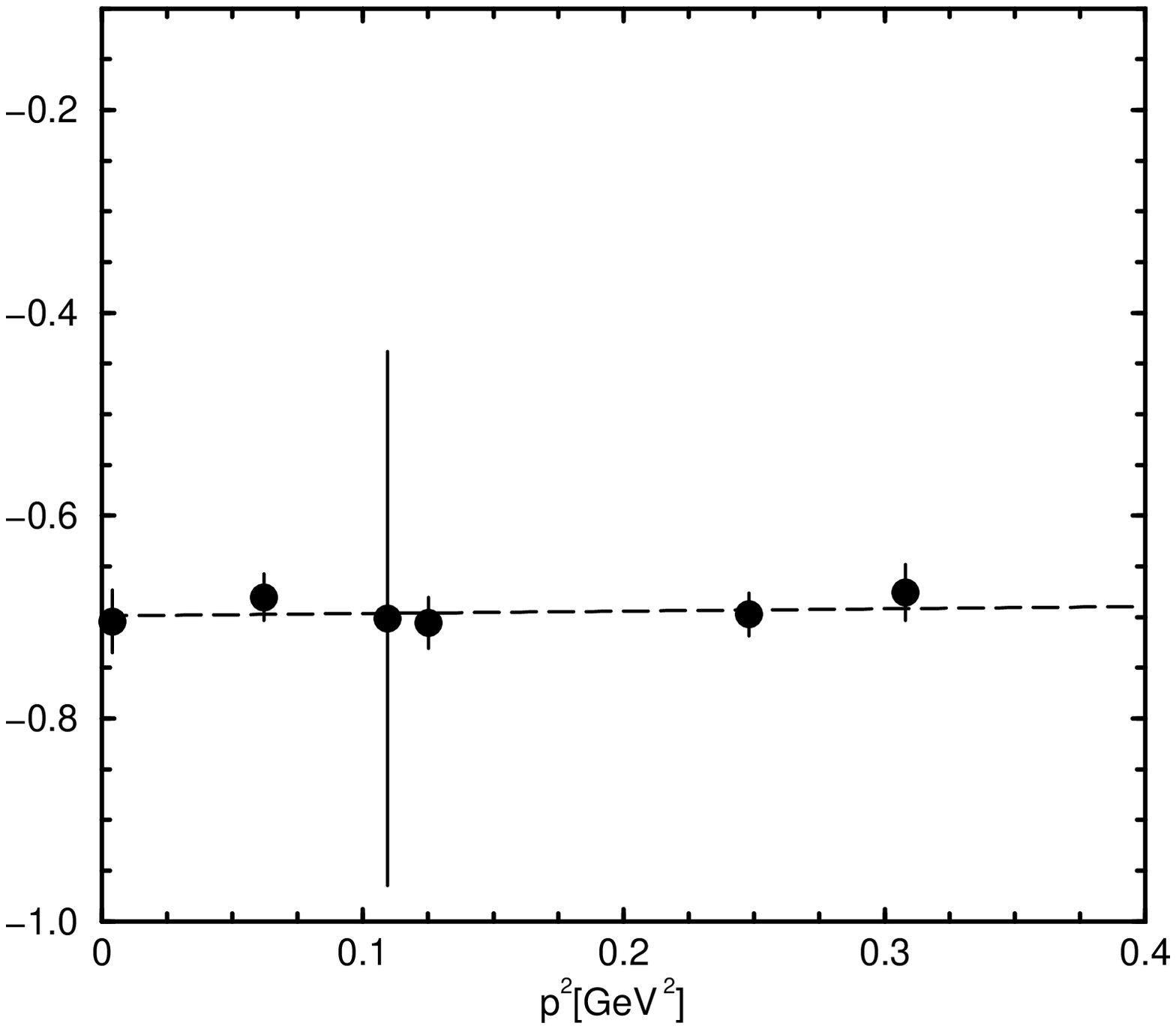}
\end{tabular}
\end{center}
\caption{
Scattering amplitude
$A( m_\pi, \overline{p} ) = \tan \delta( \overline{p} ) / \overline{p} \cdot \overline{E} / 2$.
CM$_n$ refers to the amplitude obtained from the $n$-th state in
the center of mass system, and L1$_n$ and L2$_n$ to those in the laboratory systems.
Open symbols are excluded from the global fit.
$m_\pi / m_\rho$ increases from top to bottom, 
while $\beta$ increases from left to right.
\label{fig:amp}
}
\end{figure}

The results for the scattering amplitudes $A( m_\pi, \overline{p} )$ 
defined by Eq.(\ref{eq:amplitude}) are plotted in Fig.~\ref{fig:amp}.
The CM$_n$ in the top left figure refers to the amplitude 
obtained from the $n$-th energy eigenstate state
in the CM system, 
and the L1$_n$ and L2$_n$ to those in each laboratory system. 
The ordering of the amplitude,
${\rm CM}_0$, ${\rm L1}_0$, ${\rm L2}_0$, $\cdots$, ${\rm L1}_1$, 
in the other figures is the same.
The open symbols are excluded from the following analysis.
They correspond to the data 
in which we cannot clearly observe a single exponential time behavior 
of $D_n(t)$ as noted in Sec.~\ref{Sec:Results:diag}.
The values of $A( m_\pi, \overline{p} )$
are tabulated in Appendix~\ref{tab:result}.

We obtain the scattering phase shift $\delta(\overline{p})$
at the physical pion mass for various momenta $\overline{p}$
in the continuum limit by the following procedure.
First we fit the results of $A( m_\pi, \overline{p} )$
with the following fitting assumption
at each lattice spacing,
\begin{equation}
    A( m_\pi, \overline{p} ) 
  = A_{10} m_\pi^2 
  + A_{20} m_\pi^4 
  + A_{30} m_\pi^6
  + A_{01} \overline{p}^2 
  + A_{11} m_\pi^2 \overline{p}^2 
  + A_{21} m_\pi^4 \overline{p}^2 .
\label{eq:polamp}
\end{equation}
Then we evaluate the amplitudes at the physical pion mass
for various momenta at each lattice spacing
from the constants $A_{ij}$ obtained by the fitting.
Finally the continuum extrapolation is taken for these amplitudes at each momentum.
The results of the fitting with the assumption Eq.(\ref{eq:polamp}) 
are plotted in Fig.~\ref{fig:amp},  
and the constants $A_{ij}$ are tabulated
together with $\chi^2/$d.o.f. in Table~\ref{tab:delta:fit-beta}.
The values of $\chi^2/$d.o.f. are reasonably small at all $\beta$ 
in contrast to the polynomial fit for the scattering length
where $\chi^2/$d.o.f. at $\beta=2.10$ is large 
as shown in Sec.~\ref{Sec:Results:sl}.

It is possible to make a global 
polynomial fit with the assumption Eq.(\ref{eq:polamp})
for  all momentum systems, 
so that we do not need to introduce the decay constant measured on the lattice 
as for the scattering length in the previous section.
For consistency of the analysis,
we also analyze $\hat{A}(m_\pi,\overline{p})$ defined by Eq.(\ref{eq:NOR_amplitude}).
Here we assume the same fitting assumption Eq.(\ref{eq:polamp}), 
but set $A_{20}=A_{30}=A_{21}=0$,
since the mass dependence becomes mild
by a compensation with that of the decay constant also in this case.
The fit results are tabulated in Table~\ref{tab:delta:f-fit-beta}.

\begin{figure}
\begin{center}
\begin{tabular}{c}
\leavevmode \epsfxsize=7cm\epsfbox{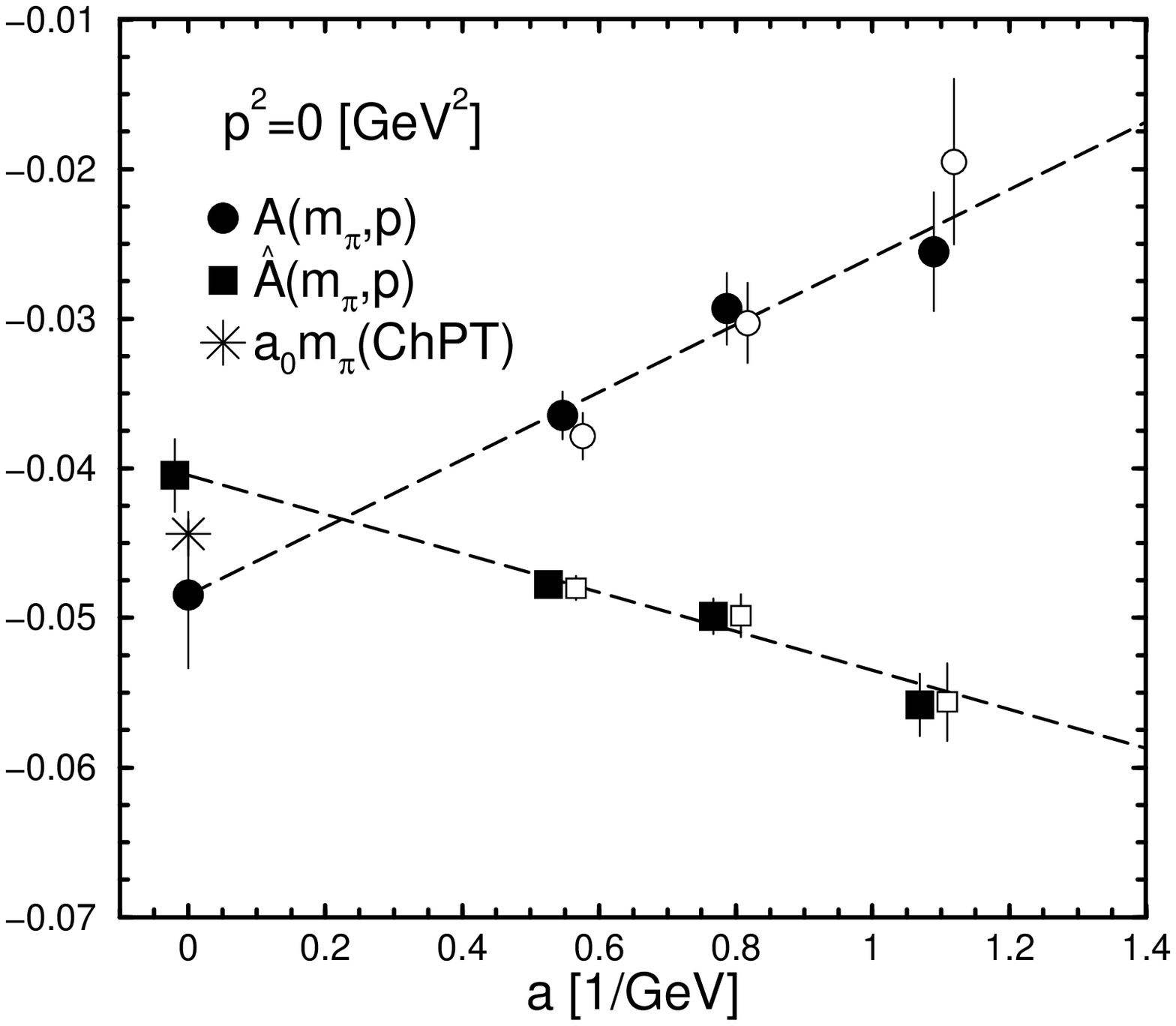} \\
\leavevmode \epsfxsize=7cm\epsfbox{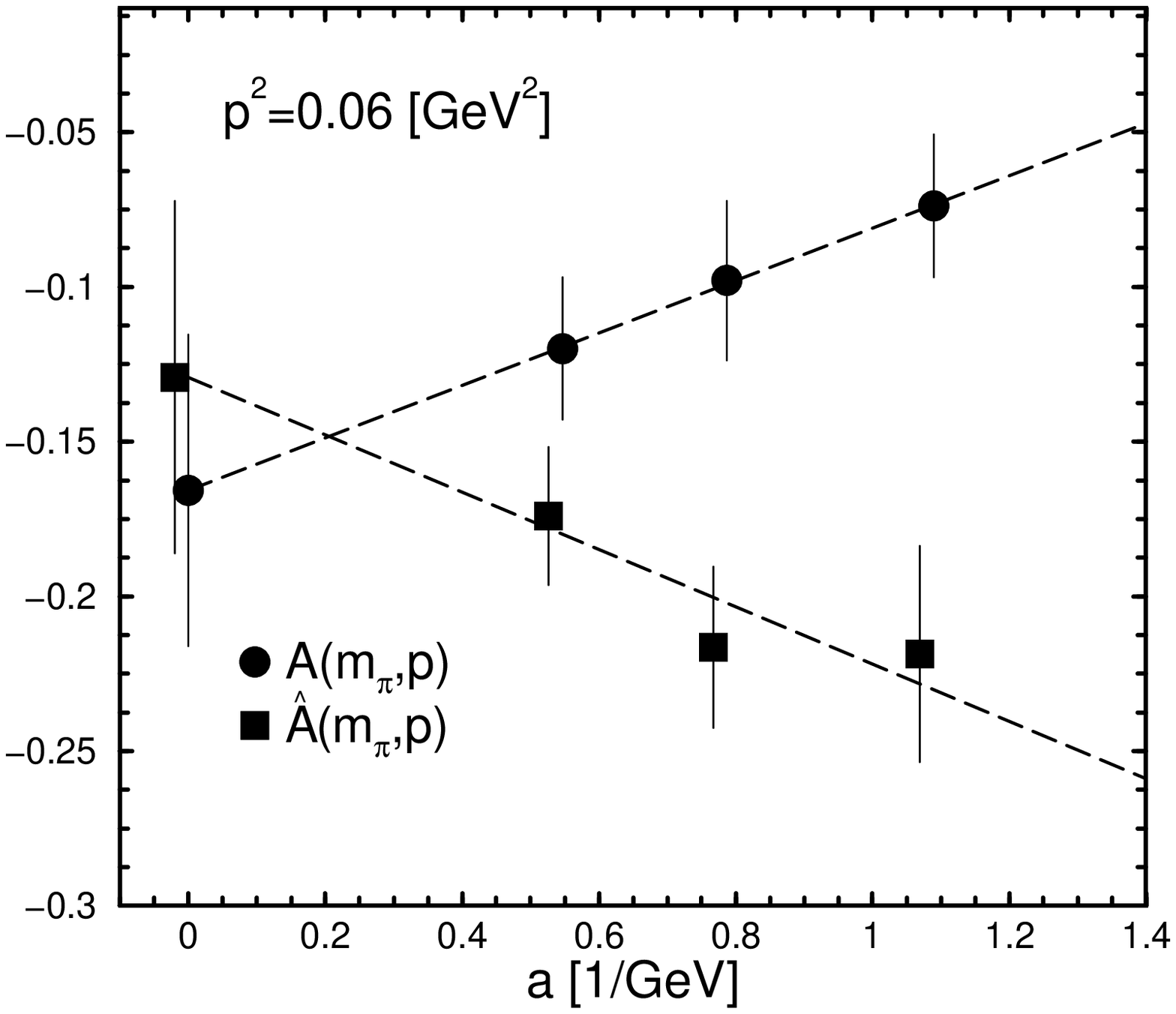} \\
\leavevmode \epsfxsize=7cm\epsfbox{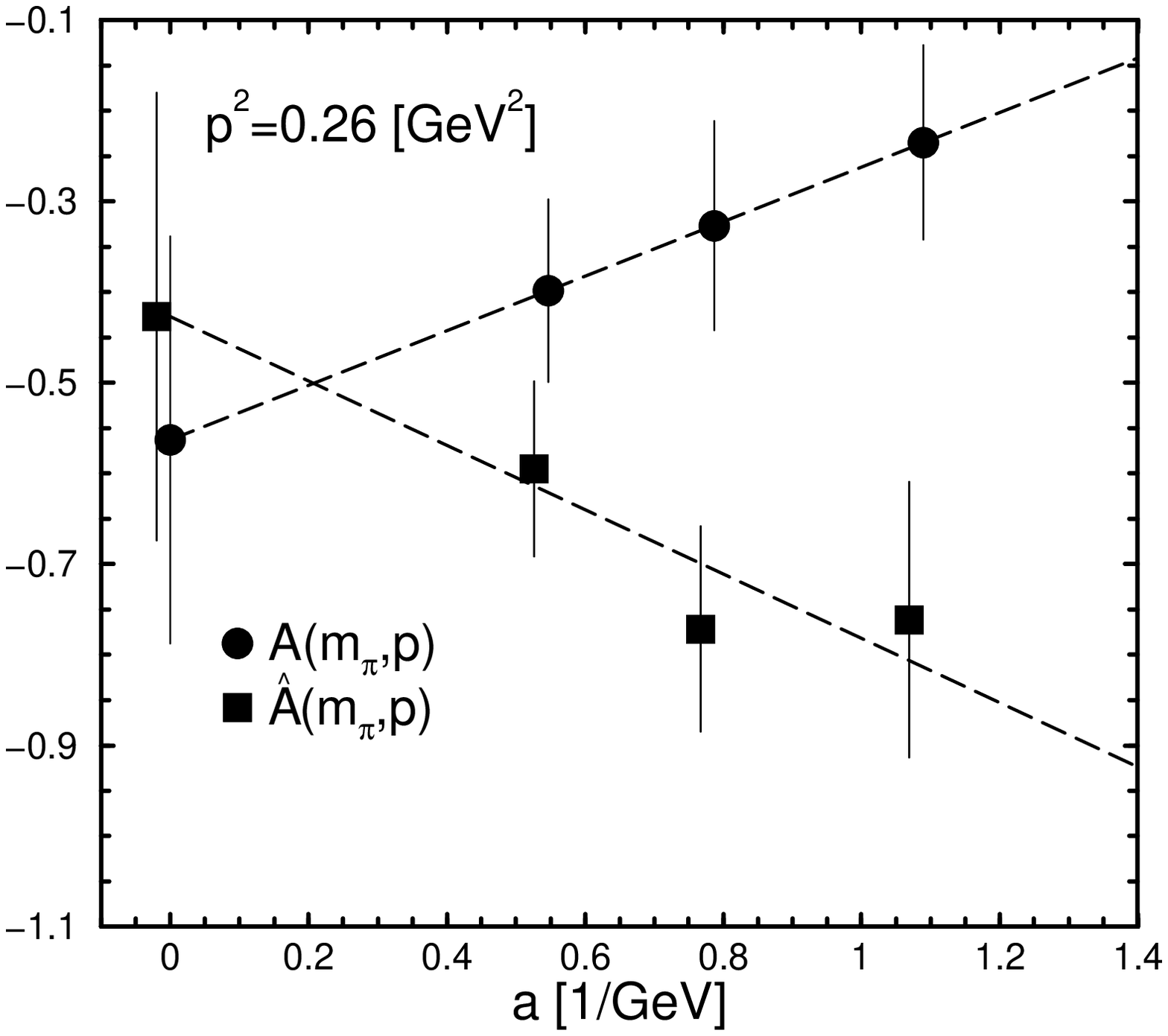}
\end{tabular}
\end{center}
\caption{
Scattering amplitudes
$A( m_\pi, \overline{p} ) = \tan \delta( \overline{p} ) / \overline{p} \cdot \overline{E} /2$ 
and 
$\hat{A}(m_\pi, \overline{p} ) = ( f_\pi^{lat} / f_\pi )^2 \cdot  A( m_\pi, \overline{p} )$
at the physical pion mass and several momenta.
The pseudoscalar decay constant $f_\pi^{lat}$ is measured on lattice and $f_\pi = 93$ MeV.
Open symbols are $a_0 m_{\pi}$ and $\hat{a}_0 m_{\pi}$ taken from Fig.~\ref{fig:a0:fit.cont}.
\label{fig:amp:cont}
}
\end{figure}

In Fig.~\ref{fig:amp:cont}
both amplitudes at the physical pion mass for various momenta
are plotted as a function of the lattice spacing, 
together with the continuum extrapolations.
Here we choose three momenta,
$\overline{p}^2=0$, $0.06$ and $0.26\ {\rm GeV}^2$ 
which are roughly equal to the momenta of CM$_0$, L1$_0$ and CM$_1$, respectively.
The result at $\overline{p}^2=0$ gives the scattering length $a_0 m_\pi$.
We also plot the scattering lengths obtained in the previous section,
which are calculated from the data only in CM$_0$.
As shown in the figure they are consistent within the statistical error 
at each lattice spacing.
There are slight differences, however, in the continuum limit between 
the results obtained from all momentum system at $\overline{p}^2=0$ given by
\begin{eqnarray}
   &&  a_0 m_{\pi}       = - 0.0484(49),  \\ 
   &&  \hat{a}_0 m_{\pi} = - 0.0404(24),
\end{eqnarray}
and the results from only the CM$_0$ data given by Eqs.(\ref{eq:a_0}) and (\ref{eq:a_0^f}).
We consider that these discrepancies represent 
the uncertainty of the continuum extrapolations 
using data at only three lattice spacings,
{\it e.g.}, the number of degrees of freedom for the extrapolation is one,
far from the continuum limit.
The difference of the two amplitudes tends to vanish in the continuum limit.
As mentioned in the previous section, however, 
there is an uncertainty 
in the determination of the decay constant on the lattice.
Thus we do not use $\hat{A}(m_\pi,\overline{p})$ 
to obtain the final results of the phase shift.

\begin{figure}
\begin{center}
\leavevmode
\epsfxsize=7cm\epsfbox{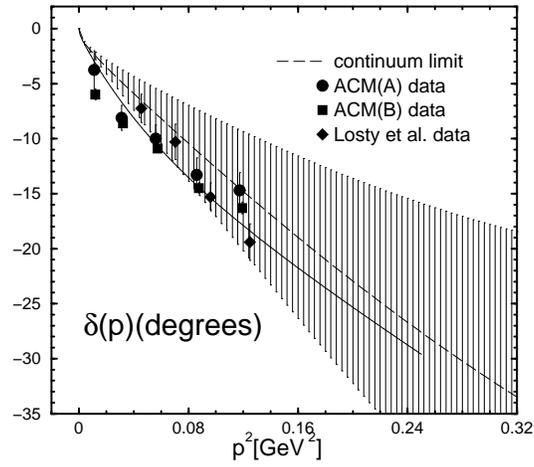}
%
\end{center}
\caption{
Results of the scattering phase shifts in the continuum limit (dashed line)
and a band of error bars.
Solid line~\protect\cite{cola} is estimated with experimental inputs using the Roy equation.
Symbols represent data of Aachen-Cern-Munich Collaboration~\protect\cite{ACM} 
and of Losty {\it et al.}~\protect\cite{Losty}.
\label{fig:delta:cont}
}
\end{figure}

In Fig.~\ref{fig:delta:cont}
the result of the phase shift obtained from the scattering amplitude 
$A(m_\pi,\overline{p})$ in the continuum limit
is presented by a dashed line, and associated by a band of error bars.
The values of the phase shift at several momenta are tabulated in Table~\ref{tab:phsh}.
Our results are compared 
with the solid curve~\cite{cola} estimated with the experimental input,
and the experimental results~\cite{ACM,Losty}.
The result in the continuum limit agrees with experiment,
although the errors of our result are large.
In order to obtain more precise results
simulations are needed
closer to the chiral and the continuum limits
with much higher statistics.
%
%
\section{Conclusions}
\label{Sec:Conclusions}
 In this article we have presented our results for the I=2 S-wave
$\pi\pi$ scattering phase shift in the continuum limit calculated with
two-flavor dynamical quark effects.  While errors are not small, it is
very encouraging to find that the phase shift in the continuum limit
shows a reasonable agreement with experiment.  

 The large errors of our final results arise from the chiral
extrapolation and the continuum extrapolation. In order to obtain more
precise results, simulations are needed closer to the chiral and the
continuum limits with much higher statistics. Technically, investigating
the correlation we found between the scattering length and the decay
constant measured on lattice, and detecting effects of chiral symmetry
breaking with Wilson fermion action are important open issues for future
work.

 In addition to being a first calculation of realistic scattering
quantity based on first principles of QCD, the importance of the present
work resides in actually showing that various technical methods, such as
diagonalization of pion four-point functions and use of laboratory
systems, necessary for practical success of the finite-volume methods
work.  Thus we can envisage a perspective toward extension of the
present work for calculations of other important scattering process,
such as $I=1$ and $I=0$ two-pion systems, systems including
unstable particles, and scattering with baryons, which are richer in
physics content. These processes are more difficult from the point of
calculations, however, and algorithmic advances are probably needed to
evaluate complicated diagrams.
%
%
\section*{Acknowledgments}
This work is supported in part by Grants-in-Aid of the Ministry of Education
(Nos.
12304011, 
12640253, 
13135204, 
13640259, 
13640260, 
14046202, 
14740173, 
15204015, 
15540251, 
15540279, 
15740134  
).
The numerical calculations have been carried out 
on the parallel computer CP-PACS.
%
%
%
\appendix
\section{ Calculation method of zeta function }
\label{appe:zeta}
In this appendix 
we introduce methods for the numerical evaluation 
of the zeta function $Z_{ 00 }^{\bf d}(s;\overline{m})$ defined in Eq.(\ref{eq:Z00d}).
This problem has already been discussed by L\"uscher in Appendix A of Ref.~\cite{FVM:L:1}
and in Appendix C of Ref.~\cite{FVM:L:2} for the center of mass system,
and by Rummukainen and Gottlieb in Sec. 5.2 of Ref.~\cite{FVM:RG}
for general systems.
Our basic idea is the same,
but our final expression for the zeta function is simpler and
more efficient for numerical evaluations.
(We find that there are some typographical errors 
in the expression of the zeta function in Ref.~\cite{FVM:RG}).
Very recently Li and Liu have reported a similar calculation method of the zeta function 
in asymmetric box~\cite{Li-Liu}.

The definition of the zeta function $Z_{ 00 }^{\bf d}(s;\overline{m})$ is
\begin{equation}
    \sqrt{ 4\pi } \cdot Z^{ \bf d }_{ 00 } ( s ; \overline{m } ) 
   = 
     \sum_{ { \bf r } \in P^{\bf d} } ( r^2 - \overline{m} )^{ -s } .
\label{eq:Z00d_appendix}
\end{equation}
The summation for ${\bf r}$ is carried out over the set 
\begin{equation}
  P^{\bf d} = \{ {\bf r} | {\bf r}= \hat{\gamma}^{ -1 } ( {\bf n} + {\bf d}/2 ), {\bf n} \in Z^3 \} .
\end{equation}
The operation $\hat{ \gamma }^{ -1 }$ is the inverse Lorentz transformation : 
$\hat{ \gamma }^{ -1 } {\bf n} = 1 / \gamma \cdot { \bf n }_{||} + {\bf n}_{\bot}$
where $ {\bf n}_{||} = ( {\bf n} \cdot {\bf d} ) {\bf d} / d^2$
is the parallel component and 
${\bf n}_{\bot} = {\bf n} - {\bf n}_{||}$ 
the perpendicular component of ${\bf n}$ in the direction ${\bf d}$.
The zeta function $Z_{ 00 }^{\bf d} ( s ; \overline{m } )$ takes a finite value
for ${\mathrm Re}\,s > 3/2 $,
and $Z_{ 00 }^{\bf d} ( 1 ; \overline{m } )$, 
which is used to obtain the scattering phase shift in Eq.(\ref{eq:FVM:RG}),
is defined by the analytic continuation from the region ${\mathrm Re}\,s > 3/2$.

First we divide the summation in $ Z_{ 00 }^{\bf d} ( s ; \overline{m} )$ into two parts as
\begin{equation}
    \sum_{\bf r } ( r^2 - \overline{m} )^{ -s }
  = \sum_{ r^2 < \overline{m} }  ( r^2 - \overline{m} )^{ -s }
  + \sum_{ r^2 > \overline{m} }  ( r^2 - \overline{m} )^{ -s },
\label{eq:Z_L:1}
\end{equation}
where the summation over ${\bf r }$ is carried out with ${\bf r }\in P^{ \bf d }$.
The second term can be written in an integral form as follows,
\begin{eqnarray}
   \sum_{ r^2 > \overline{m} } ( r^2 - \overline{m} )^{ -s }
& = &
   \frac{ 1 }{ \Gamma ( s ) }
   \sum_{ r^2 > \overline{m } } \   
   \int_0^{\infty}   \ {\rm d}t \ t^{ s - 1 } e^{ - t ( r^2 - \overline{m } ) }  \cr
& = &
   \frac{ 1 }{ \Gamma ( s ) }
   \sum_{ r^2 > \overline{m } } \ 
   \left[
   \int_0^1          \ {\rm d}t \ t^{ s - 1 } e^{ - t ( r^2 - \overline{m } ) }
 +
   \int_1^{ \infty } \ {\rm d}t \ t^{ s - 1 } e^{ - t ( r^2 - \overline{m } ) }
   \right] \cr
& = &
   \frac{ 1 }{ \Gamma ( s ) }
   \int_0^1          \ {\rm d}t \ t^{ s - 1 } e^{ t \overline{m } } \ 
   \sum_{ \bf r } e^{ -t r^2 }
 -
   \sum_{ r^2 < \overline{m } } ( r^2 - \overline{m } )^{ -s }
 +
   \sum_{ {\bf r } }  
   \frac{ e^{ -( r^2 - \overline{m } ) } }{ ( r^2 - \overline{m } )^s }.
\label{eq:Z_L:2}
\end{eqnarray}
The second term cancels out the first term in Eq.(\ref{eq:Z_L:1}).
Next we rewrite the first term in Eq.(\ref{eq:Z_L:2}) by the Poisson's summation formula
\begin{equation}
    \sum_{ {\bf n } \in Z^3 } f( {\bf n } ) 
  = \sum_{ {\bf n } \in Z^3 } \int {\rm d}^3x f( {\bf x } ) e^{ i 2 \pi { \bf n \cdot x } },
\end{equation}
and integrating over ${\bf x}$ yields,
\begin{eqnarray}
  \frac{ 1 }{ \Gamma ( s ) }
  \int_0^1 \ {\rm d}t \ t^{ s - 1 } e^{ t \overline{m } }
  \sum_{ \bf r } e^{ -t r^2 }
& = &
  \frac{ \gamma }{ \Gamma ( s ) }
  \int_0^1 \ {\rm d}t \ t^{ s - 1 } e^{ t \overline{m } }
  \left( \frac{ \pi }{ t } \right)^{ 3 / 2 } \ 
  \sum_{ {\bf n } \in Z^3 } 
     (-1)^{ \bf n \cdot d }
     e^{ \pi^2 ( \hat{ \gamma }{\bf n} )^2 / t } .
\label{eq:Z_L:3}
\end{eqnarray}
The divergence at $s=1$ comes from the ${\bf n} = {\bf 0}$ part of the integrand on the right-hand side.
We divide the integrand into a divergent part (${\bf n} = {\bf 0}$) 
and a finite part (${\bf n} \ne {\bf 0}$).
The divergent part can be evaluated for $ {\mathrm Re}\,s > 3 / 2 $ as
\begin{equation}
  \int_0^1 \ {\rm d}t \ t^{ s - 1 } 
             e^{ t \overline{m } } \left( \frac{ \pi }{ t } \right)^{ 3 / 2 }
  =
  \sum_{ l = 0 }^{ \infty } 
  \frac{ \pi^{ 3 / 2 } }{ s + l - 3 / 2 }
  \frac{ \overline{m }^l }{ l ! }.
\label{eq:Z_L:4}
\end{equation}
The right had side of this equation takes a finite value at $s=1$.

Finally by gathering all terms 
we obtain the following expression for the zeta function at $s=1$,
\begin{eqnarray}
  \sqrt{ 4\pi } \cdot Z_{ 00 }^{\bf d} ( 1 ; \overline{m } ) 
& = &
    \sum_{ \bf r }
    \frac{ e^{ -( r^2 - \overline{m } ) } }{ r^2 - \overline{m } } \cr
&&
    +
    \gamma
    \int_0^1 \ {\rm d}t \ e^{ t \overline{m } } 
    \left( \frac{ \pi }{ t } \right)^{ 3 / 2 } \ 
    \sum_{ {\bf n } \in Z^3 } \!\!\rule{0mm}{1em}^{ ' }    
      (-1)^{ {\bf n} \cdot {\bf d} }   
      e^{ \pi^2 ( \hat{ \gamma }{\bf n } )^2 / t }
+
    \gamma \sum_{ l = 0 }^{ \infty } 
    \frac{ \pi^{ 3 / 2 } }{ l - 1 / 2 }
    \frac{ \overline{m }^l }{ l ! } ,
\label{eq:Z_L:s=1}
\end{eqnarray}
where $\sum_{ {\bf n} \in Z^3 }^{'}$ is the summation without ${\bf n} = {\bf 0}$.

Substituting ${\bf d} = {\bf 0}$ and $\gamma = 1$ into the above expression,
we obtain the representation of the zeta function 
in the center of mass system appeared in Eq.(\ref{eq:FVM:L})
\begin{eqnarray}
  \sqrt{4\pi} \cdot Z_{00}( 1 ; \overline{m } )
& = &
    \sum_{ {\bf n } \in Z^3 }
    \frac{ e^{ -( n^2 - \overline{m } ) } }{ n^2 - \overline{m } }
+
    \int_0^1 \ {\rm d}t \ e^{ t \overline{m } } 
    \left( \frac{ \pi }{ t } \right)^{3/2}  \ 
    \sum_{ {\bf n } \in Z^3 } \!\!\rule{0mm}{1em}^{ ' } \
    e^{ \pi^2 n^2 / t }
+
    \sum_{ l = 0 }^{ \infty } 
    \frac{ \pi^{3/2} }{ l - 1 / 2 }
    \frac{ \overline{m }^l }{ l ! } .
\label{eq:Z_C:s=1}
\end{eqnarray}
%
%
\section{Table for Results of scattering length and scattering phase shift in each system}
\label{tab:result}
In Tables~\ref{tab:Res:CM_0} -- \ref{tab:Res:L2_1} 
we tabulate fitting ranges, energy shift $\Delta \overline{ E }_n^{\bf P}$, 
center of mass momentum $\overline{p}^2$,
Lorentz boost factor $\gamma$, 
scattering phase shift $\delta( \overline{p} )$,
scattering amplitude $A( m_\pi, \overline{p} )$ defined by Eq.(\ref{eq:amplitude}), 
and normalized scattering amplitude $\hat{A}( m_\pi, \overline{p} )$ 
defined by Eq.(\ref{eq:NOR_amplitude})
in each system for the ground $n=0 $ and first excited $n=1$ states.  
%
%
\setcounter{table}{11}
\begin{table}
\begin{center}
\begin{tabular}{lccccc}
$\beta = 1.80$ & $\kappa$              & 0.1464       & 0.1445       & 0.1430       & 0.1409       \\
$a^{-1} = 0.9176(93)$ [GeV] 
               & $m_{\pi}/m_{\rho}$    & 0.547(4)     & 0.694(2)     & 0.753(1)     & 0.807(1)     \\
               & $m_{\pi}^2$ [GeV$^2$] & 0.238(1)     & 0.571(1)     & 0.814(1)     & 1.128(1)     \\ \hline
Fitting Range  &                       & 10   --   20 & 12   --   20 & 12   --   20 & 12   --   20 \\
$\Delta \overline{E}_n^{\bf P}$        & [$ \times 10^{-3} $ GeV]
                                       & 4.61(98)     & 3.62(39)     & 2.84(37)     & 2.17(14)     \\
$\overline{p}^2_n $                    & [$\times 10^{-4}$ GeV$^2$]
                                       & 24.5(52)     & 29.9(32)     & 27.9(37)     & 25.1(17)     \\
$\delta ( \overline{p}_n )$            & [degrees]
                                       & $-$1.13(34)  & $-$1.49(22)  & $-$1.36(25)  & $-$1.17(11)  \\
$A( m_\pi, \overline{p}_n )$           & 
                                       & $-$0.196(38) & $-$0.362(35) & $-$0.406(49) & $-$0.434(26) \\
$\hat{A}( m_\pi, \overline{p}_n )$     &
                                       & $-$0.59(11)  & $-$1.66(16)  & $-$2.44(29)  & $-$3.17(19)  \\
$A( m_\pi, \overline{p}_n ) / m_\pi^2 $ & [1/GeV$^2$]
                                       & $-$0.82(16)  & $-$0.633(62) & $-$0.499(60) & $-$0.384(23) \\
$\hat{A}( m_\pi, \overline{p}_n ) / m_\pi^2$  & [1/GeV$^2$]
                                       & $-$2.47(48)  & $-$2.91(28)  & $-$3.02(33)  & $-$2.83(17)  \\ \hline\hline
%
%
$\beta = 1.95$ & $\kappa$              & 0.1410       & 0.1400       & 0.1390       & 0.1375       \\
$a^{-1} = 1.268(13)$ [GeV] 
               & $m_\pi/m_\rho$        & 0.582(3)     & 0.690(1)     & 0.752(1)     & 0.804(1)     \\
               & $m_\pi^2$ [GeV$^2$]   & 0.291(2)     & 0.573(1)     & 0.857(1)     & 1.287(1)     \\ \hline
Fitting Range  &                       & 12   --   23 & 13   --   25 & 13   --   25 & 13   --   25 \\
$\Delta \overline{E}_n^{\bf P}$        & [$ \times 10^{-3} $ GeV]
                                       & 10.95(73)    & 6.89(69)     & 5.74(30)     & 3.86(23)     \\
$\overline{p}^2_n$                     & [$\times 10^{-4}$ GeV$^2$]
                                       & 46.7(31)     & 41.2(41)     & 41.9(22)     & 34.5(21)     \\
$\delta ( \overline{p}_n )$            & [degrees]
                                       & $-$2.50(23)  & $-$2.10(29)  & $-$2.16(15)  & $-$1.65(14)  \\
$A( m_\pi, \overline{p}_n )$           & 
                                       & $-$0.348(20) & $-$0.436(38) & $-$0.540(25) & $-$0.557(30) \\
$\hat{A}( m_\pi, \overline{p}_n )$     &
                                       & $-$0.769(45) & $-$1.37(12)  & $-$2.27(10)  & $-$3.09(17)  \\
$A( m_\pi, \overline{p}_n ) / m_\pi^2$ & [1/GeV$^2$]
                                       & $-$1.195(70) & $-$0.759(67) & $-$0.630(29) & $-$0.433(23) \\
$\hat{A}( m_\pi, \overline{p}_n ) / m_\pi^2$  & [1/GeV$^2$]
                                       & $-$2.63(15)  & $-$2.39(21)  & $-$2.65(12)  & $-$2.40(13)  \\ \hline\hline
%
%
$\beta = 2.10$ & $\kappa$              & 0.1382       & 0.1374       & 0.1367       & 0.1357       \\
$a^{-1} = 1.833(22)$ [GeV] 
               & $m_{\pi}/m_{\rho}$    & 0.576(3)     & 0.691(3)     & 0.755(2)     & 0.806(1)     \\
               & $m_{\pi}^2$ [GeV$^2$] & 0.291(1)     & 0.605(2)     & 0.896(1)     & 1.332(2)     \\ \hline
Fitting Range  &                       & 18   --   35 & 18   --   35 & 18   --   35 & 18   --   35 \\
$ \Delta \overline{E}_n^{\bf P}$       & [$ \times 10^{-3} $ GeV]
                                       & 17.33(54)    & 10.48(42)    & 8.49(32)     & 6.16(30)     \\
$ \overline{p}^2_n $                   & [$\times 10^{-4}$ GeV$^2$]
                                       & 51.3(16)     & 44.6(17)     & 44.0(16)     & 38.9(19)     \\
$ \delta ( \overline{p}_n )$           & [degrees]
                                       & $-$3.17(13)  & $-$2.62(14)  & $-$2.57(13)  & $-$2.17(14)  \\
$A( m_\pi, \overline{p}_n )$           &
                                       & $-$0.421(12) & $-$0.536(18) & $-$0.643(20) & $-$7.04(30)  \\
$\hat{A}( m_\pi, \overline{p}_n )$     &
                                       & $-$0.718(20) & $-$1.471(49) & $-$2.251(72) & $-$3.12(13)  \\
$A( m_\pi, \overline{p}_n ) / m_\pi^2$ & [1/GeV$^2$]
                                       & $-$1.444(38) & $-$0.885(31) & $-$0.718(23) & $-$0.528(23) \\ 
$\hat{A}( m_\pi, \overline{p}_n ) / m_\pi^2 $  & [1/GeV$^2$]
                                       & $-$2.462(66) & $-$2.429(85) & $-$2.512(82) & $-$2.34(10)  \\
\end{tabular}
\end{center}
\caption{
Results for $n=0$ state in the center of mass system CM with energy state cut-off $N=2$.
Two scattering amplitudes are defined by
$A( m_\pi, \overline{p}_n )= \tan \delta(\overline{p}_n) / \overline{p}_n \cdot \overline{E}_n /2$
and 
$\hat{A}(m_\pi, \overline{p}_n ) = ( f_\pi^{lat} / f_\pi )^2 \cdot  A( m_\pi, \overline{p}_n )$,
where $f_\pi^{lat} $ is the pseudoscalar decay constant measured on lattice and $f_\pi = 93$ MeV.
\label{tab:Res:CM_0}
}
\end{table}
%
%
\begin{table}
\begin{center}
\begin{tabular}{lccccc}
$\beta = 1.80$ & $\kappa$              & 0.1464       & 0.1445       & 0.1430       & 0.1409       \\
$a^{-1} = 0.9176(93)$ [GeV] 
               & $m_\pi/m_\rho$        & 0.547(4)     & 0.694(2)     & 0.753(1)     & 0.807(1)     \\
               & $m_\pi^2$ [GeV$^2$]   & 0.238(1)     & 0.571(1)     & 0.814(1)     & 1.128(1)     \\ \hline
Fitting Range  &                       & 10   --   18 & 12   --   18 & 12   --   20 & 12   --   20 \\
$\Delta \overline{E}_n^{\bf P}$        & [$ \times 10^{-3} $ GeV]
                                       & 22.5(34)     & 13.2(12)     & 11.42(70)    & 8.78(40)     \\
$\overline{p}^2_n$                     & [$\times 10^{-2}$ GeV$^2$]
                                       & 24.80(26)    & 24.40(12)    & 24.380(78)   & 24.222(51)   \\
$\delta (\overline{p}_n)$              & [degrees]
                                       & $-$14.1(22)  & $-$10.7(10)  & $-$10.59(66) & $-$9.25(43)  \\
$A( m_\pi, \overline{p}_n )$           &    
                                       & $-$0.353(57) & $-$0.348(33) & $-$0.389(24) & $-$0.387(18) \\
$\hat{A}( m_\pi, \overline{p}_n ) $    & 
                                       & $-$1.06(17)  & $-$1.59(15)  & $-$2.34(14)  & $-$2.83(13)  \\ \hline\hline
%
%
$\beta = 1.95$ & $\kappa$              & 0.1410       & 0.1400       & 0.1390       & 0.1375       \\
$a^{-1} = 1.268(13)$ [GeV] 
               & $m_\pi/m_\rho$        & 0.582(3)     & 0.690(1)     & 0.752(1)     & 0.804(1)     \\
               & $m_\pi^2$ [GeV$^2$]   & 0.291(2)     & 0.573(1)     & 0.857(1)     & 1.287(1)     \\ \hline
Fitting Range  &                       & 12   --   23 & 13   --   25 & 13   --   25 & 13   --   25 \\
$\Delta \overline{E}_n^{\bf P}$        & [$ \times 10^{-3} $ GeV]
                                       & 45.7(51)     & 29.9(20)     & 22.70(95)    & 15.50(49)    \\
$\overline{p}^2_n$                     & [$\times 10^{-2}$ GeV$^2$]
                                       & 27.55(30)    & 27.02(14)    & 26.761(79)   & 26.389(48)   \\
$\delta (\overline{p}_n)$              & [degrees]
                                       & $-$20.9(24)  & $-$16.7(11)  & $-$14.63(63) & $-$11.70(38) \\
$A( m_\pi, \overline{p}_n )$           & 
                                       & $-$0.549(68) & $-$0.531(38) & $-$0.535(23) & $-$0.502(16) \\
$\hat{A}( m_\pi, \overline{p}_n ) $    & 
                                       & $-$1.21(15)  & $-$1.67(12)  & $-$2.256(99) & $-$2.787(91) \\  \hline\hline
%
%
$\beta = 2.10$ & $\kappa$              & 0.1382       & 0.1374       & 0.1367       & 0.1357       \\
$a^{-1} = 1.833(22)$ [GeV] 
               & $m_\pi/m_\rho$        & 0.576(3)     & 0.691(3)     & 0.755(2)     & 0.806(1)     \\
               & $m_\pi^2$ [GeV$^2$]   & 0.291(1)     & 0.605(2)     & 0.896(1)     & 1.332(2)     \\ \hline
Fitting Range  &                       & 18   --   35 & 18   --   35 & 18   --   35 & 18   --   35 \\
$\Delta \overline{E}_n^{\bf P}$      &  [$ \times 10^{-3} $ GeV]
                                       & 58.8(69)     & 44.1(25)     & 34.6(14)     & 26.80(78)    \\
$\overline{p}^2_n$                     & [$\times 10^{-2}$ GeV$^2$]
                                       & 25.30(28)    & 25.17(13)    & 24.969(84)   & 24.789(53)   \\
$\delta ( \overline{p}_n )$            & [degrees]
                                       & $-$19.8(24)  & $-$18.7(11)  & $-$16.99(73) & $-$15.44(46) \\
$A( m_\pi, \overline{p}_n )$           & 
                                       & $-$0.530(69) & $-$0.626(39) & $-$0.654(28) & $-$0.697(21) \\
$\hat{A}( m_\pi, \overline{p}_n )$     & 
                                       & $-$0.90(11)  & $-$1.71(10)  & $-$2.29(10)  & $-$3.097(94) \\
\end{tabular}
\end{center}
\caption{
Results for $n=1$ state in the center of mass system CM with energy state cut-off $N=2$.
Two scattering amplitudes are defined by
$A( m_\pi, \overline{p}_n )= \tan \delta(\overline{p}_n) / \overline{p}_n \cdot \overline{E}_n /2$
and 
$\hat{A}(m_\pi, \overline{p}_n ) = ( f_\pi^{lat} / f_\pi )^2 \cdot  A( m_\pi, \overline{p}_n )$,
where $f_\pi^{lat} $ is the pseudoscalar decay constant measured on lattice and $f_\pi = 93$ MeV.
\label{tab:Res:CM_1}
}
\end{table}
%
%
\begin{table}
\begin{center}
\begin{tabular}{lccccc}
$\beta = 1.80$ & $\kappa$              & 0.1464       & 0.1445       & 0.1430       & 0.1409       \\
$a^{-1} = 0.9176(93)$ [GeV] 
               & $m_\pi/m_\rho$        & 0.547(4)     & 0.694(2)     & 0.753(1)     & 0.807(1)     \\
               & $m_\pi^2$ [GeV$^2$]   & 0.238(1)     & 0.571(1)     & 0.814(1)     & 1.128(1)     \\ \hline
Fitting Range  &                       & 10   --   18 & 12   --   20 & 12   --   20 & 12   --   20 \\
$\Delta \overline{E}_n^{\bf P}$        & [$ \times 10^{-3} $ GeV]
                                       & 9.5(10)      & 6.72(50)     & 4.65(43)     & 3.70(19)     \\
$\overline{p}^2_n$                     & [$\times 10^{-3}$ GeV$^2$]
                                       & 54.16(69)    & 58.93(46)    & 59.04(45)    & 59.56(23)    \\
$\gamma$                               &
                                       & 1.09436(62)  & 1.04481(11)  & 1.032523(71) & 1.024028(25) \\
$\delta ( \overline{p}_n )$            & [degrees]
                                       & $-$7.03(76)  & $-$7.14(52)  & $-$5.81(52)  & $-$5.36(27)  \\
$A( m_\pi, \overline{p}_n )$           & 
                                       & $-$0.286(30) & $-$0.409(28) & $-$0.391(34) & $-$0.419(20) \\
$\hat{A}( m_\pi, \overline{p}_n )$     & 
                                       & $-$0.862(91) & $-$1.88(13)  & $-$2.35(20)  & $-$3.07(15)  \\ \hline\hline
%
%
$\beta = 1.95$ & $\kappa$              & 0.1410       & 0.1400       & 0.1390       & 0.1375       \\
$a^{-1} = 1.268(13)$ [GeV] 
               & $m_\pi/m_\rho$        & 0.582(3)     & 0.690(1)     & 0.752(1)     & 0.804(1)     \\
               & $m_\pi^2$ [GeV$^2$]   & 0.291(2)     & 0.573(1)     & 0.857(1)     & 1.287(1)     \\ \hline
Fitting Range  &                       & 12   --   23 & 13   --   25 & 13   --   25 & 13   --   25 \\
$\Delta \overline{E}_n^{\bf P}$        & [$ \times 10^{-3} $ GeV]
                                       & 17.3(11)     & 11.99(55)    & 9.62(35)     & 6.92(30)     \\
$\overline{p}^2_n $                    & [$\times 10^{-3}$ GeV$^2$]
                                       & 61.40(58)    & 64.47(36)    & 65.73(28)    & 65.91(28)    \\
$\gamma$                               & 
                                       & 1.08447(51)  & 1.04758(13)  & 1.033120(56) & 1.022720(32) \\
$\delta ( \overline{p}_n )$            & [degrees]
                                       & $-$9.29(59)  & $-$8.55(38)  & $-$8.20(29)  & $-$7.12(30)  \\
$A( m_\pi, \overline{p}_n )$           & 
                                       & $-$0.392(24) & $-$0.473(20) & $-$0.540(18) & $-$0.566(23) \\
$\hat{A}( m_\pi, \overline{p}_n )$     & 
                                       & $-$0.865(53) & $-$1.491(63) & $-$2.276(79) & $-$3.14(12)  \\ \hline\hline
%
%
$\beta = 2.10$ & $\kappa$              & 0.1382       & 0.1374       & 0.1367       & 0.1357       \\
$a^{-1} = 1.833(22)$ [GeV] 
               & $m_\pi/m_{\rho}$      & 0.576(3)     & 0.691(3)     & 0.755(2)     & 0.806(1)     \\
               & $m_\pi^2$ [GeV$^2$]   & 0.291(1)     & 0.605(2)     & 0.896(1)     & 1.332(2)     \\ \hline
Fitting Range  &                       & 18   --   35 & 18   --   35 & 18   --   35 & 18   --   35 \\
$\Delta \overline{E}_n^{\bf P}$        & [$ \times 10^{-3} $ GeV]
                                       & 27.8(12)     & 17.7(65)     & 14.75(48)    & 10.45(38)    \\
$\overline{p}^2_n $                    & [$\times 10^{-3}$ GeV$^2$]
                                       & 58.78(42)    & 61.02(29)    & 62.22(26)    & 61.97(25)    \\
$\gamma $                              & 
                                       & 1.07874(41)  & 1.04214(13)  & 1.029504(58) & 1.020373(39) \\
$\delta ( \overline{p}_n )$            & [degrees]
                                       & $-$11.09(46) & $-$9.66(33)  & $-$9.55(29)  & $-$8.16(28)  \\
$A( m_{\pi}, \overline{p}_n )$         &
                                       & $-$0.478(19) & $-$0.563(18) & $-$0.660(19) & $-$0.680(23) \\ 
$\hat{A}( m_{\pi}, \overline{p}_n )$   & 
                                       & $-$0.816(32) & $-$1.544(50) & $-$2.311(68) & $-$3.02(10)  \\
\end{tabular}
\end{center}
\caption{
Results for $n=0$ state in the laboratory system L1 with energy state cut-off $N=3$.
Two scattering amplitudes are defined by
$A( m_\pi, \overline{p}_n )= \tan \delta(\overline{p}_n) / \overline{p}_n \cdot \overline{E}_n /2$
and 
$\hat{A}(m_\pi, \overline{p}_n ) = ( f_\pi^{lat} / f_\pi )^2 \cdot  A( m_\pi, \overline{p}_n )$,
where $f_\pi^{lat} $ is the pseudoscalar decay constant measured on lattice and $f_\pi = 93$ MeV.
\label{tab:Res:L1_0}
}
\end{table}
%
%
\begin{table}
\begin{center}
\begin{tabular}{lccccc}
$\beta = 1.80$ & $\kappa$              & 0.1464       & 0.1445       & 0.1430       & 0.1409       \\
$a^{-1} = 0.9176(93)$ [GeV] 
               & $m_\pi/m_\rho$        & 0.547(4)     & 0.694(2)     & 0.753(1)     & 0.807(1)     \\
               & $m_\pi^2$ [GeV$^2$]   & 0.238(1)     & 0.571(1)     & 0.814(1)     & 1.128(1)     \\ \hline
Fitting Range  &                       & 10   --   18 & 12   --   18 & 12   --   20 & 12   --   20 \\
$\Delta \overline{E}_n^{\bf P}$        & [$ \times 10^{-3} $ GeV]
                                       & 28.0(63)     & 18.1(26)     & 14.6(12)     & 10.62(59)    \\
$\overline{p}^2_n $                    & [$\times 10^{-2}$ GeV$^2$]
                                       & 30.65(53)    & 30.41(27)    & 30.31(14)    & 30.061(79)   \\
$\gamma$                               & 
                                       & 1.05169(54)  & 1.03245(10)  & 1.025510(52) & 1.020014(19) \\
$\delta ( \overline{p}_n )$            & [degrees]
                                       & $-$16.9(39)  & $-$13.4(20)  & $-$12.1(10)  & $-$9.86(56)  \\
$A( m_\pi, \overline{p}_n )$           & 
                                       & $-$0.405(98) & $-$0.406(61) & $-$0.412(35) & $-$0.379(21) \\
$\hat{A}( m_\pi, \overline{p}_n )$     & 
                                       & $-$1.22(29)  & $-$1.86(28)  & $-$2.47(21)  & $-$2.77(15)  \\ \hline\hline
%
%
$\beta = 1.95$ & $\kappa$              & 0.1410       & 0.1400       & 0.1390       & 0.1375       \\
$a^{-1} = 1.268(13)$ [GeV] 
               & $m_\pi/m_\rho$        & 0.582(3)     & 0.690(1)     & 0.752(1)     & 0.804(1)     \\
               & $m_\pi^2$ [GeV$^2$]   & 0.291(2)     & 0.573(1)     & 0.857(1)     & 1.287(1)     \\ \hline
Fitting Range  &                       & 12   --   16 & 13   --   20 & 13   --   25 & 13   --   25 \\
$\Delta \overline{E}_n^{\bf P}$        & [$ \times 10^{-3} $ GeV]
                                       & 57(10)       & 34.0(37)     & 28.3(16)     & 18.98(79)    \\
$\overline{p}^2_n $                    & [$\times 10^{-2}$ GeV$^2$]
                                       & 34.21(67)    & 33.30(28)    & 33.26(14)    & 32.788(81)   \\
$\gamma$                               & 
                                       & 1.04789(53)  & 1.03371(12)  & 1.025787(45) & 1.019069(26) \\
$\delta ( \overline{p}_n )$            & [degrees]
                                       & $-$25.0(47)  & $-$17.3(19)  & $-$16.35(99) & $-$12.59(53) \\
$A( m_\pi, \overline{p}_n )$           & 
                                       & $-$0.63(13)  & $-$0.516(60) & $-$0.555(34) & $-$0.495(21) \\
$\hat{A}( m_\pi, \overline{p}_n )$     & 
                                       & $-$1.40(29)  & $-$1.62(19)  & $-$2.33(14)  & $-$2.75(11)  \\ \hline\hline
%
%
$\beta = 2.10$ & $\kappa$              & 0.1382       & 0.1374       & 0.1367       & 0.1357       \\
$a^{-1} = 1.833(22)$ [GeV] 
               & $m_\pi/m_\rho$        & 0.576(3)     & 0.691(3)     & 0.755(2)     & 0.806(1)     \\
               & $m_\pi^2$ [GeV$^2$]   & 0.291(1)     & 0.605(2)     & 0.896(1)     & 1.332(2)     \\ \hline
Fitting Range  &                       & 18   --   22 & 18   --   35 & 18   --   35 & 18   --   35 \\
$\Delta \overline{E}_n^{\bf P}$        & [$ \times 10^{-3} $ GeV]
                                       & 68(13)       & 53.9(48)     & 39.3(24)     & 32.4(12)     \\
$\overline{p}^2_n$                     & [$\times 10^{-2}$ GeV$^2$]
                                       & 31.16(58)    & 31.23(24)    & 30.82(14)    & 30.798(89)   \\
$\gamma $                              & 
                                       & 1.04646(43)  & 1.03077(10)  & 1.023545(48) & 1.017343(30) \\
$\delta ( \overline{p}_n )$            & [degrees]
                                       & $-$21.8(43)  & $-$20.7(19)  & $-$17.1(10)  & $-$16.32(64) \\
$A( m_\pi, \overline{p}_n )$           & 
                                       & $-$0.55(12)  & $-$0.651(63) & $-$0.608(39) & $-$0.675(27) \\ 
$\hat{A}( m_\pi, \overline{p}_n )$     & 
                                       & $-$0.95(20)  & $-$1.78(17)  & $-$2.13(13)  & $-$3.00(12)  \\
\end{tabular}
\end{center}
\caption{
Results for  $n=1 $ state in the laboratory system L1 with energy state cut-off $N=3$.
Two scattering amplitudes are defined by
$A( m_\pi, \overline{p}_n )= \tan \delta(\overline{p}_n) / \overline{p}_n \cdot \overline{E}_n /2$
and 
$\hat{A}(m_\pi, \overline{p}_n ) = ( f_\pi^{lat} / f_\pi )^2 \cdot  A( m_\pi, \overline{p}_n )$,
where $f_\pi^{lat} $ is the pseudoscalar decay constant measured on lattice and $f_\pi = 93$ MeV.
\label{tab:Res:L1_1}
}
\end{table}
%
%
\begin{table}
\begin{center}
\begin{tabular}{lccccc}
$\beta = 1.80$ & $\kappa$              & 0.1464       & 0.1445       & 0.1430       & 0.1409       \\
$a^{-1} = 0.9176(93)$ [GeV] 
               & $m_\pi/m_{\rho}$      & 0.547(4)     & 0.694(2)     & 0.753(1)     & 0.807(1)     \\
               & $m_\pi^2$ [GeV$^2$]   & 0.238(1)     & 0.571(1)     & 0.814(1)     & 1.128(1)     \\ \hline
Fitting Range  &                       & 10   --   18 & 12   --   18 & 12   --   18 & 12   --   20 \\
$\Delta \overline{E}_n^{\bf P}$        & [$ \times 10^{-3} $ GeV]
                                       & 7.2(21)      & 4.79(84)     & 3.81(50)     & 2.03(27)     \\
$\overline{p}^2_n $                    & [$\times 10^{-2}$ GeV$^2$]
                                       & 9.03(15)     & 10.316(82)   & 10.682(57)   & 10.820(34)   \\
$\gamma$                               & 
                                       & 1.1626(11)   & 1.08224(20)  & 1.06082(12)  & 1.045678(45) \\
$\delta ( \overline{p}_n )$            & [degrees]
                                       & $-$9.7(33)   & $-$11.4(32)  & $-$12.6(37)  & $-$6.5(14)   \\
$A( m_\pi, \overline{p}_n )$           & 
                                       & $-$0.32(11)  & $-$0.51(15)  & $-$0.65(19)  & $-$0.387(86) \\
$\hat{A}( m_\pi, \overline{p}_n )$     & 
                                       & $-$0.98(34)  & $-$2.37(69)  & $-$3.9(11)   & $-$2.83(63)  \\ \hline\hline
%
%
$\beta = 1.95$ & $\kappa$              & 0.1410       & 0.1400       & 0.1390       & 0.1375       \\
$a^{-1} = 1.268(13)$ [GeV] 
               & $m_\pi/m_\rho$        & 0.582(3)     & 0.690(1)     & 0.752(1)     & 0.804(1)     \\
               & $m_\pi^2$ [GeV$^2$]   & 0.291(2)     & 0.573(1)     & 0.857(1)     & 1.287(1)     \\ \hline
Fitting Range  &                       & 12   --   16 & 13   --   20 & 13   --   25 & 13   --   25 \\
$\Delta \overline{E}_n^{\bf P}$        & [$ \times 10^{-3} $ GeV]
                                       & 15.5(25)     & 7.5(10)      & 5.49(71)     & 3.12(37)     \\
$\overline{p}^2_n$                     & [$\times 10^{-2}$ GeV$^2$]
                                       & 10.28(14)    & 11.043(77)   & 11.473(59)   & 11.726(36)   \\ 
$\gamma$                               & 
                                       & 1.14682(99)  & 1.08709(25)  & 1.06202(10)  & 1.043336(60) \\
$\delta ( \overline{p}_n )$            & [degrees]
                                       & $-$18.3(46)  & $-$12.1(29)  & $-$12.2(32)  & $-$8.0(18)   \\
$A( m_\pi, \overline{p}_n )$           & 
                                       & $-$0.64(17)  & $-$0.53(13)  & $-$0.63(17)  & $-$0.48(11)  \\
$\hat{A}( m_\pi, \overline{p}_n )$     & 
                                       & $-$1.42(38)  & $-$1.69(41)  & $-$2.66(73)  & $-$2.70(62)  \\  \hline\hline
%
%
$\beta = 2.10$ & $\kappa$              & 0.1382       & 0.1374       & 0.1367       & 0.1357       \\
$a^{-1} = 1.833(22)$ [GeV] 
               & $m_\pi/m_\rho$        & 0.576(3)     & 0.691(3)     & 0.755(2)     & 0.806(1)     \\
               & $m_\pi^2$ [GeV$^2$]   & 0.291(1)     & 0.605(2)     & 0.896(1)     & 1.332(2)     \\ \hline
Fitting Range  &                       & 18   --   30 & 18   --   35 & 18   --   35 & 18   --   35 \\
$\Delta \overline{E}_n^{\bf P}$        & [$ \times 10^{-3} $ GeV]
                                       & 17.5(35)     & 9.3(15)      & 7.83(88)     & 4.45(61)     \\
$\overline{p}^2_n $                    & [$\times 10^{-2}$ GeV$^2$]
                                       & 9.48(13)     & 10.332(76)   & 10.746(50)   & 10.932(42)   \\
$\gamma$                               & 
                                       & 1.13877(72)  & 1.07789(27)  & 1.05563(10)  & 1.039047(74) \\
$\delta ( \overline{p}_n )$            & [degrees]
                                       & $-$14.5(41)  & $-$12.0(33)  & $-$17.3(56)  & $-$10.9(40)  \\
$A( m_\pi, \overline{p}_n )$           & 
                                       & $-$0.52(15)  & $-$0.55(16)  & $-$0.95(33)  & $-$0.70(26)  \\ 
$\hat{A}( m_\pi, \overline{p}_n )$     & 
                                       & $-$0.89(26)  & $-$1.53(44)  & $-$3.3(11)   & $-$3.1(11)   \\
\end{tabular}
\end{center}
\caption{
Results for $n=0$ state in the laboratory system L2 with energy state cut-off $N=3$.
Two scattering amplitudes are defined by
$A( m_\pi, \overline{p}_n )= \tan \delta(\overline{p}_n) / \overline{p}_n \cdot \overline{E}_n /2$
and 
$\hat{A}(m_\pi, \overline{p}_n ) = ( f_\pi^{lat} / f_\pi )^2 \cdot  A( m_\pi, \overline{p}_n )$,
where $f_\pi^{lat} $ is the pseudoscalar decay constant measured on lattice and $f_\pi = 93$ MeV.
\label{tab:Res:L2_0}
}
\end{table}
%
%
\begin{table}
\begin{center}
\begin{tabular}{lccccc}
$\beta = 1.80$ & $\kappa$              & 0.1464       & 0.1445       & 0.1430       & 0.1409       \\
$a^{-1} = 0.9176(93)$ [GeV] 
               & $m_\pi/m_\rho$        & 0.547(4)     & 0.694(2)     & 0.753(1)     & 0.807(1)     \\
               & $m_\pi^2$ [GeV$^2$]   & 0.238(1)     & 0.571(1)     & 0.814(1)     & 1.128(1)     \\ \hline
Fitting Range  &                       & 10   --   18 & 12   --   18 & 12   --   20 & 12   --   20 \\
$\Delta \overline{E}_n^{\bf P}$        & [$ \times 10^{-3} $ GeV]
                                       & 9.5(18)      & 6.80(73)     & 5.38(44)     & 4.47(24)     \\
$\overline{p}^2_n$                     & [$\times 10^{-2}$ GeV$^2$]
                                       & 12.26(13)    & 12.217(71)   & 12.152(49)   & 12.120(30)   \\
$\gamma$                               & 
                                       & 1.14900(84)  & 1.08007(19)  & 1.05990(11)  & 1.045213(44) \\
$\delta ( \overline{p}_n )$            & [degrees]
                                       & $-$11.3(18)  & $-$9.36(82)  & $-$8.12(54)  & $-$7.35(32)  \\
$A( m_\pi, \overline{p}_n )$           & 
                                       & $-$0.345(55) & $-$0.393(34) & $-$0.396(26) & $-$0.414(17) \\
$\hat{A}( m_\pi, \overline{p}_n )$     & 
                                       & $-$1.03(16)  & $-$1.80(15)  & $-$2.37(15)  & $-$3.03(13)  \\  \hline\hline
%
%
$\beta = 1.95$ & $\kappa$              & 0.1410       & 0.1400       & 0.1390       & 0.1375       \\
$a^{-1} = 1.268(13)$ [GeV] 
               & $m_\pi/m_\rho$        & 0.582(3)     & 0.690(1)     & 0.752(1)     & 0.804(1)     \\
               & $m_\pi^2$ [GeV$^2$]   & 0.291(2)     & 0.573(1)     & 0.857(1)     & 1.287(1)     \\ \hline
Fitting Range  &                       & 12   --   16 & 13   --   25 & 13   --   25 & 13   --   25 \\
$\Delta \overline{E}_n^{\bf P}$        & [$ \times 10^{-3} $ GeV]
                                       & 18.2(26)     & 13.6(11)     & 10.81(61)    & 8.71(41)     \\
$\overline{p}^2_n $                    & [$\times 10^{-2}$ GeV$^2$]
                                       & 13.49(15)    & 13.415(83)   & 13.333(51)   & 13.288(40)   \\
$\gamma$                               & 
                                       & 1.13642(69)  & 1.08428(22)  & 1.060890(97) & 1.042870(58) \\
$\delta ( \overline{p}_n )$            & [degrees]
                                       & $-$14.4(16)  & $-$12.20(83) & $-$10.63(48) & $-$9.51(36)  \\
$A( m_\pi, \overline{p}_n )$           & 
                                       & $-$0.459(53) & $-$0.496(33) & $-$0.511(23) & $-$0.547(20) \\
$\hat{A}( m_\pi, \overline{p}_n )$     & 
                                       & $-$1.01(11)  & $-$1.56(10)  & $-$2.156(98) & $-$3.04(11)  \\ \hline\hline
%
%
$\beta = 2.10$ & $\kappa$              & 0.1382       & 0.1374       & 0.1367       & 0.1357       \\
$a^{-1} = 1.833(22)$ [GeV] 
               & $m_\pi/m_\rho$        & 0.576(3)     & 0.691(3)     & 0.755(2)     & 0.806(1)     \\
               & $m_\pi^2$ [GeV$^2$]   & 0.291(1)     & 0.605(2)     & 0.896(1)     & 1.332(2)     \\ \hline
Fitting Range  &                       & 18   --   30 & 18   --   35 & 18   --   35 & 18   --   35 \\
$\Delta \overline{E}_n^{\bf P}$        & [$ \times 10^{-3} $ GeV]
                                       & 31.5(32)     & 21.2(12)     & 17.99(92)    & 15.12(66)    \\
$\overline{p}^2_n$                     & [$\times 10^{-2}$ GeV$^2$]
                                       & 12.72(13)    & 12.542(63)   & 12.522(53)   & 12.511(45)   \\
$\gamma$                               & 
                                       & 1.12865(64)  & 1.07562(21)  & 1.05469(10)  & 1.038631(69) \\
$\delta ( \overline{p}_n )$            & [degrees]
                                       & $-$17.4(14)  & $-$13.61(65) & $-$12.56(52) & $-$11.68(42) \\ 
$A( m_\pi, \overline{p}_n )$           & 
                                       & $-$0.568(48) & $-$0.584(27) & $-$0.636(26) & $-$0.705(24) \\
$\hat{A}( m_\pi, \overline{p}_n )$     & 
                                       & $-$0.969(81) & $-$1.603(75) & $-$2.226(91) & $-$3.13(11)  \\
\end{tabular}
\end{center}
\caption{
Results for $n=1$ state in the laboratory system L2 with energy state cut-off $N=3$.
Two scattering amplitudes are defined by
$A( m_\pi, \overline{p}_n )= \tan \delta(\overline{p}_n) / \overline{p}_n \cdot \overline{E}_n /2$
and 
$\hat{A}(m_\pi, \overline{p}_n ) = ( f_\pi^{lat} / f_\pi )^2 \cdot  A( m_\pi, \overline{p}_n )$,
where $f_\pi^{lat} $ is the pseudoscalar decay constant measured on lattice and $f_\pi = 93$ MeV.
\label{tab:Res:L2_1}
}
\end{table}
%
%

%
\clearpage
%
%
%
\setcounter{table}{0}
\begin{table}
\begin{center}
\begin{tabular}{ccccccccccc}
$\beta$ & $L^3 \times T$   & $c_{SW}$ & $a$ [fm]   & $La$ [fm] & $\kappa$ 
& $m_\pi/m_\rho$ & $N_{\mathrm Traj}$ & $N_{\mathrm Skip}$ & $N_{\mathrm Meas}$ \\ \hline
1.80 & $12^3 \times 24$ & 1.60 & 0.2150(22) & 2.580(26) & 0.1409 & 0.807(1) & 6530 & 10 & 645               \\
     &                  &      &            &           & 0.1430 & 0.753(1) & 5240 & 10 & 520               \\
     &                  &      &            &           & 0.1445 & 0.694(2) & 7350 & 10 & 725               \\
     &                  &      &            &           & 0.1464 & 0.547(4) & 5250 & 10 & \ 405$^{\dagger}$ \\
1.95 & $16^3 \times 32$ & 1.53 & 0.1555(17) & 2.489(27) & 0.1375 & 0.804(1) & 7000 & 10 & 595               \\
     &                  &      &            &           & 0.1390 & 0.752(1) & 7000 & 10 & 690               \\
     &                  &      &            &           & 0.1400 & 0.690(1) & 7000 & 10 & 685               \\
     &                  &      &            &           & 0.1410 & 0.582(3) & 5000 & 10 & 495               \\
2.10 & $24^3 \times 48$ & 1.47 & 0.1076(13) & 2.583(31) & 0.1357 & 0.806(1) & 4000 & 10 & 395               \\
     &                  &      &            &           & 0.1367 & 0.755(2) & 4000 & 10 & 390               \\
     &                  &      &            &           & 0.1374 & 0.691(3) & 4000 & 10 & 380               \\
     &                  &      &            &           & 0.1382 & 0.576(3) & 4000 & 5  & 640               \\
\end{tabular}
\end{center}
\caption{
Simulation parameters.
Dagger symbol means that we average two measurements on the same configuration, 
one with the temporal origin located at $t = 0$
and the other located at $t = T / 2$.
The lattice spacing $a$ is fixed by the $\rho$ meson mass 
at the physical pion mass and $m_\rho = 768.4$ MeV.
$N_{\mathrm Traj}$ is the number of all trajectories, $N_{\mathrm Skip}$ 
is the number of separation between two measurements
and $N_{\mathrm Meas}$ is the number of the configurations used the measurements.
\label{tab:param}
}
\end{table}
%
%
\begin{table}
\begin{center}
\begin{tabular}{ccccc}
CM \ ${\bf P}=$( 0, 0, 0)  &  $n=0$       &  $n=1$       &  $n=2$       &  $n=3$      \\ \hline 
${\bf p}_{1,n}$            &  ( 0, 0, 0)  &  ( 1, 0, 0)  &  ( 1, 1, 0)  &  ( 1, 1, 1) \\
${\bf p}_{2,n}$            &  ( 0, 0, 0)  &  (-1, 0, 0)  &  (-1,-1, 0)  &  (-1,-1,-1) \\ 
\hline\hline 
L1 \ ${\bf P}=$( 1, 0, 0)  &  $n=0$       &  $n=1$       &  $n=2$       &  $n=3$       \\ \hline 
${\bf p}_{1,n}$            &  ( 1, 0, 0)  &  ( 1, 1, 0)  &  ( 2, 0, 0)  &  ( 1, 1, 1)  \\
${\bf p}_{2,n}$            &  ( 0, 0, 0)  &  ( 0,-1, 0)  &  (-1, 0, 0)  &  ( 0,-1,-1)  \\
\hline\hline 
L2 \ ${\bf P}=$( 1, 1, 0)  &  $n=0$       &  $n=1$       &  $n=2$       &  $n=3$       \\ \hline 
${\bf p}_{1,n}$            &  ( 1, 1, 0)  &  ( 1, 0, 0)  &  ( 1, 1, 1)  &  ( 1, 0, 1)  \\
${\bf p}_{2,n}$            &  ( 0, 0, 0)  &  ( 0, 1, 0)  &  ( 0, 0,-1)  &  ( 0, 1,-1)  \\
\end{tabular}
\end{center}
\caption{
Momentum assignment 
for the source operator $\Omega_n^{(N_R)}(t)$ defined by Eq.(\ref{eq:Omega_source}).
Here ${\bf p}_{i,n} $ is the $i$-th pion momentum of the $n$ state
${\bf P}$ is the total momentum of two pions system ${\bf P} = {\bf p}_{1, n} + {\bf p}_{2, n}$
in units of $2 \pi / L$.
\label{tab:momsetup}
}
\end{table}
%
%
\begin{table}
\begin{center}
\begin{tabular}{cccccc}
$\beta$ & $F$ [GeV]    & $L(\mu)$    & $C_L$       & $\chi^2 / $d.o.f. & $a_0 / m_\pi$  [1/GeV$^2$] \\ 
\hline
1.80    & $0.135(20) $ & $1.74(15) $ & $0.70(77) $ & 0.04              & $-1.04(40)$            \\
1.95    & $0.1035(59)$ & $1.199(72)$ & $0.765(85)$ & 1.55              & $-1.76(20)$            \\
2.10    & $0.0909(24)$ & $0.963(28)$ & $0.702(29)$ & 4.78              & $-2.26(11)$            \\
\end{tabular}
\end{center}
\caption{
Results of fitting for the scattering length $a_0 / m_{\pi}$
obtained with the fitting function Eq.(\ref{eq:a0:fit-ChPT}),
where we set $\mu=1\ {\rm GeV}$.
The results at the physical pion mass are also tabulated.
\label{tab:a0:fit-ChPT}
}
\end{table}
%
%
\begin{table}
\begin{center}
\begin{tabular}{ccccc}
$\beta$ & $A_{00}$     
        & $A_{10}$ [1/GeV$^2$]   
        & $A_{20}$ [1/GeV$^4$]     & $\chi^2 / $d.o.f. \\ \hline
1.80    & $-0.033(89)$ 
        & $-0.78(29)$           
        & $ 0.37(20)$          & $0.09$            \\
1.95    & $-0.184(57)$ 
        & $-0.63(18)$           
        & $ 0.26(12)$          & $0.53$            \\
2.10    & $-0.273(35)$ 
        & $-0.55(12)$          
        & $ 0.169(83)$         & $0.55$            \\
\end{tabular}
\end{center}
\caption{
Results of fitting for the scattering length $a_0 / m_\pi$
obtained with a divergent form defined by Eq.(\ref{eq:a0:div}).
\label{tab:a0:fit-diverge}
}
\end{table}
%
%
\begin{table}
\begin{center}
\begin{tabular}{cccccc}
$\beta$ & $A_{10}$ [1/GeV$^2$] 
        & $A_{20}$ [1/GeV$^4$] 
        & $A_{30}$ [1/GeV$^6$] &     $\chi^2 / $d.o.f.   &   $a_0 / m_\pi$  [1/GeV$^2$] \\ 
\hline
1.80    &  $-1.01(29)$         
        &  $ 0.79(76)$           
        &  $-0.20(46)$    & $0.02$       &  $-0.99(28)$           \\
1.95    &  $-1.57(14)$         
        &  $ 1.58(33)$           
        &  $-0.54(18)$    & $2.1$        &  $-1.54(13)$           \\
2.10    &  $-1.975(82)$         
        &  $ 2.18(20)$           
        &  $-0.82(11)$    & $8.2$        & $-1.932(78)$           \\
\end{tabular}
\end{center}
\caption{
Results of fitting for the scattering length $a_0 / m_\pi$
obtained with a divergent form defined by Eq.(\ref{eq:a0:pow}).
The results at the physical pion mass are also tabulated.
\label{tab:a0:fit-pow}
}
\end{table}
%
%
\begin{table}
\begin{center}
\begin{tabular}{cccccc}
\multicolumn{2}{c}{$\beta=1.80$} &
\multicolumn{2}{c}{$\beta=1.95$} &
\multicolumn{2}{c}{$\beta=2.10$}    \\
$m_\pi^2$ [GeV$^2$] & $f^{lat}_\pi$ [GeV] & 
$m_\pi^2$ [GeV$^2$] & $f^{lat}_\pi$ [GeV] & 
$m_\pi^2$ [GeV$^2$] & $f^{lat}_\pi$ [GeV] \\ \hline
$1.128(1)$     & $0.2516(11)$ & 
$1.287(1)$     & $0.2190(16)$ & 
$1.332(2)$     & $0.1959(19)$        \\
$0.814(1)$     & $0.2279(14)$ & 
$0.857(1)$     & $0.1908(15)$ & 
$0.896(1)$     & $0.1739(21)$        \\
$0.571(1)$     & $0.1993(11)$ & 
$0.573(1)$     & $0.1650(14)$ & 
$0.605(2)$     & $0.1540(22)$        \\
$0.238(1)$     & $0.1613(15)$ & 
$0.291(2)$     & $0.1381(21)$ & 
$0.291(1)$     & $0.1214(19)$        \\
$(m_\pi^{phys})^2$   & $0.1287(33)$ & 
$(m_\pi^{phys})^2$   & $0.1054(47)$ & 
$(m_\pi^{phys})^2$   & $0.0895(45)$         \\
chiral limit & $0.1260(31)$ & 
chiral limit & $0.1032(47)$ & 
chiral limit & $0.0869(46)$       \\
\end{tabular}
\end{center}
\caption{
Pseudoscalar decay constant $f^{lat}_\pi$ measured on lattice~\protect\cite{CP-PACS:sprm}.
Here $m_{\pi}^{phys}$ is the physical pion mass.
\label{tab:f_pi}
}
\end{table}
%
%
\begin{table}
\begin{center}
\begin{tabular}{ccc}
$\beta$ & $\hat{a}_0 / m_\pi$ [1/GeV$^2$]  & $\chi^2 / $d.o.f.  \\ 
\hline
1.80    & $-2.83(13) $                     & $0.27$             \\
1.95    & $-2.543(73)$                     & $0.92$             \\
2.10    & $-2.449(40)$                     & $0.54$             \\
\end{tabular}
\end{center}
\caption{
Results of a constant fit for $\hat{a}_0 = ( f_\pi^{lat} / f_\pi )^2 \cdot a_0 / m_\pi$.
The results at the physical pion mass are also tabulated.
\label{tab:a0:fit-f_pi-a0}
}
\end{table}
%
%
\begin{table}
\begin{center}
\begin{tabular}{lccc}
                  & $A$           & $B$ [GeV]      & $\chi^2 / $d.o.f. \\ \hline
$a_0 m_\pi$       & $-0.0558(56)$ & $ 0.0328(86)$  & $0.02$            \\
$\hat{a}_0 m_\pi$ & $-0.0413(28)$ & $-0.0119(43)$  & $0.65$            \\
ChPT              & $-0.0444(10)$ & ---            & ---               \\
\end{tabular}
\end{center}
\caption{
Results of the continuum extrapolations for the two scattering lengths 
$a_0 m_\pi$ and $\hat{a}_0 m_\pi=( f_{\pi}^{lat} / f_{\pi} )^2 \cdot a_0 m_{\pi}$
as a function $A + a B$.
Here $A$ corresponds to the scattering length in the continuum limit.
\label{tab:a0:fit.cont}
}
\end{table}
%
%
\begin{table}
\begin{center}
\begin{tabular}{cccc}
$\beta$              &  1.80       &  1.95       &  2.10        \\ \hline
$A_{10}$ [1/GeV$^2$] & $-1.33(21)$ & $-1.52(12)$ & $-1.899(84)$ \\
$A_{20}$ [1/GeV$^4$] &  $1.62(53)$ &  $1.51(29)$ &  $2.00(20)$  \\
$A_{30}$ [1/GeV$^6$] & $-0.69(31)$ & $-0.52(16)$ & $-0.73(11)$  \\
$A_{01}$ [1/GeV$^2$] & $-0.83(44)$ & $-1.18(47)$ & $-1.43(40)$  \\
$A_{11}$ [1/GeV$^4$] &   $1.4(12)$ &  $1.9(11)$  &  $2.48(95)$  \\
$A_{21}$ [1/GeV$^6$] & $-0.46(83)$ & $-0.65(61)$ & $-1.04(52)$  \\
$\chi^2 / $d.o.f.    &  0.90       &  0.64       &  1.33        \\
\end{tabular}
\end{center}
\caption{
Results of a polynomial fit of $m_{\pi}^2$ and $\overline{p}^2$ for the scattering amplitude defined by
$A( m_{\pi}, \overline{p} ) = \tan \delta(\overline{p} ) / \overline{p} \cdot \overline{E} / 2$.
\label{tab:delta:fit-beta}
}
\end{table}
%
%
\begin{table}
\begin{center}
\begin{tabular}{cccc}
$\beta$                    &  1.80       &  1.95        &  2.10        \\ \hline
$\hat{A}_{10}$ [1/GeV$^2$] & $-2.84(10)$ & $-2.546(59)$ & $-2.438(38)$ \\
$\hat{A}_{01}$ [1/GeV$^2$] & $-2.78(59)$ & $-2.84(44)$  & $-2.14(37)$  \\
$\hat{A}_{11}$ [1/GeV$^4$] &  $3.67(76)$ &  $3.36(49)$  &  $2.05(43)$  \\
$\chi^2 / $d.o.f.          &  0.77       &  0.50        &  0.75        \\
\end{tabular}
\end{center}
\caption{
Results of a polynomial fit of $m_{\pi}^2$ and $\overline{p}^2$ for the normalized scattering amplitude defined by
$\hat{A}( m_{\pi}, \overline{p} ) = ( f_\pi^{lat} / f_\pi )^2 \cdot \tan \delta(\overline{p} ) / \overline{p} \cdot \overline{E} / 2$.
\label{tab:delta:f-fit-beta}
}
\end{table}
%
%
\begin{table}
\begin{center}
\begin{tabular}{ccc}
$\overline{p}^2$ [GeV$^2$] & $\sqrt{s}$ [GeV] & $\delta(\overline{p})$ [degrees] \\ \hline
0.020                      & 0.40             & $-$3.50(64)                      \\
0.072                      & 0.60             & $-$9.5(30)                       \\
0.140                      & 0.80             & $-$16.9(64)                      \\
0.232                      & 1.00             & $-$25(10)                        \\
\end{tabular}
\end{center}
\caption{
Scattering phase shift $\delta(\overline{p})$ in the continuum limit 
at the physical pion mass.
\label{tab:phsh}
}
\end{table}
%
%
\clearpage
%
%
%
%
%
%
%
%
%
%
%
%
%
%
%
%
%
%
\clearpage
%
%
%
%
%
%
%
%
%
%
%
%
\end{document}